\documentstyle[11pt,epsf,mssymb]{article}

\def\pmb#1{\setbox0=\hbox{#1}%
\kern-.025em\copy0\kern-\wd0
\kern-.05em\copy0\kern-\wd0
\kern-.025em\raise.0433em\box0}

\begin{document}

\begin {titlepage}
\title {Mechanical momentum  in nonequilibrium quantum electrodynamics} 
\author {M. de Haan \\
Service de Physique Th\'eorique et Math\'ematique\\Universit\'e
libre de Bruxelles\thanks{Campus Plaine
CP 231, Boulevard du Triomphe, 1050
Bruxelles, Belgique. email: mdehaan@ulb.ac.be}, Brussels, Belgium. }

\date{}

\maketitle

\begin{abstract}
The reformulation of field theory in which  self-energy processes are no longer present 
\cite{dH04a}, \cite{dHG03}, \cite{dHG00a} provides an adequate tool to transform Swinger-Dyson
equations into a kinetic description  outside any approximation scheme. 
Usual approaches in quantum electrodynamics (QED) are unable to cope  with the mechanical
momentum of the electron and replace it by the canonical momentum.  
The use of that unphysical momentum is responsible for the divergences that are removed by the
renormalization procedure in the $S$-matrix theory. 
The connection between distribution functions in terms of the canonical and those in terms of the
mechanical momentum is now provided by a dressing operator \cite{dH04b} that allows the elimination
of the above divergences, as the first steps are illustrated here.

\end{abstract}

\end{titlepage}

\def\theequation{\thesection.\arabic{equation}}

\section {Introduction}

The obtention of the statistical mechanical properties of interacting quantum fields is the aim of
studies in nonequilibrium  quantum fields. Their relevance in describing the early stage of the
universe in manifest but their interests lie beyond that application  to provide an unified
understanding of physics in all circonstancies.
A kinetic description (of the Boltzmann-Langevin type of equations) requires the Swinger-Dyson
equations as  starting point from which a kinetic irreversible description is derived \cite{CH00}, via a
generalization of the Bogoliubov-Born-Green-Kirkwood-Yvon (BBGKY) hierarchy. 
Different schemes can be used to perform the transition from a reversible into an irreversible 
description  \cite{CH88}. 
An appropriate truncation of the dynamics, through some molecular chaos assumption or
a Kadanoff-Baym approximation, performs usually the job.
In the framework of the scalar field, a adequate scheme of approximation enables to get rid of the
divergences through resummations ($N$-points irreductible (N-PI) approximations \cite{JB04}). 
The way of removing the infinities found in $S$-matrix theory, through the renormalisation program
initiated by S. Tomonaga, J. Schwinger and R.P. Feynman, is of no direct application since the
description in the kinetic approach lacks the asymptotic limit used in computing the collision
process.  

Our extension \cite{dHG00a} of the dynamics of correlations  \cite{RB75} provide a framework in
which irreversible equations are derived in an exact way, that contains, as particular case, the
reversible equations: the new dynamics is an enlargement from the original one that is entirely
constructed from it, without introducing new parameters for instance.
That approach enables to get rid of the self-energy processes at the level  of the equations: their
effect on the dynamics is taken exactly into account into the kinetic operator. 
We have dissociation of the approximation scheme and irreversibility. A further truncation of the
hierarchy and approximation schemes would then be introduced inside an
irreversible description and would not generate it. 
The situation is much more satisfactory on a conceptual point of view.

Equations that no longer allow self-energy processes have to contain the solution to the divergences
met in quantum field theory.
The problem of the infinities is thought to be inherited from classical physics: 
the electromagnetic field dragged along with the charge of a moving electron turns out to
provide an infinite mass to the particle.
That last point has been challenged recently \cite{dH05} using the single subdynamics approach
\cite{dHG00a}, \cite{dH04a}, \cite{dH04b}.
From an application to classical electrodynamics \cite{dH05}, using the Balescu-Poulain formulation
\cite{BP74} dealing with the mechanical momentum of the charged particles, the usual mass
divergence has been shown to be removable through resummations of classes of diagrams. 
A relation with the renormalisation group is manifest.

Considering the unexpected sucess in the classical case, a new investigation of the quantum case looks
mandatory:  does a formulation of QED exist so that no (infinite) substractions are
required to provide a finite theory?
We are convinced that a positive answer can be provided to that question.
Two new ingredients are inserted with respect to previous attempts.
The first one is the new mathematical tool constituted by the single subdynamics approach,
the second one is the reintroduction of the mechanical momentum of the electrons
(in opposition to its canonical one's) as one of the basic observables of the description.
A theory that deals with physical observables should be naturally finite!
The difference between the canonical and mechanical momentums in QED distinguishes the cases of the
quantum scalar and vectorial fields.

The relativistic interacting Dirac and electromagnetic fields necessarily involve an arbitrary large
numbers of particles, namely electrons, positrons and photons. 
It is not possible to restrict ourselves to sectors involving only a  finite number of particles.   
Therefore, it can be argued that such a system resort to quantum statistical mechanics and  that the
adequate methods are those developed in a non equilibrium statistical mechanical context \cite{RB75}.
The physical systems, classical or quantal, are described by Wigner-like reduced
distribution functions, depending on the position and the momentum of the particles (plus
obviously their nature and spin). 
The form of the infinitesimal evolution generator is dependent of the classical or
quantal nature of the system.
An interpretation in terms of usual probabilities is not possible in the quantum case: the Wigner
distribution functions are not necessarily positive everywhere.

We intend to construct a similar formalism in the high energy domain, using quantized fields.
Reduced entities, involving a finite number of degrees of freedom, are easily introduced by 
considering mean values of normally ordered creation and destruction operators , or
by taking the trace of the density operator on which destruction operators have acted on the left and
creation operators have acted on the right\footnote{Usual approaches use related $n$-points
functions, introducing differences between the non relativistic and relativistic descriptions}. 
That procedure defines a complete set of moment functions that allows the computation of the mean
value of any observable. The evolution equation of the set looks like the B.B.G.K.Y. hierarchy.
Since the creation and destruction operators of the Dirac field involve the canonical momentum, that
set is defined in terms of that momentum.
The Weyl's correspondence rule \cite{RB75} between classical and quantal observables can be then
used  when the continuous observables are the position and the canonical momentum of the particles.
Reference functions leading to a Wigner-like description are then introduced. 
Their argument involves naturally the canonical momentum of the particles. 
They constitute also a complete set for the description of the system.
Their evolution equations can be computed from the evolution equation of the set of moment
functions.
The direct application of the single subdynamics approach to the set of reference functions presents
{\it a priori} no difficulty.
When no external e.m. fied is present, a subset of reference functions, involving only the Dirac
particle, satisfies closed equations of evolutions: the reference functions functions involving the
self-field do no longer appear in the dynamics.
The analysis of the kinetic operator shows that it leads for the reference functions to the usual
(logarithmically divergent) mass correction at the second order in the charge. 

However, our aim is a description in terms of the mechanical momentum.
In Lorentz gauge, the form of the energy-momentum tensor of the electron-positron field  shows that
it involves also the electromagnetic field operators \cite{WH54}. 
That property can also be noted in the manifestly gauuge invariant formulation of quantum
electrodynamics \cite{BP75} where the distingo between the canonical and mechanical momentums is
made explicit, unlike in more recent rediscoveries \cite{BBHL96}.  
Moreover, the association between classical functions of positions and mechanical momentums and
the corresponding observables in second quantization is completely unknown and call for an
extensive study by physicists.
Two distinct roads are possible to obtain a description by Wigner functions defined in terms of the
mechanical momentums.
In the first one,  a proposal for the above association is surmised from various considerations and
the consequences of that choice examined.
In the second one, it can be argued that the searched Wigner functions can be obtained from the 
reference functions since the latters form a complete basis of the description.
The link is provided by a dressing operator acting on the reference functions \cite{dH04b}.
Inside the single subdynamics approach, when no external e.m. fied is present, a similar subset of
the  Wigner functions satisfy also closed equations of evolution governed by a dressed kinetic
operator.
It has been shown that, outside resonances processes \cite{dH04b}, a second order dressing operator
modifies the kinetic operator at the same second order.
Among the requirements for its determination, it can be required that the new kinetic operator is
finite.
Therefore, we have to deal with a description the elements of which have a clear physical meaning, for
true observables (e.g. the mechanical momentum), and so that these elements satisfy close equations
of motion. 
The set of considered observables have thus to be complete and we have to scope with all
physical situations. We can then expect that all physical entities are connected together by evolution
equations free of any divergence.
The resulting formalism is therefore rather heavy.

In order to prove the feasability of the previous scheme, we consider in this paper a naive proposal
for the above association between classical and quantal observables. 
We introduce somewhat arbitrarily a generalisation of the Weyl's correspondence rule.
We are aware that high energy processes are not correctly included into that association.
Therefore, our presentation aims to accomodate possible alternative proposals and we focus on the
structure of the new approach.
It will be illustrated how the proposal determines a dressing operator that leads, for the second order corrections to the 
kinetic  operator, to a logarithmic divergence analog to the second order mass correction, although no
complete cancellation occurs.

The usual description of the system  in terms of a state vector  in some Fock space is thus
unappropriate for our purpose and, in the second section, we treat the electromagnetic field quantized
in Lorentz gauge interacting with  a (time dependent) classical source in the density operator
formalism. That stage allows the introduction of all the required notations. 
Already at that level, the concept of superoperators  (acting on the density operator) in normal form
emerges. 
The Lorentz condition appears as a strong condition expressed in terms of  the appropriate
superoperator in normal form. Its persistence in time  is explicitly shown.  
The properties of all classically generated incident electromagnetic field are thus made explicit.

In $\S$3,  for the same system (classical source plus e.m. field), correlation functions are introduced
for the description  of the electromagnetic fied. 
They appear as the natural extension, in the density operator formalism, of the mean values of normally
ordered creation and destruction operators. 
These correlation functions obey  the equations of a B.B.G.K.Y.-like hierarchy and the solution of those
equations is provided: for a classical  source, the correlation functions simply factorize. 
The non-classical aspects of the field have to be due to the non-classical character of the sources.
Moreover, we show that the computation of the electric and magnetic fields mean values does not
require the correlation functions involving the scalar or the longitudinal photon. A description similar
to the quantization in Coulomb gauge is recovered: the fields  depend on correlation functions
involving transverse photons and on the instantaneous Coulomb potential due to the classical
source.

The fully reduced description for the Dirac and electromagnetic fields  is introduced in $\S$4. The
reduction bears not only on the electromagnetic fields variables but also on those of the Dirac field.

In the next section $\S$5,
observables relative to one electron are considered. 
The Weyl's rule of correspondence \cite{RB75}
for quantum observables corresponding to  classical dynamical functions of the
position and canonical momentum leads, from the consideration of their mean value, to our reference
functions expressed in terms of our previously introduced correlation functions.
Wigner functions dealing for a complete set of particles observables, including the mechanical
momentum, are formally introduced.
The energy-momentum tensor of the electron-positron field involves the electromagnetic field
operators and therefore, classical functions of the position and the mechanical momentum of the
particle have no unambiguously correspondence with (in second quantization) quantum operators. 
That property does not prevent to present a simple (somewhat naive) association 
provided by a generalisation of the Weyl's correspondence rule.
We are aware that high energy processes are not correctly included into that association
that allows a first exploration of the properties of the corrrespondence.
We think that the requirement of the finiteness of the final description can serve as a guide to provide
that association.

The one particle reference distribution function is further analysed in $\S$6.

In $\S$7, the evolution equations for the whole set of correlation  functions is derived. 
The number of elementary processes is of course much more important as in a
non-relativistic case but all the contributions have been given an explicit form.

The subdynamics approach is developped in the next section $\S$8, including the introduction of the
dressing operator.   
The equivalence of the single subdynamics and the usual approaches should not
prevent unexpected results using the new tool. 
The divergences met in the usual approaches may result from the use of an unappropriate
description and we investigate here the first stages of that possibility.
A finite kinetic evolution operator does not seem out of reach.
Moreover, the link between non diagonal elements of the kinetic operators and the elements of the
The dressing operatorprovides a conceptually satisfactory way of incorporating all the
assets of the renormalization program of QED.

Some conclusions are presented in the last section 
$\S$9.

\section{The density operator  description for the  electromagnetic field}   

We are concerned in this section with the electromagnetic field  quantized in Lorentz gauge  in
interaction with classical sources, i.e. sources the evolution of which is  a given function of time. This
problem is well known in Hilbert space formalism  but it is useful to reconsider it in the density
operator formalism.

Considering the undetermined number of field particles involved in the system, we use a statistical
approach and  the  state of the electromagnetic field  will be described in the next section by
correlation functions,  the evolution of which is provided by a  BBGKY-like hierarchy of equations.

The hamiltonian describing such a situation is  well-known in the second quantization formalism we need
\cite{WH54}-\cite{BS82} 
(see in particular the complement $A_{V}$ in \cite{CDG87} from which the notations are borrowed):
\begin{equation}
H=H_R+H_{Ie}
\label{2.1}
\end{equation}
$H_R$ is the field hamiltonian, decomposed in transverse ($\varepsilon,\varepsilon'$), longitudinal ($l$)
and scalar (s) modes:
\begin{equation}
H_R= \int d^3k \sum_{\lambda=\varepsilon,\varepsilon',l} 
\hbar \omega_{{\bf k}} ( a^+_{\lambda}({\bf k}) a_{\lambda}({\bf k})+\frac12) + \int d^3k 
\hbar \omega_{{\bf k}} (a^+_{s}({\bf k}) a_{s}({\bf k})+\frac12)
\label{2.2}
\end{equation}    
The interaction hamiltonian $H_{Ie}$ is:
\begin{equation}
 H_{Ie}=\int d^3r\left[-{\bf j}_e({\bf r},t).{\bf  A}({\bf r}) + c \rho_e({\bf r},t) A_s({\bf r})
\right]
\label{2.3}
\end{equation} 
The (external) charge density  $\rho_e({\bf r},t)$ and   courant ${\bf j}_e({\bf r},t)$ are  here given
functions of space and time that satisfy the  charge conservation law:
\begin{equation}
\frac{\partial}{\partial t} \rho_e({\bf r},t)+
\pmb{$\nabla$}_{{\bf r}}.{\bf j}_e({\bf r},t)=0
\label{2.4}
\end{equation} 
They  transform as a quadrivector under a Lorentz transformation.

The quantized potential operators $ {\bf  A}({\bf r})$ and $A_s({\bf r})$ admit a plane wave expansion.  
\begin{equation}
{\bf  A}({\bf r})= 
\int d^3k \sum_{\lambda} 
\sqrt{\frac{\hbar}{2\varepsilon_0 \omega_{{\bf k}} (2\pi)^3}}
 {\bf  e}_{{\bf k}\lambda}\left(  e^{ i\bf{k.r}}\,\, a_{\lambda}({\bf k})+ e^{ -i\bf{k.r}}\,
a^+_{\lambda}({\bf k})\right)  
\label{2.5}
\end{equation}  
\begin{equation}
A_s({\bf r})= \int d^3k  
\sqrt{\frac{\hbar}{2\varepsilon_0 \omega_{{\bf k}} (2\pi)^3}}
\left(  e^{ i\bf{k.r}}\,\, a_{s}({\bf k})- e^{ -i\bf{k.r}}\, a^+_{s}({\bf k})\right)  
\label{2.6}
\end{equation}  
$A_s$ is not hermitian with respect to the usual scalar product but with  respect to the scalar product
used in defining the (indefinite) metric. The infinite volume limit is used for the normalization of the
creation  and destruction operators of photons, characterized by the continuous wave  vector ${\bf k}$
and the polarization $\lambda$:
\begin{equation}
[a_{\lambda}({\bf k}),a^+_{\lambda'}({\bf k}')]=
\delta^{Kr}_{\lambda,\lambda'}
\delta({\bf k} - {\bf k}')
\label{2.7}
\end{equation}  
\begin{equation}
[a_{s}({\bf k}),a^+_{s}({\bf k}')]=
\delta({\bf k} - {\bf k}') 
\label{2.8}
\end{equation}  

The covariant quantization has been performed by using  an indefinite metric, determined by a linear
(unitary, idempotent) operator $M$ so that:
\begin{eqnarray}
&&Ma_{\lambda}M=a_{\lambda}
\label{2.10}
\nonumber\\ 
&&Ma_sM=-a_s
\label{2.11}
\end{eqnarray}
The scalar product of two vectors  $|\psi>$  and $|\phi>$  is then defined in the
indefinite metric as  $<\phi|M  |\psi>$ and 
the  adjoint $\bar A$ of an operator  $A$  through:
$<\phi|M A|\psi>=<\psi|M \bar A |\phi>^*$.
Only the adjoint of the creation and destruction operators  for the  scalar  boson are modified by the
indefinite metric and we have the link of the new adjoint  (noted with an upper bar)  with the usual
adjoint (noted  with a $+$):
$\bar a_s=-a_s^+$ and $\bar a_s^+=-a_s$. 
Therefore, the scalar potential is, as previously noted,
self-adjoint  with respect to the indefinite metric.

The mean value of an  observable, hermitian with respect to the metric $M$, is defined, when the
system is in a state  $|\psi>$ by
$<A>=<\psi|M A |\psi>$ 
The class of admissible states is the class of states for which the Lorentz  condition,  that  cannot be
satisfied in a strong sense, holds in average:
\footnote{In reference \cite{CDG87}, that condition is erroneously expressed in the complement
A${_V}$ (58) by a mean value without the  indefinite metric.  The claimed equivalence between (67) on
one hand and (66a) and the adjoint  (in some sense) relation (66b) on the other hand  shows clearly that
the Lorentz condition has to hold under the form (\ref{2.14}).}
\begin{equation}
<\psi|M \left( \pmb{$\nabla$}_{{\bf r}}.{\bf  A}({\bf r}) +\frac 1c \dot A_s\right)  |\psi>  =0
\label{2.14}
\end{equation} 
The operator $ \dot A_s$  corresponding to the time derivative  of $A_s$, is defined in  Schr\"odinger
representation by:
\begin{equation}
\dot A_s  = \frac1 {i\hbar}[ A_s ,H ]
\label{2.15}
\end{equation}

In a statistical approach, the system is described by  a density operator $\rho$ that has to be
hermitian with respect to the  metric. 
The factorized density operator $\rho_{fac} $ corresponding to a pure state
$|\psi>$ is given by:
\begin{equation}
\rho_{fac}=\frac1{<\psi|M |\psi>} |\psi><\psi|M
\label{2.16}
\end{equation} 
and the mean value of an operator $A$ is given by the trace in the usual expression:
\begin{equation}
 <A>= \rm{Tr}(A\rho)
\label{2.17}
\end{equation} 
The hamiltonian operator is thus entirely determined in the second  quantization formalism and,
as $\rho_R$,  is hermitian with respect to the indefinite metric.. 
The Liouville-von Neuman  equation determines the evolution of the  density matrix $\rho_R$ for the
radiation:
\begin{equation}
i\hbar \partial_t \rho_R = L \rho_R = [H_R + H_{Ie} ,\rho_R ].
\label{2.18}
\end{equation} 
The density operator $\rho_R$  is constructed from factorized $\rho_{fac}$ density operator based on
physical states satifying the condition (\ref{2.14}). 
Superoperators allow a direct  expression of that condition in the density operator formalism. 
They are defined as operators acting on the density operator. 
The Liouville-von Neumann (super)operator $L$ (\ref{2.18}) is an example.
With an operator $A$ acting on the left and an operator $B$ acting on the right of the density operator,
we define a factorizable superoperator $A\times B$  \cite{PGHR73} by:
\begin{equation}
\left( A\times B\right)  \rho = A\rho B
\label{2.19}
\end{equation}
In second quantization,
to each hermitian operator  $A$,  it can be associated \cite{dH91}, \cite{dH04b} in a unique way a
superoperator ${\cal A}$, written as a sum of  factorizable superoperators  in normal form (only
destruction operators act on the left and only creation operators act on the right.) so that:
\begin{equation}
<A>= \rm{Tr}(A\rho) = \rm{Tr}({\cal A}\rho)
\label{2.20}
\end{equation}   
For instance, to the scalar potential $A_s$, we associate the normal form superoperator 
${\cal A}_s$ defined as:
\begin{equation}
{\cal A}_s({\bf r})= \int d^3k  
\sqrt{\frac{\hbar}{2\varepsilon_0 \omega_{{\bf k}} (2\pi)^3}}
\left(  e^{ i\bf{k.r}}\,\left( a_{s}({\bf k})\times I\right)+ e^{ -i\bf{k.r}}\,\left(I\times \bar
a_{s}({\bf k})\right)
\right)  
\label{2.21}
\end{equation} 
and the superoperator $\dot{ {\cal A}}_s$ corresponding to $\dot A_s$ (\ref{2.15}) is constructed
in a similar way.  
The  Lorentz  condition can be  expressed  in a strong sense using the frequency positive
components of the quadrivector $A$. 
In the present formalism, we have:
\begin{equation}
\left( \pmb{$\nabla$}_{{\bf r}}.\,\pmb{${\cal A}$}({\bf r}) +\frac 1c \dot{ {\cal A}}_s\right)  
\rho_R(t)  =0
\label{2.22}
\end{equation} 
The condition  (\ref{2.14}) is   recovered by taking the trace and using (\ref{2.16}) for $\rho$.  
From (\ref{2.15}) and (\ref{2.3}), we see that the density operator has  to satisfy the following
conditions for all wave numbers ${\bf k}$ and all times:
\begin{equation}
\left( a_l({\bf k})- a_s({\bf k}) +\lambda({\bf k},t)
\right)\rho_R(t) =0
\label{2.23}
\end{equation} 
\begin{equation}
\rho_R(t) \left(\bar a_l({\bf k})-\bar a_s({\bf k}) +\lambda^*({\bf k},t) 
\right) =0
\label{2.24}
\end{equation}  
where:
\begin{equation}
\lambda({\bf k},t) =
\sqrt{\frac{c^2}{2\varepsilon_0 \hbar(2\pi)^3 \omega^3_{{\bf k}}}} \rho_e({\bf k},t)
\label{2.25}
\end{equation} 
and $\rho_e({\bf k},t)$ is the Fourier transform of the charge density.
\footnote{For the Fourier transform, we use the same convention as in \cite{RB75} in place of
the  convention of \cite{CDG87}, hence a difference in some numerical factors.}
\begin{eqnarray}
&&\rho_e({\bf r},t)=\frac1{(2\pi)^3} \int d^3k\, e^{ i\bf{k.r}}\,\rho_e({\bf k},t) 
\label{2.26} \nonumber\\ 
&&\rho_e({\bf k},t)=\int d^3r\, e^{- i\bf{k.r}}\,\rho_e({\bf r},t) 
\label{2.27}
\end{eqnarray}
The conditions (\ref{2.23}-\ref{2.24}) are mutually adjoint.  Considering the self-adjoint character of
the density operator,  they are equivalent. 
For a time independent problem, condition (\ref{2.23}) is
equivalent to condition (66a) in A${}_V$  of \cite{CDG87}.
If the conditions  (\ref{2.23}-\ref{2.24}) are satisfied
at some arbitrary  time $t$,  they remain satisfied for all times. 
Using the charge conservation (\ref{2.4}), the proof is straightforward.

\section{The correlation functions for the 
electromagnetic field}   
\setcounter{equation}{0}

In a context of statistical mechanics of large systems,  the relevant objects for its description are not
the $N$ particles distribution function or the density operator for the  whole system but the classical
or Wigner reduced distribution functions involving a limited number of particles. 
Considering the possibility of an arbitrary large number of photons  involved, we treat the
electromagnetic field along the same spirit and we have to define the appropriate reduced
description. 
Such a description can be provided  in terms a  reduced density operator \cite{dH91}, so  that the
mean value of  any observable can be evaluated for all observables  through a vacuum expectation
value.  This required that a normal form  superoperator be associated  with all  observables, as already
considered in the previous section.   
It is possible to replace the reduced operator by its matrix
elements that can be computed directly from the density operator itself and this point of view is
adopted here.  
The state of the electromagnetic field is  therefore not described   by the density
operator itself but by its  set of  correlation distribution functions  \cite{dH91}, obviously
symmetrical in each set of variables:
\begin{eqnarray}
&&
f_{nm}
( {\bf k}_1 \nu_1;{\bf k}_2 \nu_2;\dots;
{\bf k}_n \nu_n;
{\bf k}'_1 \nu'_1;{\bf k}'_2 \nu'_2;\dots;
{\bf k}'_m \nu'_m,t) 
\nonumber\\&&=
 \rm{Tr}a_{\nu_1}({\bf k}_1)
a_{\nu_2}({\bf k}_2)
\dots a_{\nu_n}({\bf k}_n) 
\rho_R(t)
a^+_{ \nu'_m}({\bf k}'_m) \dots
a^+_{ \nu'_2}({\bf k}'_2)
a^+_{ \nu'_1}({\bf k}'_1)
\nonumber\\&&
\label{3.1}
\end{eqnarray}
where the index $\nu$ takes  any of the values $s$, $l$, $\varepsilon$, or $\varepsilon'$.  Those
correlation functions  appear as mean values of particular superoperators, those in
normal form,  and depend on the diagonal, as well as the off diagonal in photon numbers elements of
$\rho$.  In the particular case where the density operator  is factorized (the so-called pure case), 
those correlation functions are nothing but mean values, in Fock space, of a normally ordered product
of creation and destruction operators.
The normalization of the density operator requires in particular that  $f_{00}=1$ for all times.
The correlation functions are not independent since they have to represent  physical states and  have
to satisfy the relations due to (\ref{2.23}) and (\ref{2.24}):
\begin{eqnarray}
&&
f_{n+1 \,m}
( {\bf k}_1 \nu_1;\dots;
{\bf k}_n \nu_n;{\bf k}_{n+1}l;
{\bf k}'_1 \nu'_1;\dots;
{\bf k}'_m \nu'_m,t)\nonumber\\ &&
-f_{n+1 \,m}
( {\bf k}_1 \nu_1;\dots;
{\bf k}_n \nu_n;{\bf k}_{n+1}s;
{\bf k}'_1 \nu'_1;\dots;
{\bf k}'_m \nu'_m,t)\nonumber\\ && +
\lambda({\bf k}_{n+1},t) f_{nm} ( {\bf k}_1 \nu_1;\dots;
{\bf k}_n \nu_n;
{\bf k}'_1 \nu'_1;\dots;
{\bf k}'_m \nu'_m,t) =0 
\label{3.2}
\end{eqnarray}
\begin{eqnarray}
&&
f_{n \,m+1}
( {\bf k}_1 \nu_1;\dots;
{\bf k}_n \nu_n;
{\bf k}'_1 \nu'_1;\dots;
{\bf k}'_m \nu'_m;{\bf k}'_{m+1}l,t) 
\nonumber\\ &&
+f_{n \,m+1}
( {\bf k}_1 \nu_1;\dots;
{\bf k}_n \nu_n;
{\bf k}'_1 \nu'_1;\dots;
{\bf k}'_m \nu'_m;{\bf k}'_{m+1}s,t)\nonumber\\ &&+
\lambda^*({\bf k}'_{m+1},t) f_{nm} ( {\bf k}_1 \nu_1;\dots;
{\bf k}_n \nu_n;
{\bf k}'_1 \nu'_1;\dots;
{\bf k}'_m \nu'_m,t) =0 
\label{3.3}
\end{eqnarray}
The  electric and magnetic field operators ${\bf  B}$ can be computed directly from the
expression  (\ref{2.5}) of the potential vector operator and from the form (\ref{2.1}) for the
Hamiltonian that is required for the evaluation of 
$\dot{{\bf  A}}({\bf r})=\frac1{i\hbar}[ {\bf  A}({\bf r}),H] $.
Their mean values involves only the moments $f_{10}({\bf k}\lambda,t)$ and
$ f_{01}({\bf k}\lambda,t)$.
The electric field operator is decomposed into 
the part ${\bf  E}_{\bot}$,
originating from the transverse component  ${\bf  A}_{\bot}({\bf r})$
of the potential operator ${\bf  A}$ 
(for $\lambda=\varepsilon,\varepsilon' $),
the part ${\bf  E}_l$,
originating from the longitudinal component 
${\bf  A}_{\Vert}({\bf r})$  
of the potential operator ${\bf  A}$ (for $\lambda=l$)  and a part
${\bf  E}_s$, originating from the scalar potential. 
Using the relations (\ref{3.2}) and (\ref{3.3}), usual manipulations lead to:
\begin{equation}
<{\bf  E}_l({\bf r})>_t+<{\bf  E}_s({\bf r})>_t= 
-\pmb{$\nabla$}_{{\bf r}} U({\bf r},t) 
\label{3.26}
\end{equation}
where $U({\bf r},t) $ is a Coulomb potential (without retardation), i.e. 
associated with the charge distribution taken at the same time $t$:
\begin{equation}
U({\bf r},t)= \frac{1}{4\pi\varepsilon_0}
\int d^3r'    
\frac{ 1}{|{\bf r}-{\bf r}'|}  
\rho_e({\bf r}',t) 
\label{3.27}
\end{equation}
The Lorentz gauge  therefore provides expressions similar 
to those  obtained in Coulomb gauge:
the fields are derived from the transverse part of the vector potential
$<{\bf  A}_{\bot}({\bf r})>_t$ and of the Coulomb potential $U({\bf r},t)$.

Let us now consider the time evolution of the correlation functions.
From the  Liouville-von Neuman equation (\ref{2.18}), we obtain immediately:
\begin{eqnarray}
&&
\frac{\partial}{\partial t}f_{nm}
( {\bf k}_1 \nu_1;{\bf k}_2 \nu_2;\dots;
{\bf k}_n \nu_n;
{\bf k}'_1 \nu'_1;{\bf k}'_2 \nu'_2;\dots;
{\bf k}'_m \nu'_m,t) 
\nonumber\\ &&=\frac{1}{i\hbar} 
 \rm{Tr}a_{\nu_1}({\bf k}_1)
a_{\nu_2}({\bf k}_2)
\dots a_{\nu_n}({\bf k}_n) 
[H_R + H_{Ie} ,\rho_R(t) ] 
\nonumber\\ &&\times
a^+_{ \nu'_m}({\bf k}'_m) \dots
a^+_{ \nu'_2}({\bf k}'_2)
a^+_{ \nu'_1}({\bf k}'_1)
\label{3.28}
\end{eqnarray}
We have to substitute in this expression the hamiltonians $H_R$ and $H_{Ie}$ 
by their expressions (\ref{2.2}) and (\ref{2.3}).
It is useful to express the interaction hamiltonian $H_{Ie}$ in terms of the
Fourier variables:
\begin{eqnarray}
&&H_{Ie}=- \int d^3k \sum_{\lambda} 
\sqrt{\frac{\hbar}{2\varepsilon_0(2\pi)^3 \omega_{{\bf k}}}}
\left( 
{\bf j}_e(-{\bf k},t).  {\bf  e}_{{\bf k}\lambda}
\, a_{\lambda}({\bf k})+
{\bf j}_e({\bf k},t).  {\bf  e}_{{\bf k}\lambda}
\, a^+_{\lambda}({\bf k})\right) 
\nonumber\\ &&
+ c \int d^3k  
\sqrt{\frac{\hbar}{2\varepsilon_0(2\pi)^3 \omega_{{\bf k}}}}
\left(  \rho_e(-{\bf k},t)\, a_{s}({\bf k})
- \rho_e({\bf k},t)\, a^+_{s}({\bf k})\right)  
\label{3.30}
\end{eqnarray}
In order to obtain a symmetrical expression, let us 
consider that the source and current densities 
define a vector with four components, labeled as the photons:
\begin{eqnarray}
&&j_{\lambda+}({\bf k},t)={\bf j}_e({\bf k},t).  {\bf  e}_{{\bf k}\lambda}
\qquad
j_{\lambda-}({\bf k},t)={\bf j}_e({\bf k},t).  {\bf  e}_{{-\bf k}\lambda}
\nonumber\\ && 
j_{s}({\bf k},t)=j_{s+}({\bf k},t)=j_{s-}({\bf k},t)=-c\rho_e({\bf k},t)
\label{3.31}
\end{eqnarray}
We have to introduce a symbol $s(\nu)$ which takes into account the 
sign difference for the scalar photon between the creation 
and destruction operators in (\ref{3.30}): $s(\nu)$ is -1 if $\nu=s$ and +1 in the
three other cases.
We have then:
\begin{eqnarray}
&&H_{Ie}=- \int d^3k \sum_{\nu} 
\sqrt{\frac{\hbar}{2\varepsilon_0(2\pi)^3 \omega_{{\bf k}}}}
\nonumber\\ && \times 
\left( 
j_{\nu-}(-{\bf k},t)\, a_{\nu}({\bf k})+ s(\nu)
j_{\nu+}({\bf k},t)\, a^+_{\nu}({\bf k})
\right) 
\label{3.32}
\end{eqnarray}
The use of the cyclic invariance of the trace leads then to:
\begin{eqnarray}
&&
\frac{\partial}{\partial t}
f_{nm}
( {\bf k}_1 \nu_1;{\bf k}_2 \nu_2;\dots;
{\bf k}_n \nu_n;
{\bf k}'_1 \nu'_1;{\bf k}'_2 \nu'_2;\dots;
{\bf k}'_m \nu'_m,t) 
\nonumber\\&& =
-i\left(\sum_{i=1, n} \omega_{ {\bf k}_i} -
\sum_{i=1,m} \omega_{ {\bf k}'_i}
\right)
\nonumber\\ && \times 
f_{nm}
( {\bf k}_1 \nu_1;{\bf k}_2 \nu_2;\dots;
{\bf k}_n \nu_n;
{\bf k}'_1 \nu'_1;{\bf k}'_2 \nu'_2;\dots;
{\bf k}'_m \nu'_m,t) 
\nonumber\\&& 
+i
\sum_{i=1, n}
\sqrt{\frac{1}{2\varepsilon_0\hbar(2\pi)^3 \omega_{{\bf k}_i}}}
s(\nu_i) j_{\nu_i+}({\bf k}_i,t) 
\nonumber\\&&  \times
f_{n-1\,m}
( {\bf k}_1 \nu_1;\dots;
{\bf k}_{i-1} \nu_{i-1};{\bf k}_{i+1} \nu_{i+1};\dots;
{\bf k}_n \nu_n;
{\bf k}'_1 \nu'_1;\dots;
{\bf k}'_m \nu'_m,t) 
\nonumber\\{}&&
-i
\sum_{i=1, m}
\sqrt{\frac{1}{2\varepsilon_0\hbar(2\pi)^3 \omega_{{\bf k}'_i}}}
j_{\nu'_i-}(-{\bf k}'_i,t)
\nonumber\\&&\times
f_{n\,m-1}
( {\bf k}_1 \nu_1;\dots;{\bf k}_n \nu_n;
{\bf k}'_1 \nu'_1;\dots;
{\bf k}'_{i-1} \nu'_{i-1};{\bf k}'_{i+1} \nu'_{i+1};\dots;
{\bf k}'_m \nu'_m,t)
\nonumber\\&&
\label{3.33}
\end{eqnarray}
This system of equations is equivalent to the Liouville-von Neumann equation
for the density operator.
It can be solved in a simple way by recurrence.
Indeed, in the r.h.s. of (\ref{3.33}), only correlation functions 
with indices equal or lower to $n$ and $m$ appear.
The equation for the first function $f_{00}$ is trivial since the r.h.s.
of (\ref{3.33}) vanishes. 
That  property corresponds to the trace conservation.
We now consider the next functions $f_{10}$ and $f_{01}$.
Let us assume that the claasical charge and currents $j_{\nu}$ vanish 
before the initial time $t_0$.
The solution of (\ref{3.33}) can then be written in a trivial way:
\begin{eqnarray}
&&f_{10}( {\bf k}_1 \nu_1,t)=i\, s(\nu_1) \sqrt{\frac{1}{2\varepsilon_0 \hbar(2\pi)^3\omega_{{\bf
k}_1}}}
\int_{t_0}^{t} d\tau_1 e^{-i\omega_{{\bf k}_1}(t-\tau_1)}      
j_{\nu_1}({\bf k}_1,\tau_1) 
\nonumber\\
&& f_{01}( {\bf k}'_1 \nu'_1,t)=-i\sqrt{\frac{1}{2\varepsilon_0 \hbar(2\pi)^3\omega_{{\bf k}'_1}}}
\int_{t_0}^{t} d\tau_1 e^{i\omega_{{\bf k}'_1}(t-\tau_1)}      
j_{\nu'_1}(-{\bf k}'_1,\tau_1) 
\label{3.35}
\end{eqnarray}
Using the charge conservation law (\ref{2.4}), a little algebra shows that the condition (\ref{3.2}) is
satisfied by our solution (\ref{3.35}). 

The element  $f_{20}$ is now considered.
The equation (\ref{3.33}) becomes for that element:
\begin{eqnarray}
&&
\frac{\partial}{\partial t}
f_{20}( {\bf k}_1 \nu_1;{\bf k}_2 \nu_2,t) 
 =
-i\left( \omega_{ {\bf k}_1} + \omega_{ {\bf k}_2}
\right)
f_{20}( {\bf k}_1 \nu_1;{\bf k}_2 \nu_2,t)
\nonumber\\&&
+i
\sqrt{\frac{1}{2\varepsilon_0\hbar(2\pi)^3 \omega_{{\bf k}_1}}}
s(\nu_1) j_{\nu_1+}({\bf k}_1,t) 
f_{10}( {\bf k}_2 \nu_2,t) 
\nonumber\\&&
+i\sqrt{\frac{1}{2\varepsilon_0\hbar(2\pi)^3 \omega_{{\bf k}_2}}}
s(\nu_2) j_{\nu_2+}({\bf k}_2,t) 
f_{10}( {\bf k}_1 \nu_1,t)
\nonumber\\&&
\label{3.38}
\end{eqnarray}
The expression (\ref{3.35})  for  $f_{10}$ can now be inserted into that 
equation and it can easily be seen that the solution of (\ref{3.38}) takes the 
factorized form:
\begin{equation}
f_{20}( {\bf k}_1 \nu_1;{\bf k}_2 \nu_2,t)
= f_{10}( {\bf k}_1 \nu_1,t)
f_{10}( {\bf k}_2 \nu_2,t)
\label{3.39}
\end{equation}
Such a factorization property holds for all the elements of $f$ in presence 
of a classical source and we have:
\begin{eqnarray}
&&f_{nm}
( {\bf k}_1 \nu_1;{\bf k}_2 \nu_2;\dots;
{\bf k}_n \nu_n;
{\bf k}'_1 \nu'_1;{\bf k}'_2 \nu'_2;\dots;
{\bf k}'_m \nu'_m,t)
\nonumber\\&&
=\left(\prod_{i=1,n}f_{10}( {\bf k}_i \nu_i,t) \right)
\left(\prod_{i=1,m}f_{01}( {\bf k}'_i \nu'_i,t) \right)
\label{3.40}
\end{eqnarray}
The state of the field characterized by such factorization relations 
is called semi-classical  or coherent.
If an observable, associated with the emitted fields at various points,
is constructed by the product of the (commuting) normal form superoperators 
associated with each component of the field, that factorization prevents a
correlation betwen the emitted fields.
The fluctuations in the emitted  field are linked to the non-classical features of the sources.

Let us examine more closely the  normal form superoperators 
$\pmb{${\cal E}$}({\bf r})$ and $\pmb{${\cal B}$}({\bf r})$ associated
with the field operators ${\bf  E}({\bf r})$ and ${\bf  B}({\bf r})$.
For those superoperators, the same decomposition into  
scalar, longitudinal and transverse components is introduced,in analogy with the operators 
themselves .
Obviously,  the magnetic field superoperator $\pmb{${\cal B}$}({\bf r})$
only contains a transverse contribution.
The electric field superoperator $\pmb{${\cal E}$}({\bf r})$
contains in addition contributions arising 
from the scalar and longitudinal components.
The sum of these two components can be given an explicit form:
\begin{eqnarray}
&&\pmb{${\cal E}$}_s({\bf r})+\pmb{${\cal E}$}_l({\bf r}) 
= ic \int d^3k  
\sqrt{\frac{\hbar}{2\varepsilon_0 \omega_{{\bf k}} (2\pi)^3}} {\bf k}
\left( 
e^{ i\bf{k.r}}\,\left[ (a_{l}({\bf k})- a_{s}({\bf k})) \times I\right]
\right.\nonumber\\&&\left.
+
e^{ -i\bf{k.r}}\,\left[ I\times (a^+_{l}({\bf k})+a^+_{s}({\bf k}))\right]
\right)  
\label{3.41}
\end{eqnarray}
Considering the conditions  (2.23-24), the density operator is an eigenoperator
of $\pmb{${\cal E}$}_s({\bf r})+\pmb{${\cal E}$}_l({\bf r}) $ at all time:
\begin{eqnarray}
&&\left(\pmb{${\cal E}$}_s({\bf r})+\pmb{${\cal E}$}_l({\bf r})\right)\rho_R(t) 
=- ic \int d^3k  
\sqrt{\frac{\hbar}{2\varepsilon_0 \omega_{{\bf k}} (2\pi)^3}} {\bf k}
\nonumber\\&&\times
\left( 
e^{ i\bf{k.r}} \lambda({\bf k},t)  + e^{ -i\bf{k.r}} \lambda(-{\bf k},t)
\right) 
\rho_R(t)  
\label{3.42}
\end{eqnarray}
Using the result (\ref{3.26}), the eigenvalue can be given a nice form:
\begin{equation}
\left(\pmb{${\cal E}$}_s({\bf r})+\pmb{${\cal E}$}_l({\bf r})\right)\rho_R(t) 
= -\pmb{$\nabla$}_{{\bf r}} U({\bf r},t) 
\rho_R(t)  
\label{3.43}
\end{equation} 
where the instantaneous Coulomb potential $ U({\bf r},t) $ is given in (\ref{3.27}) 
that is a particular case of (\ref{3.43}).
Therefore, as long as we restrict the observables to the fields, the 
scalar and longitudinal photons can be eliminated from the description 
and their effect taken into account by the instantaneous Coulomb potential.
This proves the equivalence of the quantizations in Lorentz and Coulomb
gauge.
Such a property can also be extended to the case of multiple time measurements.

\section{The reduced description for the Dirac and 
electromagnetic fields}   
\setcounter{equation}{0}

The approach of the previous section is  extended to interacting Dirac and electromagnetic fields. The
relativistic hamiltonian describing such a situation is  well-known in the second quantization
formalism we need (\cite{WH54}-\cite{BS82}, in
particular the complement $A_{V}$ (48-51) in \cite{CDG87} from which the notations are borrowed.
\begin{equation}
H=H_D+H_R+H_I
\label{4.1}
\end{equation}
where $H_D$ is the Dirac hamiltonian for the electron-positron field:
\begin{equation}
H_D=\sum_p E_p c^+_p c_p + \sum_p E_p b^+_p b_p 
\label{4.2}
\end{equation}
$c^+_p$ and $c_p$ are respectively the creation and destruction operators  associated with the
electrons characterized by the quantum number $p$ that represents the canonical momentum vector
${\bf p}$ and the spin component $\sigma$. The creation and destruction operators  $b^+_p$ and $b_p$
are associated with the positrons, with a similar sense for the indices.
$E_p$ is the relativistic energy of the free particle (characterized by a bare mass $m$) in an eigenstate
of the momentum ($E_p=\sqrt{m^2c^4+p^2c^2}$). 
$H_R$ is the electromagnetic field hamiltonian,  introduced in  previous section (\ref{2.2}). The
interaction hamiltonian $H_I$ is the sum of the  interaction hamiltonian $H_{Ie}$ with external sources,
considered in the two previous sections,  and of the interaction hamiltonian 
$H_{ID}$, involving (in normal form) the operators associated with  the electron-positron field:
\begin{equation}
H_{ID}=\int d^3r\left[-{\bf j}_D({\bf r}).{\bf  A}({\bf r})
+ c \rho_D({\bf r}) A_s({\bf r})\right] 
\label{4.3}
\end{equation}
The Dirac field charge density operator $\rho_D({\bf r})$ and  current ${\bf j}_D({\bf r})$ are now
operators given  in second quantization (in terms of the Dirac spinor $\psi$,  its adjoint
$\tilde \psi^+$ and the Dirac matrices $\pmb{$\alpha$}$.) ($\psi^+$ is a spinor, the elements of which are
the adjoint of the  elements of $\psi$, and its transposed $\tilde \psi^+$ is the adjoint of $\psi$):
\begin{eqnarray}
&&\rho_D({\bf r})=q_e \frac12[\tilde \psi^+({\bf r}) \psi({\bf r})-\tilde\psi({\bf r}) \psi^+({\bf r}) ]
\label{4.4} \\
&&{\bf j}_D({\bf r})=q_e c\frac12[ \tilde \psi^+({\bf r})\,\pmb{$\alpha$}\, \psi({\bf r})-
\tilde\psi({\bf r})\,\pmb{$\alpha$}\, \psi^+({\bf r})]
\label{4.5}
\end{eqnarray}
wher $q_e$ is the electronic charge.
The quantized Dirac field spinor $\psi({\bf r})$  and its adjoint $\tilde \psi^+({\bf r})$ are given by:
\begin{eqnarray}
&&\psi({\bf r})=\sum_p \left[ c_p u_p({\bf r})  +   b^+_p v_{\bar p}({\bf r})  \right] 
\label{4.6} \nonumber\\
&&\tilde \psi^+({\bf r})=\sum_p \left[ c^+_p u^+_p({\bf r})  +   b_p v^+_{\bar p}({\bf r})  \right] 
\label{4.7}
\end{eqnarray}
The $u_p({\bf r})$ are the positive energy plane waves spinors, 
eigenstates of the  momentum and  spin (along the momentum ${\bf p}$) operators
and of the Dirac Hamiltonian ${\cal H}_D$ for the free electron.
The index $p$ represents the set of quantum numbers
(${\bf p}$, $\mu$) ($\mu$ is the helicity  index).
The index $\bar p$ represents the set of quantum numbers
(-${\bf p}$, -$\mu$) that are opposed to those of $p$.
The wave function $v_{\bar p}({\bf r})$ represents a negative energy state 
and is therefore associated with the positron (to which is 
associated a positive energy).

The Dirac matrices  are chosen to be given in terms of  the Pauli spin matrices by:
\begin{equation}
\pmb{$\alpha$}=\left(
\matrix{  
0          &  \pmb{$\sigma$}\cr
\pmb{$\sigma$}& 0\cr
}
\right) 
\qquad 
\beta= \left(\matrix{  
I & 0 \cr
0 & -I\cr}\right) \qquad 
\pmb{$\Sigma$}=
\left(\matrix {  
\pmb{$\sigma$}& 0 \cr 0 &\pmb{$\sigma$}\cr}\right)\qquad
\delta= \left(\matrix {  
0&  I \cr
I  & 0\cr}\right) 
\label{4.8}
\end{equation}
The $\pmb{$\Sigma$}$ and $\delta$ matrices are given for future reference.
We will also need the 16 hermitian matrices that form a basis.
If we define $\sigma_0$ as the unity matrix, that basis can be represented by
$B_{\alpha\beta}=\sigma_\alpha\otimes \sigma_\beta$
where $\otimes $ represents the direct product of the Pauli matrices.
In these notations, the component $i$ of the vectorial $\pmb{$\alpha$}$ matrix is $B_{1i}$.
 
Fot the hamiltonians  (\ref{4.1}-\ref{4.2}),
we  use a continuum basis. The  anticommutators
of the creation and destruction operators are then  expressed in terms of the Dirac delta functions.
The quantized Dirac field takes  then the form, taking into account the normalization of 
$u_{{\bf p},\mu}({\bf r})  $ and $v_{{\bf p},\mu}({\bf r}) $:
\begin{eqnarray}
&&\psi({\bf r})=\frac1{\hbar^3(2\pi)^{\frac32}}
\sum_{\mu}\int d^3p 
\left[ c_{\mu}({\bf p}) u_{{\bf p},\mu}({\bf r})  
+   b^+_{\mu}({\bf p}) v_{-{{\bf p}},-\mu}({\bf r})  \right] 
\label{4.9}
\end{eqnarray}
Those operators $c$, $c^+$, $b$ and $b^+$ obey the usual fermion anticommutation relations and we
have for instance:
\begin{eqnarray}
&&\left[c_{ \mu}({\bf p}),c^+_{ \mu'}({\bf p}')\right]_{+}= \delta^{Kr}_{\mu,\mu'}\,
\delta(\frac{{\bf p} -{\bf p}'}{\hbar}) 
\label{4.12}
\end{eqnarray}

The explicit form of the normalized spinors we use can be given.
The canonical basis for the spinors is constructed from the common eigenvectors
to $\beta$ and  $\pmb{$\Sigma$}_z$. 
Those vectors are noted $|\lambda \mu>$ where $\lambda$ corresponds to the
eigenvalue of $\beta$ and $\mu$ to that of $\pmb{$\Sigma$}_z$.
The link with the states in (\ref{4.9}) 
is $|11>=u_{0+1}$, $|1-1>=u_{0-1}$,
$|-11>=v_{0+1}$, $|-1-1>=v_{0-1}$. 
The  plane waves spinors $u_{{\bf p},\mu}$,
$v_{{\bf p},\mu}$ can be obtained by the action of a unitary transformation 
$T({\bf p})$ \cite{CDG87} on those basis spinors:
\begin{eqnarray}
&&u_{{\bf p},\pm1}=T({\bf p}) u_{0\pm1} 
\qquad
v_{{\bf p},\pm1}=T({\bf p}) v_{0\pm1}
\label{4.14}
\end{eqnarray}
where the unitary matrix is ($\hat {{\bf p}}$ represents an unit vector aligned along 
${\bf p}$.):
\begin{equation}
T({\bf p})=\left[ \cos\frac{ \theta_p}2 -
\left( \beta \pmb{$\alpha$}.\hat {{\bf p}}
\right)\sin \frac{ \theta_p}2
\right] \exp \frac{i{\bf p}.{\bf r}}{\hbar} 
\label{4.15}
\end{equation}
where the angle $\theta_p$ is given by
$\theta_p={\rm{Arc tg}}(\frac{p}{mc} )$.
In terms of the relativistic energy $E_p$ and the bare electronic mass,
we have:
\begin{eqnarray}
&&\cos\frac{ \theta_p}2=\sqrt{\frac{E_p+mc^2}{2E_p}} 
\qquad
\sin\frac{ \theta_p}2=\sqrt{\frac{E_p-mc^2}{2E_p}} 
\label{4.17}
\end{eqnarray}
The  ${\bf r}$ independent part of $T$ is the unitary matrix $\tau({\bf p})$ 
($\tau^+({\bf p})=\tau(-{\bf p})$): 
\begin{equation}
\tau({\bf p})=\left[ \cos\frac{ \theta_p}2 -
\left( \beta \pmb{$\alpha$}.\hat {{\bf p}}
\right)\sin \frac{ \theta_p}2
\right].
\label{4.18a}
\end{equation}
The interaction hamiltonian can still be written in Fourier variables 
as  (\ref{3.32})
with new expressions for the charge and current density:
\begin{eqnarray}
&&j_{\lambda+}({\bf k},t) =\left({\bf j}_e({\bf k},t)+{\bf j}_D({\bf k},t)\right).  
{\bf  e}_{{\bf k}\lambda}
\nonumber\\ 
&&j_{\lambda-}({\bf k},t) =\left({\bf j}_e({\bf k},t)+{\bf j}_D({\bf k},t)\right).  
{\bf  e}_{{-\bf k}\lambda}
\nonumber\\ 
&&j_{s}({\bf k},t)=j_{s+}({\bf k},t)=j_{s-}({\bf k},t)=-c\left(\rho_e({\bf k},t)+\rho_D({\bf k},t)\right)
\label{4.20}
\end{eqnarray}
The explicit form of the charge an current operators $j_{D\nu\pm}$ (\ref{4.20}) in 
Fourier variables can now be given.
\begin{eqnarray}
&&j_{D\nu+}({\bf k})=\sum_{\mu \mu'}\int d^3p
\left( \alpha_{\nu+\,\mu \,\mu'}^{11}({\bf p},{\bf k}) 
c^+_{\mu}({\bf p}-\frac12\hbar {\bf k}) c_{ \mu'}({\bf p}+\frac12\hbar {\bf k}) 
\right.\nonumber\\ &&\left.
+ \alpha_{\nu+\,\mu \,\mu'}^{\bar1\bar1}({\bf p},{\bf k})
b^+_{\mu}({\bf p}-\frac12\hbar {\bf k}) b_{ \mu'}({\bf p}+\frac12\hbar {\bf k})
\right.\nonumber\\ &&\left.
+\alpha_{\nu+\,\mu \,\mu'}^{1\bar1}({\bf p},{\bf k})
c^+_{\mu}({\bf p}-\frac12\hbar {\bf k}) b^+_{ \mu'}(-{\bf p}-\frac12\hbar {\bf k})
\right.\nonumber\\ &&\left.
+\alpha_{\nu+\,\mu \,\mu'}^{\bar1 1}({\bf p},{\bf k})
b_{ \mu}(-{\bf p}+\frac12\hbar {\bf k})c_{\mu'}({\bf p}+\frac12\hbar {\bf k}) 
\right)  
\label{4.21}
\end{eqnarray}
and similar expressions for a minus subscript in place of a +.
The detailed expressions of the coefficients $\alpha$ are provided 
in the appendix A.
The charge and current operators are known \cite{CDG87} to satisfy the charge conservation equation  that
implies a  link between the scalar density and the longitudinal part of the current.

The hamiltonian operator is thus entirely determined in the second 
quantization formalism and we can now consider the equation for the 
density matrix $\rho$ that satisfies the Liouville-von Neuman 
equation:
\begin{equation}
i\hbar \partial_t \rho = L \rho =
[H_D + H_R + H_I ,\rho ]
\label{4.22}
\end{equation} 
The density operator $\rho$ has to involve only physical states. 
Therefore, it has to satisfy the following conditions (\ref{2.23}-\ref{2.24}) 
for all wave number ${\bf k}$  (c.f.  \cite{CDG87} (67) p.425) where
$\lambda({\bf k},t)$ is now a self-adjoint operator in the Dirac field:
\begin{equation}
\lambda({\bf k},t) =
\sqrt{\frac{c^2}{2\varepsilon_0 \hbar \omega^3_{{\bf k}}}} 
\left(\rho_e({\bf k},t)+\rho_D({\bf k})\right)
\label{4.25}
\end{equation}
where $\rho_D({\bf k})$ is now the Fourier transform of the  charge density operator.
As for external sources, if these conditions (\ref{2.23}-\ref{2.24})  hold at some given
time, they hold for all times as a consequence of the charge conservation.

The basic entities for our description of the interacting electron-positron
and  electromagnetic fields are  defined, as in (\ref{3.1}), by correlation functions obtained by mean
values  of particular superoperators in normal form:
\begin{eqnarray}
&&
f_{ss'\tilde s\tilde s'nn'}
^{\mu_1\dots\mu_s\mu'_{s'}\dots \mu'_1
\tilde \mu_1\dots\tilde \mu_{\tilde s}\tilde \mu'_{\tilde s'}\dots\tilde  \mu'_1}
({\bf p}_1 ;{\bf p}_2 \dots{\bf p}_s;
{\bf p}'_1 ;{\bf p}'_2 \dots{\bf p}'_s;
{\tilde{{\bf p}_1}} ;{\tilde{{\bf p}}}_2 \dots{\tilde{{\bf p}}}_{\tilde{s}};
\nonumber\\&&
{\tilde{{\bf p}}}'_1 ;{\tilde{{\bf p}}}'_2 \dots{\tilde{{\bf p}}}'_{\tilde{s}};
{\bf k}_1 \nu_1;{\bf k}_2 \nu_2;\dots;
{\bf k}_n \nu_n;
{\bf k}'_1 \nu'_1;{\bf k}'_2 \nu'_2;\dots;
{\bf k}'_{n'} \nu'_{n'};t) \nonumber\\&&  =
\hbar^{-\frac32(s+s'+\tilde s+\tilde s')} \rm{Tr}a_{\nu_1}({\bf k}_1)
a_{\nu_2}({\bf k}_2)
\dots a_{\nu_n}({\bf k}_n) 
\nonumber\\&&\times
b_{\tilde \mu_1}(\tilde{{\bf p}}_1) 
\dots b_{\tilde \mu_{\tilde s}}(\tilde{{\bf p}}_{\tilde s}) 
c_{ \mu_1}({\bf p}_1)
\dots c_{ \mu_s}({\bf p}_s) 
\rho(t) 
c^+_{ \mu'_{s'}}({\bf p}'_{s'})
\dots 
c^+_{ \mu'_1}({\bf p}'_1)
\nonumber\\ &&\times
b^+_{\tilde \mu'_{\tilde s'}}(\tilde{{\bf p}}'_{\tilde s'}) 
\dots 
b^+_{\tilde \mu'_1}(\tilde{{\bf p}}'_1) 
a^+_{ \nu'_{n'}}({\bf k}'_{n'}) \dots
a^+_{ \nu'_2}({\bf k}'_2)
a^+_{ \nu'_1}({\bf k}'_1)
\label{4.26}
\end{eqnarray}
The variables with an upper index ``tilde" are associated with the positrons.
The variables with a ``prime" are associated with the creation operators, as in 
(\ref{3.1}).
The element  $f_{000000}$ corresponds to the normalization of the 
density operator and takes the value 1.
When the arguments of the $f_{ss'\tilde s\tilde s'nn'}$ functions
are labelled as in the definition (\ref{4.26}), they will not
written explicitly.

It is useful to introduce substitution operators $S$ for the  variables  of
electromagnetic or electron-positron fields.
Acting on a function of ${\bf k}_i$, $\nu_i$, the operator
$S({\bf k}, \nu;{\bf k}_i, \nu_i)$  replaces the variables  
${\bf k}_i$, $\nu_i$ by new variables  ${\bf k}$, $\nu$, and a similar
definition for the matter variables.
Those substitution operators allow to avoid to write explicitly the 
variables of the $f_{ss'\tilde s\tilde s'nn'}$ functions: only the
modifications with respect to those of the definition (\ref{4.26}) have to be 
given in terms of the substitution operators.

These moments (\ref{4.26})  are not independent since they have to represent 
physical states and  have to satisfy relations due to (\ref{2.23}) and (\ref{2.24})
besides those induced by quantum symmetry of the Dirac field \cite{RB75}.
From(\ref{2.23}), the following relations have to hold:
\begin{eqnarray}
&&
\hbar^{-\frac32(s+s'+\tilde s+\tilde s')} \rm{Tr}a_{\nu_1}({\bf k}_1)
a_{\nu_2}({\bf k}_2)
\dots a_{\nu_n}({\bf k}_n) 
b_{\tilde \mu_1}(\tilde{{\bf p}}_1) 
\dots b_{\tilde \mu_{\tilde s}}(\tilde{{\bf p}}_{\tilde s})
\nonumber\\ &&\times
c_{ \mu_1}({\bf p}_1)
\dots c_{ \mu_s}({\bf p}_s) 
\left[ a_l({\bf k})- a_s({\bf k}) +
\sqrt{\frac{c^2}{2\varepsilon_0 \hbar \omega^3_{{\bf k}}}} 
\left(\rho_e({\bf k},t)+:\rho_D({\bf k}):\right) 
\right]\rho(t) 
\nonumber\\ &&\times 
c^+_{ \mu'_{s'}}({\bf p}'_{s'})
\dots 
c^+_{ \mu'_1}({\bf p}'_1)
b^+_{\tilde \mu'_{\tilde s'}}(\tilde{{\bf p}}'_{\tilde s'}) 
\dots 
b^+_{\tilde \mu'_1}(\tilde{{\bf p}}'_1) 
a^+_{ \nu'_{n'}}({\bf k}'_{n'}) \dots
a^+_{ \nu'_2}({\bf k}'_2)
a^+_{ \nu'_1}({\bf k}'_1)
\nonumber\\ && =0
\label{4.27}
\end{eqnarray}
from which, using the cyclic invariance of the trace, relations for 
$f_{ss'\tilde s\tilde s'nn'} $ that generalize (\ref{3.2}-\ref{3.3})
can be deduced.

For a classical external source, 
the longitudinal and scalar components do not appear in
the computation of the mean value of the  electric field (\ref{3.26}). 
A similar property holds for the combination of quantum and classical sources.
If we neglect for the moment the time dependence in the classical source,  to the electric field
observable, we can associate a superoperator in normal form.
Its transverse  components are given by analogy with the external source case.
The sum of the longitudinal and transverse components is still given by (\ref{3.41}).
The use of the relations (\ref{2.23}-\ref{2.24}) in (\ref{3.41}) imposes a reorganization  
to obtain an operator in normal form: 
\begin{eqnarray}
&&\pmb{${\cal E}$}_s({\bf r})+\pmb{${\cal E}$}_l({\bf r}) 
=-\pmb{$\nabla$}_{{\bf r}} U_e({\bf r}) 
+i\frac1{(2\pi)^{\frac32}}\frac{c}{2\varepsilon_0}  \int d^3k  
\frac{{\bf k}}{ \omega^2_{{\bf k}}}
\left( 
e^{ i\bf{k.r}}-e^{ -i\bf{k.r}}\right)
\nonumber\\ &&  \times
\sum_{\mu \mu'}\int d^3p
\left\{ \alpha_{s\,\mu \,\mu'}^{11}({\bf p},{\bf k}) 
\left[c_{ \mu'}({\bf p}+\frac12\hbar {\bf k}) \times
c^+_{\mu}({\bf p}-\frac12\hbar {\bf k})\right]  
\right.\nonumber\\ &&\left.
+ \alpha_{s\,\mu \,\mu'}^{\bar1\bar1}({\bf p},{\bf k})
\left[b_{ \mu'}({\bf p}+\frac12\hbar {\bf k}) \times 
b^+_{\mu}({\bf p}-\frac12\hbar {\bf k})\right] 
\right.\nonumber\\ &&\left.
+\alpha_{s\,\mu \,\mu'}^{1\bar1}({\bf p},{\bf k}) \left[ I\times 
c^+_{\mu}({\bf p}-\frac12\hbar {\bf k}) 
b^+_{ \mu'}(-{\bf p}-\frac12\hbar {\bf k})
\right] 
\right.\nonumber\\ &&\left.
+\alpha_{s\,\mu \,\mu'}^{\bar1 1}({\bf p},{\bf k})
\left[ b_{ \mu}(-{\bf p}+\frac12\hbar {\bf k})
c_{\mu'}({\bf p}+\frac12\hbar {\bf k}) 
\times I\right]
\right\} 
\label{4.32}
\end{eqnarray}
For all types of sources, the mean values of the electric field operator can therefore be easily 
expressed in terms of the correlation functions that involve neither  the longitudinal nor the scalar
photons.

\section{The observables of  the Dirac field}
\setcounter{equation}{0}

The continuous observables associated with the Dirac field are the positions, the canonical momentum
and the mechanical momentum,  that can be combined with spins and nature (electron-positron)
properties. 
The mechanical momentum is more physically relevant than the canonical momentum
but, for commodity reasons, the correlation functions (\ref{4.26}) are more easily defined in terms of
the canonical momentum. 
For a physical interpretation, we have however to keep in mind that
only the mechanical momentum is meaningful and that the canonical momentum has to be
considered as a useful intermediate variable.

In non relativistic quantum theory,
to a classical function $O_{cl}({\bf r}, {\bf p})$ of the position and momentum of
the particle, is associated a quantum observable, $\hat O$ defined through the
operator (Weyl's correspondence rule) \cite{RB75}:
\begin{equation}
\hat O=\frac1{(2\pi)^6}\int d^3K\,d^3X\,\tilde O_{cl}({\bf  K},{\bf  X})
e^{i({\bf  K}.\hat {{\bf r}}+\frac1\hbar{\bf  X}.\hat{{\bf p}})}
\label{4.33}
\end{equation}
where $\tilde O_{cl}({\bf  K}, {\bf  X})$ is the Fourier transform of $O_{cl}({\bf r}, {\bf p})$:
\begin{equation}
\tilde O_{cl}({\bf  K}, {\bf  X})=\frac1{\hbar^6}
\int d^3r\,d^3p\,O_{cl}({\bf r}, {\bf p})e^{-i({\bf  K}.{\bf r}+\frac1\hbar{\bf  X}.{\bf p})}
\label{4.34}
\end{equation}
The second quantization form $O_{[2Q]}$ of an  operator $\hat O$ is
constructed {\it a priori} \cite{RB75}
 as:
\begin{equation}
O_{[2Q]} =\int d^3r\, \psi^+({\bf r})  \hat O \psi({\bf r}) 
\label{4.35}
\end{equation}
The present relativistic approach requires the generalization of those concepts.
No straightforward procedure is directly available.
Indeed, the expression of the mechanical momentum, derived for instance from the energy
momentum tensor \cite{WH54} p419  requires also the fields operators describing creation and
destruction of particles,  besides the operators dealing with the field. 

However, for the canonical momentum, no such problem arises.
Therefore, we consider the Weyl's procedure for that momentum 
and possible extension to the mechanical momentum will be presented.

The joint characteristic operator $\Xi^{(0)}(\hat {{\bf r}},\hat{\pmb{$\pi$}},\alpha,\beta)$,
involving the position and the canonical momentum of the
electron, and  the hermitian operator $B_{\alpha\beta}$ previously introduced
($B_{\alpha\beta}\equiv\sigma_\alpha\otimes \sigma_\beta$), is defined through an expression that
respects the symmetry between electron and positron. Due to the presence of a general
$B_{\alpha\beta}$ matrix, the simple substraction procedure present in (\ref{4.4}), (\ref{4.5}) cannot
be generalized and the normal ordering of the creation and destruction operators has to be restored
by hand, without considering the contribution due to the anticommutator. The existence of that
procedure is  noted by a prime on the bracket. For one particle, we have (The generalization to an
arbitrary numbers of particle is staightforward and allows to introduce a complete set of
observables.):
\begin{eqnarray}
&&
\Xi^{(0)}(\hat {{\bf r}},\hat{\pmb{$\pi$}},\alpha,\beta)=
\left.
e^{i({\bf  K}.\hat {{\bf r}}+\frac1\hbar{\bf  X}.\hat {\bf p})}
B_{\alpha\beta}
\right\vert_{[2Q]}=\int d^3r\,
\left[\tilde \psi^+({\bf r})
e^{{\bf  X}.\nabla+i{\bf  K}.{\bf r}}
B_{\alpha\beta}
\psi({\bf r}) 
 \right]'
\nonumber\\&&
\label{4.37}
\end{eqnarray}

The Weyl's procedure cannot be applied directly in the case of the mechanical momentum.
A naive proposal based on it provides nevertheless clues on the structure and potentialities of the
approach.
In a non relativistic approach, the mechanical momentum of the electron $\hat{\pmb{$\pi$}}$ is
linked with the canonical's one ${\bf p}$ by the relation :
\begin{equation}
\hat{\pmb{$\pi$}}=\hat{{\bf p}}-q_e \hat{{\bf A}}_{\bot}({\bf r})
\label{4.36}
\end{equation}
where the transverse vectorial operator $ \hat{{\bf A}}_{\bot}({\bf r})$ can be deduced from
(\ref{2.5}). 
The use of (\ref{4.36}) in a relativistic context in place of  the true momentum tensor \cite{WH54} p419
provides a limitation in specific results.

$ \hat{{\bf A}}_{\bot}({\bf r})$  contains creation and destruction operator associated with the field.
The use of a similar expression in the Dirac description of the electron-positron particles requires
some adaptation.
Indeed, the description involves simultaneously both kinds of particles and we have to provide a
mechanism that enables the correct sign of the associated charge that plaus a role in (\ref{4.36}).
A free electron is represented by the positive energy plane waves spinor $u_p({\bf r})$ 
($ \vert u_p>$ in Dirac notation). 
We can therefore introduce a projector 
$P_e$ ($P_e=\sum_p  \vert u_p><u_p\vert$) on these states and a similar projector $P_p$ for the
positrons. 

The basic observables for the Dirac field are the position and the mechanical momentum of the
electron-positrons, and  the hermitian operator $B_{\alpha\beta}$ introduced previously
We have to propose a form for the joint characteristic operator involving these basis observables.

The hermitian operator $B_{\alpha\beta}$, is defined
through an expression that respects the symmetry between electron and positron.
The difference between the canonical and mechanical momentum involves the transverse field
operator 
${\bf A}_{\bot}({\bf r})$  and the operators $P_e$ and $P_p$.
The $\psi({\bf r})$ operator is not symmetrical with respect to the electron-positron
symmetry.
Indeed, the destruction operator associated with the electron would be on the right of the
generalisation of (\ref{4.35}), as the creation operator for the positron.
In presence of a general $B_{\alpha\beta}$ matrix,  the normal
ordering of the creation and destruction operators of the particle has to be restored by hand, without
considering the contribution due to the contractions between similar fields. 
The existence of that procedure is  noted by a prime on the bracket.
We want a similar association for both kinds of particle.
We separate  the positive frequency part ${\bf A}^{(+)}_{\bot}({\bf r})$ (with the
destruction operator) and negative frequency part  ${\bf A}_{\bot}^{(-)}({\bf r})$ of
${\bf A}_{\bot}({\bf r})$.
If ${\bf A}_{\bot}^{(+)}({\bf r})$ has to be associated with the electron destruction operator, 
${\bf A}_{\bot}^{(-)}({\bf r})$ has to be associated with the positron creation operator.
For the term linear in ${\bf A}_{\bot}({\bf r})$, it will be manifest that the field operator has to be
outside the first quantisation operator $e^{i({\bf K}.\hat {{\bf r}}+\frac1\hbar{\bf  X}.\hat{{\bf p}})}$
where 
$\hat{{\bf p}}$ will stands for $\frac \hbar i\nabla$.
The operator to be placed at the right will be noted $\bar {{\bf A}}_{\bot r} ({\bf r})$ and the
operator to  be placed at the left will be noted $\bar {{\bf A}}_{\bot l} ({\bf r})$.
We have 
\begin{equation}
\bar{{\bf A}}_{\bot r} ({\bf r})={\bf A}_{\bot}^{(+)}({\bf r})P_e-{\bf A}_{\bot}^{(-)}({\bf r})P_p
\label{4.36a}
\end{equation}
\begin{equation}
\bar{{\bf A}}_{\bot l} ({\bf r})={\bf A}_{\bot}^{(-)}({\bf r})P_e-{\bf A}_{\bot}^{(+)}({\bf r})P_p
\label{4.36b}
\end{equation}
The  proposal for the joint characteristic operator 
$\Xi(\hat {{\bf r}},\hat{\pmb{$\pi$}},\alpha,\beta)$ is:
\begin{eqnarray}
&&
\left.
e^{i({\bf K}.\hat {{\bf r}}+\frac1\hbar{\bf  X}.\hat{\pmb{$\pi$}})}
B_{\alpha\beta}
\right\vert_{[2Q]}=
\int d^3r\,
\left[\tilde \psi^+({\bf r})e^{-\frac{iq_e}\hbar{\bf  X}.\bar{{\bf A}}_{\bot l} ({\bf r})}
e^{{\bf  X}.\nabla+i{\bf K}.{\bf r}}
B_{\alpha\beta}
\right.\nonumber\\&&\left.\times
e^{-\frac{iq_e}\hbar{\bf  X}.\bar{{\bf A}}_{\bot r} ({\bf r})}
\psi({\bf r}) 
 \right]'
\label{4.37a}
\end{eqnarray}
The choice of the correspondence  has to be justified later on by the properties of the
description that it generates.
We anticipate by noting simply that if we interchange 
$\bar{{\bf A}}_{\bot l} ({\bf r})$ and $\bar{{\bf A}}_{\bot r} ({\bf r})$ in (\ref{4.37a}), no contribution
analogous to the second order mass correction is obtained.

As we have seen in quantum optics \cite{dH04b}, a dressing operator can be introduced  in the single
subdynamics approach.
That dressing operator will be able to connect all the descriptions arising from different proposals
for the connection between classical and quantum observables.
In the present study, we will simply consider that proposal (\ref{4.37a}) 
and examine mainly its consequences
at the level of the finiteness of the evolution operator.
Relativistic corrections may be required to the present proposal based on the non relativistic Weyl's
rules.
For a proposal that is not based on the energy momentum tensor, two different gauges may provide
inequivalent representation: we have no reason to suspect a gauge invariance of our first proposal
that is formulated without caring about that point but aims to ilustrate the feasibility of the approach .

The transverse component (also the longitudinal one) of the field operator 
${\bf A}_{\bot}({\bf r})$ does not commute with powers of the operator
$\nabla$ and that expression (\ref{4.37a}) is defined through the perturbation expansion of the
exponential.

As we have seen in quantum optics \cite{dH04b}, a dressing operator can be introduced  in the
single subdynamics approach.
The dressing operator  \cite{dH04b} is a tool to perform the transition between the distribution
functions in terms of positions and  canonical momentums towards distribution functions in terms of
positions and mechanical momentums.
It is entirely determined by the proposal (\ref{4.37a}).
That dressing operator is able to connect all the descriptions arising from different proposals
for the correspondence between classical and quantum observables.

For the mechanical momentum $\hat{\pmb{$\pi$}}$, 
the simple form (\ref{4.37a}) is not expected to hold in general and, for all proposals, 
we would obtain, for the joint characteristic operator 
$\Xi(\hat {{\bf r}},\hat{\pmb{$\pi$}},\alpha,\beta)$,
the following structure ($k_1$ is here a shorthand writing for the set of variables ${\bf k}_1$,
$\nu_1$): 
\begin{eqnarray}
&&
\Xi(\hat {{\bf r}},\hat{\pmb{$\pi$}},\alpha,\beta) = 
\sum_{n,n'}
 \sum_{\mu\mu'}\int d^3p\,\int d^3p'\,
\int d^3k_1\,\dots\int d^3k_n\,
\int d^3k'_1\,\dots\int d^3k'_{n'}\,
\nonumber\\&&\times
\sum_{\nu_1,\dots\nu_n}\sum_{\nu'_1,\dots\nu'_{n'}}
\left( \gamma_{\alpha\beta}^{1\mu1\mu'nn'}({\bf p},{\bf p}',k_1,\dots k_n;k'_1\dots k'_{n'}) 
c^+_{\mu'}({\bf p}') c_{ \mu}({\bf p}) 
\right.\nonumber\\&&\left.\times
a^+_{ \nu'_{n'}}({\bf k}'_{n'}) \dots
a^+_{ \nu'_2}({\bf k}'_2)
a^+_{ \nu'_1}({\bf k}'_1)
a_{\nu_1}({\bf k}_1)
a_{\nu_2}({\bf k}_2)
\dots a_{\nu_n}({\bf k}_n) 
\right.\nonumber\\ &&\left.
+ \gamma_{\alpha\beta}^{\bar1\mu\bar1\mu'nn'}({\bf p},{\bf p}',k_1,\dots k_n;k'_1\dots k'_{n'})
b^+_{\mu'}({\bf p}') b_{ \mu}({\bf p})
\right.\nonumber\\&&\left.\times
a^+_{ \nu'_{n'}}({\bf k}'_{n'}) \dots
a^+_{ \nu'_2}({\bf k}'_2)
a^+_{ \nu'_1}({\bf k}'_1)
a_{\nu_1}({\bf k}_1)
a_{\nu_2}({\bf k}_2)
\dots a_{\nu_n}({\bf k}_n) 
\right.\nonumber\\ &&\left.
+\gamma_{\alpha\beta}^{1\mu\bar1\mu'nn'}({\bf p},{\bf p}',k_1,\dots k_n;k'_1\dots k'_{n'})
c^+_{\mu}({\bf p}) b^+_{ \mu'}({\bf p}')
\right.\nonumber\\&&\left.\times
a^+_{ \nu'_{n'}}({\bf k}'_{n'}) \dots
a^+_{ \nu'_2}({\bf k}'_2)
a^+_{ \nu'_1}({\bf k}'_1)
a_{\nu_1}({\bf k}_1)
a_{\nu_2}({\bf k}_2)
\dots a_{\nu_n}({\bf k}_n) 
\right.\nonumber\\ &&\left.
+\gamma_{\alpha\beta}^{\bar1\mu 1\mu'nn'}({\bf p},{\bf p}',k_1,\dots k_n;k'_1\dots k'_{n'})
b_{ \mu'}({\bf p}')c_{\mu}({\bf p}) 
\right.\nonumber\\&&\left.\times
a^+_{ \nu'_{n'}}({\bf k}'_{n'}) \dots
a^+_{ \nu'_2}({\bf k}'_2)
a^+_{ \nu'_1}({\bf k}'_1)
a_{\nu_1}({\bf k}_1)
a_{\nu_2}({\bf k}_2)
\dots a_{\nu_n}({\bf k}_n) 
+\dots
\right) 
\label{4.39}
\end{eqnarray}
The dots refer to terms involving more particles creation and destruction operators.
These $\gamma$ functions define the joint characteristic operator for one particle.
Their dependence in ${\bf  K}$ and ${\bf  X}$ is implicit.
The knowledge of these functions is still a challenging problem for physicists.
Even in ordinary quantum mechanics, the correspondence between a function of classical
observables, so that their quantum operators do not commute, and the associated function of
quantum operators is not settled: Weyll correspondence rules is only one useful proposal.
In the relativistic case, the mechanical momentum of one electron can be obtained throught the
energy-momentum tensor but at our knowledge, no proposal for products of that momentum or the
joint distribution of product of the moment and the position operator has ever be emitted.
Imposing the finiteness of QED involves constraints to the $\gamma$ functions.
We think that these constraints are natural and we link in fact the finiteness of the description with
the general correspondence rules between a classical function and the associated quantum operator.
Gauge invariance provides of course also constraints on the choice of the $\gamma$'s.

To the form (\ref{4.37}) correspond a limited number of non-vanishing  $\gamma$ functions 
(both $n$ and $n'$ vanish and no supplementary terms are present) that can be
easily computed.
They will constitute our reference description.
The form (\ref{4.37a}) defines non trivial  $\gamma$ functions 
and enables to illustrate the use and properties of  (\ref{4.39}).

The mean value of the characteristic operator for the canonical momentum
$\Xi^{(0)}(\hat {{\bf r}},\hat{\pmb{$\pi$}},\alpha,\beta)$
 involves the trace of the product of  that operator with the density operator
and can be written in terms of the moments that we have previously introduced (\ref{4.26}), 
with the same
convention for the name for their argument:
\begin{eqnarray}
&&
\Xi^{(0)}(\hat {{\bf r}},\hat{\pmb{$\pi$}},\alpha,\beta) =
\hbar^3 
\left( 
\sum_{\mu\mu'}\int d^3p\,\int d^3p'\,
\gamma_{\alpha\beta}^{1\mu1\mu' 00}({\bf p},{\bf p}') 
 f_{1100 00}^{\mu  \mu'}
\right.\nonumber\\ &&\left.
+\sum_{\tilde\mu\tilde\mu'}\int d^3\tilde p\,\int d^3\tilde p'\, 
\gamma_{\alpha\beta}^{\bar1\tilde\mu\bar1\tilde\mu' 00}(\tilde{{\bf p}},\tilde{{\bf p}}') 
f_{001100}^{\tilde \mu\tilde  \mu'}
\right.\nonumber\\ &&\left.
+\sum_{\mu'\tilde\mu'}\int d^3 p'\,\int d^3\tilde p'\, 
\gamma_{\alpha\beta}^{1\mu'\bar1\tilde\mu' nn'}({\bf p}',\tilde{{\bf p}}') 
f_{010100}^{\mu' \tilde\mu'}
\right.\nonumber\\ &&\left.
+\sum_{\mu\tilde\mu}\int d^3 p\,\int d^3\tilde p\, 
\gamma_{\alpha\beta}^{\bar1\mu 1\tilde\mu00}({\bf p},\tilde{{\bf p}})
f_{101000'}^{\mu \tilde\mu}
\right) 
\label{4.40}
\end{eqnarray}
If we consider the quantum operator ${\hat O}B_{\alpha\beta}$ associated with a classical function 
$O_{cl}({\bf  K}, {\bf  X})$ and the quantum operator  $B$, its  mean value is
provided by
\begin{equation}
<{\hat O}B_{\alpha\beta}>=
\frac1{(2\pi)^6}\int d^3K\,d^3X\,\tilde O_{cl}({\bf  K},{\bf  X})
<\Xi(\hat {{\bf r}},\hat{\pmb{$\pi$}},\alpha,\beta)>
\label{4.41}
\end{equation}
where $\hat{\pmb{$\pi$}}$ represents the mechanical momentum of the electron-positron field.
We can define the generalization $f_{\alpha\beta}^W$ of the Wigner distribution functions  by imposing
\begin{equation}
<{\hat O}B_{\alpha\beta}>=
\frac1{(2\pi)^6}\int d^3K\,d^3X\,\tilde O_{cl}({\bf  K},{\bf  X})
f_{\alpha\beta}^W(-{\bf  K},-{\bf  X})
\label{4.42}
\end{equation}
and by identifying the two previous expressions (\ref{4.41}) and (\ref{4.42}).
The use of $\Xi^{(0)}(\hat {{\bf r}},\hat{\pmb{$\pi$}},\alpha,\beta)$ in (\ref{4.41}) and a similar
identification provide the definition of the reference functions $f_{\alpha\beta}^{(0)}$.

The 16 dressed distribution functions $f_{\alpha\beta}^W$ are the natural variables to deal with the
electron-positron field for observables involving only one particle. 
They are completely determined by the gamma's 
and expressed in terms of the bare distribution
functions $f_{ss'\tilde s{\tilde s}' nn'}^{\mu \mu'}$ for all values of $n$ and $n'$.
If the observables involve more than one particle, we need also joint Wigner distribution functions
that can be defined in a similar way.

These relations (\ref{4.40})-(\ref{4.42}) and the similar ones for more that one particle
(for describing for instance a collision process) form a basis
for studying the time evolution of the joint dressed distribution functions $f^W$, describing all
physical observables associated with the particles. 

Although we deal with a complete set of quantum observables for the particles,
the set of Wigner distribution functions for the particles is not closed
{\it a priori} by the evolution equation:
indeed, distribution functions involving the em field appear naturally.
When no incident field is involved, 
our reformulation of quantum field without self energy parts 
allows to express those functions in terms of the distribution functions involving the particles only
through a so-called creation operator.
Therefore, in that approach, closed kinetic equations can be obtained for the Wigner distribution
functions when no incident field is needed.
That incident field represents of course an external field and not the self-field generated by the
particles: the effect of that self-field is already taken into account in the kinetic equations.

Once the gamma's (\ref{4.39}) are known, the link provided by the creation operators can be used to
express
$<\Xi(\hat {{\bf r}},\hat{\pmb{$\pi$}},\alpha,\beta)>$ in a way similar
to the right hand side of (\ref{4.42}): only distribution functions of the particles are present.
They can be expressed in terms of the reference distribution functions $f^{0}_{\alpha\beta}$
describing the canonical momentum.
That process is equivalent to the determination of the  set of Wigner distribution functions 
$f_{\alpha\beta}^W$ in terms of the reference distribution functions $f^{(0)}_{\alpha\beta}$, relative
to the canonical momentum, obtained in using (\ref{4.40}) in place of (\ref{4.41}).
It corresponds thus to the determination of a dressing operator.
We know \cite{dH04b} that a dressing operator of the second order (in the charge) will modify the
kinetic equations at the same order, outside resonances.
We will see in a next section that the second order kinetic operator contains the usual divergent
mass correction for the reference distribution functions $f^{(0)}_{\alpha\beta}$.
The supplementary contributions due to the dressing could compensate that divergence.
Our attempt (\ref{4.37a}) that aims to replace exponential of the canonical momentum $\hat{{\bf
p}}$ by some combination of exponentials of $\hat{{\bf p}}$ and $q_e\hat{{\bf  A}}({\bf r})$ in
(\ref{4.33}) does not provide the solution since high energy processes are not correctly described,
due to the lack of fermions pairs creation and destruction. That attemp proves nevertheless
that new terms, presenting the same kind of divergences, appear although no complete cancellation
occurs.

\section{The one particle reference distribution function}  
\setcounter{equation}{0}

The explicit evaluation of the first terms in the joint characteristic operator (\ref{4.37}) is the object
of this section. In place of (\ref{4.37}), we can use a more symmetric form with respect to the
electrons-positrons field using ($\nabla_{op}$ is a derivative operator acting on the left cf.
\cite{WH54}). The final result do not depend on the form chosen  
\begin{eqnarray}
&&\left.
\Xi(\hat {{\bf r}},\hat{\pmb{$\pi$}},\alpha,\beta)
\right\vert^{(0)}=\left.
e^{i({\bf  K}.\hat {{\bf r}}+\frac1\hbar{\bf  X}.\hat{{\bf p}})}
B_{\alpha\beta}
\right\vert_{[2Q]}
\nonumber\\&&
=\int d^3r\,
\left[\tilde \psi^+({\bf r})
e^{{\bf  X}.(\frac12(\nabla-\nabla_{op})+i{\bf  K}.{\bf r}}
B_{\alpha\beta}
\psi({\bf r}) 
 \right]'
\label{5.1}
\end{eqnarray}
We see that expression as the zero${}^{th}$ order in the charge of $\Xi$ (\ref{4.39}), hence the upper index.

The well known relations:
\begin{equation}
e^{{\bf  X}.\nabla+i{\bf  K}.{\bf r}}=e^{i\frac12{\bf  K}.{\bf  X}}
e^{i{\bf  K}.{\bf r}}e^{{\bf  X}.\nabla}
\label{5.2}
\end{equation}
\begin{equation}
e^{{\bf  X}.\nabla_{op}+i{\bf  K}.{\bf r}}=
e^{i{\bf  K}.{\bf r}}e^{{\bf  X}.\nabla_{op}}e^{i\frac12{\bf  K}.{\bf  X}}
\label{5.2a}
\end{equation}
will be useful for the computation of (\ref{5.1}).
Since $[\frac12(\nabla-\nabla_{op}),{\bf r}]=[\nabla,{\bf r}]=I$, and  
$[\nabla,\nabla_{op}]=0$
we have also
\begin{equation}
e^{{\bf  X}.\frac12(\nabla-\nabla_{op})+i{\bf  K}.{\bf r}}=e^{i\frac12{\bf  K}.{\bf  X}}
e^{i{\bf  K}.{\bf r}}e^{{\bf  X}.\frac12(\nabla-\nabla_{op})}
=e^{-{\bf  X}.\frac12(\nabla_{op})}
e^{i{\bf  K}.{\bf r}}
e^{{\bf  X}.\frac12(\nabla)}
\label{5.2b}
\end{equation}
Using an integration by parts, we obtain at zero order in the charge:
\begin{eqnarray}
\left.
e^{i({\bf  K}.\hat {{\bf r}}+\frac1\hbar{\bf  X}.\hat{\pmb{$\pi$}})}
B_{\alpha\beta}
\right\vert^{(0)}_{[2Q]}&&=e^{i\frac12{\bf  K}.{\bf  X}}\int d^3r\,
\left[\tilde \psi^+({\bf r})
e^{i{\bf  K}.{\bf r}}
B_{\alpha\beta}
\psi({\bf r}+{\bf  X}) 
\right]'
\label{5.3}
\end{eqnarray}
Using the expression (\ref{4.9}) for $\psi({\bf r})$ and the relations (\ref{4.15}), (\ref{4.14}) 
and (\ref{4.15}), we obtain
\begin{equation}
\psi({\bf r})=\frac1{\hbar^3(2\pi)^{\frac32}}
\sum_{\mu}\int d^3p 
\left[ c_{\mu}({\bf p}) T({\bf p}) u_{0\mu} ({\bf r})  
+   b^+_{\mu}({\bf p}) T(-{\bf p}) v_{0-\mu}({\bf r})  \right] 
\label{5.4}
\end{equation}
We note $T({\bf p})$ given in (\ref{4.15}) as
\begin{equation}
T({\bf p}) =\tau({\bf p})\exp \frac{i{\bf p}.{\bf r}}{\hbar} 
\label{5.5}
\end{equation}
where the unitary matrix $\tau({\bf p})$, given by (\ref{4.18a}), takes the form
\begin{equation}
\tau({\bf p})=\left[ \cos\frac{ \theta_p}2 \left(\otimes I\right) -
i\sin \frac{ \theta_p}2\left( \sigma_2\otimes  \pmb{$\sigma$}.\hat {{\bf p}}
\right)
\right]
\label{5.6}
\end{equation}
and $\tau^+({\bf p})=\tau(-{\bf p})$.
We use also the previously introduced notation $\bar \mu=-\mu$.
Therefore,
\begin{equation}
\psi({\bf r})=\frac1{\hbar^3(2\pi)^{\frac32}}
\sum_{\mu}\int d^3p 
\left[ c_{\mu}({\bf p}) e^{\frac{i{\bf p}.{\bf r}}{\hbar}}\tau({\bf p}) u_{0\mu}
+   b^+_{\mu}({\bf p}) e^{ \frac{-i{\bf p}.{\bf r}}{\hbar}}\tau(-{\bf p})  v_{0\bar\mu} \right] 
\label{5.7}
\end{equation}
\begin{equation}
\tilde\psi^+({\bf r})=\frac1{\hbar^3(2\pi)^{\frac32}}
\sum_{\mu}\int d^3p 
\left[ c^+_{\mu}({\bf p})e^{\frac{-i{\bf p}.{\bf r}}{\hbar}} \tilde u_{0\mu}   \tau(-{\bf p}) 
+   b_{\mu}({\bf p})e^{ \frac{i{\bf p}.{\bf r}}{\hbar}}  \tilde v_{0\bar\mu}\tau({\bf p})  \right] 
\label{5.8}
\end{equation}
$\gamma_{\alpha\beta}^{1\mu1\mu'00}({\bf p},{\bf p}') $ is given by 
\begin{eqnarray}
&&
\gamma_{\alpha\beta}^{1\mu1\mu'00}({\bf p},{\bf p}')
= 
\frac1{\hbar^6(2\pi)^3}
e^{i\frac12{\bf  K}.{\bf  X}}
\int d^3r\,
\left[
e^{\frac{-i{\bf p}'.{\bf r}}{\hbar}} \tilde u_{0\mu'}  \tau(-{\bf p}') 
e^{i{\bf  K}.{\bf r}}
B_{\alpha\beta}
\right.\nonumber\\ &&\left.\times
 e^{\frac{i{\bf p}.({\bf r}+{\bf  X})}{\hbar}}\tau({\bf p}) u_{0\mu}
\right]
\label{5.9}
\end{eqnarray}
The integration over ${\bf r}$ can be performed and provides a Dirac delta function.
We introduce a notation for the geometrical factors:
\begin{equation}
\delta_{\alpha\beta}^{\xi'\mu'\xi\mu}({\bf p}', {\bf p})
=<\xi'\mu'\vert\tau^+({\bf p}') B_{\alpha\beta}\tau({\bf p}) \vert\xi\mu>
\label{5.13}
\end{equation}
The links between the states $\vert\xi\mu>$ on one hand and $u_{0\mu}$, $v_{0\mu}$ on the other hand
has been provided previously.
Therefore,
\begin{equation}
\gamma_{\alpha\beta}^{1\mu1\mu'00}({\bf p},{\bf p}')
=\frac1{\hbar^3}
e^{i\frac12{\bf  K}.{\bf  X}}
\delta({\bf p}-{\bf p}'+\hbar{\bf  K})
 e^{\frac{i{\bf p}.{\bf  X}}{\hbar}}
\delta_{\alpha\beta}^{1\mu'1\mu}({\bf p}', {\bf p})
\label{5.15}
\end{equation}
Similarly,
\begin{equation}
\gamma_{\alpha\beta}^{\bar1\mu\bar1\mu'00}({\bf p},{\bf p}')
=-\frac1{\hbar^3}
e^{i\frac12{\bf  K}.{\bf  X}}
\delta({\bf p}-{\bf p}'+\hbar{\bf  K})
 e^{\frac{-i{\bf p}'.{\bf  X}}{\hbar}}
\delta_{\alpha\beta}^{\bar1\bar\mu\bar1\bar\mu'}(-{\bf p}, -{\bf p}')
\label{5.17}
\end{equation}
\begin{equation}
\gamma_{\alpha\beta}^{1\mu\bar1\mu'00}({\bf p},{\bf p}')
=\frac1{\hbar^3}
e^{i\frac12{\bf  K}.{\bf  X}}
\delta({\bf p}+{\bf p}'-\hbar{\bf  K})
 e^{\frac{-i{\bf p}'.{\bf  X}}{\hbar}}
\delta_{\alpha\beta}^{1\mu\bar1\bar\mu'}({\bf p}, -{\bf p}')
\label{5.19}
\end{equation}
\begin{equation}
\gamma_{\alpha\beta}^{\bar1\mu1\mu'00}({\bf p},{\bf p}')
=\frac1{\hbar^3}
e^{i\frac12{\bf  K}.{\bf  X}}
\delta({\bf p}+{\bf p}'+\hbar{\bf  K})
e^{\frac{i{\bf p}.{\bf  X}}{\hbar}}
\delta_{\alpha\beta}^{1\mu\bar1\bar\mu'}(-{\bf p}', {\bf p})
\label{5.21}
\end{equation}
The gamma's corresponding to the reference one particle distribution function is thus explicitly
known. They allow to express the reference functions
in terms of the correlation functions (\ref{4.26}) that do not involve the field.
The two sets of functions provide an equivalent description.
The gamma's corresponding to the proposal (\ref{4.37a}) can be
found in appendix B.
The connection between the $\alpha$'s (\ref{4.21}) and the $\delta$'s (\ref{5.13}) can be made explicit.
For $\nu\not=s$, we have the relevant values:
\begin{eqnarray}
&&
 \alpha_{\nu+\,\mu \,\mu'}^{11}({\bf p},{\bf k}) 
=
q_e c\frac1{\hbar^3}\sum_{i=1,3}
({\bf e}_\nu)_i
\delta_{1i}^{1\mu1\mu'}({\bf p}-\frac12\hbar {\bf k}, {\bf p}+\frac12\hbar {\bf k})
\nonumber\\ &&
\alpha_{\nu+\,\mu \,\mu'}^{\bar1\bar1}({\bf p},{\bf k})
=
-q_e c\frac1{\hbar^3}\sum_{i=1,3}
({\bf e}_\nu)_i
 \delta_{1i}^{\bar1\bar\mu \bar1\bar\mu'}(-{\bf p}-\frac12\hbar{\bf k},
-{\bf p}+\frac12\hbar {\bf k})
\nonumber\\ &&
\alpha_{\nu+\,\mu \,\mu'}^{1\bar1}({\bf p},{\bf k})
=
q_e c\frac1{\hbar^3}\sum_{i=1,3}
({\bf e}_\nu)_i
\delta_{1i}^{1\mu\,\bar1\bar\mu'}
({\bf p}-\frac12\hbar {\bf k}, {\bf p}+\frac12\hbar{\bf k}) 
\nonumber\\ &&
\alpha_{\nu+\,\mu \,\mu'}^{\bar1 1}({\bf p},{\bf k})
= 
q_e c\frac1{\hbar^3}\sum_{i=1,3}
({\bf e}_\nu)_i
\delta_{1i}^{\bar1\bar\mu \,1\mu'}
({\bf p}-\frac12\hbar{\bf k}, {\bf p}+\frac12\hbar {\bf k}) 
\label{5.39}
\end{eqnarray}

The next section is devoted to the time evolution of the correlation 
functions.

\section{The evolution equations for the correlation functions}  
\setcounter{equation}{0}

The evolution equations for the correlation functions (\ref{4.26}) can be obtained
from the Liouville-von Neumann equation (\ref{4.22}), using the cyclic invariance of
the trace.
The structure of the equation is the following:
\begin{equation}
\frac{\partial}{\partial t}  f_{ss'\tilde s\tilde s'nn'}
=\sum_{\underline s\,\underline s'\underline {\tilde s}\,
\underline {\tilde s}'\underline n\,\underline n'}
 <ss'\tilde s\tilde s'nn'|{\cal L} 
|\underline s\,\underline s'\underline {\tilde s}\,\underline{\tilde s}'
\underline n\,\underline n'>
f_{\underline s\,\underline s'\underline {\tilde s}\,\underline {\tilde s}'
\underline n\,\underline n'}
\label{6.1}
\end{equation}
The evolution operator ${\cal L}$ admits the same decomposition as $L$ (\ref{4.22})
or $H$ (\ref{4.1}).
From the explicit form of the Dirac hamiltonian (\ref{4.2}), written in the 
continuous basis as:
\begin{equation}
H_D=\frac1{\hbar^3}
\sum_{\mu} \int d^3p E_p\left( c^+_{\mu}({\bf p})c_{\mu}({\bf p})
+ b^+_{\mu}({\bf p})b_{\mu}({\bf p})\right)
\label{6.2}
\end{equation}
it can easily seen that the  contribution of ${\cal L}_D$ is diagonal and its 
only non-vanishing elements are:
\begin{eqnarray}
&&<ss'\tilde s\tilde s'nn'|{\cal L}_D |ss'\tilde s\tilde s'nn' > 
=\frac1{i\hbar}\left(\sum_{i=1,s}E_{p_i}+\sum_{i=1,\tilde s}E_{\tilde p_i}  
-\sum_{i=1,s'}E_{p'_i}-\sum_{i=1,\tilde s'}E_{\tilde p'_i}
\right)
\nonumber\\ &&
\label{6.3}
\end{eqnarray}
The contribution of ${\cal L}_R$ is also purely diagonal and corresponds to the
first term of (\ref{3.33}):
\begin{equation}
<ss'\tilde s\tilde s'nn'|{\cal L}_D |ss'\tilde s\tilde s'nn' > 
=-i\left(\sum_{i=1, n} \omega_{ {\bf k}_i} -
\sum_{i=1,n'} \omega_{ {\bf k}'_i}
\right)
\label{6.4}
\end{equation}
The contributions due to ${\cal L}_I$ are not diagonal. 
The non-vanishing contributions arising from the external sources 
can be inferred from (\ref{3.33}):
\begin{eqnarray}
&&<ss'\tilde s\tilde s'nn'|{\cal L}_{Ie} |ss'\tilde s\tilde s'\,n-1\,n' >
=
i\sum_{i=1, n}
\sqrt{\frac{1}{2\varepsilon_0\hbar(2\pi)^3 \omega_{{\bf k}_i}}}
s(\nu_i) j_{e\nu_i+}({\bf k}_i,t) 
\nonumber\\ &&\times
\left(\prod_{j=i,n-1}S({\bf k}_{j+1},\nu_{j+1};{\bf k}_j,\nu_j)
\right)
\label{6.5}
\end{eqnarray}
\begin{eqnarray}
&&<ss'\tilde s\tilde s'nn'|{\cal L}_{Ie} |ss'\tilde s\tilde s'n\,n'-1 >
=-i
\sum_{i=1, n'}
\sqrt{\frac{1}{2\varepsilon_0\hbar(2\pi)^3 \omega_{{\bf k}_i}}}
\nonumber\\ &&  \times 
s(\nu_i) j_{e\nu_i-}(-{\bf k}'_i,t) 
\left(\prod_{j=i,n'-1}S({\bf k}'_{j+1},\nu'_{j+1};{\bf k}'_j,\nu'_j)
\right)
\label{6.6}
\end{eqnarray}
The contributions due to ${\cal L}_{ID}$ are more complicated since they involve
changes both in the e.m. field and the Dirac field variables.
In order to visualize those contributions, it is useful to introduce a 
diagrammatic representation for them.
Let us represent the correlation function $f_{ss'\tilde s\tilde s'nn'}$
by $s$ left orientated straight lines, bearing a plus sign, 
$s'$ right orientated straight lines, bearing a plus sign,
$\tilde s$ left orientated straight lines, bearing a minus sign, 
$\tilde s'$ right orientated straight lines, bearing a minus sign, 
$n$ left orientated wavy lines and 
$n'$ right orientated wavy lines.
The different elements of ${\cal L}_{ID}$ can then be represented by vertices 
inducing changes in the corresponding lines.
These vertices involve at most one photon.
Each occupation  numbers $s$, $s'$, $\tilde s$ and $\tilde s'$ 
can change by at most one unit and the modifications are restricted to 
a couple of them, leaving unchanged the parity of 
$s+s'+\tilde s+\tilde s'$.
Moreover, the modifications leave invariant $S$, given by the expression 
$S=s-s'-\tilde s+\tilde s'$. 
$S$ corresponds to the difference of the charge number $N_L=s-\tilde s$ 
of the state at the left of the density matrix with the 
charge number $N_R=s-\tilde s$  of the state at the right.
Each charge number is no longer individually conserved, 
as in the  Hilbert space
or in the density operator description, but only their difference   is
invariant in the correlation functions formalism.
Each value of  $S$ defines therefore a sector and the evolution is completely 
decoupled between those sectors.
The equation (\ref{6.1}) can therefore be analysed in each subdynamics defined
by those sectors.

Those equations (\ref{6.1}), with the explicit values of the vertices (given in appendix C), 
together with the expression of
the mean value of the physical observables  (\ref{4.42}), form the basis for a further analysis of the
evolution of the coupled Dirac  and electromagnetic fields.

\section{The single subdynamics approach}
\setcounter{equation}{0}

\subsection{The extension of dynamics and the choice of the subdynamics} 

We intend to describe a Dirac field in interaction with the electromagnetic field.
The structure of our description is the same as in the classical case \cite{dH05} 
and we proceed in a very similar way.
From our hierarchy equations of motions for the correlations functions (\ref{4.26}), 
we derive a description in which we can distinguish whether a field is incident, emitted or dressing
while including the original description. 
Of course, such an extension is largely arbitrary:
the criterion may indeed be defined in various ways.
Since we do not depart from equivalence conditions with the original formulation,
all choices are equivalent and are simply determined by commodity reasons for the physical situation
in mind.
In the classical case,  the distinction between incident, emitted or dressing field could easily be based
in reference with each classical charge \cite{dH05}.
Considering the quantal undistinguishability, leading to relations induced by the quantum symmetry
\cite{RB75}, it is convenient to define the criterion with respect to the whole set of Dirac field. 
Each photon is therefore labeled by a supplementary index $\eta$ that describes its status,
either as incident photon $\eta=i$, outgoing photon $\eta=o$ or dressing photon (self-interaction)
$\eta=s$\footnote{The procedure is reminescent of the in and out states usual in $S$-matrix theory
but no aymptotic limit is required here for the validity of the approach, hence the
consideration of unstable states does not prevent any peculiar difficulty}.   
The self interaction includes here transfert of photons from a charge to ``another" one. Since our main
interest is to investigate the presence of infinities due to the self interaction, the above criterion is
suitable for our purpose.

The equation of motions for the new set of correlation functions 
\begin{eqnarray}
&&
\tilde f_{ss'\tilde s\tilde s'nn'}
^{\mu_1\dots\mu_s\mu'_{s'}\dots \mu'_1
\tilde \mu_1\dots\tilde \mu_{\tilde s}\tilde \mu'_{\tilde s'}\dots\tilde  \mu'_1}
({\bf p}_1 ;{\bf p}_2 \dots{\bf p}_s;
{\bf p}'_1 ;{\bf p}'_2 \dots{\bf p}'_s;
{\tilde{{\bf p}_1}} ;{\tilde{{\bf p}}}_2 \dots{\tilde{{\bf p}}}_{\tilde{s}};
\nonumber\\&& 
{\tilde{{\bf p}}}'_1 ;{\tilde{{\bf p}}}'_2 \dots{\tilde{{\bf p}}}'_{\tilde{s}};
{\bf k}_1 \nu_1\eta_1;{\bf k}_2 \nu_2\eta_2;\dots;
{\bf k}_n \nu_n\eta_n;
{\bf k}'_1 \nu'_1\eta'_1;{\bf k}'_2 \nu'_2\eta'_2;\dots;
{\bf k}'_{n'} \nu'_{n'}\eta'_{n'};t)
\nonumber\\&& 
\label{2.a1}
\end{eqnarray}
is constructed in the usual way: the non-vanishing matrix elements are numerically the same as in
$f_{ss'\tilde s\tilde s'nn'} ^{\mu_1\dots\mu_s\mu'_{s'}\dots \mu'_1
\tilde \mu_1\dots\tilde \mu_{\tilde s}\tilde \mu'_{\tilde s'}\dots\tilde  \mu'_1}$ but certain matrix
elements vanish and reflect the nature of the photon absorbed or emitted: for instance, an incident
photon is never emitted by the interaction and a outgoing photon  cannot be absorbed.
The constitutive equations are similar to the classical case:
\begin{eqnarray}
&&
f_{ss'\tilde s\tilde s'nn'} ^{\mu_1\dots\mu_s\mu'_{s'}\dots \mu'_1
\tilde \mu_1\dots\tilde \mu_{\tilde s}\tilde \mu'_{\tilde s'}\dots\tilde  \mu'_1}
({\bf p}_1 ;{\bf p}_2 \dots{\bf p}_s;
{\bf p}'_1 ;{\bf p}'_2 \dots{\bf p}'_s;
{\tilde{{\bf p}_1}} ;{\tilde{{\bf p}}}_2 \dots{\tilde{{\bf p}}}_{\tilde{s}};
\nonumber\\&&
{\tilde{{\bf p}}}'_1 ;{\tilde{{\bf p}}}'_2 \dots{\tilde{{\bf p}}}'_{\tilde{s}};
 {\bf k}_1 \nu_1;{\bf k}_2 \nu_2;\dots;
{\bf k}_n \nu_n;
{\bf k}'_1 \nu'_1;{\bf k}'_2 \nu'_2;\dots;
{\bf k}'_{n'} \nu'_{n'};t)
\nonumber\\
&&=\sum'_{\{\eta\}}\tilde f_{ss'\tilde s\tilde s'nn'}
^{\mu_1\dots\mu_s\mu'_{s'}\dots \mu'_1
\tilde \mu_1\dots\tilde \mu_{\tilde s}\tilde \mu'_{\tilde s'}\dots\tilde  \mu'_1}
({\bf p}_1 ;{\bf p}_2 \dots{\bf p}_s;
{\bf p}'_1 ;{\bf p}'_2 \dots{\bf p}'_s;
{\tilde{{\bf p}_1}} ;{\tilde{{\bf p}}}_2 \dots{\tilde{{\bf p}}}_{\tilde{s}};
\nonumber\\&&
{\tilde{{\bf p}}}'_1 ;{\tilde{{\bf p}}}'_2 \dots{\tilde{{\bf p}}}'_{\tilde{s}};
{\bf k}_1 \nu_1\eta_1;{\bf k}_2 \nu_2\eta_2;\dots;
{\bf k}_n \nu_n\eta_n;
{\bf k}'_1 \nu'_1\eta'_1;{\bf k}'_2 \nu'_2\eta'_2;\dots;
{\bf k}'_{n'} \nu'_{n'}\eta'_{n'};t)
\nonumber\\&&
\label{2.a2}
\end{eqnarray}
where the prime on the sum means that only the values of $\eta$ =$i$ or $o$ have to be
included.
Providing that the $\tilde f_{ss'\tilde s\tilde s'nn'}
^{\mu_1\dots\mu_s\mu'_{s'}\dots \mu'_1
\tilde \mu_1\dots\tilde \mu_{\tilde s}\tilde \mu'_{\tilde s'}\dots\tilde  \mu'_1}$ functions keep 
their value upon substitution of the value $o$ of any $\eta$ by the value $s$ (and
{\it vice-versa}), the set of equations for $\tilde f$ provide for $f$, via the constitutive relations, the
original set of equations. Of course, the new set is richer since the link between the initial conditions is
not yet determined. 
In the classical case \cite{dH05}, that link has been  provided by the condition of belonging to the single
subdynamics and that condition holds also here.

We have to define now our subdynamics and define the ``vacuum"
 of correlations.
As in previous papers, the vacuum is defined by all the ``states" that do not include  a self-field:
a $\tilde f$ function belongs to the vacuum if all the indices $\eta$ are of the $i$ or $o$ type.
The set of equations is not close {\it a priori} for the corresponding elements, if the criterion of
belonging to the vacuum bears only to the self-field. 
Indeed, in the kinetic equation, the one particle Wigner distribution function is driven by the value of
all possible Wigner distribution functions involving an arbitrary large number of Dirac particles.
Indeed, all  those functions are considered as independent and we deal with the equivalent of a
BBGKY hierachy.

Other choices  may be more appropriate acccording to the physical situation in mind, e.g. if the aim is
not the description of an assembly of electrons-positrons, with an arbitrary initial condition, but the
description of only a few specified Dirac particles, for which the initial condition for the Wigner
distribution functions is known.   
Through another choice of the vacuum,  
a solution of the hierarchy equation can be obtained so that it is driven by the elements describing 
those specified Dirac particles, while the other elements, also present, are determined as
time-independent functionals of the relevant one's. 
They describe the ``cloud" of virtual particles, determined by the ``real" ones.
For the elements of the subdynamics approach
that will be considered here,  the specific choice is in fact irrelevant.

Let us present our scheme for the determination of the closed equations for the Wigner distribution
functions.
It will then be made explicit.

A new generator of evolution $\Theta$ determines the time dependence of the elements of the vacuum
while the correlated elements are determined by a creation operator $C$ acting on the elements of the
vacuum. 
The computational rules of these operators are well known and only the relevant part for our
purpose wil be recalled later on.

Since the emitted photons cannot interact with the electron (by construction), the kinetic equation for  
 $\tilde f_{ss'\tilde s\tilde s'00}
^{\mu_1\dots\mu_s\mu'_{s'}\dots \mu'_1
\tilde \mu_1\dots\tilde \mu_{\tilde s}\tilde \mu'_{\tilde s'}\dots\tilde  \mu'_1}$ 
involve only elements $\tilde f_{ss'\tilde s\tilde s'nn'}
^{\mu_1\dots\mu_s\mu'_{s'}\dots \mu'_1
\tilde \mu_1\dots\tilde \mu_{\tilde s}\tilde \mu'_{\tilde s'}\dots\tilde  \mu'_1}$ 
(with possible other values for the indices) where the
$\eta$ indices are of the incident $i$ type. 
We label by an index $inc$ such elements and 
since by definition, the field described in 
$\tilde f_{ss'\tilde s\tilde s'nn'}
^{\mu_1\dots\mu_s\mu'_{s'}\dots \mu'_1
\tilde \mu_1\dots\tilde \mu_{\tilde s}\tilde \mu'_{\tilde s'}\dots\tilde  \mu'_1}$ 
has never been in interaction with the Dirac particles, we can factorize the particle and field
dependence in these functions:
\begin{equation}
\tilde f_{ss'\tilde s\tilde s'nn'}
^{\mu_1\dots\mu_s\mu'_{s'}\dots \mu'_1
\tilde \mu_1\dots\tilde \mu_{\tilde s}\tilde \mu'_{\tilde s'}\dots\tilde  \mu'_1}(t)=
\tilde f_{ss'\tilde s\tilde s'00}
^{\mu_1\dots\mu_s\mu'_{s'}\dots \mu'_1
\tilde \mu_1\dots\tilde \mu_{\tilde s}\tilde \mu'_{\tilde s'}\dots\tilde  \mu'_1}(t)
\tilde f_{nn'inc}(t)
\label{2.a3}
\end{equation}
$\tilde f_{nn'inc}(t)$ describes the incident field and can be of  quasi classical or arbitrary, as
considered in previous sections.
Its time evolution is independent of the evolution of the Dirac field and corresponds to the
propagation of a free field.  Of course, the resulting field depends, via the constitutive relation
(\ref{2.a2}), also of the emitted field and of their interference.
Keeping that factorization (\ref{2.a3}) in mind, the kinetic equations describes now a closed
equation for the Dirac correlation functions, in canonical momentum variables.
The relation (\ref{4.40}) and the similar ones for more than one particle allows to transform these
kinetic equations into kinetic equations for the reference distribution functions, with the help of
relations (\ref{5.15}-\ref{5.21}).

From (\ref{4.39}) and (\ref{4.41}), we know the link between the Wigner distribution functions (in
variables position and mechanical momentum) in terms of the correlation functions in
the other set of particles variables and the electromagnetic field.
Through the creation operator, the correlation functions involving the self-field
are determined by the vacuum elements only.
Using the factorisation (\ref{2.a3}),  
the Wigner distribution functions can finally be expressed in terms of the reference distribution
functions.
Therefore,  the closed evolution equation in one set of variables (the reference functions) can be
transferred into a closed evolution equation for the Wigner functions in the other set of variables.
If the gamma's (\ref{4.39}) are known, the kinetic equation for the Wigner functions  is completely
determined (and is dependent of the free incident field that has to be considered as given) and it is
possible to check if divergences are present.

\subsection{Explicit link between  the mechanical and canonical momentum distributions}

We make explicit the link introduced in the previous subsection between the Wigner distribution
function and the correlation functions for the Dirac particles.

From (\ref{4.39}),
the doubly Fourier transform one article Wigner distribution functions 
$f_{\alpha\beta}^W(-{\bf  K},-{\bf  X})$ and the correlation functions are connected by:
\begin{eqnarray}
&&
f_{\alpha\beta}^W(-{\bf  K},-{\bf  X}) =
\hbar^3 
\sum_{n,n'}
\int d^3k_1\,\dots\int d^3k_n\,
\int d^3k'_1\,\dots\int d^3k'_{n'}\,
\sum_{\nu_1,\dots\nu_n}\sum_{\nu'_1,\dots\nu'_{n'}}
\nonumber\\&&\times
\left( 
\sum_{\mu\mu'}\int d^3p\,\int d^3p'\,
\gamma_{\alpha\beta}^{1\mu1\mu' nn'}({\bf p},{\bf p}',k_1,\dots k_n;k'_1\dots k'_{n'}) 
 f_{1100 nn'}^{\mu  \mu'}
\right.\nonumber\\ &&\left.
+\sum_{\tilde\mu\tilde\mu'}\int d^3\tilde p\,\int d^3\tilde p'\, 
\gamma_{\alpha\beta}^{\bar1\tilde\mu\bar1\tilde\mu' nn'}(\tilde{{\bf p}},\tilde{{\bf p}}',k_1,\dots
k_n;k'_1\dots k'_{n'}) 
f_{0011nn'}^{\tilde \mu\tilde  \mu'}
\right.\nonumber\\ &&\left.
+\sum_{\mu'\tilde\mu'}\int d^3 p'\,\int d^3\tilde p'\, 
\gamma_{\alpha\beta}^{1\mu'\bar1\tilde\mu' nn'}({\bf p}',\tilde{{\bf p}}',k_1,\dots k_n;k'_1\dots
k'_{n'}) 
f_{0101nn'}^{\mu' \tilde\mu'}
\right.\nonumber\\ &&\left.
+\sum_{\mu\tilde\mu}\int d^3 p\,\int d^3\tilde p\, 
\gamma_{\alpha\beta}^{\bar1\tilde\mu 1\mu  nn'}({\bf p},\tilde{{\bf p}},k_1,\dots
k_n;k'_1\dots k'_{n'})
f_{1010nn'}^{\mu \tilde\mu}
+\dots
\right) 
\label{3.a1}
\end{eqnarray}
The  $\gamma$'s are defined through (\ref{4.39}) in Lorentz gauge. The $f$'s functions are defined in
(\ref{4.26}) with a convention that the arguments are those of (\ref{2.a1}) when they are not
explicitely written.

We use the constitutive relations (\ref{2.a2}) to obtain for instance:
\begin{eqnarray}
&&
f_{1100nn'}^{\mu  \mu'}({\bf p},{\bf p}';{\bf k}_1 \nu_1;{\bf k}_2 \nu_2;\dots;
{\bf k}_n \nu_n;
{\bf k}'_1 \nu'_1;{\bf k}'_2 \nu'_2;\dots;
{\bf k}'_{n'} \nu'_{n'};t)
\nonumber\\&&=
\sum'_{\{\eta\}}
\tilde f_{1100nn'}^{\mu  \mu'}({\bf p},{\bf p}';{\bf k}_1 \nu_1\eta_1;{\bf k}_2 \nu_2\eta_2;\dots;
{\bf k}_n \nu_n\eta_n;
\nonumber\\&&
{\bf k}'_1 \nu'_1\eta'_1;{\bf k}'_2 \nu'_2\eta'_2;\dots;
{\bf k}'_{n'} \nu'_{n'}\eta'_{n'};t)
\label{3.a2}
\end{eqnarray}
Considering the invariance of the distribution and of the relation (\ref{3.1}) 
with respect to the permutation of photons, in each term of the sum $\sum'_{\{\eta\}}$, 
we can place all photons of the same type ($i$) in
the last indices, with a counting factor. 
Obvious conventions are in use for  values of  $m$ and $m'$ near their limits.
\begin{eqnarray}
&&
f_{1100nn'}^{\mu  \mu'}({\bf p},{\bf p}';{\bf k}_1 \nu_1;{\bf k}_2 \nu_2;\dots;
{\bf k}_n \nu_n;
{\bf k}'_1 \nu'_1;{\bf k}'_2 \nu'_2;\dots;
{\bf k}'_{n'} \nu'_{n'};t)
\nonumber\\&&
=\sum_{m,m'}\frac{n!}{m!(n-m)!}\frac{n'!}{m'!(n'-m')!}
\nonumber\\&&\times
\tilde f_{1100nn'}^{\mu  \mu'}({\bf p},{\bf p}';{\bf k}_1 \nu_1o;\dots{\bf k}_m \nu_no;{\bf k}_{m+1}\nu_{m+1}i\dots;
{\bf k}_n \nu_ni;
\nonumber\\&&\qquad
{\bf k}'_1 \nu'_1o;\dots{\bf k}'_{m'} \nu'_{m'}o;{\bf k}'_{m'+1} \nu'_{m'+1}i\dots;
{\bf k}'_{n'} \nu'_{n'}i;t)
\label{3.a3}
\end{eqnarray}
Considering the equivalence conditions, the distribution functions with an index $o$ are numerically
equal   with the distribution functions where the index $o$ is replaced by $s$.
That operation has to be considered as realized when (\ref{3.a2}) is reported in (\ref{3.a1}).
We have then the factorization similar to (\ref{2.a3}) between the incident field and the electron with its
dressing photons. The index $i$ is useless for the function $\tilde f_{nn'inc}$ where all photons are of that
type.
\begin{eqnarray}
&&
f_{1100nn'}^{\mu  \mu'}({\bf p},{\bf p}';{\bf k}_1 \nu_1;{\bf k}_2 \nu_2;\dots;
{\bf k}_n \nu_n;
{\bf k}'_1 \nu'_1;{\bf k}'_2 \nu'_2;\dots;
{\bf k}'_{n'} \nu'_{n'};t)
\nonumber\\&&
=\sum_{m,m'}\frac{n!}{m!(n-m)!}\frac{n'!}{m'!(n'-m')!}
\nonumber\\&&\times
\tilde f_{1100mm'}^{\mu  \mu'}({\bf p},{\bf p}';{\bf k}_1 \nu_1s;\dots{\bf k}_m \nu_ms;
{\bf k}'_1 \nu'_1s;\dots{\bf k}'_{m'} \nu'_{m'}s;t)
\nonumber\\&&\times
\tilde f_{(n-m)(n'-m')inc}({\bf k}_{m+1}\nu_{m+1}\dots;
{\bf k}_n \nu_n;
{\bf k}'_{m'+1} \nu'_{m'+1}\dots;
{\bf k}'_{n'} \nu'_{n'};t)
\nonumber\\&&
\label{3.a4}
\end{eqnarray}
In the subdynamics, the correlated components (involving e.g. the self-field) are obtained by the action
of a creation operator $C$ on the elements of the vacuum and we have:
\begin{eqnarray}
&&
\tilde f_{1100mm'}^{\mu  \mu'}({\bf p},{\bf p}';{\bf k}_1 \nu_1s;\dots{\bf k}_m \nu_ms;
{\bf k}'_1 \nu'_1s;\dots{\bf k}'_{m'} \nu'_{m'}s;t)
\nonumber\\&&=\sum_{ss'\tilde s\tilde s'}<1100mm'\mu  \mu'\vert C\vert ss'\tilde s\tilde s'> \tilde f_{ss'\tilde
s\tilde s'}
\label{3.a5}
\end{eqnarray}
where the spin indices have not been written.

Dynamical restrictions are present in the summation $\sum_{ss'\tilde s\tilde s'}$: the parity of the
sum $s+s'+\tilde s+\tilde s'$ and the value of $S=s-s'-\tilde s+\tilde s'$ remain invariant under the
evolution.  We place accordingly a prime on the summation symbol to remember that property.

Therefore, combining the two previous relations, we obtain:
\begin{eqnarray}
&&
f_{1100nn'}^{\mu  \mu'}({\bf p},{\bf p}';{\bf k}_1 \nu_1;{\bf k}_2 \nu_2;\dots;
{\bf k}_n \nu_n;
{\bf k}'_1 \nu'_1;{\bf k}'_2 \nu'_2;\dots;
{\bf k}'_{n'} \nu'_{n'};t)
\nonumber\\&&
=\sum_{m,m'}\frac{n!}{m!(n-m)!}\frac{n'!}{m'!(n'-m')!}
\sum'_{ss'\tilde s\tilde s'}
<1100mm'\mu  \mu'\vert C\vert ss'\tilde s\tilde s'> \tilde f_{ss'\tilde s\tilde s'00}
\nonumber\\&&\times
\tilde f_{(n-m)(n'-m')inc}({\bf k}_{m+1}\nu_{m+1}\dots;
{\bf k}_n \nu_n;
{\bf k}'_{m'+1} \nu'_{m'+1}\dots;
{\bf k}'_{n'} \nu'_{n'};t)
\label{3.a6}
\end{eqnarray}
Inserting that relation into (\ref{3.a1}) leads to:
\begin{eqnarray}
&&
f_{\alpha\beta}^W(-{\bf  K},-{\bf  X}) =
\hbar^3 \sum'_{ss'\tilde s\tilde s'}
\sum_{n,n'}
\int d^3k_1\,\dots\int d^3k_n\,
\int d^3k'_1\,\dots\int d^3k'_{n'}
\nonumber\\&&\times
\sum_{\nu_1,\dots\nu_n}\sum_{\nu'_1,\dots\nu'_{n'}}
\left( 
\sum_{\mu\mu'}\int d^3p\,\int d^3p'\,
\gamma_{\alpha\beta}^{1\mu1\mu' nn'}({\bf p},{\bf p}',k_1,\dots k_n;k'_1\dots k'_{n'}) 
\right.\nonumber\\ &&\left.\times
 \sum_{m,m'}\frac{n!}{m!(n-m)!}\frac{n'!}{m'!(n'-m')!}
<1100mm'\mu  \mu'\vert C\vert ss'\tilde s\tilde s'> \tilde f_{ss'\tilde s\tilde s'00}
\right.
\nonumber\\ &&\left. \times
\tilde f_{(n-m)(n'-m')inc}({\bf k}_{m+1}\nu_{m+1}\dots;
{\bf k}_n \nu_n;
{\bf k}'_{m'+1} \nu'_{m'+1}\dots;
{\bf k}'_{n'} \nu'_{n'};t)
\right.
\nonumber\\ &&\left.
+\sum_{\tilde\mu\tilde\mu'}\int d^3\tilde p\,\int d^3\tilde p'\, 
\gamma_{\alpha\beta}^{\bar1\tilde\mu\bar1\tilde\mu' nn'}(\tilde{{\bf p}},\tilde{{\bf p}}',k_1,\dots
k_n;k'_1\dots k'_{n'}) 
\right.\nonumber\\ &&\left.\times
\sum_{m,m'}\frac{n!}{m!(n-m)!}\frac{n'!}{m'!(n'-m')!}
<0011mm'\tilde\mu\tilde\mu'\vert C\vert ss'\tilde s\tilde s'> \tilde f_{ss'\tilde s\tilde s'00}
\right.\nonumber\\ &&\left.\times
\tilde f_{(n-m)(n'-m')inc}({\bf k}_{m+1}\nu_{m+1}\dots;
{\bf k}_n \nu_n;
{\bf k}'_{m'+1} \nu'_{m'+1}\dots;
{\bf k}'_{n'} \nu'_{n'};t)
\right.\nonumber\\ &&\left.
+\sum_{\mu'\tilde\mu'}\int d^3 p'\,\int d^3\tilde p'\, 
\gamma_{\alpha\beta}^{1\mu'\bar1\tilde\mu' nn'}({\bf p}',\tilde{{\bf p}}',k_1,\dots k_n;k'_1\dots
k'_{n'}) 
\right.\nonumber\\ &&\left.\times
\sum_{m,m'}\frac{n!}{m!(n-m)!}\frac{n'!}{m'!(n'-m')!}
<0101mm'\mu'\tilde\mu'\vert C\vert ss'\tilde s\tilde s'> \tilde f_{ss'\tilde s\tilde s'00}
\right.\nonumber\\&& \left.\times
\tilde f_{(n-m)(n'-m')inc}({\bf k}_{m+1}\nu_{m+1}\dots;
{\bf k}_n \nu_n;
{\bf k}'_{m'+1} \nu'_{m'+1}\dots;
{\bf k}'_{n'} \nu'_{n'};t)
\right.\nonumber\\ &&\left.
+\sum_{\mu\tilde\mu}\int d^3 p\,\int d^3\tilde p\, 
\gamma_{\alpha\beta}^{\bar1\tilde\mu 1\mu  nn'}({\bf p},\tilde{{\bf p}},k_1,\dots
k_n;k'_1\dots k'_{n'})
\right.\nonumber\\ &&\left.\times
\sum_{m,m'}\frac{n!}{m!(n-m)!}\frac{n'!}{m'!(n'-m')!}
<1010mm'\mu\tilde\mu\vert C\vert ss'\tilde s\tilde s'> \tilde f_{ss'\tilde s\tilde s'00}
\right.\nonumber\\ &&\left.\times
\tilde f_{(n-m)(n'-m')inc}({\bf k}_{m+1}\nu_{m+1}\dots;
{\bf k}_n \nu_n;
{\bf k}'_{m'+1} \nu'_{m'+1}\dots;
{\bf k}'_{n'} \nu'_{n'};t)
\right) 
\nonumber\\ &&
+\dots
\label{3.a7}
\end{eqnarray}

\subsection{The dressing operator}

Similar relations can be written for all the joint Wigner distribution functions involving an arbitrary
number of particles.
They can be symbolized with the use of a dressing operator $\chi^{-1}(\tilde f_{inc})$
\cite{dH04b}, depending on the incident field distribution functions, as  ($r$ is the number of particles
described in
$f^W$) (the definition is general although we use it for one particle only):
\begin{equation}
f_r^W=\sum'_{ss'\tilde s\tilde s'}<r\vert \chi^{-1}(\tilde f_{inc})\vert {ss'\tilde s\tilde s'}>
\tilde f_{ss'\tilde s\tilde s'00}
\label{3.a8}
\end{equation}
The elements of $\chi^{-1}(\tilde f_{inc})$ for $r=1$ can be read directly on expression (\ref{3.a7}) for the
choice of variables used in that expression.
Since the correlation functions that do not involve the field can be replaced in terms of the reference
function (see section 6), the dressing operator could also be defined as the link between the reference
and Wigner functions.

That relation (\ref{3.a8}) can be inverted, 
for the relevant values of the indices noted by a $R$, to provide:
\begin{equation}
\tilde f_{ss'\tilde s\tilde s'00 R}=\sum_r<ss'\tilde s\tilde s' R\vert \chi\vert r> f_r^W
\label{3.a9}
\end{equation}
From now on, we will leave out the writing of that index $R$: a restriction on the relevant
values of $ss'\tilde s\tilde s'$ is implicit.

The dressing operators $\chi^{-1}$, $\chi$ considered above present new features with respect to the
dressing operators considered up to now (\cite{dH04b}  and references within).
First of all, it is a time dependent operator since the incident field distribution $\tilde f_{inc}$ evolves
freely with time.
This point should not appear as unexpected.
Indeed, the canonical and mechanical momentums of the electrons-positrons are linked in a naive
approach by transverse components of the potential
vector ${\bf  A}$ that evolves with time and contains obviously also the external field and not only the
self-field.

A second difference is that, in no coupling limit ($q_e\to 0$) of the dressing operator, such as it is
defined in (\ref{3.a8}) and (\ref{3.a9}) with respect to the correlation functins, the dressing
operator does not reduce to the identity.
If we define it with respect to the reference function, it reduces to the identity in the no coupling
limit.

In \cite{dH04b}, in a quantum optics context, we have constructed the
dressing operator by imposing a form of the kinetic operator so that some observables are
naturally defined in the physical representation, among them, the observable linked with the
probability of finding the atom in the upper state. 
Here, the dressing is necessary to deal with the physical variables in place of the auxiliary ones
(canonical momentums) that are required for the quantization of the Dirac field. 
The two points of view are closely connected.
In the present approach, we may begin with an {\it a priori} proposal for the gamma's (\ref{4.39})
or obtain conditions on them by imposing the finiteness of the evolution operator.

\subsection{The evolution equation of the Wigner distribution functions} 

We know that $\tilde f_{ss'\tilde s\tilde s'00}(t)$ satisfies a close kinetic equation in the subdynamics:
\begin{equation}
\partial_t\tilde f_{ss'\tilde s\tilde s'00}(t)=
\sum'_{rr'\tilde r\tilde r'}< ss'\tilde s\tilde s'\vert\Theta(\tilde f_{inc})\vert rr'\tilde r\tilde r'>
\tilde f_{rr'\tilde r\tilde r'00}(t)
\label{4.a1}
\end{equation}
where the $\Theta$ operator depends on time via the incident field.
The dependence of the operator in the spin  variables is not written explicitly.
From the connection (\ref{3.a8}), we have successively
\begin{eqnarray}
&&\partial_t f_r^W(t)=\sum'_{ss'\tilde s\tilde s'}<r\vert \partial_t\chi^{-1}(\tilde f_{inc}(t))\vert
{ss'\tilde s\tilde s'}>
\tilde f_{ss'\tilde s\tilde s'00}(t)
\nonumber\\&&
+\sum'_{ss'\tilde s\tilde s'}<r\vert \chi^{-1}(\tilde f_{inc}(t))\vert {ss'\tilde s\tilde s'}>
\partial_t\tilde f_{ss'\tilde s\tilde s'00}
\nonumber\\&&
=\sum'_{ss'\tilde s\tilde s'}<r\vert\partial_t \chi^{-1}(\tilde f_{inc}(t))\vert {ss'\tilde s\tilde s'}>
\tilde f_{ss'\tilde s\tilde s'00}(t)
\nonumber\\&&
+\sum'_{ss'\tilde s\tilde s'}<r\vert \chi^{-1}(\tilde f_{inc}(t))\vert {ss'\tilde s\tilde s'}>
\nonumber\\&&\times
\sum'_{rr'\tilde r\tilde r'}< ss'\tilde s\tilde s'\vert\Theta(\tilde f_{inc}(t))\vert rr'\tilde r\tilde r'>
\tilde f_{rr'\tilde r\tilde r'00}(t)
\nonumber\\&&
=\sum_{u}
\sum'_{ss'\tilde s\tilde s'}<r\vert\partial_t \chi^{-1}(\tilde f_{inc}(t))\vert {ss'\tilde s\tilde s'}>
<ss'\tilde s\tilde s' \vert \chi(\tilde f_{inc}(t))\vert u> f_u^W(t)
\nonumber\\&&
+\sum_{u}
\sum'_{ss'\tilde s\tilde s'}<r\vert \chi^{-1}(\tilde f_{inc}(t))\vert {ss'\tilde s\tilde s'}>
\sum'_{rr'\tilde r\tilde r'}< ss'\tilde s\tilde s'\vert\Theta(\tilde f_{inc}(t))\vert rr'\tilde r\tilde r'>
\nonumber\\&&\times
< rr'\tilde r\tilde r' \vert \chi(\tilde f_{inc}(t))\vert u> f_u^W(t)
\label{4.a2}
\end{eqnarray}
Therefore,
\begin{equation}
\partial_t f_r^W(t)=\sum_{u}<r\vert\Phi(\tilde f_{inc}(t))\vert u>f_u^W(t)
\label{4.a3}
\end{equation}
with
\begin{eqnarray}
&&<r\vert\Phi(\tilde f_{inc}(t))\vert u>=<r\vert\left(\partial_t \chi^{-1}(\tilde f_{inc}(t))\right)
 \chi(\tilde f_{inc}(t))\vert u>
\nonumber\\&&
+\sum_{u}
<r\vert \chi^{-1}(\tilde f_{inc}(t))\Theta(\tilde f_{inc}(t)) \chi(\tilde f_{inc}(t))\vert u> 
\label{4.a4}
\end{eqnarray}
That set of equations (\ref{4.a3}) is the basis of the dynamical equations of QED in the physical
variables..

As will be established soon, the kinetic operator $\Theta(\tilde f_{inc}(t))$ presents the usual ultraviolet
logarithmic divergence. Our aim in this paper is to examine whether $\Phi(\tilde f_{inc}(t))$
presents the same illness. 
We will not proceed to a systematic analysis of that kinetic operator  $\Phi(\tilde f_{inc}(t))$ but we
focus on the contributions that are divergent in $\Theta(\tilde f_{inc}(t))$.
They appear already at the second order in the charge via the self interaction.
Although our approach is based on an expansion in powers of the charge, we do not really assume the
analycity of the operator. 
We are aware that resummations of class of contributions may be required, as
well known in statistical mechanics in similar context \cite{RB75,dH05} for dealing with non-analycity.
For the first orders we treat here, no problem should arise.

For examining the divergence property, we may restrict ourselves to contributions that do not involve
the incident field.
Indeed, the incident field provides a natural cut-off at high energy in its wave number expansion.
Therefore, we consider the system of equations (\ref{4.a3}) for initial conditions where no incident field is
present.
The influence of the self-field on the spreading of the one particle distribution function can still be
studied under these initial conditions.
Such kind of problem is not usually considered in QED while it is treated as a first application of the
Schr\"odinger equation in non relativistic theory.

\subsection{The zero${}^{th}$ and first order evolution operator}

The zero${}^{th}$ order $<1\vert\Phi^{(0)}\vert 1>$ operator is provided by:
\begin{equation}
<1\vert\Phi^{(0)}\vert 1>=<1\vert\chi^{-1(0)}\Theta^{(0)}\chi^{(0)}\vert 1>
\label{4.a5}
\end{equation}
The free motion operator $\Theta^{(0)}$ is identified with ${\cal L}_D$ given in (6.3). 
It does not involve transitions in the numbers $ss'\tilde s\tilde s$ of the distribution function and is a
diagonal operator in the canonical momentum representation . 

The zero${}^{th}$ order of the operator 
$<1\vert(\chi^{-1})^{(0)}\vert ss'\tilde s\tilde s'00>$ can be read
in (\ref{3.a7}) by considering  the values $n=n'=m=m'=0$.
The relation will be inverted, by computing the inverse of a $16\times 16$ matrix, to provide the
operator $<ss'\tilde s\tilde s'00\vert\chi^{(0)}\vert 1>$ (see later on (\ref{4.a23})).

The first order $<1\vert\Phi^{(1)}\vert 1>$ operator involves the incident field and is not considered
here.
Such contributions are present for describing the absorption of the incident free field but the
contributions of self-energy retained here requires at least a the second order in the charge.
To evaluate the order of the contributions in equations (\ref{4.a4}), we have to take into account that
the creation operators $C$ in (\ref{3.a7}) are at least of order $m+m'$ in the charge.
Since for instance 
$\gamma_{\alpha\beta\mu \,\mu'}^{11mm'}$  involves 
a factor $q_e^{m+m'}$, the first correction to $\chi^{-1}$ is of
the second order and arises from the contributions where $m=1$, $m'=0$ or $m=0$, $m'=1$.
$(\chi^{-1})^{(2)}$ is thus obtained from (\ref{3.a7}) by considering these two contributions while taking the
creation operator at the lowest order.

\subsection{The second order evolution operator}

The second order $<1\vert\Phi^{(2)}\vert 1>$ is thus provided by:
\begin{eqnarray}
&&<1\vert\Phi^{(2)}\vert 1>=<1\vert(\chi^{-1})^{(2)}\Theta^{(0)}\chi^{(0)}\vert 1>
+<1\vert(\chi^{-1})^{(0)}\Theta^{(2)}\chi^{(0)}\vert 1>
\nonumber\\&&
+<1\vert(\chi^{-1})^{(0)}\Theta^{(0)}\chi^{(2)}\vert 1>
\label{4.a7}
\end{eqnarray}
We determine first of all $\chi^{(2)}$ in terms of $(\chi^{-1})^{(2)}$.
The relation $\chi\chi^{-1}=I$ provides the condition
$\chi^{(2)}(\chi^{-1})^{(0)}+\chi^{(0)}(\chi^{-1})^{(2)}=0$, hence we have
\begin{equation}
\chi^{(2)}=-\chi^{(0)}(\chi^{-1})^{(2)}\chi^{(0)}
\label{4.a8}
\end{equation}
The kinetic operator $\Phi^{(2)}$ can thus be written as:
\begin{eqnarray}
&&\Phi^{(2)}=(\chi^{-1})^{(0)}\Theta^{(2)}\chi^{(0)}+(\chi^{-1})^{(2)}\Theta^{(0)}\chi^{(0)}
-(\chi^{-1})^{(0)}\Theta^{(0)}\chi^{(0)}(\chi^{-1})^{(2)}\chi^{(0)}
\nonumber\\&&
\label{4.a9}
\end{eqnarray}
In order to analyse the possible singularities of $\Phi^{(2)}$, we can consider the operator
$\chi^{(0)}\Phi^{(2)}(\chi^{-1})^{(0)}$ for which we have:
\begin{equation}
\chi^{(0)}\Phi^{(2)}(\chi^{-1})^{(0)}=\Theta^{(2)}+[\chi^{(0)} (\chi^{-1})^{(2)},\Theta^{(0)}]
\label{4.a10}
\end{equation}
The operator $\Theta^{(2)}$ is an operator in variables position and canonical momentum.
From the subdynamics, we know that ($P$ and $Q=1-P$ are the projectors respectively on the states of
the vacuum and of the correlations, the usual $L$ is $i\hbar {\cal L}$, with a similar link for 
$P\Theta^{(2)}P$ with the usual
collision operator $\psi^{(2)}$ \cite{PGHR73}
($P\Theta^{(2)}P=\frac1{i\hbar}\psi^{(2)}$) due to the lack of factors in the definition (\ref{4.a1}))
\begin{equation}
i\hbar P\Theta^{(2)}P=i\hbar P{\cal L}'QC^{(1)}P
\label{4.a11}
\end{equation}
The explicit computation of a relevant element of $\Theta^{(2)}$ can be found in appendix D.
It will be further analysed later on.
It contains the usual logarithmic divergence for its part diagonal in variables ${\bf  K}$ and
${\bf  P}$.
That divergence has to be compensated by the commutator in the
second term of (\ref{4.a10}) to provide a finite result.
That compensation can be studied as well on (\ref{4.a7}) as on (\ref{4.a10}).
The advantage of (\ref{4.a10}) is that we work in the original variables and we can study separately the
consequences of the dressing.

\subsection{Expliciting the zero${}^{th}$  order dressing operator $(\chi^{-1})^{(0)}$}

We focus now on the second term in (\ref{4.a10}) and determine first of all the $(\chi^{-1})^{(0)}$, $\chi^{(0)}$
operators.
If we call $f_{\alpha\beta}^{W(0)}(-{\bf  K},-{\bf  X})(\equiv f_{\alpha\beta}^{(0)}(-{\bf  K},-{\bf  X}))$
the links hand side of (\ref{3.a7}) when no incident field is present and the $(\chi^{-1})$ operator is
replaced by
$(\chi^{-1})^{(0)}$, we have:
\begin{eqnarray}
&&
f_{\alpha\beta}^{W(0)}(-{\bf  K},-{\bf  X}) =
\hbar^3 \sum_{\mu\mu'}\int d^3p\,\int d^3p'\,
\gamma_{\alpha\beta}^{1\mu 1\mu'00}({\bf p},{\bf p}') 
 \tilde f_{110000}^{\mu\mu'}
\nonumber\\&&
+\hbar^3\sum_{\tilde\mu\tilde\mu'}\int d^3\tilde p\,\int d^3\tilde p'\, 
\gamma_{\alpha\beta}^{\bar1\tilde\mu \bar1\tilde\mu'00}(\tilde{{\bf p}},\tilde{{\bf p}}') 
 \tilde f_{001100}^{\tilde\mu \,\tilde\mu'}
\nonumber\\&&
+\hbar^3\sum_{\mu'\tilde\mu'}\int d^3 p'\,\int d^3\tilde p'\, 
\gamma_{\alpha\beta}^{1\mu '\bar1\tilde\mu'00}({\bf p}',\tilde{{\bf p}}') 
 \tilde f_{010100}^{\mu '\,\tilde\mu'}
\nonumber\\&&
+\hbar^3\sum_{\mu\tilde\mu}\int d^3 p\,\int d^3\tilde p\, 
\gamma_{\alpha\beta}^{\bar1\tilde\mu 1\mu00}({\bf p},\tilde{{\bf p}})
 \tilde f_{101000}^{\mu \,\tilde\mu}
\label{4.a12}
\end{eqnarray}
$f_{\alpha\beta}^{W(0)}(-{\bf  K},-{\bf  X})$ can be identified with the reference function
$f_{\alpha\beta}^{(0)}(-{\bf  K},-{\bf  X}))$. 
We take the value of the relevant $\gamma$'s in the
expressions  (\ref{5.15}), (\ref{5.17}), (\ref{5.19}) and
(\ref{5.21}).
In order to obtain an algebraic relation in place of (\ref{4.a12}), we replace the $\gamma$'s by their
expression and take the Fourier transform with respect to the variable ${\bf  X}$.
All the dependence in ${\bf  X}$ lies in exponential and the  Fourier transform provides Dirac delta
functions.
We multiply both sides by $\frac1{(2\pi)^3} e^{\frac{-i{\bf  P}.{\bf  X}}{\hbar}}$ and integrate over
${\bf  X}$ to obtain:
\begin{eqnarray}
&&
f_{\alpha\beta}^{W(0)}(-{\bf  K},{\bf  P}) =
\hbar^3\sum_{\mu\mu'}\int d^3p\,\int d^3p'\,
\delta({\bf p}-{\bf p}'+\hbar{\bf  K})
\delta({\bf  P}-{\bf p}-\frac{\hbar}{2}{\bf  K})
\nonumber\\&&\times
\delta_{\alpha\beta}^{1\mu'1\mu}({\bf p}', {\bf p})
 \tilde f_{110000}^{\mu\mu'}({\bf p}, {\bf p}')
\nonumber\\&&
-\hbar^3\sum_{\tilde\mu\tilde\mu'}\int d^3\tilde p\,\int d^3\tilde p'\, 
\delta(\tilde{{\bf p}}-\tilde{{\bf p}'}+\hbar{\bf  K})
\delta({\bf  P}+\tilde {{\bf p}}'-\frac{\hbar}{2}{\bf  K})
\nonumber\\&&\times
\delta_{\alpha\beta}^{\bar1\bar{\tilde\mu}\bar1\bar{\tilde\mu}'}(-\tilde{{\bf p}}, -\tilde{{\bf p}}')
 \tilde f_{001100}^{\tilde\mu \,\tilde\mu'}(\tilde{{\bf p}},\tilde{{\bf p}}')
\nonumber\\&&
+\hbar^3\sum_{\mu'\tilde\mu'}\int d^3 p'\,\int d^3\tilde p'\, 
\delta({\bf p}'+\tilde{{\bf p}}'-\hbar{\bf  K})
\delta({\bf  P}+\tilde{{\bf p}}'-\frac{\hbar}{2}{\bf  K})
\nonumber\\&&\times
\delta_{\alpha\beta}^{1\mu'\bar1\bar{\tilde\mu'}}({\bf p}', -\tilde{{\bf p}}')
 \tilde f_{010100}^{\mu '\,\tilde\mu'}({{\bf p}}',\tilde{{\bf p}}')
\nonumber\\&&
+\hbar^3\sum_{\mu\tilde\mu}\int d^3 p\,\int d^3\tilde p\, 
\delta({\bf p}+\tilde{{\bf p}}+\hbar{\bf  K})
\delta({\bf  P}-{\bf p}-\frac{\hbar}{2}{\bf  K})
\nonumber\\&&\times
\delta_{\alpha\beta}^{\bar1\bar{\tilde\mu}1\mu}(-\tilde{{\bf p}}, {\bf p})
 \tilde f_{101000}^{\mu \,\tilde\mu}({{\bf p}},\tilde{{\bf p}})
\label{4.a18}
\end{eqnarray}
Performing the trivial integrations, we obtain
\begin{equation}
f_{\alpha\beta}^{W(0)}({\bf  K},{\bf  P})=\sum_{\xi\mu\xi'\mu'}
\delta_{\alpha\beta}^{\xi'\mu'\xi\mu}({\bf  P}-\frac{\hbar}{2}{\bf  K}, 
{\bf  P}+\frac{\hbar}{2}{\bf  K})
\tilde f^{\xi\mu\xi'\mu'}({\bf  K},{\bf  P})
\label{4.a20}
\end{equation}
where the matrix $\delta_{\alpha\beta}$ is given in (\ref{5.13}) and the non vanishing elements of the 
one particle distribution function $\tilde f^{\xi\mu\xi'\mu'}({\bf  K},{\bf  P})$ are by definition:
\begin{eqnarray}
&&\tilde f^{1\mu1\mu'}({\bf  K},{\bf  P})=\hbar^3\tilde f_{110000}^{\mu\mu'}
({\bf  P}+\frac{\hbar}{2}{\bf  K}, {\bf  P}-\frac{\hbar}{2}{\bf  K})
\nonumber\\
&&\tilde f^{\bar1\mu\bar1\mu'}({\bf  K},{\bf  P})=-\hbar^3\tilde f_{001100}^{\bar{\mu}'\bar{\mu}}
(-{\bf  P}+\frac{\hbar}{2}{\bf  K},-{\bf  P}-\frac{\hbar}{2}{\bf  K})
\nonumber\\
&&\tilde f^{1\mu\bar1\mu'}({\bf  K},{\bf  P})=\hbar^3\tilde f_{101000}^{\mu \bar\mu'}
({\bf  P}+\frac{\hbar}{2}{\bf  K},-{\bf  P}+\frac{\hbar}{2}{\bf  K})
\nonumber\\
&&\tilde f^{\bar1\mu1\mu'}({\bf  K},{\bf  P})=\hbar^3\tilde f_{010100}^{\mu' \bar\mu}
({\bf  P}-\frac{\hbar}{2}{\bf  K},-{\bf  P}-\frac{\hbar}{2}{\bf  K}
\label{4.a21}
\end{eqnarray}
From (\ref{4.a20}), $\tilde f^{1\mu\bar1\mu'}({\bf  K},{\bf  P})$ is indeed associated with 
$\delta_{\alpha\beta}^{\bar1\mu'1\mu}$ and (\ref{4.a18}) implies the association with 
$\tilde f_{101000}^{\mu\bar\mu'}$.
Similarly, $\tilde f^{\bar1\mu1\mu'}({\bf  K},{\bf  P})$ is associated with 
$\delta_{\alpha\beta}^{1\mu'\bar1\mu}$ and (\ref{4.a18}) implies the association with 
$\tilde f_{010100}^{\mu'\bar\mu}$.

This new set of functions \{$\tilde f^{\xi\mu\xi'\mu'}({\bf  K},{\bf  P})$\} can replace everywhere the set
of functions present in the right hand side of the previous expression and the elements of the
dressing operator can be defined as the connection between
$f_{\alpha\beta}^{W}({\bf  K},{\bf  P})$ and $\tilde f^{\xi\mu\xi'\mu'}({\bf  K},{\bf  P})$.
The relation (\ref{4.a20}) allows to identify in these variables the new elements of the $(\chi^{-1})^{(0)}$
dressing operator with those of the $\delta$ matrix:
\begin{equation}
(\chi^{-1})^{(0)\xi\mu\xi'\mu'}_{\alpha\beta}=
\delta_{\alpha\beta}^{\xi'\mu'\xi\mu}({\bf  P}-\frac{\hbar}{2}{\bf  K}, 
{\bf  P}+\frac{\hbar}{2}{\bf  K})
\label{4.a22}
\end{equation}

\subsection{Computing the zero${}^{th}$  order dressing operator $\chi^{(0)}$}

The dressing operator $(\chi^{-1})^{(0)}$, linking $f_{\alpha\beta}^{W(0)}({\bf  K},{\bf  P})$ and
$\tilde f^{\xi\mu\xi'\mu'}({\bf  K},{\bf  P})$ (and thus also the four functions in the right hand side of
(\ref{4.a21}),  is thus diagonal in the variables ${\bf  K}$, ${\bf  P}$
and this property has to be shared by $\chi^{(0)}$.
We can determine the form of $\chi^{(0)}$.
Using the expression of $\delta$, the unitarity of $\tau$, the invariance of the trace, the explicit
form of our basis matrices $B_{\alpha\beta}$ as direct product 
and the properties of the Pauli matrices,
$\chi^{(0)}$ is easily seen to be:
\begin{equation}
\chi^{(0)\xi\mu\xi'\mu'}_{\alpha\beta}=\frac14
\delta_{\alpha\beta}^{\xi\mu\xi'\mu'}({\bf  P}+\frac{\hbar}{2}{\bf  K}, 
{\bf  P}-\frac{\hbar}{2}{\bf  K})
\label{4.a23}
\end{equation}

\subsection{Free motion operator $\Theta^{(0)}$}

Let us provide the one  particle free motion operator $\Theta^{(0)}$ (\ref{4.a5}), (\ref{6.3}) 
in the  representation provided by (\ref{4.a21}) for the  Wigner distribution
function $\tilde f^W({\bf  K},{\bf  P})$.
The diagonal operator $\Theta^{(0)}$ has the following non vanishing elements:
\begin{eqnarray}
&&<1\mu1\mu'{\bf  K}{\bf  P}\vert\Theta^{(0)}\vert 1\mu1\mu'{\bf  K}{\bf P}>
=\frac1{i\hbar}\left(E_{{\bf  P}+\frac{\hbar}{2}{\bf  K}} -E_{{\bf  P}-\frac{\hbar}{2}{\bf  K}}
\right)
\nonumber\\
&&<\bar1\mu\bar1\mu'{\bf  K}{\bf  P}\vert\Theta^{(0)}\vert \bar1\mu\bar1\mu'{\bf  K}{\bf  P}>
=\frac1{i\hbar}\left(E_{-{\bf  P}+\frac{\hbar}{2}{\bf  K}} -E_{-{\bf  P}-\frac{\hbar}{2}{\bf  K}}
\right)
\nonumber\\
&&<1\mu\bar1\mu'{\bf  K}{\bf  P}\vert\Theta^{(0)}\vert 1\mu\bar1\mu'{\bf  K}{\bf  P}>
=\frac1{i\hbar}\left(E_{{\bf  P}+\frac{\hbar}{2}{\bf  K}}
+E_{-{\bf  P}+\frac{\hbar}{2}{\bf  K}}
\right)
\nonumber\\
&&<1\mu\bar1\mu'{\bf  K}{\bf  P}\vert\Theta^{(0)}\vert 1\mu\bar1\mu'{\bf  K}{\bf  P}>
=\frac1{i\hbar}\left(-E_{{\bf  P}-\frac{\hbar}{2}{\bf  K}}
-E_{-{\bf  P}-\frac{\hbar}{2}{\bf  K}}
\right)
\label{4.a25}
\end{eqnarray}

\subsection{Second order diagonal motion operator $\Theta^{(2)}$}
From the appendix D, we have for the four contributions corresponding to the four diagrams
(the lower indices refer to the vertices of appendix C involved):
\begin{eqnarray}
&&<{\bf  K} {\bf  P} |\theta_{M\ref{C.2b}/\ref{C.4}}^{(2)}|{\bf  K} {\bf  P}>  
=i\int d^3k
\frac{q^2_ec^2}{2\varepsilon_0(2\pi)^3 \omega_{{\bf k}}} 
\nonumber\\&&\times
\left\{
\cos^2\frac{ \theta_{|\bf P+\frac12\hbar\bf K|}}2
\cos ^2\frac{ \theta_{|\bf P+\frac12\hbar\bf K-\hbar\bf k|}}2 
+\sin^2\frac{ \theta_{|\bf P+\frac12\hbar\bf K-\hbar\bf k|}}2
\sin^2 \frac{ \theta_{|\bf P+\frac12\hbar\bf K|}}2
\nonumber\right.\\&&\left.\quad
-3\cos^2\frac{ \theta_{|\bf P+\frac12\hbar\bf K|}}2
\sin^2\frac{ \theta_{|\bf P+\frac12\hbar\bf K-\hbar\bf k|}}2
-3\cos ^2\frac{ \theta_{|\bf P+\frac12\hbar\bf K-\hbar\bf k|}}2 
\sin^2 \frac{ \theta_{|\bf P+\frac12\hbar\bf K|}}2
\nonumber\right.\\&&\left.\quad
+\left( 
\frac {\bf P+\frac12\hbar\bf K-\hbar\bf k}
{|\bf P+\frac12\hbar\bf K-\hbar\bf k|}
.\frac{\bf P+\frac12\hbar\bf K}{|\bf P+\frac12\hbar\bf K|} 
\right)
\sin  \theta_{|\bf P+\frac12\hbar\bf K|} 
\sin  \theta_{|\bf P+\frac12\hbar\bf K-\hbar\bf k|} 
\right\}
\nonumber\\ &&\times
\frac1{- E_{|{\bf  P}+\frac12\hbar {\bf  K}-\hbar{\bf k}|}
+ E_{|{\bf  P}+\frac12\hbar {\bf  K}|}-\hbar\omega_k} 
\label{5.224}
\end{eqnarray}
For the second diagram, we have:
\begin{eqnarray}
&&<{\bf  K} {\bf  P} |\theta_{M\ref{C.5b}/\ref{C.1}}^{(2)}|{\bf  K} {\bf  P}>  
=i\int d^3k
\frac{q^2_ec^2}{2\varepsilon_0(2\pi)^3 \omega_{{\bf k}}} 
\nonumber\\&& \times 
\left\{
\cos^2\frac{ \theta_{|\bf P-\frac12\hbar\bf K|}}2
\cos ^2\frac{ \theta_{|\bf P-\frac12\hbar\bf K-\hbar\bf k|}}2 
+\sin^2\frac{ \theta_{|\bf P-\frac12\hbar\bf K-\hbar\bf k|}}2
\sin^2 \frac{ \theta_{|\bf P-\frac12\hbar\bf K|}}2
\nonumber\right.\\&&\left.\quad
-3\cos^2\frac{ \theta_{|\bf P-\frac12\hbar\bf K|}}2
\sin^2\frac{ \theta_{|\bf P-\frac12\hbar\bf K-\hbar\bf k|}}2
-3\cos ^2\frac{ \theta_{|\bf P-\frac12\hbar\bf K-\hbar\bf k|}}2 
\sin^2 \frac{ \theta_{|\bf P-\frac12\hbar\bf K|}}2
\nonumber\right.\\&&\left.\quad
+\left( 
\frac {\bf P-\frac12\hbar\bf K-\hbar\bf k}
{|\bf P-\frac12\hbar\bf K-\hbar\bf k|}
.\frac{\bf P+\frac12\hbar\bf K}{|\bf P+\frac12\hbar\bf K|} 
\right)
\sin  \theta_{|\bf P-\frac12\hbar\bf K|} 
\sin  \theta_{|\bf P-\frac12\hbar\bf K-\hbar\bf k|} 
\right\}
\nonumber\\ &&\times
\frac1{ E_{|{\bf  P}-\frac12\hbar {\bf  K}-\hbar{\bf k}|}
- E_{|{\bf  P}-\frac12\hbar {\bf  K}|}+\hbar\omega_k} 
\label{5.225}
\end{eqnarray}
For the third diagram, we have:
\begin{eqnarray}
&&<{\bf  K} {\bf  P} |\theta_{M\ref{C.20}/\ref{C.28}}^{(2)}|{\bf  K} {\bf  P}>  
=-i
\int d^3k
\frac{q^2_ec^2}{2\varepsilon_0(2\pi)^3 \omega_{{\bf k}}} 
\nonumber\\ &&  \times 
\left\{
3\cos^2\frac{ \theta_{|\bf P+\frac12\hbar\bf K|}}2
\cos ^2\frac{ \theta_{|\bf P+\frac12\hbar\bf K+\hbar\bf k|}}2 
+3\sin^2\frac{ \theta_{|\bf P+\frac12\hbar\bf K+\hbar\bf k|}}2
\sin^2 \frac{ \theta_{|\bf P+\frac12\hbar\bf K|}}2
\nonumber\right.\\&&\left.\quad
-\cos^2\frac{ \theta_{|\bf P+\frac12\hbar\bf K|}}2
\sin^2\frac{ \theta_{|\bf P+\frac12\hbar\bf K+\hbar\bf k|}}2
-\cos ^2\frac{ \theta_{|\bf P+\frac12\hbar\bf K+\hbar\bf k|}}2 
\sin^2 \frac{ \theta_{|\bf P+\frac12\hbar\bf K|}}2
\nonumber\right.\\&&\left.\quad
+\left( 
\frac {\bf P+\frac12\hbar\bf K+\hbar\bf k}
{|\bf P+\frac12\hbar\bf K+\hbar\bf k|}
.\frac{\bf P+\frac12\hbar\bf K}{|\bf P+\frac12\hbar\bf K|} 
\right)
\sin  \theta_{|\bf P+\frac12\hbar\bf K|} 
\sin  \theta_{|\bf P+\frac12\hbar\bf K+\hbar\bf k|} 
\right\}
\nonumber\\ &&\times  
\frac1{ E_{|{\bf  P}+\frac12\hbar {\bf  K}+\hbar{\bf k}|}
+ E_{|{\bf  P}+\frac12\hbar {\bf  K}|}+\hbar\omega_k} 
\label{5.226}
\end{eqnarray}
For the fourth diagram, we have:
\begin{eqnarray}
&&<{\bf  K} {\bf  P} |\theta_{M\ref{C.27}/\ref{C.21}}^{(2)}|{\bf  K} {\bf  P}>  
=-i\int d^3k
\frac{q^2_ec^2}{2\varepsilon_0(2\pi)^3 \omega_{{\bf k}}} 
\nonumber\\ && \times 
\left\{
3\cos^2\frac{ \theta_{|\bf P-\frac12\hbar\bf K|}}2
\cos ^2\frac{ \theta_{|\bf P-\frac12\hbar\bf K+\hbar\bf k|}}2 
+3\sin^2\frac{ \theta_{|\bf P-\frac12\hbar\bf K+\hbar\bf k|}}2
\sin^2 \frac{ \theta_{|\bf P-\frac12\hbar\bf K|}}2
\nonumber\right.\\&&\left.\quad
-\cos^2\frac{ \theta_{|\bf P-\frac12\hbar\bf K|}}2
\sin^2\frac{ \theta_{|\bf P-\frac12\hbar\bf K+\hbar\bf k|}}2
-\cos ^2\frac{ \theta_{|\bf P-\frac12\hbar\bf K+\hbar\bf k|}}2 
\sin^2 \frac{ \theta_{|\bf P-\frac12\hbar\bf K|}}2
\nonumber\right.\\&&\left.\quad
+\left( 
\frac {\bf P-\frac12\hbar\bf K+\hbar\bf k}
{|\bf P-\frac12\hbar\bf K+\hbar\bf k|}
.\frac{\bf P-\frac12\hbar\bf K}{|\bf P-\frac12\hbar\bf K|} 
\right)
\sin  \theta_{|\bf P-\frac12\hbar\bf K|} 
\sin  \theta_{|\bf P-\frac12\hbar\bf K+\hbar\bf k|} 
\right\}\nonumber\\ && \times 
\frac1{- E_{|{\bf  P}-\frac12\hbar {\bf  K}+\hbar{\bf k}|}
- E_{|{\bf  P}-\frac12\hbar {\bf  K}|}-\hbar\omega_k}  
\label{5.227}
\end{eqnarray}
From (\ref{4.17}), we have $\sin{\theta_p}=\frac{pc}{E_p}$.
Let us consider first the problem of convergence of the integral 
over ${\bf k}$.
The problem of convergence could arise for large wave numbers ${\bf k}$.
In that limit, we may use the properties:
\begin{equation}
\lim_{k\to \infty}
\sin ^2\frac{ \theta_{|{\bf  P}-\frac12\hbar{\bf  K}+\hbar{\bf k}|}}2 
=\lim_{k\to \infty}
\cos^2\frac{ \theta_{|{\bf  P}-\frac12\hbar{\bf  K}+\hbar{\bf k}|}}2=\frac12
\label{A.97}
\end{equation}
arising from
\begin{eqnarray}
&&\cos^2\frac{ \theta_p}2=\frac{E_p+mc^2}{2E_p}=\frac12+\frac{mc^2}{2E_p} 
\label{A.100} \\
&&\sin^2\frac{ \theta_p}2=\frac{E_p-mc^2}{2E_p}=\frac12-\frac{mc^2}{2E_p} 
\label{A.101}
\end{eqnarray}
We look in appendix E how the various integrands behave in that limit.
Combining partial results, we obtain
\begin{eqnarray}
&&
\left.I(\omega_k)_{M}\right\vert_{Pr} 
+\left.I(\omega_k)_{M}\right\vert_{F} 
+\left.I(\omega_k)_{M}\right\vert_{Tr} 
\nonumber\\&&=-i\frac{q^2_ec^2}{2\varepsilon_0(2\pi)^3 \omega_{{\bf k}}} 
\frac{m^2c^4}{2\hbar^2\omega^2_{k}}
\left(\frac1{E_{|{\bf  P}+\frac12\hbar {\bf  K}|}}
-\frac1{E_{|{\bf  P}-\frac12\hbar {\bf  K}|}}
\right)  
\label{5.249}
\end{eqnarray}
For $P$ and $K$ in the non relativistic domain, we obtain
\begin{eqnarray}
&&
\left.I(\omega_k)_{M}\right\vert_{PrNR} 
+\left.I(\omega_k)_{M}\right\vert_{FNR} 
+\left.I(\omega_k)_{M}\right\vert_{TrNR} 
\nonumber\\&&=-i3\frac{q^2_ec^2}{2\varepsilon_0(2\pi)^3 \omega_{{\bf k}}}
\frac{1}{ 2\hbar^2\omega_{{\bf k}}^2} 
\frac{ -\hbar{\bf  K}.{\bf  P}}{m}
\nonumber\\&&
=-i\frac{{\bf  K}.{\bf  P}}{m}(-)3
\frac{q^2_e}{4\pi\varepsilon_0}c^2
\frac{1}{ 8\pi^2\hbar\omega_{{\bf k}}^3} 
\label{5.250}
\end{eqnarray}
and the mass correction $\delta m$ is
\begin{equation}
\delta m=\int d^3k\,3
\frac{q^2_e}{4\pi\varepsilon_0}c^2
\frac{1}{ 8\pi^2\hbar\omega_{{\bf k}}^3} 
=
3
\frac{q^2_e}{4\pi\varepsilon_0}
\frac{1}{ 2\pi\hbar c}
\int_0^{\infty}dk\,k^2\frac1{k^3}
\label{5.251}
\end{equation}
%That is the expression found in \cite{WH54} p302.

We examine whether the second term in the right hand side of (\ref{4.a10}) provides similar
divergences for the proposal (\ref{4.37a}).
Details of the computation are provided in the appendixes.
The result can be synthesized as follows:
in the non relativistic domain, small $\bf K$ and  $\bf P$, we obtain obviously a behaviour in
${\bf K}.{\bf P}$ with a coefficient that diverges also logarithmically with a cut-off for large wave
numbers: the same kind of behaviour as for the collision operator $\Theta^{(2)}$. Therefore, we can
ascertain that the transition to a description in terms of the mechanical momentum provides new
terms that are comparable to the divergent part for the description in terms of the canonical
momentum.

\section{Conclusions}  

This paper treats the interacting Dirac and electromagnetic fields as a statistical system in the
sense of a $N$-body system, $N$ arbitrary large. Therefore, we have introduced an analogue of the
reduced distribution functions for the determination of the mean values of the observables.
Our main preoccupation has been the obtention of the distribution functions $f_{\alpha\beta}^W$
describing the variables position and mechanical momentum associated with the electron.
These functions can be interpreted as the physical representation \cite{dH04b} describing the
observables and these functions have to replace the original description in terms of the canonical
momentum. The transition between the two sets of functions is ensured by a dressing operator
previously introduced in the context of quantum optics.

A main requirement on the dressing operator is that it leads for the physical representation to
evolution equations free from the usual divergences.
The usual trick to eliminate of these divergences is through counterterms to provide mass (charge,
\dots) renormalization, assuming an infinite bare mass for the charged particle.
However, we think that the origin of the divergences lies in the use of unphysical variables (canonical
momentums) that have to be considered only as useful intermediate entities.
Progress  along these lines requires progress in the association of operators describing  physical
observables with classical functions of position and mechanical momentum.
The present paper is a prerequisite to tackle those questions still pending.
The two aspects (choice of the association and finiteness) can enlight each other.
No contradiction with the renormalization program is expected: the latter may serve as a useful guide
for the determination of the dressing operator.

The main differences of the present work with usual QED lie in interpretating differently the origin
of the divergences and in providing a more satisfactory conceptual framework. We hope that it will
induce further work on the subject.

\appendix
\section{The charge and current operators}  
\def\theequation{\thesection.\arabic{equation}}

\setcounter{equation}{0}
The explicit computation of the charge and current operators  requires the evaluation of the product
of the spinors
$u_{{\bf p},\mu}$ and $v_{{\bf p},\mu}$ by the adjoints. 
For instance, the coefficient of $c^+_{\mu}({\bf p}) c_{
\mu'}({\bf p}')$  in the expression of $\rho_D$ is given by:
\begin{equation}
\tilde u^*_{{\bf p},\mu} u_{{\bf p}',\mu'}=
\tilde u_{0\mu}  \tilde  T^*({\bf p}) T({\bf p}')  u_{0\mu'} =\left(T^+({\bf p}) T({\bf p}')\right)_{1\mu
1\mu'} 
\label{A.1}
\end{equation} 
The explicit evaluation of all terms is lengthy but straightforward.
In terms of elements of the Pauli matrices and of the angle 
$\varphi_{\hat {{\bf p}},\hat {{\bf p}}'}$ that satisfies the fundamental  trigonometric relation on the 
sphere:
\begin{equation}
\cos \varphi_{ {{\bf p}}, {{\bf p}}'}=
\cos\frac{ \theta_p}2 \cos\frac{ \theta_{p'}}2+
\sin \frac{ \theta_p}2 \sin \frac{ \theta_{p'}}2 
\hat {{\bf p}}.\hat {{\bf p}}',
\label{A.5}
\end{equation}
we get for the coefficients $\alpha$: 
\begin{eqnarray}
&&\alpha_{s\,\mu \,\mu'}^{11}({\bf p},{\bf k})  =- \frac{q_e c}{\hbar^3} \left(
\cos \varphi_{{{\bf p}-\frac12\hbar{\bf k}} , {{\bf p}+\frac12\hbar{\bf k}}}  I_{\mu\mu'} 
\right. \nonumber\\ && \left.   
+ i\frac{\hbar({{\bf p}}\pmb{$\times$}{{\bf k}})}
 {|{\bf p}-\frac12\hbar{\bf k}||{\bf p}+\frac12\hbar{\bf k}|} .
\pmb{$\sigma$}_{\mu\mu'}
\sin\frac{ \theta_{|{\bf p}+\frac12\hbar{\bf k}|}}2
\sin \frac{ \theta_{|{\bf p}-\frac12\hbar{\bf k}|}}2  
\right)
\label{A.19}\\ 
&&\alpha_{s\,\mu \,\mu'}^{\bar1\bar1}({\bf p},{\bf k})  = \frac{q_e c}{\hbar^3} \left(
\cos \varphi_{{{\bf p}-\frac12\hbar{\bf k}} , {{\bf p}+\frac12\hbar{\bf k}}}  I_{\mu\mu'} 
\right. \nonumber\\ && \left.   
+ i\frac{\hbar({{\bf p}}\pmb{$\times$}{{\bf k}})} 
{|{\bf p}-\frac12\hbar{\bf k}||{\bf p}+\frac12\hbar{\bf k}|} .
\pmb{$\sigma$}_{\bar\mu'\bar\mu} 
\sin\frac{ \theta_{|{\bf p}+\frac12\hbar{\bf k}|}}2
\sin \frac{ \theta_{|{\bf p}-\frac12\hbar{\bf k}|}}2 
\right)
\label{A.20}\\ 
&&\alpha_{s\,\mu \,\mu'}^{1\bar1}({\bf p},{\bf k})   =\frac{q_e c}{\hbar^3}\left[
\left(\pmb{$\sigma$}_{\mu\bar\mu'}.
\frac {{\bf p}+\frac12\hbar{\bf k}}{|{\bf p}+\frac12\hbar{\bf k}|}
\right)
\cos\frac{ \theta_{|{\bf p}-\frac12\hbar{\bf k}|}}2
\sin \frac{ \theta_{|{\bf p}+\frac12\hbar{\bf k}|}}2  
\right. \nonumber\\ && \left.\qquad  
-\left( \pmb{$\sigma$}_{\mu\bar\mu'}.
\frac{{\bf p}-\frac12\hbar{\bf k}}{|{\bf p}-\frac12\hbar{\bf k}|} 
\right)
\cos\frac{ \theta_{|{\bf p}+\frac12\hbar{\bf k}|}}2
\sin \frac{ \theta_{|{\bf p}-\frac12\hbar{\bf k}|}}2
\right] 
\label{A.21}\\ 
&&\alpha_{s\,\mu \,\mu'}^{\bar1 1}({\bf p},{\bf k})  =-\alpha_{s\,\bar\mu \,\bar\mu'}^{1\bar1}({\bf
p},{\bf k})   
\label{A.22}
\end{eqnarray}
\begin{eqnarray} 
&&\alpha_{\lambda+\,\mu \,\mu'}^{11}({\bf p},{\bf k})  =\frac{q_e c}{\hbar^3}{\bf  e}_{{\bf k}\lambda}.
\left[
\left( \frac {{\bf p}+\frac12\hbar {\bf k}}{|{\bf p}+\frac12\hbar {\bf k}|}
\cos\frac{ \theta_{|{\bf p}-\frac12\hbar {\bf k}|}}2   
\sin \frac{ \theta_{|{\bf p}+\frac12\hbar {\bf k}|}}2 
\right.\right. \nonumber\\ && \left. \left. +\frac {{\bf p}-\frac12\hbar {\bf k}}{|{\bf p}-\frac12\hbar {\bf k}|}    
\sin \frac{ \theta_{|{\bf p}-\frac12\hbar {\bf k}|}}2 
\cos\frac{ \theta_{{{\bf p}+\frac12\hbar {\bf k}}}}2 
\right)
 I_{\mu\, \mu'}
\right. \nonumber\\ && \left.   
+i\left(
\left( \frac{{\bf p}+\frac12\hbar {\bf k}}{|{\bf p}+\frac12\hbar {\bf k}|} 
\pmb{$\times$}   \pmb{$\sigma$}_{\mu\, \mu'}
\right)
\cos\frac{ \theta_{|{\bf p}-\frac12\hbar {\bf k}|}}2  
\sin \frac{ \theta_{|{{\bf p}+\frac12\hbar {\bf k}}|}}2
\right.\right. \nonumber\\ && \left. \left. + \left( \pmb{$\sigma$}_{\mu\, \mu'}\pmb{$\times$}
\frac {{\bf p}-\frac12\hbar {\bf k}}{|{\bf p}-\frac12\hbar {\bf k}|}   
\right) 
\sin \frac{ \theta_{|{\bf p}-\frac12\hbar {\bf k}|}}2 
\cos\frac{ \theta_{|{{\bf p}+\frac12\hbar {\bf k}}|}}2 
\right)
\right] 
\label{A.23}
\end{eqnarray}
\begin{eqnarray}
&&\alpha_{\lambda+\,\mu \,\mu'}^{\bar1\bar1}({\bf p},{\bf k})  
=- \frac{q_e c}{\hbar^3}{\bf  e}_{{\bf k}\lambda}.
\left[
\left( \frac {{\bf p}+\frac12\hbar {\bf k}}{|{\bf p}+\frac12\hbar {\bf k}|}
\cos\frac{ \theta_{|{\bf p}-\frac12\hbar {\bf k}|}}2   
\sin \frac{ \theta_{|{\bf p}+\frac12\hbar {\bf k}|}}2 
\right.\right. \nonumber\\ && \left. \left. +\frac {{\bf p}-\frac12\hbar {\bf k}}{|{\bf p}-\frac12\hbar {\bf k}|}    
\sin \frac{ \theta_{|{\bf p}-\frac12\hbar {\bf k}|}}2 
\cos\frac{ \theta_{{{\bf p}+\frac12\hbar {\bf k}}}}2 
\right)
 I_{\mu\, \mu'}
\right. \nonumber\\ && \left.   
-i\left(
\left( \frac{{\bf p}+\frac12\hbar {\bf k}}{|{\bf p}+\frac12\hbar {\bf k}|} 
\pmb{$\times$}   \pmb{$\sigma$}_{\bar\mu'\,\bar \mu}
\right)
\cos\frac{ \theta_{|{\bf p}-\frac12\hbar {\bf k}|}}2  
\sin \frac{ \theta_{|{{\bf p}+\frac12\hbar {\bf k}}|}}2
\right.\right. \nonumber\\ && \left. \left. + \left( \pmb{$\sigma$}_{\bar\mu'\, \bar\mu}\pmb{$\times$}
\frac {{\bf p}-\frac12\hbar {\bf k}}{|{\bf p}-\frac12\hbar {\bf k}|}   
\right) 
\sin \frac{ \theta_{|{\bf p}-\frac12\hbar {\bf k}|}}2 
\cos\frac{ \theta_{|{{\bf p}+\frac12\hbar {\bf k}}|}}2 
\right)
\right] 
\label{A.24}
\end{eqnarray}
\begin{eqnarray}
&&\alpha_{\lambda+\,\mu \,\mu'}^{1\bar1}({\bf p},{\bf k})  
=\frac{q_e c}{\hbar^3}{\bf  e}_{{\bf k}\lambda}.
\left[
\pmb{$\sigma$}_{\mu\,\bar\mu'} 
\cos\varphi_{{{\bf p}-\frac12\hbar{\bf k}} , {{\bf p}+\frac12\hbar{\bf k}}} 
\right.\nonumber\\&& \left.  -iI_{\mu\,\bar\mu'}
\left(\frac {\hbar{\bf k}}{|{\bf p}+\frac12\hbar{\bf k}|}\pmb{$\times$}
\frac {{\bf p}}{|{\bf p}-\frac12\hbar{\bf k}|}\right)
\sin \frac{ \theta_{|{\bf p}-\frac12\hbar{\bf k}|}}2 
\sin \frac{ \theta_{|{\bf p}+\frac12\hbar{\bf k}|}}2  
\right.\nonumber\\&& \left.   
- \left[\left(\pmb{$\sigma$}_{\mu\,\bar\mu'}.
\frac {{\bf p}-\frac12\hbar{\bf k}}{|{\bf p}-\frac12\hbar{\bf k}|}
\right)
\frac {{\bf p}+\frac12\hbar{\bf k}}{|{\bf p}+\frac12\hbar{\bf k}|} 
+\left(\pmb{$\sigma$}_{\mu\,\bar\mu'}.
\frac {{\bf p}+\frac12\hbar{\bf k}}{|{\bf p}+\frac12\hbar{\bf k}|} 
\right)
\frac {{\bf p}-\frac12\hbar{\bf k}}{|{\bf p}-\frac12\hbar{\bf k}|}
\right]
\right.\nonumber\\&& \left. \quad \times 
\sin \frac{ \theta_{|{\bf p}-\frac12\hbar{\bf k}|}}2 
\sin \frac{ \theta_{|{\bf p}+\frac12\hbar{\bf k}|}}2 
\right] 
\label{A.25}
\end{eqnarray}

\begin{eqnarray}
&&\alpha_{\lambda+\,\mu \,\mu'}^{\bar1 1}({\bf p},{\bf k})  
= \frac{q_e c}{\hbar^3}{\bf  e}_{{\bf k}\lambda}.
\left[
\pmb{$\sigma$}_{\bar\mu\,\mu'} 
\cos\varphi_{{{\bf p}-\frac12\hbar{\bf k}} , {{\bf p}+\frac12\hbar{\bf k}}}
\right.\nonumber\\&& \left. 
 -iI_{\bar\mu\,\mu'}
\left(\frac {\hbar{\bf k}}{|{\bf p}+\frac12\hbar{\bf k}|}\pmb{$\times$}
\frac {{\bf p}}{|{\bf p}-\frac12\hbar{\bf k}|}\right)  
\sin \frac{ \theta_{|{\bf p}-\frac12\hbar{\bf k}|}}2 
\sin \frac{ \theta_{|{\bf p}+\frac12\hbar{\bf k}|}}2  
\right.\nonumber\\&& \left.  -
\left[\left(\pmb{$\sigma$}_{\bar\mu\,\mu'}.
\frac {{\bf p}-\frac12\hbar{\bf k}}{|{\bf p}-\frac12\hbar{\bf k}|}
\right)
\frac {{\bf p}+\frac12\hbar{\bf k}}{|{\bf p}+\frac12\hbar{\bf k}|} 
+\left(\pmb{$\sigma$}_{\bar\mu\,\mu'}.
\frac {{\bf p}+\frac12\hbar{\bf k}}{|{\bf p}+\frac12\hbar{\bf k}|} 
\right)
\frac {{\bf p}-\frac12\hbar{\bf k}}{|{\bf p}-\frac12\hbar{\bf k}|}
\right]
\right.\nonumber\\&& \left. \times
\sin \frac{ \theta_{|{\bf p}-\frac12\hbar{\bf k}|}}2 
\sin \frac{ \theta_{|{\bf p}+\frac12\hbar{\bf k}|}}2 
\right] 
\label{A.26}
\end{eqnarray}
\begin{equation}
\alpha_{\lambda\,\mu \,\mu'}^{1\bar 1}({\bf p},{\bf k}) =
\alpha_{\lambda\,\mu \,\mu'}^{\bar1 1}({\bf p},{\bf k}) 
\label{A.27}
\end{equation}
The elements of $\alpha$ with a subscript minus can be btained by replacing in the above expressions
${\bf  e}_{{\bf k}\lambda}$ by ${\bf  e}_{{-\bf k}\lambda}$

\section{Elements of the joint characteristic operator}  
\def\theequation{\thesection.\arabic{equation}}
\setcounter{equation}{0}

The computation of the elements 
$\gamma_{\alpha\beta}^{\xi\mu\xi'\mu'10}({\bf p},{\bf p}')$ and 
$\gamma_{\alpha\beta}^{\xi\mu\xi'\mu'01}({\bf p},{\bf p}')$ is straightforward.
We need the first contribution in the expansion of (\ref{5.1}) in powers of $q_e$.
We have obviously:
\begin{eqnarray}
&&\left.e^{-\frac{iq_e}\hbar
{\bf  X}.\bar{{\bf A}}_{\bot l} ({\bf r})}
e^{{\bf  X}.\nabla+i{\bf K}.{\bf r}}
B_{\alpha\beta}
e^{-\frac{iq_e}\hbar
{\bf  X}.\bar{{\bf A}}_{\bot r} ({\bf r})}
\right\vert^{(1)}
\nonumber\\&&=
-\frac{iq_e}\hbar
{\bf  X}.\bar{{\bf A}}_{\bot l}({\bf r})
e^{{\bf  X}.\nabla+i{\bf K}.{\bf r}}
B_{\alpha\beta}
\nonumber\\&&
+ 
e^{{\bf  X}.\nabla+i{\bf K}.{\bf r}}
B_{\alpha\beta}
(-)\frac{iq_e}\hbar
{\bf  X}.\bar{{\bf A}}_{\bot r}r({\bf r})
\label{5.22}
\end{eqnarray}
Introducing the relation (\ref{5.2}), we obtain
\begin{eqnarray}
&&\left.
e^{-\frac{iq_e}\hbar
{\bf  X}.\bar{{\bf A}}_{\bot l} ({\bf r})}
e^{{\bf  X}.\nabla+i{\bf K}.{\bf r}}
B_{\alpha\beta}
e^{-\frac{iq_e}\hbar
{\bf  X}.\bar{{\bf A}}_{\bot r} ({\bf r})}
\right\vert^{(1)}
\nonumber\\&&\\&&=
-\frac{iq_e}\hbar
{\bf  X}.\bar{{\bf A}}_{\bot l} ({\bf r})
e^{i\frac12{\bf K}.{\bf  X}}
e^{i{\bf K}.{\bf r}}e^{{\bf  X}.\nabla}
B_{\alpha\beta}
\nonumber\\&&
-\frac{iq_e}\hbar
B_{\alpha\beta}
{\bf  X}.\bar{{\bf A}}_{\bot r}({\bf r}+{\bf  X})
e^{i\frac12{\bf K}.{\bf  X}}
e^{i{\bf K}.{\bf r}}e^{{\bf  X}.\nabla}
\label{5.23}
\end{eqnarray}
$\gamma_{\alpha\beta}^{1\mu1\mu'10}({\bf p},{\bf p}') $ is given by 
\begin{eqnarray}
&&\gamma_{\alpha\beta}^{1\mu1\mu'10}({\bf p},{\bf p}')
= 
\frac1{\hbar^6(2\pi)^3}
\frac{-iq_e}\hbar\sqrt{\frac{\hbar}{2\varepsilon_0  (2\pi)^3}}
\int d^3r\,
e^{\frac{-i{\bf p}'.{\bf r}}{\hbar} } \tilde u_{0\mu'}  \tau(-{\bf p}') 
\nonumber\\&&\times
e^{i\frac12{\bf K}.{\bf  X}}
e^{i{\bf K}.{\bf r}}
e^{{\bf  X}.\nabla}
\frac1{\sqrt{\omega_{{\bf k}_1}}}{\bf e}_{{\bf k}_1 \nu_1}.
{\bf  X} e^{ i\bf{k_1.r}}
B_{\alpha\beta}
 e^{\frac{i{\bf p}.{\bf r}}{\hbar} }\tau({\bf p}) u_{0\mu}
\label{5.26}
\end{eqnarray}
Acting with the displacement operators $e^{{\bf  X}.\nabla}$, we  obtain:
\begin{eqnarray}
&&\gamma_{\alpha\beta}^{1\mu1\mu'10}({\bf p},{\bf p}')
= 
\frac1{\hbar^6(2\pi)^3}
\frac{-iq_e}\hbar\sqrt{\frac{\hbar}{2\varepsilon_0  (2\pi)^3}}
\int d^3r\,
e^{\frac{-i{\bf p}'.{\bf r}}{\hbar} } \tilde u_{0\mu'}  \tau(-{\bf p}') 
\nonumber\\&&\times
e^{i\frac12{\bf K}.{\bf  X}}
e^{i{\bf K}.{\bf r}}
\frac1{\sqrt{\omega_{{\bf k}_1}}}{\bf e}_{{\bf k}_1 \nu_1}.
{\bf  X} e^{ i{\bf k}_1.({\bf r}+{\bf  X})}
B_{\alpha\beta}
 e^{\frac{i{\bf p}.({\bf r}+{\bf  X})}{\hbar} }\tau({\bf p}) u_{0\mu}
\label{5.27}
\end{eqnarray}
The integration over ${\bf r}$ can be performed and provides a Dirac delta function.
\begin{eqnarray}
&&\gamma_{\alpha\beta}^{1\mu1\mu'10}({\bf p},{\bf p}')
= 
\frac1{\hbar^3}
\frac{-iq_e}\hbar\sqrt{\frac{\hbar}{2\varepsilon_0  (2\pi)^3}}
\delta({\bf p}-{\bf p}'+{\bf K}+{\bf k}_1)
e^{i\frac12{\bf K}.{\bf  X}}
\nonumber\\&&\times
\frac1{\sqrt{\omega_{{\bf k}_1}}} {\bf e}_{{\bf k}_1 \nu_1}.
{\bf  X} e^{ i{\bf k}_1.{\bf  X}}
e^{\frac{i{\bf p}.{\bf  X}}{\hbar} }
\delta_{\alpha\beta}^{1\mu'1\mu}({\bf p}', {\bf p})
\label{5.29}
\end{eqnarray}
\begin{eqnarray}
&&\gamma_{\alpha\beta}^{1\mu1\mu'01}({\bf p},{\bf p}')
= 
\frac1{\hbar^3}
\frac{-iq_e}\hbar\sqrt{\frac{\hbar}{2\varepsilon_0  (2\pi)^3}}
\delta({\bf p}-{\bf p}'+\hbar{\bf K}-\hbar{\bf k}'_1)
e^{i\frac12{\bf K}.{\bf  X}}
\nonumber\\&&\times
\frac1{\sqrt{\omega_{{\bf k}'_1}}}{\bf e}_{{\bf k}'_1 \nu_1}.{\bf  X} 
e^{\frac{i{\bf p}.{\bf  X}}{\hbar} }
\delta_{\alpha\beta}^{1\mu'1\mu}({\bf p}', {\bf p})
\label{5.30}
\end{eqnarray}
\begin{eqnarray}
&&\gamma_{\alpha\beta}^{\bar1\mu\bar1\mu'10}({\bf p},{\bf p}')
= 
\frac1{\hbar^3}
\frac{-iq_e}\hbar\sqrt{\frac{\hbar}{2\varepsilon_0  (2\pi)^3}}
\delta({\bf p}-{\bf p}'+\hbar{\bf K}+\hbar{\bf k}_1)
e^{i\frac12{\bf K}.{\bf  X}}
\nonumber\\&&\times
\frac1{\sqrt{\omega_{{\bf k}_1}}}{\bf e}_{{\bf k}_1 \nu_1}.
{\bf  X} 
e^{\frac{-i{\bf p}'.{\bf  X}}{\hbar} }
\delta_{\alpha\beta}^{\bar1\bar\mu\bar1\bar\mu'}(-{\bf p}, -{\bf p}')
\label{5.31}
\end{eqnarray}
\begin{eqnarray}
&&\gamma_{\alpha\beta}^{\bar1\mu\bar1\mu'01}({\bf p},{\bf p}')
= 
\frac1{\hbar^3}
\frac{-iq_e}\hbar\sqrt{\frac{\hbar}{2\varepsilon_0  (2\pi)^3}}
\delta({\bf p}-{\bf p}'+\hbar{\bf K}-\hbar{\bf k}'_1)
e^{i\frac12{\bf K}.{\bf  X}}
\nonumber\\&&\times
\frac1{\sqrt{\omega_{{\bf k}'_1}}}{\bf e}_{{\bf k}'_1 \nu_1}.
{\bf  X} 
e^{- i{\bf k}'_1.{\bf  X}}
e^{\frac{-i{\bf p}'.{\bf  X}}{\hbar} }
\delta_{\alpha\beta}^{\bar1\bar\mu\bar1\bar\mu'}(-{\bf p}, -{\bf p}')
\label{5.32}
\end{eqnarray}
\begin{eqnarray}
&&\gamma_{\alpha\beta}^{1\mu\bar1\mu'10}({\bf p},{\bf p}')
= 0
\label{5.33}
\end{eqnarray}
\begin{eqnarray}
&&\gamma_{\alpha\beta}^{1\mu\bar1\mu'01}({\bf p},{\bf p}')
= 
\frac1{\hbar^3}
\frac{-iq_e}\hbar\sqrt{\frac{\hbar}{2\varepsilon_0  (2\pi)^3}}
\delta(-{\bf p}-{\bf p}'+\hbar{\bf K}-\hbar{\bf k}'_1)
e^{i\frac12{\bf K}.{\bf  X}}
\nonumber\\&&\times
\frac1{\sqrt{\omega_{{\bf k}'_1}}}{\bf e}_{{\bf k}'_1 \nu_1}.
{\bf  X} 
(1-e^{- i{\bf k}'_1.{\bf  X}})
e^{\frac{-i{\bf p}'.{\bf  X}}{\hbar} }
\delta_{\alpha\beta}^{1\mu\bar1\bar\mu'}({\bf p}, -{\bf p}')
\label{5.34}
\end{eqnarray}
\begin{eqnarray}
&&\gamma_{\alpha\beta}^{\bar1\mu'1\mu10}({\bf p},{\bf p}')
= 
\frac1{\hbar^3}
\frac{-iq_e}\hbar\sqrt{\frac{\hbar}{2\varepsilon_0  (2\pi)^3}}
\delta({\bf p}+{\bf p}'+\hbar{\bf K}+\hbar{\bf k}_1)
e^{i\frac12{\bf K}.{\bf  X}}
\nonumber\\&&\times
\frac1{\sqrt{\omega_{{\bf k}_1}}}{\bf e}_{{\bf k}_1 \nu_1}.
{\bf  X} 
(e^{ i{\bf k}_1.{\bf  X}}-1)
e^{\frac{i{\bf p}.{\bf  X}}{\hbar} }
\delta_{\alpha\beta}^{\bar1\bar\mu'1\mu}(-{\bf p}', {\bf p})
\label{5.35}
\end{eqnarray}
\begin{eqnarray}
&&\gamma_{\alpha\beta}^{\bar1\mu1\mu'01}({\bf p},{\bf p}')
= 0
\label{5.36}
\end{eqnarray}

\section{The evolution equations for the correlation functions}  
\def\theequation{\thesection.\arabic{equation}}
\setcounter{equation}{0}

This appendix provides all the  non-vanishing contributions due to the 
part ${\cal L}_{ID}$ of the evolution operator for the correlation functions.
Let us first consider the contributions due to the part ${\cal L}_{ID1}$  
of ${\cal L}_{ID}$ involving the product of creation and destruction operators 
$c^+ca$.
The contributions  diagonal in the 
occupation numbers $ ss'\tilde s\tilde s'$ can be written as: 
\begin{eqnarray}
&&<ss'\tilde s\tilde s'nn'|{\cal L}_{ID1} |ss'\tilde s\tilde s'\,n\,n'-1 >
\nonumber\\&&
=\frac{1}{i\hbar}\hbar^3\sum_{\mu}
\sum_{i=1, s'} \sum_{j=1,n'}
\sqrt{\frac{\hbar}{2\varepsilon_0(2\pi)^3 \omega_{{\bf k}'_{j}}}}
\alpha^{11}_{\nu'_{j} -\mu \mu'_i}
({\bf p}'_i+\frac12\hbar {\bf k}'_j,-{\bf k}'_j)
\nonumber\\ && \times 
S({\bf p}'_i+\hbar {\bf k}'_j,\mu;{\bf p}'_i,\mu'_i)
\left(\prod_{r=j,n'-1}S({\bf k}'_{r+1},\nu'_{r+1};{\bf k}'_r,\nu'_r)
\right)  
\label{C.1}
\end{eqnarray}
Two contributions correspond to the same matrix element of ${\cal L}_{ID1}$:
\begin{eqnarray}
&&
<ss'\tilde s\tilde s'nn'|{\cal L}_{ID1} |ss'\tilde s\tilde s'\,n+1\,n' >_a
\nonumber\\ && 
=\frac{1}{i\hbar}\hbar^3\sum_{\nu_{n+1}}\sum_{\mu}\int d^3k_{n+1}
\sum_{i=1, s'}
\sqrt{\frac{\hbar}{2\varepsilon_0(2\pi)^3 \omega_{{\bf k}_{n+1}}}}
\nonumber\\ &&\times 
\alpha^{11}_{\nu_{n+1}- \mu \mu'_i}({\bf p}'_i+\frac12\hbar {\bf k}_{n+1},
-{\bf k}_{n+1})
S({\bf p}'_i+\hbar {\bf k}_{n+1},\mu;{\bf p}'_i,\mu'_i)
\label{C.2a}
\end{eqnarray}
\begin{eqnarray}
&&
<ss'\tilde s\tilde s'nn'|{\cal L}_{ID1} |ss'\tilde s\tilde s'\,n+1\,n' >_b
\nonumber\\ && 
=-\frac{1}{i\hbar}\hbar^3\sum_{\nu_{n+1}}\sum_{\mu'}\int d^3k_{n+1}
\sum_{i=1, s}
\sqrt{\frac{\hbar}{2\varepsilon_0(2\pi)^3 \omega_{{\bf k}_{n+1}}}}
\nonumber\\ &&  \times 
\alpha^{11}_{\nu_{n+1} -\mu_i \mu'}({\bf p}_i-\frac12\hbar {\bf k}_{n+1},
-{\bf k}_{n+1})
S({\bf p}_i-\hbar {\bf k}_{n+1},\mu';{\bf p}_i,\mu_i)  
\label{C.2b}
\end{eqnarray}
The contribution involving a change  in the 
occupation numbers $ ss'\tilde s\tilde s'$ can be written as: 
\begin{eqnarray}
&&
<ss'\tilde s\tilde s'nn'|{\cal L}_{ID1} |s+1\,s'+1\,\tilde s\tilde s'\,n\,n'-1 >
\nonumber\\ && 
=\frac{1}{i\hbar}\hbar^3\sum_{\mu\mu'} (-1)^{s+s'+\tilde s+\tilde s'}
\sum_{j=1,n'} 
\sqrt{\frac{\hbar}{2\varepsilon_0(2\pi)^3 \omega_{{\bf k}'_{j}}}}
\nonumber\\ &&  \times 
\int d^3p \,
\alpha^{11}_{\nu'_{j} -\mu \mu'}({\bf p},-{\bf k}'_j)
S({\bf p}-\frac12\hbar {\bf k}'_j,\mu';{\bf p}_{s+1},\mu_{s+1})
\nonumber\\ &&\times 
S({\bf p}+\frac12\hbar {\bf k}'_j,\mu;{\bf p}'_{s'+1},\mu'_{s'+1})
\left(\prod_{r=j,n'-1}S({\bf k}'_{r+1},\nu'_{r+1};{\bf k}'_r,\nu'_r)
\right)  
\label{C.3}
\end{eqnarray}
We now consider the contributions due to the part ${\cal L}_{ID2}$  
of ${\cal L}_{ID}$ involving the product of creation and destruction operators 
$c^+ca^+$.
The contributions  diagonal in the 
occupation numbers $ ss'\tilde s\tilde s'$ can be written as: 
\begin{eqnarray}
&&
<ss'\tilde s\tilde s'nn'|{\cal L}_{ID2} |ss'\tilde s\tilde s'\,n-1\,n' >
\nonumber\\&&
=-\frac{1}{i\hbar}\hbar^3\sum_{\mu'}
\sum_{i=1, s} \sum_{j=1,n} s(\nu_{j})
\sqrt{\frac{\hbar}{2\varepsilon_0(2\pi)^3 \omega_{{\bf k}_{j}}}}
\alpha^{11}_{\nu_{j}+ \mu_i \mu'}
({\bf p}_i+\frac12\hbar {\bf k}_j,{\bf k}_j)
\nonumber\\ && \times 
S({\bf p}_i+\hbar {\bf k}_j,\mu';{\bf p}_i,\mu_i)
\left(\prod_{r=j,n-1}S({\bf k}_{r+1},\nu_{r+1};{\bf k}_r,\nu_r)
\right)  
\label{C.4}
\end{eqnarray}
Two contributions correspond to the same matrix element of ${\cal L}_{ID2}$:
\begin{eqnarray}
&&
<ss'\tilde s\tilde s'nn'|{\cal L}_{ID2} |ss'\tilde s\tilde s'n\,n'+1 >_a
\nonumber\\ &&
=-\frac{1}{i\hbar}\hbar^3\sum_{\nu'_{n'+1}} s(\nu'_{n'+1})
\sum_{\mu'}\int d^3k'_{n'+1}
\sum_{i=1, s}
\sqrt{\frac{\hbar}{2\varepsilon_0(2\pi)^3 \omega_{{\bf k}'_{n'+1}}}}
\nonumber\\ && \times 
\alpha^{11}_{\nu'_{n'+1}+ \mu_i \mu'}({\bf p}_i+\frac12\hbar {\bf k}'_{n'+1},
{\bf k}'_{n'+1})
S({\bf p}_i+\hbar {\bf k}'_{n'+1},\mu';{\bf p}_i,\mu_i)  
\label{C.5a}
\end{eqnarray}
\begin{eqnarray}
&&
<ss'\tilde s\tilde s'nn'|{\cal L}_{ID2} |ss'\tilde s\tilde s'n\,n'+1 >_b
\nonumber\\ && 
=\frac{1}{i\hbar}\hbar^3\sum_{\nu'_{n'+1}} s(\nu'_{n'+1})
\sum_{\mu}\int d^3k'_{n'+1}
\sum_{i=1, s'}
\sqrt{\frac{\hbar}{2\varepsilon_0(2\pi)^3 \omega_{{\bf k}'_{n'+1}}}}
\nonumber\\ &&\times 
\alpha^{11}_{\nu'_{n'+1} +\mu \mu'_i}({\bf p}'_i-\frac12\hbar {\bf k}'_{n'+1},
{\bf k}'_{n'+1})
S({\bf p}'_i-\hbar {\bf k}'_{n'+1},\mu;{\bf p}'_i,\mu'_i)
\label{C.5b}
\end{eqnarray}
The contribution involving a change  in the 
occupation numbers $ ss'\tilde s\tilde s'$ can be written as: 
\begin{eqnarray}
&&
<ss'\tilde s\tilde s'nn'|{\cal L}_{ID2} |s+1\,s'+1\,\tilde s\tilde s'\,n-1\,n' >
\nonumber\\ &&
=-\frac{1}{i\hbar}\hbar^3\sum_{\mu\mu'} (-1)^{s+s'+\tilde s+\tilde s'}
\sum_{j=1,n} s(\nu_j)
\sqrt{\frac{\hbar}{2\varepsilon_0(2\pi)^3 \omega_{{\bf k}_{j}}}}
\nonumber\\ &&  \times 
\int d^3p \,
\alpha^{11}_{\nu_{j}+ \mu \mu'}({\bf p},{\bf k}_j)
S({\bf p}+\frac12\hbar {\bf k}_j,\mu';{\bf p}_{s+1},\mu_{s+1})
\nonumber\\ &&\times 
S({\bf p}-\frac12\hbar {\bf k}_j,\mu;{\bf p}'_{s'+1},\mu'_{s'+1})
\left(\prod_{r=j,n-1}S({\bf k}_{r+1},\nu_{r+1};{\bf k}_r,\nu_r)
\right)  
\label{C.6}
\end{eqnarray}
We now consider the contributions due to the part ${\cal L}_{ID3}$  
of ${\cal L}_{ID}$ involving the product of creation and destruction operators 
$b^+ba$.
The contributions  diagonal in the 
occupation numbers $ ss'\tilde s\tilde s'$ can be written as: 
\begin{eqnarray}
&&
<ss'\tilde s\tilde s'nn'|{\cal L}_{ID3} |ss'\tilde s\tilde s'\,n\,n'-1 >
\nonumber\\&&
=\frac{1}{i\hbar}\hbar^3\sum_{\mu}
\sum_{i=1, \tilde{s}'} \sum_{j=1,n'}
\sqrt{\frac{\hbar}{2\varepsilon_0(2\pi)^3 \omega_{{\bf k}'_{j}}}}
\alpha^{\bar1\bar1}_{\nu'_{j} -\mu \tilde{\mu}'_i}
(\tilde{{\bf p}}'_i+\frac12\hbar {\bf k}'_j,-{\bf k}'_j)
\nonumber\\ &&\times 
S(\tilde{{\bf p}}'_i+\hbar {\bf k}'_j,\mu;\tilde{{\bf p}}'_i,\tilde{\mu}'_i)
\left(\prod_{r=j,n'-1}S({\bf k}'_{r+1},\nu'_{r+1};{\bf k}'_r,\nu'_r)
\right)  
\label{C.7}
\end{eqnarray}
Two contributions correspond to the same matrix element of ${\cal L}_{ID3}$:
\begin{eqnarray}
&&
<ss'\tilde s\tilde s'nn'|{\cal L}_{ID3} |ss'\tilde s\tilde s'\,n+1\,n' >_a
\nonumber\\ &&
=\frac{1}{i\hbar}\hbar^3\sum_{\nu_{n+1}}\sum_{\mu}\int d^3k_{n+1}
\sum_{i=1, \tilde{s}'}
\sqrt{\frac{\hbar}{2\varepsilon_0(2\pi)^3 \omega_{{\bf k}_{n+1}}}}
\nonumber\\ &&\times 
\alpha^{\bar1\bar1}_{\nu_{n+1}- \mu \tilde{\mu}'_i}
(\tilde{{\bf p}}'_i+\frac12\hbar {\bf k}_{n+1},
-{\bf k}_{n+1})
S(\tilde{{\bf p}}'_i+\hbar {\bf k}_{n+1},\mu;\tilde{{\bf p}}'_i,\tilde{\mu}'_i)
\label{C.8a}
\end{eqnarray}
\begin{eqnarray}
&&
<ss'\tilde s\tilde s'nn'|{\cal L}_{ID3} |ss'\tilde s\tilde s'\,n+1\,n' >_b
\nonumber\\ &&
=-\frac{1}{i\hbar}\hbar^3\sum_{\nu_{n+1}}\sum_{\mu'}\int d^3k_{n+1}
\sum_{i=1, \tilde{s}}
\sqrt{\frac{\hbar}{2\varepsilon_0(2\pi)^3 \omega_{{\bf k}_{n+1}}}}
\nonumber\\ && \times 
\alpha^{\bar1\bar1}_{\nu_{n+1} -\tilde{\mu_i} \mu'}(\tilde{{\bf p}}_i-\frac12\hbar {\bf k}_{n+1},
-{\bf k}_{n+1})
S(\tilde{{\bf p}}_i-\hbar {\bf k}_{n+1},\mu';\tilde{{\bf p}}_i,\tilde{\mu}_i)  
\label{C.8b}
\end{eqnarray}
The contribution involving a change  in the 
occupation numbers $ ss'\tilde s\tilde s'$ can be written as: 
\begin{eqnarray}
&&
<ss'\tilde s\tilde s'nn'|{\cal L}_{ID3} |ss'\,\tilde s+1\,\tilde s'+1\,n\,n'-1 >
\nonumber\\ && 
=\frac{1}{i\hbar}\hbar^3\sum_{\mu\mu'} (-1)^{\tilde s+\tilde s'}
\sum_{j=1,n'} 
\sqrt{\frac{\hbar}{2\varepsilon_0(2\pi)^3 \omega_{{\bf k}'_{j}}}}
\nonumber\\ &&  \times 
\int d^3p \,
\alpha^{\bar1\bar1}_{\nu'_{j}- \mu \mu'}({\bf p},-{\bf k}'_j)
S({\bf p}-\frac12\hbar {\bf k}'_j,\mu';
\tilde{{\bf p}}_{\tilde s+1},\tilde{\mu}_{\tilde s+1})
\nonumber\\ &&\times 
S({\bf p}+\frac12\hbar {\bf k}'_j,\mu;
\tilde{{\bf p}}'_{\tilde s'+1},\tilde{\mu}'_{\tilde s'+1})
\left(\prod_{r=j,n'-1}S({\bf k}'_{r+1},\nu'_{r+1};{\bf k}'_r,\nu'_r)
\right)
\nonumber\\ &&  
\label{C.9}
\end{eqnarray}
We now consider the contributions due to the part ${\cal L}_{ID4}$  
of ${\cal L}_{ID}$ involving the product of creation and destruction operators 
$b^+ba^+$.
The contributions  diagonal in the 
occupation numbers $ ss'\tilde s\tilde s'$ can be written as: 
\begin{eqnarray}
&&
<ss'\tilde s\tilde s'nn'|{\cal L}_{ID4} |ss'\tilde s\tilde s'\,n-1\,n' >
\nonumber\\&&
=-\frac{1}{i\hbar}\hbar^3\sum_{\mu'}
\sum_{i=1,\tilde{s}} \sum_{j=1,n} s(\nu_{j})
\sqrt{\frac{\hbar}{2\varepsilon_0(2\pi)^3 \omega_{{\bf k}_{j}}}}
\alpha^{\bar1\bar1}_{\nu_{j}+ \tilde{\mu}_i \mu'}
(\tilde{{\bf p}}_i+\frac12\hbar {\bf k}_j,{\bf k}_j)
\nonumber\\ &&  \times 
S(\tilde{{\bf p}}_i+\hbar {\bf k}_j,\mu';\tilde{{\bf p}}_i,\tilde{\mu}_i)
\left(\prod_{r=j,n-1}S({\bf k}_{r+1},\nu_{r+1};{\bf k}_r,\nu_r)
\right)  
\label{C.10}
\end{eqnarray}
Two contributions correspond to the same matrix element of ${\cal L}_{ID4}$:
\begin{eqnarray}
&&
<ss'\tilde s\tilde s'nn'|{\cal L}_{ID4} |ss'\tilde s\tilde s'n\,n'+1 >_a
\nonumber\\ && 
=-\frac{1}{i\hbar}\hbar^3\sum_{\nu'_{n'+1}} s(\nu'_{n'+1})
\sum_{\mu'}\int d^3k'_{n'+1}
\sum_{i=1, \tilde{s}}
\sqrt{\frac{\hbar}{2\varepsilon_0(2\pi)^3 \omega_{{\bf k}'_{n'+1}}}}
\nonumber\\ &&\times 
\alpha^{\bar1\bar1}_{\nu'_{n'+1} +\tilde{\mu}_i \mu'}
(\tilde{{\bf p}}_i+\frac12\hbar {\bf k}'_{n'+1},
{\bf k}'_{n'+1})
S(\tilde{{\bf p}}_i+\hbar {\bf k}'_{n'+1},\mu';
\tilde{{\bf p}}_i,\tilde{\mu}_i)  
\label{C.11a}
\end{eqnarray}
\begin{eqnarray}
&&
<ss'\tilde s\tilde s'nn'|{\cal L}_{ID4} |ss'\tilde s\tilde s'n\,n'+1 >_b
\nonumber\\ && 
=\frac{1}{i\hbar}\hbar^3\sum_{\nu'_{n'+1}} s(\nu'_{n'+1})
\sum_{\mu}\int d^3k'_{n'+1}
\sum_{i=1, \tilde{s}'}
\sqrt{\frac{\hbar}{2\varepsilon_0(2\pi)^3 \omega_{{\bf k}'_{n'+1}}}}
\nonumber\\ &&\times 
\alpha^{\bar1\bar1}_{\nu'_{n'+1}+ \mu \tilde{\mu}'_i}
(\tilde{{\bf p}}'_i-\frac12\hbar {\bf k}'_{n'+1},
{\bf k}'_{n'+1})
S(\tilde{{\bf p}}'_i-\hbar {\bf k}'_{n'+1},\mu;
\tilde{{\bf p}}'_i,\tilde{\mu}'_i)
\label{C.11b}
\end{eqnarray}
The contribution involving a change  in the 
occupation numbers $ ss'\tilde s\tilde s'$ can be written as: 
\begin{eqnarray}
&&
<ss'\tilde s\tilde s'nn'|{\cal L}_{ID4} |ss'\,\tilde s+1\,\tilde s'+1\,n-1\,n' >
\nonumber\\ &&
=-\frac{1}{i\hbar}\hbar^3\sum_{\mu\mu'} (-1)^{\tilde s+\tilde s'}
\sum_{j=1,n} s(\nu_j)
\sqrt{\frac{\hbar}{2\varepsilon_0(2\pi)^3 \omega_{{\bf k}_{j}}}}
\nonumber\\ && \times 
\int d^3p \,
\alpha^{\bar1\bar1}_{\nu_{j} +\mu \mu'}({\bf p},{\bf k}_j)
S({\bf p}+\frac12\hbar {\bf k}_j,\mu';
\tilde{{\bf p}}_{\tilde{s}+1},\tilde{\mu}_{\tilde{s}+1})
\nonumber\\ && \times 
S({\bf p}-\frac12\hbar {\bf k}_j,\mu;
\tilde{{\bf p}}'_{\tilde{s}'+1},\tilde{\mu}'_{\tilde{s}'+1})
\left(\prod_{r=j,n-1}S({\bf k}_{r+1},\nu_{r+1};{\bf k}_r,\nu_r)
\right)  
\nonumber\\ && 
\label{C.12}
\end{eqnarray}
We now consider the contributions due to the part ${\cal L}_{ID5}$  
of ${\cal L}_{ID}$ involving the product of creation and destruction operators 
$c^+b^+a$.
No contribution is diagonal in the 
occupation numbers $ ss'\tilde s\tilde s'$. 
The term corresponding to the 
diminution of two occupation numbers is: 
\begin{eqnarray}
&&
<ss'\tilde s\tilde s'nn'|{\cal L}_{ID5} |s-1\,s'\,\tilde s-1\,\tilde s'\,n+1\,n'>
\nonumber\\ && 
=-\frac{1}{i\hbar}\sum_{\nu_{n+1}}
\sum_{i=1, s}\sum_{j=1,\tilde s}  (-1)^{i-1+\tilde s-j}
\sqrt{\frac{\hbar}{2\varepsilon_0(2\pi)^3 
\omega_{\frac{-1}{\hbar}({\bf p}_i+\tilde{{\bf p}}_j)}}}
\nonumber\\ &&  \times 
\alpha^{1\bar1}_{\nu_{n+1} -\mu_i \tilde{\mu}_j}
(\frac12({\bf p}_i-\tilde{{\bf p}}_j),
-\frac1{\hbar}({\bf p}_i+\tilde{{\bf p}}_j))
\nonumber\\ &&\times 
\left(\prod_{r=i,s-1}
S({\bf p}_{r+1},\mu_{r+1};{\bf p}_r,\mu_r)\right) 
\left(\prod_{\tilde{r}=j,\tilde{s}-1}
S(\tilde{{\bf p}}_{{\tilde{r}}+1},\tilde{\mu}_{{\tilde{r}}+1};
\tilde{{\bf p}}_{\tilde{r}},\tilde{\mu_{\tilde{r}}})\right)  
\nonumber\\ &&  \times 
S(\frac1{\hbar}({\bf p}_i+\tilde{{\bf p}}_j),\nu_{n+1};{\bf k}_{n+1},\nu_{n+1})
\label{C.13}
\end{eqnarray}
The terms corresponding to a transfert of occupation numbers are:
\begin{eqnarray}
&&
<ss'\tilde s\tilde s'nn'|{\cal L}_{ID5} |s-1\,s'\,\tilde s\,\tilde s'+1\,n+1\,n'>
\nonumber\\ &&
=-\frac{1}{i\hbar}\hbar^3\sum_{\nu_{n+1}}  \int d^3k_{n+1}   \sum_{\mu'}
\sum_{i=1, s}  (-1)^{i-1+\tilde s+\tilde s'}
\sqrt{\frac{\hbar}{2\varepsilon_0(2\pi)^3 
\omega_{{\bf k}_{n+1}}}}
\nonumber\\ &&  \times 
\alpha^{1\bar1}_{\nu_{n+1} -\mu_i \mu'}
({\bf p}_i-\frac12\hbar{{\bf k}}_{n+1},
-{\bf k}_{n+1})
\left(\prod_{r=i,s-1}
S({\bf p}_{r+1},\mu_{r+1};{\bf p}_r,\mu_r)\right) 
\nonumber\\ &&  \times 
S({-{\bf p}}_{i}+\hbar{{\bf k}}_{n+1},{\mu}';
\tilde{{\bf p}}'_{\tilde s'+1},\tilde{\mu}'_{\tilde s'+1})  
\label{C.14}
\end{eqnarray}
\begin{eqnarray}
&&
<ss'\tilde s\tilde s'nn'|{\cal L}_{ID5} |s\,s'+1\,\tilde s-1\,\tilde s'\,n+1\,n'>
\nonumber\\ && 
=-\frac{1}{i\hbar}\hbar^3\sum_{\nu_{n+1}}  \int d^3k_{n+1}   \sum_{\mu}
\sum_{i=1, \tilde s}  (-1)^{i-1+ s'+\tilde s'}
\sqrt{\frac{\hbar}{2\varepsilon_0(2\pi)^3 
\omega_{{\bf k}_{n+1}}}}
\nonumber\\ &&  \times 
\alpha^{1\bar1}_{\nu_{n+1} -\mu \tilde{\mu}_i}
(-\tilde{{\bf p}}_i+\frac12\hbar{{\bf k}}_{n+1},
-{\bf k}_{n+1})
\left(\prod_{r=i,\tilde{s}-1}
S(\tilde{{\bf p}}_{r+1},\tilde{\mu}_{r+1};
\tilde{{\bf p}}_r,\tilde{\mu}_r)\right) 
\nonumber\\ &&\times 
S(-\tilde{{\bf p}}_{i}+\hbar{{\bf k}}_{n+1},{\mu};
{{\bf p}}'_{ s'+1},{\mu'_{ s'+1}})  
\label{C.15}
\end{eqnarray}
The term corresponding to the 
majoration  of two occupation numbers is: 
\begin{eqnarray}
&&
<ss'\tilde s\tilde s'nn'|{\cal L}_{ID5} |s\,s'+1\,\tilde s\,\tilde s'+1\,n\,n'-1 >
\nonumber\\ &&
=\frac{1}{i\hbar}\hbar^3\sum_{\mu\mu'} (-1)^{s'}
\sum_{j=1,n'} 
\sqrt{\frac{\hbar}{2\varepsilon_0(2\pi)^3 \omega_{{\bf k}'_{j}}}}
\nonumber\\ && \times 
\int d^3p \,
\alpha^{1\bar1}_{\nu'_{j} -\mu \mu'}({\bf p},-{\bf k}'_j)
S(-{\bf p}+\frac12\hbar {\bf k}'_j,\mu';
\tilde{{\bf p}}'_{\tilde{s}'+1},\tilde{\mu}'_{\tilde{s}'+1})
\nonumber\\ &&\times 
S({\bf p}+\frac12\hbar {\bf k}'_j,\mu;{\bf p}'_{s'+1},\mu'_{s'+1})
\left(\prod_{r=j,n'-1}S({\bf k}'_{r+1},\nu'_{r+1};{\bf k}'_r,\nu'_r)
\right)  
\nonumber\\ &&
\label{C.16}
\end{eqnarray}
We now consider the contributions due to the part ${\cal L}_{ID6}$  
of ${\cal L}_{ID}$ involving the product of creation operators 
$c^+b^+a^+$.
No contribution is diagonal in the 
occupation numbers $ ss'\tilde s\tilde s'$. 
The terms corresponding to the 
diminution of two occupation numbers are: 
\begin{eqnarray}
&&
<ss'\tilde s\tilde s'nn'|{\cal L}_{ID6} |s-1\,s'\,\tilde s-1\,\tilde s'\,n-1\,n'>
\nonumber\\ &&
=-\frac{1}{i\hbar}
\sum_{i=1, s}\sum_{j=1,\tilde s}\sum_{l=1,n}s(\nu_l)  (-1)^{i-1+\tilde s-j}
\sqrt{\frac{\hbar}{2\varepsilon_0(2\pi)^3 
\omega_{{\bf k}_l}}}
\alpha^{1\bar1}_{\nu_{l} +\mu_i \tilde{\mu}_j}
(\frac12({\bf p}_i-\tilde{{\bf p}}_j),
{\bf k}_l)
\nonumber\\ && \times 
\delta\left({\bf k}_l+\frac1{\hbar}({\bf p}_i+\tilde{{\bf p}}_j)\right)
\left(\prod_{r=i,s-1}
S({\bf p}_{r+1},\mu_{r+1};{\bf p}_r,\mu_r)\right) 
\nonumber\\ && \times 
\left(\prod_{\tilde{r}=j,\tilde{s}-1}
S(\tilde{{\bf p}}_{\tilde{r}+1},\tilde{\mu}_{\tilde{r}+1};
\tilde{{\bf p}}_{\tilde r},\tilde{\mu}_{\tilde r})\right)  
\left(\prod_{r=l,n-1}S({\bf k}_{r+1},\nu_{r+1};{\bf k}_r,\nu_r)
\right)
\label{C.17}
\end{eqnarray}
\begin{eqnarray}
&&
<ss'\tilde s\tilde s'nn'|{\cal L}_{ID6} |s-1\,s'\,\tilde s-1\,\tilde s'\,n\,n'+1>
\nonumber\\ &&
=-\frac{1}{i\hbar}\sum_{\nu'_{n'+1}} s(\nu'_{n'+1})
\sum_{i=1, s}\sum_{j=1,\tilde s}  (-1)^{i-1+\tilde s-j}
\sqrt{\frac{\hbar}{2\varepsilon_0(2\pi)^3 
\omega_{\frac{-1}{\hbar}({\bf p}_i+\tilde{{\bf p}}_j)}}}
\nonumber\\ &&\times
\alpha^{1\bar1}_{\nu'_{n'+1}+ \mu_i \tilde{\mu}_j}
(\frac12({\bf p}_i-\tilde{{\bf p}}_j),
-\frac1{\hbar}({\bf p}_i+\tilde{{\bf p}}_j))
\left(\prod_{r=i,s-1}
S({\bf p}_{r+1},\mu_{r+1};{\bf p}_r,\mu_r)\right) 
\nonumber\\ && \times
\left(\prod_{\tilde{r}=j,\tilde{s}-1}
S(\tilde{{\bf p}}_{\tilde{r}+1},\tilde{\mu}_{\tilde{r}+1};
\tilde{{\bf p}}_{\tilde{r}},\tilde{\mu}_{\tilde{r}})\right)  
S(-\frac1{\hbar}({\bf p}_i+\tilde{{\bf p}}_j),\nu'_{n'+1};
{\bf k}'_{n'+1},\nu'_{n'+1})
\nonumber\\ &&
\label{C.18}
\end{eqnarray}
The terms corresponding to a transfert of occupation numbers are:
\begin{eqnarray}
&&
<ss'\tilde s\tilde s'nn'|{\cal L}_{ID6} |s-1\,s'\,\tilde s\,\tilde s'+1\,n-1\,n'>
\nonumber\\ &&
=-\frac{1}{i\hbar}\hbar^3\sum_{l=1,n} s(\nu_l)    
\sum_{\mu'}
\sum_{i=1, s}  (-1)^{i-1+\tilde s+\tilde s'}
\sqrt{\frac{\hbar}{2\varepsilon_0(2\pi)^3 
\omega_{{\bf k}_l}}}
\nonumber\\ && \times
\alpha^{1\bar1}_{\nu_{l} +\mu_i \mu'}
({\bf p}_i+\frac12\hbar{{\bf k}}_{l},
{\bf k}_{l})
\left(\prod_{r=i,s-1}
S({\bf p}_{r+1},\mu_{r+1};{\bf p}_r,\mu_r)\right) 
\nonumber\\ && \times
S(-{{\bf p}}_{i}-\hbar{{\bf k}}_{l},{\mu}';
\tilde{{\bf p}}'_{\tilde s'+1},\tilde{\mu}'_{\tilde s'+1})  
\left(\prod_{r=l,n-1}S({\bf k}_{r+1},\nu_{r+1};{\bf k}_r,\nu_r)
\right)
\nonumber\\ &&
\label{C.19}
\end{eqnarray}
\begin{eqnarray}
&&
<ss'\tilde s\tilde s'nn'|{\cal L}_{ID6} |s-1\,s'\,\tilde s\,\tilde s'+1\,n\,n'+1>
\nonumber\\ &&
=-\frac{1}{i\hbar}\hbar^3\sum_{\nu'_{n'+1}} s(\nu'_{n'+1}) \int d^3k'_{n'+1}   
\sum_{\mu'}
\sum_{i=1, s}  (-1)^{i-1+\tilde s+\tilde s'}
\sqrt{\frac{\hbar}{2\varepsilon_0(2\pi)^3 
\omega_{{\bf k}'_{n'+1}}}}
\nonumber\\ && \times
\alpha^{1\bar1}_{\nu'_{n'+1}+ \mu_i \mu'}
({\bf p}_i+\frac12\hbar{{\bf k}}'_{n'+1},
{\bf k}'_{n'+1})
\left(\prod_{r=i,s-1}
S({\bf p}_{r+1},\mu_{r+1};{\bf p}_r,\mu_r)\right) 
\nonumber\\ && \times
S(-{{\bf p}}_{i}-\hbar{{\bf k}}'_{n'+1},{\mu}';
\tilde{{\bf p}}'_{\tilde s'+1},\tilde{\mu}'_{\tilde s'+1})  
\label{C.20}
\end{eqnarray}
\begin{eqnarray}
&&
<ss'\tilde s\tilde s'nn'|{\cal L}_{ID6} |s\,s'+1\,\tilde s-1\,\tilde s'\,n-1\,n'>
\nonumber\\ &&
=-\frac{1}{i\hbar}\hbar^3\sum_{l=1,n} s(\nu_l)    
\sum_{\mu}
\sum_{i=1, \tilde s}  (-1)^{i-1+ s'+\tilde s'}
\sqrt{\frac{\hbar}{2\varepsilon_0(2\pi)^3 
\omega_{{\bf k}_l}}}
\nonumber\\ && \times
\alpha^{1\bar1}_{\nu_{l}+ \mu \tilde\mu_i}
(-\tilde{{\bf p}}_i-\frac12\hbar{{\bf k}}_{l},
{\bf k}_{l})
\left(\prod_{r=i,\tilde{s}-1}
S(\tilde{{\bf p}}_{r+1},\tilde{\mu}_{r+1};
\tilde{{\bf p}}_r,\tilde{\mu}_r)\right) 
\nonumber\\ &&\times
S(-\tilde{{\bf p}}_{i}-\hbar{{\bf k}}_{l},{\mu};
{{\bf p}}'_{ s'+1},{\mu}'_{ s'+1})  
\left(\prod_{r=l,n-1}S({\bf k}_{r+1},\nu_{r+1};{\bf k}_r,\nu_r)
\right)
\nonumber\\ &&
\label{C.21}
\end{eqnarray}
\begin{eqnarray}
&&
<ss'\tilde s\tilde s'nn'|{\cal L}_{ID6} |s\,s'+1\,\tilde s-1\,\tilde s'\,n\,n'+1>
\nonumber\\ &&
=-\frac{1}{i\hbar}\hbar^3\sum_{\nu'_{n'+1}} s(\nu'_{n'+1}) \int d^3k'_{n'+1}   
\sum_{\mu}
\sum_{i=1, \tilde{s}}  (-1)^{i-1+ s'+\tilde s'}
\sqrt{\frac{\hbar}{2\varepsilon_0(2\pi)^3 
\omega_{{\bf k}'_{n'+1}}}}
\nonumber\\ &&\times
\alpha^{1\bar1}_{\nu'_{n'+1} +\mu \tilde\mu_i}
(-\tilde{{\bf p}}_i-\frac12\hbar{{\bf k}}'_{n'+1},
{\bf k}'_{n'+1})
\left(\prod_{r=i,\tilde{s}-1}
S(\tilde{{\bf p}}_{r+1},\tilde{\mu}_{r+1};
\tilde{{\bf p}}_r,\tilde{\mu}_r)\right) 
\nonumber\\ &&\times
S(-\tilde{{\bf p}}_{i}-\hbar{{\bf k}}'_{n'+1},{\mu};
{{\bf p}}'_{ s'+1},{\mu}'_{ s'+1})  
\label{C.22}
\end{eqnarray}
The term corresponding to the 
majoration  of two occupation numbers is: 
\begin{eqnarray}
&&
<ss'\tilde s\tilde s'nn'|{\cal L}_{ID6} |s\,s'+1\,\tilde s\,\tilde s'+1\,n-1\,n'>
\nonumber\\ &&
=-\frac{1}{i\hbar}\hbar^3\sum_{\mu\mu'} (-1)^{s'}
\sum_{j=1,n} s(\nu_j)
\sqrt{\frac{\hbar}{2\varepsilon_0(2\pi)^3 \omega_{{\bf k}_{j}}}}
\nonumber\\ && \times 
\int d^3p \,
\alpha^{1\bar1}_{\nu_{j} +\mu \mu'}({\bf p},{\bf k}_j)
S(-{\bf p}-\frac12\hbar {\bf k}_j,\mu';
\tilde{{\bf p}}'_{\tilde{s}'+1},\tilde{\mu}'_{\tilde{s}'+1})
\nonumber\\ &&\times 
S({\bf p}-\frac12\hbar {\bf k}_j,\mu;{\bf p}'_{s'+1},\mu'_{s'+1})
\left(\prod_{r=j,n-1}S({\bf k}_{r+1},\nu_{r+1};{\bf k}_r,\nu_r)
\right) 
\nonumber\\ &&
\label{C.23}
\end{eqnarray}
We now consider the contributions due to the part ${\cal L}_{ID7}$  
of ${\cal L}_{ID}$ involving the product of destruction operators 
$bca$.
No contribution is diagonal in the 
occupation numbers $ ss'\tilde s\tilde s'$. 
The terms corresponding to the 
diminution of two occupation numbers are: 
\begin{eqnarray}
&&
<ss'\tilde s\tilde s'nn'|{\cal L}_{ID7} |s\,s'-1\,\tilde s\,\tilde s'-1\,n\,n'-1>
\nonumber\\ &&
=\frac{1}{i\hbar}
\sum_{i=1, s'}\sum_{j=1,\tilde s'}\sum_{l=1,n'}  (-1)^{i-1+\tilde s'-j}
\sqrt{\frac{\hbar}{2\varepsilon_0(2\pi)^3 
\omega_{{\bf k}'_l}}}
\alpha^{\bar11}_{\nu'_{l}  -\tilde{\mu}'_j\mu'_i}
(\frac12({\bf p}'_i-\tilde{{\bf p}}'_j),
-{\bf k}'_l)
\nonumber\\ &&\times 
\delta\left({\bf k}'_l+\frac1{\hbar}({\bf p}'_i+\tilde{{\bf p}}'_j)\right)
\left(\prod_{r=i,s'-1}
S({\bf p}'_{r+1},\mu'_{r+1};{\bf p}'_r,\mu'_r)\right) 
\nonumber\\ &&\times 
\left(\prod_{\tilde{r}=j,\tilde{s}'-1}
S(\tilde{{\bf p}}'_{\tilde{r}+1},\tilde{\mu}'_{\tilde{r}+1};
\tilde{{\bf p}}'_{\tilde{r}},\tilde{\mu}'_{\tilde{r}})\right)  
\left(\prod_{r=l,n-1}S({\bf k}'_{r+1},\nu'_{r+1};{\bf k}'_r,\nu'_r)
\right)
\label{C.24}
\end{eqnarray}
\begin{eqnarray}
&&
<ss'\tilde s\tilde s'nn'|{\cal L}_{ID7} |s\,s'-1\,\tilde s\,\tilde s'-1\,n+1\,n'>
\nonumber\\ &&
=\frac{1}{i\hbar}\sum_{\nu_{n+1}} 
\sum_{i=1, s'}\sum_{j=1,\tilde s'}  (-1)^{i-1+\tilde s'-j}
\sqrt{\frac{\hbar}{2\varepsilon_0(2\pi)^3 
\omega_{\frac{-1}{\hbar}({\bf p}'_i+\tilde{{\bf p}}'_j)}}}
\nonumber\\ &&\times
\alpha^{\bar11}_{\nu_{n+1} - \tilde{\mu}'_j\mu'_i}
(\frac12({\bf p}'_i-\tilde{{\bf p}}'_j),
\frac1{\hbar}({\bf p}_i+\tilde{{\bf p}}_j))
\left(\prod_{r=i,s'-1}
S({\bf p}'_{r+1},\mu'_{r+1};{\bf p}'_r,\mu'_r)\right) 
\nonumber\\ &&\times
\left(\prod_{\tilde{r}=j,\tilde{s}'-1}
S(\tilde{{\bf p}}'_{\tilde{r}+1},\tilde{\mu}'_{\tilde{r}+1};
\tilde{{\bf p}}'_{\tilde{r}},\tilde{\mu}'_{\tilde{r}})\right)  
S(-\frac1{\hbar}({\bf p}'_i+\tilde{{\bf p}}'_j),\nu_{n+1};
{\bf k}_{n+1},\nu_{n+1})
\nonumber\\ &&
\label{C.25}
\end{eqnarray}
The terms corresponding to a transfert of occupation numbers are:
\begin{eqnarray}
&&
<ss'\tilde s\tilde s'nn'|{\cal L}_{ID7} |s\,s'-1\,\tilde s+1\,\tilde s'\,n\,n'-1>
\nonumber\\ &&
=\frac{1}{i\hbar}\hbar^3\sum_{l=1,n'}     
\sum_{\mu'}
\sum_{i=1, s'}  (-1)^{i-1+\tilde s+\tilde s'}
\sqrt{\frac{\hbar}{2\varepsilon_0(2\pi)^3 
\omega_{{\bf k}'_l}}}
\nonumber\\ &&\times
\alpha^{\bar11}_{\nu'_{l}-\mu' \mu'_i}
({\bf p}'_i+\frac12\hbar{{\bf k}}'_{l},
-{\bf k}'_{l})
\left(\prod_{r=i,s'-1}
S({\bf p}'_{r+1},\mu'_{r+1};{\bf p}'_r,\mu'_r)\right) 
\nonumber\\ &&\times
S(-{{\bf p}}'_{i}-\hbar{{\bf k}}'_{l},{\mu}';
\tilde{{\bf p}}_{\tilde s+1},\tilde{\mu}_{\tilde s+1})  
\left(\prod_{r=l,n'-1}S({\bf k}'_{r+1},\nu'_{r+1};{\bf k}'_r,\nu'_r)
\right)
\nonumber\\ &&
\label{C.26}
\end{eqnarray}
\begin{eqnarray}
&&
<ss'\tilde s\tilde s'nn'|{\cal L}_{ID7} |s\,s'-1\,\tilde s+1\,\tilde s'\,n+1\,n'>
\nonumber\\ &&
=\frac{1}{i\hbar}\hbar^3\sum_{\nu_{n+1}}  \int d^3k_{n+1}   
\sum_{\mu'}
\sum_{i=1, s'}  (-1)^{i-1+\tilde s+\tilde s'}
\sqrt{\frac{\hbar}{2\varepsilon_0(2\pi)^3 
\omega_{{\bf k}_{n+1}}}}
\nonumber\\ &&\times
\alpha^{\bar11}_{\nu_{n+1}-\mu' \mu'_i}
({\bf p}'_i+\frac12\hbar{{\bf k}}_{n+1},
-{\bf k}_{n+1})
\left(\prod_{r=i,s'-1}
S({\bf p}'_{r+1},\mu'_{r+1};{\bf p}'_r,\mu'_r)\right) 
\nonumber\\ &&\times
S(-{{\bf p}}'_{i}-\hbar{{\bf k}}_{n+1},{\mu}';
\tilde{{\bf p}}_{\tilde s+1},\tilde{\mu}_{\tilde s+1})  
\label{C.27}
\end{eqnarray}
\begin{eqnarray}
&&
<ss'\tilde s\tilde s'nn'|{\cal L}_{ID7} |s+1\,s'\,\tilde s\,\tilde s'-1\,n\,n'-1>
\nonumber\\ &&
=\frac{1}{i\hbar}\hbar^3\sum_{l=1,n'}     
\sum_{\mu}
\sum_{i=1, \tilde s'}  (-1)^{i-1+ s+\tilde s}
\sqrt{\frac{\hbar}{2\varepsilon_0(2\pi)^3 
\omega_{{\bf k}'_l}}}
\nonumber\\ &&\times
\alpha^{\bar11}_{\nu_{l} -\tilde{\mu}'_i\mu}
(-\tilde{{\bf p}}'_i-\frac12\hbar{{\bf k}}'_{l},
-{\bf k}'_{l})
\left(\prod_{r=i,\tilde{s}'-1}
S(\tilde{{\bf p}}'_{r+1},\tilde{\mu}'_{r+1};
\tilde{{\bf p}}'_r,\tilde{\mu}'_r)\right) 
\nonumber\\ &&\times
S(-\tilde{{\bf p}}'_{i}-\hbar{{\bf k}}'_{l},{\mu};
{{\bf p}}_{ s+1},{\mu}_{ s+1})  
\left(\prod_{r=l,n'-1}S({\bf k}'_{r'+1},\nu'_{r'+1};{\bf k}'_r,\nu'_r)
\right)
\nonumber\\ &&
\label{C.28}
\end{eqnarray}
\begin{eqnarray}
&&
<ss'\tilde s\tilde s'nn'|{\cal L}_{ID7} |s+1\,s'\,\tilde s\,\tilde s'-1\,n+1\,n'>
\nonumber\\ &&
=\frac{1}{i\hbar}\hbar^3\sum_{\nu_{n+1}}  \int d^3k_{n+1}   
\sum_{\mu}
\sum_{i=1, \tilde{s}'}  (-1)^{i-1+ s+\tilde s}
\sqrt{\frac{\hbar}{2\varepsilon_0(2\pi)^3 
\omega_{{\bf k}_{n+1}}}}
\nonumber\\ &&\times
\alpha^{\bar11}_{\nu_{n+1} -\tilde{\mu}'_i \mu}
(-\tilde{{\bf p}}'_i-\frac12\hbar{{\bf k}}_{n+1},
-{\bf k}_{n+1})
\left(\prod_{r=i,\tilde{s}'-1}
S(\tilde{{\bf p}}'_{r+1},\tilde{\mu}'_{r+1};
\tilde{{\bf p}}'_r,\tilde{\mu}'_r)\right) 
\nonumber\\ &&\times
S(-\tilde{{\bf p}}'_{i}-\hbar{{\bf k}}_{n+1},{\mu};
{{\bf p}}_{ s+1},{\mu}_{ s+1})  
\label{C.29}
\end{eqnarray}
The term corresponding to the 
majoration  of two occupation numbers is: 
\begin{eqnarray}
&&
<ss'\tilde s\tilde s'nn'|{\cal L}_{ID7} |s+1\,s'\,\tilde s+1\,\tilde s'\,n\,n'-1>
\nonumber\\ &&
=\frac{1}{i\hbar}\hbar^3\sum_{\mu\mu'} (-1)^{s}
\sum_{j=1,n'} 
\sqrt{\frac{\hbar}{2\varepsilon_0(2\pi)^3 \omega_{{\bf k}'_{j}}}}
\nonumber\\ &&\times 
\int d^3p \,
\alpha^{\bar11}_{\nu'_{j} -\mu \mu'}({\bf p},-{\bf k}'_j)
S(-{\bf p}-\frac12\hbar {\bf k}'_j,\mu;
\tilde{{\bf p}}_{\tilde{s}+1},\tilde{\mu}_{\tilde{s}+1})
\nonumber\\ &&\times 
S({\bf p}-\frac12\hbar {\bf k}'_j,\mu';{\bf p}_{s+1},\mu_{s+1})
\left(\prod_{r=j,n'-1}S({\bf k}'_{r+1},\nu'_{r+1};{\bf k}'_r,\nu'_r)
\right)  
\nonumber\\ &&
\label{C.30}
\end{eqnarray}
We now consider the contributions due to the part ${\cal L}_{ID8}$  
of ${\cal L}_{ID}$ involving the product of creation and destruction operators 
$bca^+$.
No contribution is diagonal in the 
occupation numbers $ ss'\tilde s\tilde s'$. 
The term corresponding to the 
diminution of two occupation numbers is: 
\begin{eqnarray}
&&
<ss'\tilde s\tilde s'nn'|{\cal L}_{ID8} |s\,s'-1\,\tilde s\,\tilde s'-1\,n\,n'+1>
\nonumber\\ &&
=\frac{1}{i\hbar}\sum_{\nu'_{n'+1}} s(\nu'_{n'+1})
\sum_{i=1, s'}\sum_{j=1,\tilde s'}  (-1)^{i-1+\tilde s'-j}
\sqrt{\frac{\hbar}{2\varepsilon_0(2\pi)^3 
\omega_{\frac{1}{\hbar}({\bf p}'_i+\tilde{{\bf p}}'_j)}}}
\nonumber\\ &&\times 
\alpha^{\bar11}_{\nu'_{n'+1} +\tilde{\mu}'_j\mu'_i}
(\frac12({\bf p}'_i-\tilde{{\bf p}}'_j),
\frac1{\hbar}({\bf p}'_i+\tilde{{\bf p}}'_j))
\nonumber\\ &&\times 
\left(\prod_{r=i,s'-1}
S({\bf p}'_{r+1},\mu'_{r+1};{\bf p}'_r,\mu'_r)\right) 
\left(\prod_{\tilde{r}=j,\tilde{s}'-1}
S(\tilde{{\bf p}}'_{{\tilde{r}}+1},\tilde{\mu}'_{{\tilde{r}}+1};
\tilde{{\bf p}}'_{\tilde{r}},\tilde{\mu'_{\tilde{r}}})\right)  
\nonumber\\ &&\times 
S(\frac1{\hbar}({\bf p}'_i+\tilde{{\bf p}}'_j),\nu'_{n'+1};
{\bf k}'_{n'+1},\nu'_{n'+1})
\label{C.31}
\end{eqnarray}
The terms corresponding to a transfert of occupation numbers are:
\begin{eqnarray}
&&
<ss'\tilde s\tilde s'nn'|{\cal L}_{ID8} |s\,s'-1\,\tilde s+1\,\tilde s'\,n\,n'+1>
\nonumber\\ &&
=\frac{1}{i\hbar}\hbar^3\sum_{\nu'_{n'+1}} s(\nu'_{n'+1}) \int d^3k'_{n'+1}   
\sum_{\mu}
\sum_{i=1, s'}  (-1)^{i-1+\tilde s+\tilde s'}
\sqrt{\frac{\hbar}{2\varepsilon_0(2\pi)^3 
\omega_{{\bf k}'_{n'+1}}}}
\nonumber\\ &&\times 
\alpha^{\bar11}_{\nu'_{n'+1}+\mu \mu'_i}
({\bf p}'_i-\frac12\hbar{{\bf k}}'_{n'+1},
{\bf k}'_{n'+1})
\left(\prod_{r=i,s'-1}
S({\bf p}'_{r+1},\mu'_{r+1};{\bf p}'_r,\mu'_r)\right) 
\nonumber\\ &&\times 
S({-{\bf p}}'_{i}+\hbar{{\bf k}}'_{n'+1},{\mu};
\tilde{{\bf p}}_{\tilde s+1},\tilde{\mu}_{\tilde s+1})  
\label{C.32}
\end{eqnarray}
\begin{eqnarray}
&&
<ss'\tilde s\tilde s'nn'|{\cal L}_{ID8} |s+1\,s'\,\tilde s\,\tilde s'-1\,n\,n'+1>
\nonumber\\ &&
=\frac{1}{i\hbar}\hbar^3\sum_{\nu'_{n'+1}} s(\nu'_{n'+1}) \int d^3k'_{n'+1}   
\sum_{\mu'}
\sum_{i=1, \tilde s'}  (-1)^{i-1+ s+\tilde s}
\sqrt{\frac{\hbar}{2\varepsilon_0(2\pi)^3 
\omega_{{\bf k}'_{n'+1}}}}
\nonumber\\ &&\times 
\alpha^{\bar11}_{\nu'_{n'+1} +\tilde{\mu}'_i\mu'}
(-\tilde{{\bf p}}'_i+\frac12\hbar{{\bf k}}'_{n'+1},
{\bf k}'_{n'+1})
\left(\prod_{r=i,\tilde{s}'-1}
S(\tilde{{\bf p}}'_{r+1},\tilde{\mu}'_{r+1};
\tilde{{\bf p}}'_r,\tilde{\mu}'_r)\right) 
\nonumber\\ &&\times 
S(-\tilde{{\bf p}}'_{i}+\hbar{{\bf k}}'_{n'+1},{\mu'};
{{\bf p}}_{ s+1},{\mu_{ s+1}})  
\label{C.33}
\end{eqnarray}
The term corresponding to the 
majoration  of two occupation numbers is: 
\begin{eqnarray}
&&
<ss'\tilde s\tilde s'nn'|{\cal L}_{ID8} |s+1\,s'\,\tilde s+1\,\tilde s'\,n-1\,n'>
\nonumber\\ &&
=-\frac{1}{i\hbar}\hbar^3\sum_{\mu\mu'} (-1)^{s}
\sum_{j=1,n} s(\nu_j)
\sqrt{\frac{\hbar}{2\varepsilon_0(2\pi)^3 \omega_{{\bf k}_{j}}}}
\nonumber\\ &&\times 
\int d^3p \,
\alpha^{\bar11}_{\nu_{j} +\mu \mu'}({\bf p},{\bf k}_j)
S(-{\bf p}+\frac12\hbar {\bf k}_j,\mu;
\tilde{{\bf p}}_{\tilde{s}+1},\tilde{\mu}_{\tilde{s}+1})
\nonumber\\ &&\times 
S({\bf p}+\frac12\hbar {\bf k}_j,\mu';{\bf p}_{s+1},\mu_{s+1})
\left(\prod_{r=j,n-1}S({\bf k}_{r+1},\nu_{r+1};{\bf k}_r,\nu_r)
\right)  
\label{C.34}
\end{eqnarray}

\section{The diagonal second order kinetic operator $\Theta^{(2)}$ for one electron}  

\def\theequation{\thesection.\arabic{equation}}

\setcounter{equation}{0}

The operator $\Theta^{(2)}$ is the operator in variables position and canonical momentum 
(\ref{4.a10}).
We will be interested in its elements that connect state $|110000>$ to itself, with the same value for
the arguments of the functions.
These terms allow to illustrate obviously the problems associated with the self-energy interactions.
The basis for their computation is the relation (\ref{4.a11}).
The terms of interest to us are:
\begin{equation}
<110000 |\Theta^{(2)}|110000> =\frac1{i\hbar}<110000 |i\hbar P{\cal L}'QC^{(1)}|110000>
\label{A.a1}
\end{equation} 
The explicit forms of the last vertices are found in appendix C and have to be combined
with the expression of the $C^{(1)}$ operator.
As for all elements of the subdynamics operators, we begin from:
\begin{equation}
e^{{\cal L} t}=\frac{-1}{2\pi i} \int'_c dz\,e^{-izt}\frac1{z-i{\cal L}}
\label{4.1a}
\end{equation}
and we have to compute the residue arising from the vacuum states only to obtain $\Sigma(t)$.

Since at first order,
\begin{equation}
\left(\frac1{z-i{\cal L}}\right)^{(1)}
=\frac1{z-i{\cal L}_0}i{\cal L}'\frac1{z-i{\cal L}_0},
\label{4.2a}
\end{equation}
the first order creation operator has a very
simple form:
\begin{eqnarray}
&&<110010\mu_1 \mu'_1|C^{(1)} |110000\mu_2\mu'_2 >
\nonumber\\&&=
\frac{-1}{2\pi i} \int'_c dz\,
<110010\mu_1 \mu'_1|\frac1{z-i{\cal L}_0}i{\cal L}_{ID2}\frac1{z-i{\cal L}_0}
|110000\mu_2\mu'_2 >
\nonumber\\&&
\label{4.a62}
\end{eqnarray}
where the prime on the integral sign means that only the contribution of the pole corresponding to a
vacuum state has to be included in the integration path (here, only the second propagator satisfies
the criterion).
Only ${\cal L}_{ID2}$ provides a non-vanishing contribution.

Let us list all the required matrix elements:
The combination of vertices which provide an $\psi$ diagonal in the variables
are: (\ref{C.2b}/\ref{C.4}), (\ref{C.5b}/\ref{C.1}), (\ref{C.20}/\ref{C.28}), (\ref{C.27}/\ref{C.21}).
To each combination is associated in the usual way a diagram \cite{RB75}.
If we explicit the variables (in the ${\bf  K}$, ${\bf  P}$ representation), we need the following
expressions, written in Lorentz gauge, of appendix D  ($\nu$ takes four values: 
three values for $\lambda$ and  one value $s$).
From (\ref{C.2b}), (\ref{C.4}), (\ref{C.5b}), (\ref{C.1}), (\ref{C.20}), (\ref{C.28}), 
(\ref{C.27}), (\ref{C.21}), we
obtain
\begin{eqnarray}
&&
<{\bf  P}+\frac12\hbar {\bf  K},\mu;{\bf  P}-\frac12\hbar {\bf  K},\mu';0;0;0;0|
{\cal L}_{ID1}
\nonumber\\ &&
 |{\bf  P}+\frac12\hbar {\bf  K}-\hbar{\bf k},\bar\mu;
{\bf  P}-\frac12\hbar {\bf  K},\mu';0;0;{\bf k}, \nu;0>
\nonumber\\ &&
=-\frac{1}{i\hbar}\hbar^3
\sqrt{\frac{\hbar}{2\varepsilon_0(2\pi)^3 \omega_{{\bf k}}}}
\alpha^{11}_{\nu -\mu \bar\mu}
({\bf  P}+\frac12\hbar {\bf  K}-\frac12\hbar {\bf k},-{\bf k})
\label{A.a13}
\end{eqnarray}
\begin{eqnarray}
&&
<{\bf  P}+\frac12\hbar {\bf  K}-\hbar{\bf k},\bar\mu;
{\bf  P}-\frac12\hbar {\bf  K},\mu';0;0;{\bf k}, \nu;0|{\cal L}_{ID2} 
\nonumber\\ &&
|{\bf  P}+\frac12\hbar {\bf  K},\mu_1;{\bf  P}-\frac12\hbar {\bf  K},\mu';0;0;0;0>
\nonumber\\&&
=\frac{-1}{i\hbar}\hbar^3
s(\nu)
\sqrt{\frac{\hbar}{2\varepsilon_0(2\pi)^3 \omega_{{\bf k}}}}
\alpha^{11}_{\nu +\bar\mu \mu_1}
({\bf  P}+\frac12\hbar {\bf  K}-\frac12\hbar {\bf k},{\bf k})
\label{A.a14}
\end{eqnarray}
\begin{eqnarray}
&&
<{\bf  P}+\frac12\hbar {\bf  K},\mu;{\bf  P}-\frac12\hbar {\bf  K},\mu';0;0;0;0|
{\cal L}_{ID2}
\nonumber\\ && |
{\bf  P}+\frac12\hbar {\bf  K},\mu;
{\bf  P}-\frac12\hbar {\bf  K}-\hbar{\bf k},\bar\mu;0;0;0;{\bf k}, \nu>_b
\nonumber\\ && \qquad
=\frac{1}{i\hbar}\hbar^3 s(\nu)
\sqrt{\frac{\hbar}{2\varepsilon_0(2\pi)^3 \omega_{{\bf k}}}}
\alpha^{11}_{\nu+\bar\mu \mu'}
({\bf  P}-\frac12\hbar {\bf  K}-\frac12\hbar {\bf k},
{\bf k})
\label{A.a15}
\end{eqnarray}
\begin{eqnarray}
&&
<{\bf  P}+\frac12\hbar {\bf  K},\mu;
{\bf  P}-\frac12\hbar {\bf  K}-\hbar{\bf k},\bar\mu;0;0;0;{\bf k}, \nu
|{\cal L}_{ID1}
\nonumber\\ && |
{\bf  P}+\frac12\hbar {\bf  K},\mu;{\bf  P}-\frac12\hbar {\bf  K},\mu'_1;0;0;0;0>
\nonumber\\&&
=\frac{1}{i\hbar}\hbar^3
\sqrt{\frac{\hbar}{2\varepsilon_0(2\pi)^3 \omega_{{\bf k}}}}
\alpha^{11}_{\nu -\mu'_1 \bar\mu}
({\bf  P}-\frac12\hbar {\bf  K} -\frac12\hbar {\bf k},-{\bf k})
\label{A.a16}
\end{eqnarray}
\begin{eqnarray}
&&
<{\bf  P}+\frac12\hbar {\bf  K},\mu;
{\bf  P}-\frac12\hbar {\bf  K},\mu';0;0;0;0
|{\cal L}_{ID6}
\nonumber\\ && |
0;{\bf  P}-\frac12\hbar {\bf  K},\mu';0;
-{\bf  P}-\frac12\hbar {\bf  K}-\hbar{\bf k},\bar\mu;0;{\bf k}, \nu>
\nonumber\\ &&
=-\frac{1}{i\hbar}\hbar^3 s(\nu')    
\sqrt{\frac{\hbar}{2\varepsilon_0(2\pi)^3 
\omega_{{\bf k}}}}
\alpha^{1\bar1}_{\nu' +\mu \bar\mu}
({\bf  P}+\frac12\hbar {\bf  K}+\frac12\hbar{{\bf k}},
{\bf k})
\label{A.a17}
\end{eqnarray}
\begin{eqnarray}
&&
<0;{\bf  P}-\frac12\hbar {\bf  K},\mu';0;
-{\bf  P}-\frac12\hbar {\bf  K}-\hbar{\bf k},\bar\mu;0;{\bf k}, \nu|
{\cal L}_{ID7} 
\nonumber\\ &&
|{\bf  P}+\frac12\hbar {\bf  K},\mu_1;
{\bf  P}-\frac12\hbar {\bf  K},\mu';0;0;0;0 >
\nonumber\\ && \qquad
=\frac{1}{i\hbar}\hbar^3     
\sqrt{\frac{\hbar}{2\varepsilon_0(2\pi)^3 
\omega_{{\bf k}}}}
\alpha^{\bar11}_{\nu -\bar\mu\mu_1}
({\bf  P}+\frac12\hbar {\bf  K}+\frac12\hbar{{\bf k}},
-{\bf k})
\label{A.a18}
\end{eqnarray}
\begin{eqnarray}
&&
<{\bf  P}+\frac12\hbar {\bf  K},\mu;
{\bf  P}-\frac12\hbar {\bf  K},\mu';0;0;0;0 
|{\cal L}_{ID7}
\nonumber\\ && |
{\bf  P}+\frac12\hbar {\bf  K},\mu;0;
-{\bf  P}+\frac12\hbar {\bf  K}-\hbar{\bf k},\bar\mu;0;{\bf k}, \nu;0>
\nonumber\\ &&
=\frac{1}{i\hbar}\hbar^3   
\sqrt{\frac{\hbar}{2\varepsilon_0(2\pi)^3 
\omega_{{\bf k}}}}
\alpha^{\bar11}_{\nu-\bar\mu \mu'}
({\bf  P}-\frac12\hbar {\bf  K}+\frac12\hbar{{\bf k}},
-{\bf k})
\label{A.a19}
\end{eqnarray}
\begin{eqnarray}
&&
<{\bf  P}+\frac12\hbar {\bf  K},\mu;0;
-{\bf  P}+\frac12\hbar {\bf  K}-\hbar{\bf k},\bar\mu;0;{\bf k}, \nu;0
|{\cal L}_{ID6}
\nonumber\\ && |
{\bf  P}+\frac12\hbar {\bf  K},\mu;
{\bf  P}-\frac12\hbar {\bf  K},\mu'_1;0;0;0;0  >
\nonumber\\ && \qquad
=-\frac{1}{i\hbar}\hbar^3 s(\nu)    
\sqrt{\frac{\hbar}{2\varepsilon_0(2\pi)^3 
\omega_{{\bf k}}}}
\alpha^{1\bar1}_{\nu +\mu'_1 \bar\mu}
({\bf  P}-\frac12\hbar {\bf  K}+\frac12\hbar{{\bf k}},
{\bf k})
\label{A.a20}
\end{eqnarray}
Equations (5.39) provide the relations between the $\alpha$'s and the $\delta$'s, for
$\nu\not=s$

The denominators which appear in the four diagrams are, after the integration over $z$ has been
performed:
\begin{eqnarray}
&&P_{\ref{C.2b}/\ref{C.4}}=\frac1{i\varepsilon- E_{|{\bf  P}+\frac12\hbar {\bf  K}-\hbar{\bf k}|}
+ E_{|{\bf  P}+\frac12\hbar {\bf  K}|}-\hbar\omega_k}
\label{A.a21a}\\
&&P_{\ref{C.5b}/\ref{C.1}}=\frac1{i\varepsilon+ E_{|{\bf  P}-\frac12\hbar {\bf  K}-\hbar{\bf k}|}
- E_{|{\bf  P}-\frac12\hbar {\bf  K}|}+\hbar\omega_k} 
\label{A.a22a}\\
&&P_{\ref{C.20}/\ref{C.28}}=\frac1{i\varepsilon+ E_{|{\bf  P}+\frac12\hbar {\bf  K}+\hbar{\bf k}|}
+ E_{|{\bf  P}+\frac12\hbar {\bf  K}|}+\hbar\omega_k}   
\label{A.a23a}r\\
&&P_{\ref{C.27}/\ref{C.21}}=\frac1{i\varepsilon- E_{|{\bf  P}-\frac12\hbar {\bf  K}+\hbar{\bf k}|}
- E_{|{\bf  P}-\frac12\hbar {\bf  K}|}-\hbar\omega_k}   
\label{A.a24a}
\end{eqnarray}
In all propagators, the $i\varepsilon$'s play no role since the 
denominators never vanish.

Inserting the value of the different elements, we obtain for the expression of the relevant elements
of $C^{(1)}$:
\begin{eqnarray}
&&<110010\mu_1 \mu'_1|C^{(1)} |110000\mu_2\mu'_2 >
\nonumber\\&&=
(-)\hbar^2
\sqrt{\frac{\hbar}{2\varepsilon_0(2\pi)^3 \omega_{{\bf k}_1} }}
\alpha^{11}_{\lambda_1+ \mu_1 \mu_2}
({\bf p}+\frac12\hbar {\bf k}_1,{\bf k}_1)
\delta^{Kr}_{\mu'_1,\mu'_2}
\nonumber\\&&\times
\frac{-1}{2\pi i} \int'_c dz\,
\frac1{z-\frac1\hbar E_{{\bf p}}+\frac1\hbar E_{{\bf p}'}-\omega_{k_1}}
S({\bf p}+\hbar {\bf k}_1;{\bf p})
\frac1{z-\frac1\hbar E_{{\bf p}}+\frac1\hbar E_{{\bf p}'}}
\nonumber\\&&=
(-)\hbar^2
\sqrt{\frac{\hbar}{2\varepsilon_0(2\pi)^3 \omega_{{\bf k}_1} }}
\alpha^{11}_{\lambda_1+ \mu_1 \mu_2}
({\bf p}+\frac12\hbar {\bf k}_1,{\bf k}_1)
\delta^{Kr}_{\mu'_1,\mu'_2}
\nonumber\\&&\times
\frac{-1}{2\pi i} \int'_c dz\,
\frac1{z-\frac1\hbar E_{{\bf p}}+\frac1\hbar E_{{\bf p}'}-\omega_{k_1}}
\frac1{z-\frac1\hbar E_{{\bf p}+\hbar {\bf k}_1}+\frac1\hbar E_{{\bf p}'}}
S({\bf p}+\hbar {\bf k}_1;{\bf p})
\nonumber\\&&=
-\hbar^3
\sqrt{\frac{\hbar}{2\varepsilon_0(2\pi)^3 \omega_{{\bf k}_1} }}
\alpha^{11}_{\lambda_1+ \mu_1 \mu_2}
({\bf p}+\frac12\hbar {\bf k}_1,{\bf k}_1)
\delta^{Kr}_{\mu'_1,\mu'_2}
\nonumber\\&&\times
\frac1{E_{{\bf p}+\hbar {\bf k}_1}-E_{{\bf p}}-\hbar\omega_{k_1}}
S({\bf p}+\hbar {\bf k}_1;{\bf p})
\label{4.a63}
\end{eqnarray}
It can easily be realized that $E_{{\bf p}+\hbar {\bf k}_1}-E_{{\bf p}}-\hbar\omega_{k_1}$ cannot
vanish, so that no prescription involving a $i\varepsilon$ is required.
Similarly,
\begin{eqnarray}
&&<110001\mu_1 \mu'_1|C^{(1)} |110000\mu_2\mu'_2 >
\nonumber\\&&=
\hbar^3
\sqrt{\frac{\hbar}{2\varepsilon_0(2\pi)^3 \omega_{{\bf k}_1} }}
\alpha^{11}_{\lambda'_1- \mu'_2 \mu'_1}
({\bf p}'+\frac12\hbar {\bf k}'_1,-{\bf k}'_1)
\delta^{Kr}_{\mu_1,\mu_2}
\nonumber\\&&\times
\frac1{-E_{{\bf p}'+\hbar {\bf k}'_1}+E_{{\bf p}'}+\hbar\omega_{k'_1}}
S({\bf p}'+\hbar {\bf k}'_1;{\bf p}')
\label{4.a64}
\end{eqnarray}
\begin{eqnarray}
&&<010101\mu_1 \mu'_1|C^{(1)} |110000\mu_2\mu'_2>
\nonumber\\&&=
\hbar^3
\sqrt{\frac{\hbar}{2\varepsilon_0(2\pi)^3 \omega_{{\bf k}_1} }}
\alpha^{\bar11}_{\lambda'_1- \mu'_1\mu_2 }
(-\tilde{{\bf p}}'-\frac12\hbar{{\bf k}}'_1,
-{\bf k}'_1)
\delta^{Kr}_{\mu_1,\mu'_2}
\nonumber\\&&\times
\frac1{E_{-\tilde{{\bf p}}'-\hbar{{\bf k}}'_1}+ E_{\tilde{{\bf p}}'}+\hbar\omega_{k'_1}}
S(-\tilde{{\bf p}}'-\hbar{{\bf k}}'_1;{{\bf p}})  
\label{4.a65}
\end{eqnarray}
\begin{eqnarray}
&&<101010\mu_1 \mu'_1|C^{(1)} |110000\mu_2\mu'_2>
\nonumber\\&&=
-\hbar^3
\sqrt{\frac{\hbar}{2\varepsilon_0(2\pi)^3 \omega_{{\bf k}_1} }}
\alpha^{1\bar1}_{\lambda_1+ \mu'_2\mu'_1}
(-\tilde{{\bf p}}-\frac12\hbar{{\bf k}}_1,{\bf k}_1)
\delta^{Kr}_{\mu_1,\mu_2}
\nonumber\\&&\times
\frac1{-E_{-\tilde{{\bf p}}-\hbar{{\bf k}}_1}-E_{\tilde{{\bf p}}}-\hbar\omega_{k_1}}
S(-\tilde{{\bf p}}-\hbar{{\bf k}}_1;{{\bf p}}')  
\label{4.a66}
\end{eqnarray}

Let us  now focus on the spin dependence of the diagrams which is provided
by the product of the vertices only: the propagators play no role 
in that respect. 
We take into account the expression of the $\alpha$'s 
for the computation of the  spin dependence
of the diagrams and we  introduce new functions  $S$'s.
Their definition takes into account the compensation of the factor $\frac1{i\hbar}$ in 
(\ref{A.a13}-\ref{A.a20})
by the factor $i\hbar$ present in (\ref{A.a1}).
In a similar way, the factor $\hbar^3$ in (\ref{A.a13}-\ref{A.a20}) is compensated by the factor
$\frac1{\hbar^3}$ in the $\alpha$'s.
The factors arising from the $\alpha$'s are  taken into account 
in the expressions of the $S$'s.
We keep explicit the $s(\nu)$ factor.
\begin{eqnarray}
&&S_{\ref{C.2b}/\ref{C.4}}=\sum_{\bar \mu\nu} s(\nu)
(a_{\ref{C.2b}\nu} I_{\mu \bar\mu} +{\bf  b}_{\ref{C.2b}\nu}.\pmb{$\sigma$}_{\mu \bar\mu})
(a_{\ref{C.4}\nu}' I_{\bar\mu \mu_1} +{\bf  b}_{\ref{C.4}\nu}'.\pmb{$\sigma$}_{\bar\mu \mu_1})
\label{A.a28} \nonumber\\ &&
= \sum_{\nu} s(\nu)\left[
(a_{2b\nu} a_{4\nu}'+{\bf  b}_{2b\nu}.{\bf  b}_{4\nu}') I_{\mu \mu_1}
+(a_{\ref{C.2b}\nu}{\bf  b}_{\ref{C.4}\nu}' +a_{\ref{C.4}\nu}'{\bf  b}_{\ref{C.2b}\nu} 
\right.\nonumber\\ &&\left.
+i{\bf  b}_{\ref{C.2b}\nu}\pmb{$\times$} {\bf  b}_{\ref{C.4}\nu}').
\pmb{$\sigma$}_{\mu \mu_1} \right]
\label{A.a29}
\end{eqnarray}
The coefficient of  the spin identity operator will be called $S_{\ref{C.2b}/\ref{C.4}}^s$
and that of the Pauli matrix  ${\bf  S}_{\ref{C.2b}/\ref{C.4}}^v$. 
In a similar way, we have:
\begin{eqnarray}
&&S_{\ref{C.5b}/\ref{C.1}}=\sum_{\bar \mu\nu} s(\nu)
(a_{\ref{C.5b}\nu} I_{ \bar\mu\mu'} +{\bf  b}_{\ref{C.5b}\nu}.\pmb{$\sigma$}_{ \bar\mu\mu'})
(a_{\ref{C.1}\nu}' I_{\mu'_1\bar\mu} +{\bf  b}_{\ref{C.1}\nu}'.\pmb{$\sigma$}_{\mu'_1\bar\mu})
\label{A.a30} \nonumber\\ &&
= \sum_{\nu} s(\nu) \left[
(a_{\ref{C.5b}\nu} a_{\ref{C.1}\nu}'+{\bf  b}_{\ref{C.5b}\nu}.{\bf  b}_{\ref{C.1}\nu}') I_{\mu'_1 \mu'}
\right.\nonumber\\ &&\left.
+(a_{\ref{C.5b}\nu}{\bf  b}_{1\nu}' +a_{\ref{C.1}\nu}'{\bf  b}_{\ref{C.5b}\nu} 
-i{\bf  b}_{\ref{C.5b}\nu}\pmb{$\times$} {\bf  b}_{\ref{C.1}\nu}').
\pmb{$\sigma$}_{\mu'_1 \mu'} \right]
\label{A.a31}
\end{eqnarray}
For the last two diagrams (\ref{C.20}/\ref{C.28}) and (\ref{C.27}/\ref{C.21}), it ought a priori to be taken 
into account that the label of the argument present 
in the spin operator (present in the expression of the $\alpha$'s) 
requires a spin component opposite  to the index of $\alpha$.
Since we have to perform an summation over the spin variables, 
this point does not play a role and the contribution of the two last diagrams
is similar to that of the first two diagrams. 

We will see that only the scalar parts  $S_{\ref{C.2b}/\ref{C.4}}^s$,  $S_{\ref{C.5b}/\ref{C.1}}^s$,
$S_{\ref{C.20}/\ref{C.28}}^s$,  $S_{\ref{C.27}/\ref{C.21}}^s$ of the $S$'s do not vanish.   
From the expression of the coefficients $\alpha$, we obtain by direct 
identification, taking into account the arguments required in  (\ref{A.a13}-\ref{A.a20}):
(we shall consider successively the expression with the scalar and with the 
vectorial photon.)
\begin{eqnarray}
&&\alpha_{s\,\mu \,\mu'}^{11}({\bf p},{\bf k}) 
=-\frac{q_e c}{\hbar^3} \left(
\cos \varphi_{{{\bf p}-\frac12\hbar{\bf k}} , {{\bf p}+\frac12\hbar{\bf k}}} 
I_{\mu\mu'}
\right.\nonumber\\ &&\left.
+ i\frac{\hbar( {{\bf p}}\pmb{$\times$}{{\bf k}})}
{|{\bf p}-\frac12\hbar{\bf k}||{\bf p}+\frac12\hbar{\bf k}|}
. \pmb{$\sigma$}_{\mu\mu'}
\sin\frac{ \theta_{|{\bf p}-\frac12\hbar{\bf k}|}}2
\sin \frac{ \theta_{|{\bf p}+\frac12\hbar{\bf k}|}}2  
\right)
\nonumber\\
&&\label{A.a32}\nonumber\\ 
&&\alpha_{s\,\mu \,\mu'}^{1\bar1}({\bf p},{\bf k})  
=\frac{q_e c}{\hbar^3}\left[
\left(\pmb{$\sigma$}_{\mu\bar\mu'}.
\frac {{\bf p}+\frac12\hbar{\bf k}}{|{\bf p}+\frac12\hbar{\bf k}|}
\right)
\cos\frac{ \theta_{|{\bf p}-\frac12\hbar{\bf k}|}}2
\sin \frac{ \theta_{|{\bf p}+\frac12\hbar{\bf k}|}}2  
\right. \nonumber\\ && \left.\qquad 
-\left( \pmb{$\sigma$}_{\mu\bar\mu'}.
\frac{{\bf p}-\frac12\hbar{\bf k}}{|{\bf p}-\frac12\hbar{\bf k}|} 
\right)
\cos\frac{ \theta_{|{\bf p}+\frac12\hbar{\bf k}|}}2
\sin \frac{ \theta_{|{\bf p}-\frac12\hbar{\bf k}|}}2
\right] 
\label{A.a33}\\ 
&&\alpha_{s\,\mu \,\mu'}^{\bar1 1}({\bf p},{\bf k}) 
=-\alpha_{s\,\bar\mu \,\bar\mu'}^{1\bar1}({\bf p},{\bf k})   
\label{A.a34}
\end{eqnarray}
with the convention $\bar\mu'=-\mu'$.
From those expressions, we deduce, taking into account the value of 
the arguments of the $\alpha$'s in (\ref{A.a13}-\ref{A.a20}):
\begin{eqnarray}
&&a_{\ref{C.2b}s}  
=-q_ec 
\cos \varphi_{{{\bf  P}+\frac12\hbar{\bf  K}-\hbar{\bf k}} , 
{{\bf  P}+\frac12\hbar{\bf  K}}} 
\label{A.a35}\\&&
{\bf  b}_{\ref{C.2b}s}
=-iq_ec   
 \frac{\hbar(( {{\bf  P}+\frac12\hbar{\bf  K}})\pmb{$\times$}{(-{\bf k})})}
{|{\bf  P}+\frac12\hbar{\bf  K}-\hbar{\bf k}||{\bf  P}+\frac12\hbar{\bf  K}|}
\sin\frac{ \theta_{|{\bf  P}+\frac12\hbar{\bf  K}-\hbar{\bf k}|}}2
\sin \frac{ \theta_{|{\bf  P}+\frac12\hbar{\bf  K}|}}2
\nonumber\\ &&  
\label{A.a36}
\end{eqnarray}
\begin{eqnarray}
&&a'_{\ref{C.4} s}  
=a_{\ref{C.2b}s} 
\label{A.a37}\\
&&{\bf  b}'_{\ref{C.4} s}
=- {\bf  b}_{\ref{C.2b}s}
\label{A.a38}
\end{eqnarray}
The contribution to the scalar part $S_{\ref{C.2b}/\ref{C.4}}^s$  becomes therefore:
\begin{eqnarray}
&&(a_{\ref{C.2b}s} a_{\ref{C.4}s}'+{\bf  b}_{\ref{C.2b}s}.{\bf  b}_{\ref{C.4}s}')
=q^2_ec^2 
\cos^2 \varphi_{{{\bf  P}+\frac12\hbar{\bf  K}-\hbar{\bf k}} , 
{{\bf  P}+\frac12\hbar{\bf  K}}}
\nonumber\\ &&
+q^2_ec^2 \hbar^2 
\frac{
k^2 |{\bf  P}+\frac12\hbar{\bf  K}|^2
-(( {{\bf  P}+\frac12\hbar{\bf  K}}).{{\bf k}})^2}
{|{\bf  P}+\frac12\hbar{\bf  K}-\hbar{\bf k}|^2|{\bf  P}+\frac12\hbar{\bf  K}|^2}
\sin^2\frac{ \theta_{|{\bf  P}+\frac12\hbar{\bf  K}-\hbar{\bf k}|}}2
\sin^2\frac{ \theta_{|{\bf  P}+\frac12\hbar{\bf  K}|}}2
\nonumber\\ &&
\label{A.a39}
\end{eqnarray}
while the contribution to the vectorial part clearly vanishes considering
(\ref{A.a37}-\ref{A.a38}).

The corresponding contribution to the scalar part  $S_{\ref{C.5b}/\ref{C.1}}^s$  
can be obtained by changing the sign of ${\bf  K}$:
\begin{eqnarray}
&&(a_{\ref{C.5b}s} a_{\ref{C.1}s}'+{\bf  b}_{\ref{C.5b}s}.{\bf  b}_{\ref{C.1}s}')
=q^2_ec^2 
\cos^2 \varphi_{{{\bf  P}-\frac12\hbar{\bf  K}-\hbar{\bf k}} , 
{{\bf  P}-\frac12\hbar{\bf  K}}}
\nonumber\\ &&
+q^2_ec^2 \hbar^2 
\frac{
k^2 |{\bf  P}-\frac12\hbar{\bf  K}|^2
-(( {{\bf  P}-\frac12\hbar{\bf  K}}).{{\bf k}})^2}
{|{\bf  P}-\frac12\hbar{\bf  K}-\hbar{\bf k}|^2|{\bf  P}-\frac12\hbar{\bf  K}|^2}
\sin^2\frac{ \theta_{|{\bf  P}-\frac12\hbar{\bf  K}-\hbar{\bf k}|}}2
\sin^2\frac{ \theta_{|{\bf  P}-\frac12\hbar{\bf  K}|}}2 
\nonumber\\ &&\label{A.a40}
\end{eqnarray}
while the contribution to the vectorial part clearly vanishes.

The computation of the contributions due to the combination of 
vertices (\ref{C.20}/\ref{C.28}) and (\ref{C.27}/\ref{C.21}) requires the expressions:
\begin{equation}
a_{\ref{C.20}s}=a_{\ref{C.27}s}=a'_{\ref{C.28}s}=a'_{\ref{C.21}s}=0
\label{A.a41}
\end{equation}
and:
\begin{eqnarray}
&&{\bf  b}_{\ref{C.20}s}=q_ec
\left[
\frac {{\bf  P}+\frac12\hbar{\bf  K}+\hbar{\bf k}}
{|{\bf  P}+\frac12\hbar{\bf  K}+\hbar{\bf k}|}
\cos\frac{ \theta_{|{\bf  P}+\frac12\hbar{\bf  K}|}}2
\sin \frac{ \theta_{|{\bf  P}+\frac12\hbar{\bf  K}+\hbar{\bf k}|}}2  
\right. \nonumber\\ && \left.\qquad 
-\frac{{\bf  P}+\frac12\hbar{\bf  K}}{|{\bf  P}+\frac12\hbar{\bf  K}|} 
\cos\frac{ \theta_{|{\bf  P}+\frac12\hbar{\bf  K}+\hbar{\bf k}|}}2
\sin \frac{ \theta_{|{\bf  P}+\frac12\hbar{\bf  K}|}}2
\right]
\label{A.a42}
\end{eqnarray}
\begin{equation}
{\bf  b}'_{\ref{C.28}s} ={\bf  b}_{\ref{C.20}s}
\label{A.a43}
\end{equation}
\begin{eqnarray}
&&{\bf  b}_{27s}=q_ec\left[
\frac {{\bf  P}-\frac12\hbar{\bf  K}+\hbar{\bf k}}
{|{\bf  P}-\frac12\hbar{\bf  K}+\hbar{\bf k}|}
\cos\frac{ \theta_{|{\bf  P}-\frac12\hbar{\bf  K}|}}2
\sin \frac{ \theta_{|{\bf  P}-\frac12\hbar{\bf  K}+\hbar{\bf k}|}}2  
\right. \nonumber\\ && \left.\qquad 
-\frac{{\bf  P}-\frac12\hbar{\bf  K}}{|{\bf  P}-\frac12\hbar{\bf  K}|} 
\cos\frac{ \theta_{|{\bf  P}-\frac12\hbar{\bf  K}+\hbar{\bf k}|}}2
\sin \frac{ \theta_{|{\bf  P}-\frac12\hbar{\bf  K}|}}2
\right]
\label{A.a44}
\end{eqnarray}
\begin{equation}
{\bf  b}'_{\ref{C.21}s} ={\bf  b}_{\ref{C.27}s}
\label{A.a45}
\end{equation}
The contribution to the scalar part of $S_{\ref{C.20}/\ref{C.28}}^s$ and $S_{\ref{C.27}/\ref{C.21}}^s$ 
requires the computation of:
\begin{eqnarray}
&&{\bf  b}_{\ref{C.20}s}.{\bf  b}'_{\ref{C.28}s}=q_e^2c^2\left( 
\cos^2\frac{ \theta_{|{\bf  P}+\frac12\hbar{\bf  K}|}}2
\sin ^2\frac{ \theta_{|{\bf  P}+\frac12\hbar{\bf  K}+\hbar{\bf k}|}}2
\right.\nonumber\\&&\left. 
+\cos^2\frac{ \theta_{|{\bf  P}+\frac12\hbar{\bf  K}+\hbar{\bf k}|}}2
\sin^2 \frac{ \theta_{|{\bf  P}+\frac12\hbar{\bf  K}|}}2
-\frac12\frac {{\bf  P}+\frac12\hbar{\bf  K}+\hbar{\bf k}}
{|{\bf  P}+\frac12\hbar{\bf  K}+\hbar{\bf k}|}
.\frac{{\bf  P}+\frac12\hbar{\bf  K}}{|{\bf  P}+\frac12\hbar{\bf  K}|} 
\right.\nonumber\\&&\left.\times
\sin  \theta_{|{\bf  P}+\frac12\hbar{\bf  K}|} 
\sin  \theta_{|{\bf  P}+\frac12\hbar{\bf  K}+\hbar{\bf k}|} 
\right)
\label{A.a46}
\end{eqnarray}
\begin{eqnarray}
&&{\bf  b}_{\ref{C.27}s}.{\bf  b}'_{\ref{C.21}s}=q_e^2c^2\left( 
\cos^2\frac{ \theta_{|{\bf  P}-\frac12\hbar{\bf  K}|}}2
\sin ^2\frac{ \theta_{|{\bf  P}-\frac12\hbar{\bf  K}+\hbar{\bf k}|}}2
\right.\nonumber\\&&\left. 
+\cos^2\frac{ \theta_{|{\bf  P}-\frac12\hbar{\bf  K}+\hbar{\bf k}|}}2
\sin^2 \frac{ \theta_{|{\bf  P}-\frac12\hbar{\bf  K}|}}2
-\frac12\frac {{\bf  P}-\frac12\hbar{\bf  K}+\hbar{\bf k}}
{|{\bf  P}-\frac12\hbar{\bf  K}+\hbar{\bf k}|}
.\frac{{\bf  P}-\frac12\hbar{\bf  K}}{|{\bf  P}+\frac12\hbar{\bf  K}|} 
\right.\nonumber\\&&\left.\times
\sin  \theta_{|{\bf  P}-\frac12\hbar{\bf  K}|} 
\sin  \theta_{|{\bf  P}-\frac12\hbar{\bf  K}+\hbar{\bf k}|} 
\right)
\label{A.a47}
\end{eqnarray}
Anew, the contribution to the vectorial parts of $S$ vanishes.

We now consider the contributions involving the vectorial photon.
It is useful to write the expressions of the $\alpha$ in the following way,
in order to allow a more direct identification of the $a$'s 
and the ${\bf  b}$'s:

\begin{eqnarray}
&&\alpha_{\lambda+\,\mu \,\mu'}^{11}({\bf  P},{\bf k}) 
=\frac{q_e c}{\hbar^3}{\bf  e}_{{\bf k}\lambda}.
\left[
\left( \frac {{\bf  P}+\frac12\hbar {\bf k}}{|{\bf  P}+\frac12\hbar {\bf k}|}
\cos\frac{ \theta_{|{\bf  P}-\frac12\hbar {\bf k}|}}2   
\sin \frac{ \theta_{|{\bf  P}+\frac12\hbar {\bf k}|}}2 
\right.\right. \nonumber\\ && \left. \left.
+\frac {{\bf  P}-\frac12\hbar {\bf k}}{|{\bf  P}-\frac12\hbar {\bf k}|}    
\sin \frac{ \theta_{|{\bf  P}-\frac12\hbar {\bf k}|}}2 
\cos\frac{ \theta_{{{\bf  P}+\frac12\hbar {\bf k}}}}2 
\right)
 I_{\mu\, \mu'}
\right] \nonumber\\ &&
+i\frac{q_e c}{\hbar^3} \,\pmb{$\sigma$}_{\mu\, \mu'}.\left[  
\left({\bf  e}_{{\bf k}\lambda} 
\pmb{$\times$}
\frac{{\bf  P}+\frac12\hbar {\bf k}}{|{\bf  P}+\frac12\hbar {\bf k}|} 
\right)
\cos\frac{ \theta_{|{\bf  P}-\frac12\hbar {\bf k}|}}2  
\sin \frac{ \theta_{|{{\bf  P}+\frac12\hbar {\bf k}}|}}2
\right. \nonumber\\ &&  \left.
- \left( {\bf  e}_{{\bf k}\lambda}\pmb{$\times$}
\frac {{\bf  P}-\frac12\hbar {\bf k}}{|{\bf  P}-\frac12\hbar {\bf k}|}   
\right) 
\sin \frac{ \theta_{|{\bf  P}-\frac12\hbar {\bf k}|}}2 
\cos\frac{ \theta_{|{{\bf  P}+\frac12\hbar {\bf k}}|}}2 
\right] 
\label{A.a48}
\end{eqnarray}
\begin{eqnarray}
&&\alpha_{\lambda+\,\mu \,\mu'}^{1\bar1}({\bf  P},{\bf k}) 
=-i\frac{q_e c}{\hbar^3} I_{\mu\,\bar\mu'} {\bf  e}_{{\bf k}\lambda}.
\left(\frac {\hbar{\bf k}}{|{\bf  P}+\frac12\hbar{\bf k}|}\pmb{$\times$}
\frac {{\bf  P}}{|{\bf  P}-\frac12\hbar{\bf k}|}\right)
\nonumber\\ &&\times
\sin \frac{ \theta_{|{\bf  P}-\frac12\hbar{\bf k}|}}2 
\sin \frac{ \theta_{|{\bf  P}+\frac12\hbar{\bf k}|}}2
\nonumber\\ &&
+\frac{q_e c}{\hbar^3}\, \pmb{$\sigma$}_{\mu\,\bar\mu'}.
\left[ 
{\bf  e}_{{\bf k}\lambda}
\cos\varphi_{{{\bf  P}-\frac12\hbar{\bf k}} , {{\bf  P}+\frac12\hbar{\bf k}}} 
\right.\nonumber\\&& \left.
- 
\left[\left({\bf  e}_{{\bf k}\lambda}.
\frac {{\bf  P}-\frac12\hbar{\bf k}}{|{\bf  P}-\frac12\hbar{\bf k}|}
\right)
\frac {{\bf  P}+\frac12\hbar{\bf k}}{|{\bf  P}+\frac12\hbar{\bf k}|} 
+\left({\bf  e}_{{\bf k}\lambda}.
\frac {{\bf  P}+\frac12\hbar{\bf k}}{|{\bf  P}+\frac12\hbar{\bf k}|} 
\right)
\frac {{\bf  P}-\frac12\hbar{\bf k}}{|{\bf  P}-\frac12\hbar{\bf k}|}
\right]
\right.\nonumber\\&& \left.\times 
\sin \frac{ \theta_{|{\bf  P}-\frac12\hbar{\bf k}|}}2 
\sin \frac{ \theta_{|{\bf  P}+\frac12\hbar{\bf k}|}}2 
\right] 
\label{A.a49}
\end{eqnarray}
\begin{eqnarray}
&&\alpha_{\lambda+\,\mu \,\mu'}^{\bar1 1}({\bf  P},{\bf k}) 
=-i\frac{q_e c}{\hbar^3}I_{\bar\mu\,\mu'} {\bf  e}_{{\bf k}\lambda}.
\left(\frac {\hbar{\bf k}}{|{\bf  P}+\frac12\hbar{\bf k}|}\pmb{$\times$}
\frac {{\bf  P}}{|{\bf  P}-\frac12\hbar{\bf k}|}\right)
\nonumber\\ &&\times
\sin \frac{ \theta_{|{\bf  P}-\frac12\hbar{\bf k}|}}2 
\sin \frac{ \theta_{|{\bf  P}+\frac12\hbar{\bf k}|}}2
\nonumber\\ &&
+\frac{q_e c}{\hbar^3}\, \pmb{$\sigma$}_{\bar\mu\,\mu'}.
\left[ 
{\bf  e}_{{\bf k}\lambda}
\cos\varphi_{{{\bf  P}-\frac12\hbar{\bf k}} , {{\bf  P}+\frac12\hbar{\bf k}}} 
\right.\nonumber\\&& \left.
- 
\left[\left({\bf  e}_{{\bf k}\lambda}.
\frac {{\bf  P}-\frac12\hbar{\bf k}}{|{\bf  P}-\frac12\hbar{\bf k}|}
\right)
\frac {{\bf  P}+\frac12\hbar{\bf k}}{|{\bf  P}+\frac12\hbar{\bf k}|} 
+\left({\bf  e}_{{\bf k}\lambda}.
\frac {{\bf  P}+\frac12\hbar{\bf k}}{|{\bf  P}+\frac12\hbar{\bf k}|} 
\right)
\frac {{\bf  P}-\frac12\hbar{\bf k}}{|{\bf  P}-\frac12\hbar{\bf k}|}
\right]
\right.\nonumber\\&& \left.\times 
\sin \frac{ \theta_{|{\bf  P}-\frac12\hbar{\bf k}|}}2 
\sin \frac{ \theta_{|{\bf  P}+\frac12\hbar{\bf k}|}}2 
\right]
\label{A.a50}
\end{eqnarray}
with the relation (\ref{A.5})

From those expressions, we derive:
\begin{eqnarray}
&&a_{\ref{C.2b}\lambda}=q_ec{\bf  e}_{{\bf k}\lambda}.
\left( \frac {{\bf  P}+\frac12\hbar {\bf  K}}{|{\bf  P}+\frac12\hbar {\bf  K}|}
\cos\frac{ \theta_{|{\bf  P}+\frac12\hbar {\bf  K}-\hbar {\bf k}|}}2   
\sin \frac{ \theta_{|{\bf  P}+\frac12\hbar {\bf  K}|}}2 
\right. \nonumber\\ &&  \left.
+\frac {{\bf  P}+\frac12\hbar {\bf  K}-\hbar {\bf k}}
{|{\bf  P}+\frac12\hbar {\bf  K}-\hbar {\bf k}|}    
\sin \frac{ \theta_{|{\bf  P}+\frac12\hbar {\bf  K}-\hbar {\bf k}|}}2 
\cos\frac{ \theta_{{{\bf  P}+\frac12\hbar {\bf  K}}}}2 
\right)
\label{A.a51}
\end{eqnarray}
\begin{equation}
a'_{\ref{C.4}\lambda}= a_{\ref{C.2b}\lambda}
\label{A.a52}
\end{equation}
and therefore, the corresponding contribution to the first term of 
$S_{\ref{C.2b}/\ref{C.4}}^s$ can be obtained as
\begin{eqnarray}
&&\sum_{\lambda} a_{\ref{C.2b}\lambda}a'_{\ref{C.4}\lambda}=
q_e^2c^2 \left(
\cos^2\frac{ \theta_{{{\bf  P}+\frac12\hbar {\bf  K}}}}2 
\sin ^2\frac{ \theta_{|{\bf  P}+\frac12\hbar {\bf  K}-\hbar {\bf k}|}}2 
 \right.\nonumber\\ &&  \left.
+\sin^2\frac{ \theta_{{{\bf  P}+\frac12\hbar {\bf  K}}}}2 
\cos ^2\frac{ \theta_{|{\bf  P}+\frac12\hbar {\bf  K}-\hbar {\bf k}|}}2 
+\frac12\frac {\bf P+\frac12\hbar \bf K-\hbar \bf k}
{|\bf P+\frac12\hbar \bf K-\hbar \bf k|}. 
\frac {\bf P+\frac12\hbar \bf K}{|\bf P+\frac12\hbar \bf K|}   
 \right.\nonumber\\ &&  \left.\times
\sin \theta_{{|\bf P+\frac12\hbar \bf K|}} 
\sin  \theta_{|\bf P+\frac12\hbar \bf K-\hbar \bf k|} 
\right)
\label{A.a53}
\end{eqnarray}
The computation of the contribution of the second term of  $S_{\ref{C.2b}/\ref{C.4}}^s$
requires the expressions:
\begin{eqnarray}
&&{\bf  b}_{\ref{C.2b}\lambda}=iq_e c
\left[  
\left({\bf  e}_{{\bf k}\lambda} 
\pmb{$\times$}
\frac{{\bf  P}+\frac12\hbar {\bf  K}-\hbar{\bf k}}
{|{\bf  P}+\frac12\hbar {\bf  K}-\hbar{\bf k}|} 
\right)
\cos\frac{ \theta_{|{\bf  P}+\frac12\hbar {\bf  K}|}}2  
\sin \frac{ \theta_{|{{\bf  P}+\frac12\hbar {\bf  K}-\hbar{\bf k}}|}}2
\right. \nonumber\\ &&  \left.
- \left( {\bf  e}_{{\bf k}\lambda}\pmb{$\times$}
\frac {{\bf  P}+\frac12\hbar {\bf  K}}{|{\bf  P}+\frac12\hbar {\bf  K}|}   
\right) 
\sin \frac{ \theta_{|{\bf  P}+\frac12\hbar {\bf  K}|}}2 
\cos\frac{ \theta_{|{{\bf  P}+\frac12\hbar {\bf  K}-\hbar{\bf k}}|}}2 
\right] 
\label{A.a54}
\end{eqnarray}
\begin{equation}
{\bf  b}'_{\ref{C.4}\lambda}=-{\bf  b}_{2b\lambda}
\label{A.a55}
\end{equation}
Using the  relation:
\begin{eqnarray}
&&
\sum_{\lambda} \left(\bf e_{\bf k\lambda}\pmb{$\times$}
\bf c\right)
.\left(\bf e_{\bf k\lambda}\pmb{$\times$}\bf c'\right)
 =
2\bf c.\bf c'
\label{A.a56}
\end{eqnarray}
we obtain:
\begin{eqnarray}
&&\sum_{\lambda} {\bf  b}_{\ref{C.2b}\lambda} .{\bf  b}'_{\ref{C.4}\lambda} =
2q_e^2c^2
\left\vert  
\left(
\frac{\bf P+\frac12\hbar \bf K-\hbar\bf k}
{|\bf P+\frac12\hbar \bf K-\hbar\bf k|} 
\right)
\cos\frac{ \theta_{|\bf P+\frac12\hbar \bf K|}}2  
\sin \frac{ \theta_{|{\bf P+\frac12\hbar \bf K-\hbar\bf k}|}}2
\right. \nonumber\\&&   \left.\quad
- \left(
\frac {\bf P+\frac12\hbar \bf K}{|\bf P+\frac12\hbar \bf K|}   
\right) 
\sin \frac{ \theta_{|\bf P+\frac12\hbar \bf K|}}2 
\cos\frac{ \theta_{|{\bf P+\frac12\hbar \bf K-\hbar\bf k}|}}2 
\right\vert^2
\label{A.a58}
\end{eqnarray}
The expressions (\ref{A.a53}) and (\ref{A.a58}) determine the contribution due the 
vectorial photon  to the scalar part $S_{\ref{C.2b}/\ref{C.4}}^s$ while the contribution
to the vectorial part vanishes considering (\ref{A.a52})-(\ref{A.a55}).

We now consider the contribution of the vectorial virtual photon to the 
second diagram (\ref{C.1}/\ref{C.5b}). From
\begin{eqnarray}
&&a_{\ref{C.5b}\lambda}=q_ec{\bf  e}_{{\bf k}\lambda}.
\left( \frac {{\bf  P}-\frac12\hbar {\bf  K}}{|{\bf  P}+\frac12\hbar {\bf  K}|}
\cos\frac{ \theta_{|{\bf  P}-\frac12\hbar {\bf  K}-\hbar {\bf k}|}}2   
\sin \frac{ \theta_{|{\bf  P}-\frac12\hbar {\bf  K}|}}2 
\right. \nonumber\\ &&  \left.\quad
+\frac {{\bf  P}-\frac12\hbar {\bf  K}-\hbar {\bf k}}
{|{\bf  P}-\frac12\hbar {\bf  K}-\hbar {\bf k}|}    
\sin \frac{ \theta_{|{\bf  P}-\frac12\hbar {\bf  K}-\hbar {\bf k}|}}2 
\cos\frac{ \theta_{{{\bf  P}-\frac12\hbar {\bf  K}}}}2 
\right)
\label{A.a59}
\end{eqnarray}
\begin{equation}
a'_{\ref{C.1}\lambda}= a_{\ref{C.5b}\lambda}
\label{A.a60}
\end{equation}
the corresponding contribution to the first term of 
$S_{\ref{C.2b}/\ref{C.4}}^s$ can be obtained as 
\begin{eqnarray}
&&\sum_{\lambda} a_{\ref{C.5b}\lambda}a'_{\ref{C.1}\lambda}=
q_e^2c^2 \left(
\cos^2\frac{ \theta_{{{\bf P}-\frac12\hbar {\bf  K}}}}2 
\sin ^2\frac{ \theta_{|{\bf P}-\frac12\hbar {\bf  K}-\hbar {\bf k}|}}2 
 \right.\nonumber\\ &&  \left.
+\sin^2\frac{ \theta_{{{\bf P}-\frac12\hbar {\bf  K}}}}2 
\cos ^2\frac{ \theta_{|{\bf P}-\frac12\hbar {\bf  K}-\hbar {\bf k}|}}2 
+\frac12\frac {\bf P-\frac12\hbar \bf K-\hbar \bf k}
{|\bf P-\frac12\hbar \bf K-\hbar \bf k|}. 
\frac {\bf P-\frac12\hbar \bf K}{|\bf P-\frac12\hbar \bf K|}   
 \right.\nonumber\\ &&  \left.\times
\sin \theta_{{|\bf P-\frac12\hbar \bf K|}} 
\sin  \theta_{|\bf P-\frac12\hbar \bf K-\hbar \bf k|} 
\right)
\label{A.a61}
\end{eqnarray}
In a similar way,  the computation of the contribution  to 
the second part of $S_{\ref{C.5b}/\ref{C.1}}^s$ requires:
\begin{eqnarray}
&&{\bf  b}_{\ref{C.5b}\lambda}=-iq_e c
\left[  
\left(
\left({\bf  e}_{{\bf k}\lambda} 
\pmb{$\times$}
\frac{{\bf  P}-\frac12\hbar {\bf  K}-\hbar{\bf k}}
{|{\bf  P}-\frac12\hbar {\bf  K}-\hbar{\bf k}|} 
\right)
\cos\frac{ \theta_{|{\bf  P}-\frac12\hbar {\bf  K}|}}2  
\sin \frac{ \theta_{|{{\bf  P}-\frac12\hbar {\bf  K}-\hbar{\bf k}}|}}2
\right.\right. \nonumber\\ && \left. \left.
- \left( {\bf  e}_{{\bf k}\lambda}\pmb{$\times$}
\frac {{\bf  P}-\frac12\hbar {\bf  K}}{|{\bf  P}-\frac12\hbar {\bf  K}|}   
\right) 
\sin \frac{ \theta_{|{\bf  P}-\frac12\hbar {\bf  K}|}}2 
\cos\frac{ \theta_{|{{\bf  P}-\frac12\hbar {\bf  K}-\hbar{\bf k}}|}}2 
\right)
\right] 
\label{A.a62}
\end{eqnarray}
from which we derive:
\begin{equation}
{\bf  b}'_{\ref{C.1}\lambda}=-{\bf  b}_{\ref{C.5b}\lambda}
\label{A.a63}
\end{equation} 
\begin{eqnarray}
&&\sum_{\lambda} {\bf  b}_{\ref{C.5b}\lambda} .{\bf  b}'_{\ref{C.1}\lambda} 
=
2q_e^2c^2
\left\vert  
\left(
\frac{\bf P-\frac12\hbar \bf K-\hbar\bf k}
{|\bf P-\frac12\hbar \bf K-\hbar\bf k|} 
\right)
\cos\frac{ \theta_{|\bf P-\frac12\hbar \bf K|}}2  
\sin \frac{ \theta_{|{\bf P-\frac12\hbar \bf K-\hbar\bf k}|}}2
\right. \nonumber\\ &&  \left.\quad
- \left(
\frac {\bf P-\frac12\hbar \bf K}{|\bf P-\frac12\hbar \bf K|}   
\right) 
\sin \frac{ \theta_{|\bf P-\frac12\hbar \bf K|}}2 
\cos\frac{ \theta_{|{\bf P-\frac12\hbar \bf K-\hbar\bf k}|}}2 
\right\vert^2
\nonumber\\&&
=
2q_e^2c^2
\left(
\cos^2\frac{ \theta_{|\bf P-\frac12\hbar \bf K|}}2  
\sin^2 \frac{ \theta_{|{\bf P-\frac12\hbar \bf K-\hbar\bf k}|}}2
\right. \nonumber\\ &&  \left.
+\sin^2 \frac{ \theta_{|\bf P-\frac12\hbar \bf K|}}2 
\cos^2\frac{ \theta_{|{\bf P-\frac12\hbar \bf K-\hbar\bf k}|}}2 
\right. \nonumber\\ &&  \left.
-\frac12
\left(
\frac{\bf P-\frac12\hbar \bf K-\hbar\bf k}
{|\bf P-\frac12\hbar \bf K-\hbar\bf k|} .
\frac {\bf P-\frac12\hbar \bf K}{|\bf P-\frac12\hbar \bf K|}   
\right) 
\sin \theta_{|\bf P-\frac12\hbar \bf K|} 
\sin  \theta_{|{\bf P-\frac12\hbar \bf K-\hbar\bf k}|}
\right)
\label{A.a65}
\end{eqnarray}
The contribution to $S_{\ref{C.5b}/\ref{C.1}}^s$ can be obtained from (\ref{A.a61}) and (\ref{A.a65}).
It can be obtained also from the expressions for the diagram (\ref{C.2b}/\ref{C.4}) 
by changing the sign of ${\bf  K}$.
Anew, the contribution to the vectorial part  ${\bf  S}_{\ref{C.5b}/\ref{C.1}}^v$ vanishes.

We now consider the contribution of the vectorial virtual photon to the 
third diagram (\ref{C.20}/\ref{C.28}). The computation is similar and we obtain:  
\begin{eqnarray}
&&a_{\ref{C.20}\lambda}=-iq_ec {\bf  e}_{{\bf k}\lambda}.
\left(\frac {\hbar{\bf k}}{|{\bf  P}+\frac12\hbar{\bf  K}+\hbar{\bf k}|}
\pmb{$\times$}
\frac {{\bf  P}+\frac12\hbar{\bf  K}}{|{\bf  P}+\frac12\hbar{\bf  K}|}\right)
\nonumber\\&&\times
\sin \frac{ \theta_{|{\bf  P}+\frac12\hbar{\bf  K}|}}2 
\sin \frac{ \theta_{|{\bf  P}+\frac12\hbar{\bf  K}+\hbar{\bf k}|}}2 
\label{A.a66}
\end{eqnarray}
\begin{equation}
a'_{\ref{C.28}\lambda}=-a_{\ref{C.20}\lambda}
\label{A.a67}
\end{equation}
from which we deduce:
\begin{eqnarray}
&&
\sum_{\lambda} a_{\ref{C.20}\lambda} a'_{\ref{C.28}\lambda-}=
q_e^2c^2\hbar^2
\frac{k^2|{\bf  P}+\frac12\hbar{\bf  K}|^2-|{\bf k}.({\bf  P}+\frac12\hbar{\bf  K})|^2}
{|{\bf  P}+\frac12\hbar{\bf  K}+\hbar{\bf k}|^2
|{\bf  P}+\frac12\hbar{\bf  K}|^2}
\nonumber\\&&
\times
\sin^2 \frac{ \theta_{|\bf P+\frac12\hbar \bf K|}}2 
\sin^2 \frac{ \theta_{|\bf P+\frac12\hbar \bf K+\hbar{\bf k}|}}2 
\label{A.a68}
\end{eqnarray}
The second part of $S_{\ref{C.20}/\ref{C.28}}^s$ requires the evaluation of
\begin{eqnarray}
&&{\bf  b}_{\ref{C.20}\lambda}=q_ec\left[ 
{\bf  e}_{{\bf k}\lambda}
\cos\varphi_{{{\bf  P}+\frac12\hbar{\bf  K}} , 
{{\bf  P}+\frac12\hbar{\bf  K}+\hbar{\bf k}}} 
\right.\nonumber\\&& \left.
- \left[\left({\bf  e}_{{\bf k}\lambda}.
\frac {{\bf  P}+\frac12\hbar{\bf  K}}{|{\bf  P}+\frac12\hbar{\bf  K}|}
\right)
\frac {{\bf  P}+\frac12\hbar{\bf  K}+\hbar{\bf k}}
{|{\bf  P}+\frac12\hbar{\bf  K}+\hbar{\bf k}|} 
\right.\right.\nonumber\\&& \left.\left.
+\left({\bf  e}_{{\bf k}\lambda}.
\frac {{\bf  P}+\frac12\hbar{\bf  K}+\hbar{\bf k}}
{|{\bf  P}+\frac12\hbar{\bf  K}+\hbar{\bf k}|} 
\right)
\frac {{\bf  P}+\frac12\hbar{\bf  K}}{|{\bf  P}+\frac12\hbar{\bf  K}|}
\right]
\right.\nonumber\\&& \left.\qquad\qquad \times 
\sin \frac{ \theta_{|{\bf  P}+\frac12\hbar{\bf  K}|}}2 
\sin \frac{ \theta_{|{\bf  P}+\frac12\hbar{\bf  K}+\hbar{\bf k}|}}2 
\right]  
\label{A.a69}
\end{eqnarray}
\begin{eqnarray}
&&{\bf  b}'_{\ref{C.28}\lambda}=q_ec\left[ 
{\bf  e}_{{\bf k}\lambda}
\cos\varphi_{{{\bf  P}+\frac12\hbar{\bf  K}} , 
{{\bf  P}+\frac12\hbar{\bf  K}+\hbar{\bf k}}} 
\right.\nonumber\\&& \left.
-\left[\left({\bf  e}_{{\bf k}\lambda}.
\frac {{\bf  P}+\frac12\hbar{\bf  K}}{|{\bf  P}+\frac12\hbar{\bf  K}|}
\right)
\frac {{\bf  P}+\frac12\hbar{\bf  K}+\hbar{\bf k}}
{|{\bf  P}+\frac12\hbar{\bf  K}+\hbar{\bf k}|} 
\right.\right.\nonumber\\&& \left.\left.
+\left({\bf  e}_{{\bf k}\lambda}.
\frac {{\bf  P}+\frac12\hbar{\bf  K}+\hbar{\bf k}}
{|{\bf  P}+\frac12\hbar{\bf  K}+\hbar{\bf k}|} 
\right)
.\frac {{\bf  P}+\frac12\hbar{\bf  K}}{|{\bf  P}+\frac12\hbar{\bf  K}|}
\right]
\right.\nonumber\\&& \left.\qquad \times 
\sin \frac{ \theta_{|{\bf  P}+\frac12\hbar{\bf  K}|}}2 
\sin \frac{ \theta_{|{\bf  P}+\frac12\hbar{\bf  K}+\hbar{\bf k}|}}2 
\right]  
\label{A.a70}
\end{eqnarray}
and therefore, we obtain:
\begin{eqnarray}
&&\sum_{\lambda}{\bf  b}_{\ref{C.20}\lambda}{\bf  b}'_{\ref{C.28}\lambda}=
3q_e^2c^2 \cos^2\varphi_{{\bf P+\frac12\hbar\bf K} , 
{\bf P+\frac12\hbar\bf K+\hbar\bf k}} 
\nonumber\\&& 
-4q_e^2c^2 
\cos\varphi_{{\bf P+\frac12\hbar\bf K} , 
{\bf P+\frac12\hbar\bf K+\hbar\bf k}} 
\left(
\frac {(\bf P+\frac12\hbar\bf K+\hbar\bf k)}{|\bf P+\frac12\hbar\bf K+\hbar\bf k|} .
\frac {\bf P+\frac12\hbar\bf K}{|\bf P+\frac12\hbar\bf K|}
\right)
\nonumber\\&& \times
\sin \frac{ \theta_{|\bf P+\frac12\hbar\bf K|}}2 
\sin \frac{ \theta_{|\bf P+\frac12\hbar\bf K+\hbar\bf k|}}2 
\nonumber\\&& 
+ q_e^2c^2\sum_\lambda
 \left\vert\left({\bf  e}_{{\bf k}\lambda}.
\frac {\bf P+\frac12\hbar\bf K}{|\bf P+\frac12\hbar\bf K|}
\right)
\frac {\bf P+\frac12\hbar\bf K+\hbar\bf k}
{|\bf P+\frac12\hbar\bf K+\hbar\bf k|} 
\right.\nonumber\\&&  \left.
+\left({\bf  e}_{{\bf k}\lambda}.
\frac {\bf P+\frac12\hbar\bf K+\hbar\bf k}
{|\bf P+\frac12\hbar\bf K+\hbar\bf k|} 
\right)
\frac {\bf P+\frac12\hbar\bf K}{|\bf P+\frac12\hbar\bf K|}
\right\vert^2
\nonumber\\&&\times
\sin^2 \frac{ \theta_{|\bf P+\frac12\hbar\bf K|}}2 
\sin^2 \frac{ \theta_{|\bf P+\frac12\hbar\bf K+\hbar\bf k|}}2 
\nonumber\\&& 
=
3q_e^2c^2 \cos^2\varphi_{{\bf P+\frac12\hbar\bf K} , 
{\bf P-\frac12\hbar\bf K+\hbar\bf k}} 
\nonumber\\&& 
-4q_e^2c^2 
\cos\varphi_{{\bf P+\frac12\hbar\bf K} , 
{\bf P+\frac12\hbar\bf K+\hbar\bf k}} 
\left(
\frac {(\bf P+\frac12\hbar\bf K+\hbar\bf k)}{|\bf P+\frac12\hbar\bf K+\hbar\bf k|} .
\frac {\bf P+\frac12\hbar\bf K}{|\bf P+\frac12\hbar\bf K|}
\right)
\nonumber\\&& \times
\sin \frac{ \theta_{|\bf P+\frac12\hbar\bf K|}}2 
\sin \frac{ \theta_{|\bf P+\frac12\hbar\bf K+\hbar\bf k|}}2 
\nonumber\\&& 
+2 q_e^2c^2
 \left\vert
1+
\left(
\frac {\bf P+\frac12\hbar\bf K}{|\bf P+\frac12\hbar\bf K|}.
\frac {\bf P+\frac12\hbar\bf K+\hbar\bf k}
{|\bf P+\frac12\hbar\bf K+\hbar\bf k|}
\right)^2 \right\vert
\nonumber\\&&\times
\sin^2 \frac{ \theta_{|\bf P+\frac12\hbar\bf K|}}2 
\sin^2 \frac{ \theta_{|\bf P+\frac12\hbar\bf K+\hbar\bf k|}}2 
\label{A.a72}
\end{eqnarray}
The contribution due to the vectorial photon to $S_{\ref{C.20}/\ref{C.28}}^s$ is obtained 
from (\ref{A.a68}) and (\ref{A.a72}) while the contribution to the vectorial part vanishes.
               
We now consider the contribution of the vectorial virtual photon to the 
fourth diagram (\ref{C.27}/\ref{C.21}). From: 
\begin{eqnarray}
&&a_{\ref{C.27}\lambda}=iq_ec {\bf  e}_{-{\bf k}\lambda}.
\left(\frac {\hbar{\bf k}}{|{\bf  P}-\frac12\hbar{\bf  K}+\hbar{\bf k}|}
\pmb{$\times$}
\frac {{\bf  P}-\frac12\hbar{\bf  K}}{|{\bf  P}-\frac12\hbar{\bf  K}|}\right)
\nonumber\\&&\times 
\sin \frac{ \theta_{|{\bf  P}-\frac12\hbar{\bf  K}|}}2 
\sin \frac{ \theta_{|{\bf  P}-\frac12\hbar{\bf  K}+\hbar{\bf k}|}}2 
\label{A.a73}
\end{eqnarray}
\begin{equation}
a'_{\ref{C.21}\lambda}=-a_{\ref{C.27}\lambda}
\label{A.a74}
\end{equation}
we obtain:
\begin{eqnarray}
&&\sum_{\lambda} a_{\ref{C.27}\lambda} a'_{\ref{C.21}\lambda}=
q_e^2c^2\hbar^2
\frac{k^2|{\bf  P}-\frac12\hbar{\bf  K}|^2-|{\bf k}.({\bf  P}-\frac12\hbar{\bf  K})|^2}
{
|{\bf  P}-\frac12\hbar{\bf  K}+\hbar{\bf k}|^2
|{\bf  P}-\frac12\hbar{\bf  K}|^2
}
\nonumber\\&&
\times
\sin^2 \frac{ \theta_{|\bf P-\frac12\hbar \bf K|}}2 
\sin^2 \frac{ \theta_{|\bf P-\frac12\hbar \bf K+\hbar{\bf k}|}}2 
\label{A.a75}
\end{eqnarray}
From:
\begin{eqnarray}
&&{\bf  b}_{\ref{C.27}\lambda}=q_ec\left[ 
{\bf  e}_{{\bf k}\lambda}
\cos\varphi_{{{\bf  P}-\frac12\hbar{\bf  K}} , 
{{\bf  P}-\frac12\hbar{\bf  K}+\hbar{\bf k}}} 
\right.\nonumber\\&& \left.
-\left[\left({\bf  e}_{{\bf k}\lambda}.
\frac {{\bf  P}-\frac12\hbar{\bf  K}}{|{\bf  P}-\frac12\hbar{\bf  K}|}
\right)
\frac {{\bf  P}-\frac12\hbar{\bf  K}+\hbar{\bf k}}
{|{\bf  P}-\frac12\hbar{\bf  K}+\hbar{\bf k}|} 
\right.\right.\nonumber\\&& \left.\left.
+\left({\bf  e}_{{\bf k}\lambda}.
\frac {{\bf  P}-\frac12\hbar{\bf  K}+\hbar{\bf k}}
{|{\bf  P}-\frac12\hbar{\bf  K}+\hbar{\bf k}|} 
\right)
\frac {{\bf  P}-\frac12\hbar{\bf  K}}{|{\bf  P}-\frac12\hbar{\bf  K}|}
\right]
\right.\nonumber\\&& \left.\times 
\sin \frac{ \theta_{|{\bf  P}-\frac12\hbar{\bf  K}|}}2 
\sin \frac{ \theta_{|{\bf  P}-\frac12\hbar{\bf  K}+\hbar{\bf k}|}}2 
\right]  
\label{A.a76}
\end{eqnarray}
\begin{eqnarray}
&&{\bf  b}'_{\ref{C.21}\lambda}=q_ec\left[ 
{\bf  e}_{{\bf k}\lambda}
\cos\varphi_{{{\bf  P}-\frac12\hbar{\bf  K}} , 
{{\bf  P}-\frac12\hbar{\bf  K}+\hbar{\bf k}}} 
\right.\nonumber\\&& \left. 
-\left[\left({\bf  e}_{{\bf k}\lambda}.
\frac {{\bf  P}-\frac12\hbar{\bf  K}}{|{\bf  P}-\frac12\hbar{\bf  K}|}
\right)
\frac {{\bf  P}-\frac12\hbar{\bf  K}+\hbar{\bf k}}
{|{\bf  P}-\frac12\hbar{\bf  K}+\hbar{\bf k}|} 
\right.\right.\nonumber\\&& \left.\left.
+\left({\bf  e}_{{\bf k}\lambda}.
\frac {{\bf  P}-\frac12\hbar{\bf  K}+\hbar{\bf k}}
{|{\bf  P}-\frac12\hbar{\bf  K}+\hbar{\bf k}|} 
\right)
\frac {{\bf  P}-\frac12\hbar{\bf  K}}{|{\bf  P}-\frac12\hbar{\bf  K}|}
\right]
\right.\nonumber\\&& \left. \times 
\sin \frac{ \theta_{|{\bf  P}-\frac12\hbar{\bf  K}|}}2 
\sin \frac{ \theta_{|{\bf  P}-\frac12\hbar{\bf  K}+\hbar{\bf k}|}}2 
\right]  
\label{A.a77}
\end{eqnarray}
we obtain:
\begin{eqnarray}
&&\sum_{\lambda}{\bf  b}_{\ref{C.27}\lambda-}{\bf  b}'_{\ref{C.21}\lambda+}
=3q_e^2c^2 \cos^2\varphi_{{{\bf P}-\frac12\hbar{\bf K}} , 
{{\bf P}-\frac12\hbar{\bf K}+\hbar{\bf k}}} 
\nonumber\\&& 
-4q_e^2c^2 
\cos\varphi_{{{\bf P}-\frac12\hbar{\bf K}} , 
{{\bf P}-\frac12\hbar{\bf K}+\hbar{\bf k}}} 
\left(
\frac {({\bf P}-\frac12\hbar{\bf K}+\hbar{\bf k})}{|{\bf P}-\frac12\hbar{\bf K}+\hbar{\bf k}|} .
\frac {{\bf P}-\frac12\hbar{\bf K}}{|{\bf P}-\frac12\hbar{\bf K}|}
\right)
\nonumber\\&& \times
\sin \frac{ \theta_{|{\bf P}-\frac12\hbar{\bf K}|}}2 
\sin \frac{ \theta_{|{\bf P}-\frac12\hbar{\bf K}+\hbar{\bf k}|}}2 
\nonumber\\&& 
+2 q_e^2c^2
 \left\vert
1+
\left(
\frac {{\bf P}-\frac12\hbar{\bf K}}{|{\bf P}-\frac12\hbar{\bf K}|}.
\frac {{\bf P}-\frac12\hbar{\bf K}+\hbar{\bf k}}
{|{\bf P}-\frac12\hbar{\bf K}+\hbar{\bf k}|}
\right)^2 \right\vert
\nonumber\\&&\times
\sin^2 \frac{ \theta_{|{\bf P}-\frac12\hbar{\bf K}|}}2 
\sin^2 \frac{ \theta_{|{\bf P}-\frac12\hbar{\bf K}+\hbar{\bf k}|}}2 
\label{A.a78}
\end{eqnarray}
The contribution due to the vectorial photon to $S_{\ref{C.27}/\ref{C.21}}^s$ is obtained 
from (A.a75) and (A.a78) while the contribution to the vectorial part vanishes.

We now collect our partial result to obtain the complete contribution for 
each diagram. For the first one, we have to add the contributions of (\ref{A.a39}),
(\ref{A.a53}) and (\ref{A.a58}) and we obtain, for the scalar part $S^s_{\ref{C.2b}/\ref{C.4}}$ of
$S_{\ref{C.2b}/\ref{C.4}}$ (A.a29), taking into account the value of $s(\nu)$:
\begin{eqnarray}
&&S^s_{\ref{C.2b}/\ref{C.4}}=-q^2_ec^2 
\cos^2 \varphi_{{{\bf  P}+\frac12\hbar{\bf  K}-\hbar{\bf k}} , 
{{\bf  P}+\frac12\hbar{\bf  K}}}
\nonumber\\ &&
-q^2_ec^2 \hbar^2 
\frac{
k^2 |{\bf  P}+\frac12\hbar{\bf  K}|^2
-(( {{\bf  P}+\frac12\hbar{\bf  K}}).{{\bf k}})^2}
{|{\bf  P}+\frac12\hbar{\bf  K}-\hbar{\bf k}|^2|{\bf  P}+\frac12\hbar{\bf  K}|^2}
\sin^2\frac{ \theta_{|{\bf  P}+\frac12\hbar{\bf  K}-\hbar{\bf k}|}}2
\sin^2\frac{ \theta_{|{\bf  P}+\frac12\hbar{\bf  K}|}}2
\nonumber\\ &&
+q^2_ec^2  
\left(
\cos^2\frac{ \theta_{|\bf P+\frac12\hbar \bf K|}}2  
\sin^2 \frac{ \theta_{|\bf P+\frac12\hbar \bf K-\hbar{\bf k}|}}2 
+\cos^2 \frac{ \theta_{|\bf P+\frac12\hbar \bf K-\hbar{\bf k}|}}2 
\sin^2\frac{ \theta_{|\bf P+\frac12\hbar \bf K|}}2  
\right.\nonumber\\&& \left.
+\frac12
\frac {{\bf P}+\frac12\hbar{\bf K}}{|{\bf P}+\frac12\hbar{\bf K}|}.
\frac {{\bf P}+\frac12\hbar{\bf K}-\hbar{\bf k}}
{|{\bf P}+\frac12\hbar{\bf K}-\hbar{\bf k}|}
\sin{ \theta_{|{\bf  P}+\frac12\hbar{\bf  K}-\hbar{\bf k}|}}
\sin{ \theta_{|{\bf  P}+\frac12\hbar{\bf  K}|}} 
\right)
\nonumber\\ &&
+2q_e^2c^2 
\left(
\cos^2\frac{ \theta_{|\bf P+\frac12\hbar \bf K|}}2  
\sin^2 \frac{ \theta_{|\bf P+\frac12\hbar \bf K-\hbar{\bf k}|}}2 
+\cos^2 \frac{ \theta_{|\bf P+\frac12\hbar \bf K-\hbar{\bf k}|}}2 
\sin^2\frac{ \theta_{|\bf P+\frac12\hbar \bf K|}}2  
\right.\nonumber\\&& \left.
-\frac12
\frac {{\bf P}+\frac12\hbar{\bf K}}{|{\bf P}+\frac12\hbar{\bf K}|}.
\frac {{\bf P}+\frac12\hbar{\bf K}-\hbar{\bf k}}
{|{\bf P}+\frac12\hbar{\bf K}-\hbar{\bf k}|}
\sin{ \theta_{|{\bf  P}+\frac12\hbar{\bf  K}-\hbar{\bf k}|}}
\sin{ \theta_{|{\bf  P}+\frac12\hbar{\bf  K}|}} 
\right)
\nonumber\\&&
=-q^2_ec^2 
\cos^2 \varphi_{{{\bf  P}+\frac12\hbar{\bf  K}-\hbar{\bf k}} , 
{{\bf  P}+\frac12\hbar{\bf  K}}}
\nonumber\\ &&
-q^2_ec^2 \hbar^2 
\frac{
k^2 |{\bf  P}+\frac12\hbar{\bf  K}|^2
-(( {{\bf  P}+\frac12\hbar{\bf  K}}).{{\bf k}})^2}
{|{\bf  P}+\frac12\hbar{\bf  K}-\hbar{\bf k}|^2|{\bf  P}+\frac12\hbar{\bf  K}|^2}
\sin^2\frac{ \theta_{|{\bf  P}+\frac12\hbar{\bf  K}-\hbar{\bf k}|}}2
\sin^2\frac{ \theta_{|{\bf  P}+\frac12\hbar{\bf  K}|}}2
\nonumber\\ &&
+q^2_ec^2
\left(
3\cos^2\frac{ \theta_{|\bf P+\frac12\hbar \bf K|}}2  
\sin^2 \frac{ \theta_{|\bf P+\frac12\hbar \bf K-\hbar{\bf k}|}}2 
+3\cos^2 \frac{ \theta_{|\bf P+\frac12\hbar \bf K-\hbar{\bf k}|}}2 
\sin^2\frac{ \theta_{|\bf P+\frac12\hbar \bf K|}}2  
\right.\nonumber\\&& \left.
-\frac12
\frac {{\bf P}+\frac12\hbar{\bf K}}{|{\bf P}+\frac12\hbar{\bf K}|}.
\frac {{\bf P}+\frac12\hbar{\bf K}-\hbar{\bf k}}
{|{\bf P}+\frac12\hbar{\bf K}-\hbar{\bf k}|}
\sin{ \theta_{|{\bf  P}+\frac12\hbar{\bf  K}-\hbar{\bf k}|}}
\sin{ \theta_{|{\bf  P}+\frac12\hbar{\bf  K}|}} 
\right)
\label{A.a79}
\end{eqnarray}

Using the explicit form of 
$\cos \varphi_{{{\bf  P}},{{\bf  P}'}} $, we obtain:
\begin{eqnarray}
&&S^s_{\ref{C.2b}/\ref{C.4}}=-q^2_ec^2 \left(
\cos^2\frac{ \theta_{|{\bf  P}+\frac12\hbar{\bf  K}|}}2
\cos ^2\frac{ \theta_{|{\bf  P}+\frac12\hbar{\bf  K}-\hbar{\bf k}|}}2 
\right.\nonumber\\&&\left.
+\left|\frac {{\bf  P}+\frac12\hbar{\bf  K}-\hbar{\bf k}}
{|{\bf  P}+\frac12\hbar{\bf  K}-\hbar{\bf k}|}
.\frac{{\bf  P}+\frac12\hbar{\bf  K}}{|{\bf  P}+\frac12\hbar{\bf  K}|} 
\right|^2
\sin^2\frac{ \theta_{|{\bf  P}+\frac12\hbar{\bf  K}-\hbar{\bf k}|}}2
\sin^2 \frac{ \theta_{|{\bf  P}+\frac12\hbar{\bf  K}|}}2
\right.\nonumber\\&&\left.
+\frac12\frac {{\bf  P}+\frac12\hbar{\bf  K}-\hbar{\bf k}}
{|{\bf  P}+\frac12\hbar{\bf  K}-\hbar{\bf k}|}
.\frac{{\bf  P}+\frac12\hbar{\bf  K}}{|{\bf  P}+\frac12\hbar{\bf  K}|} 
\sin  \theta_{|{\bf  P}+\frac12\hbar{\bf  K}|} 
\sin  \theta_{|{\bf  P}+\frac12\hbar{\bf  K}-\hbar{\bf k}|}
\right)
\nonumber\\ &&
-q^2_ec^2 \hbar^2 
\frac{
k^2 |{\bf  P}+\frac12\hbar{\bf  K}|^2
-(( {{\bf  P}+\frac12\hbar{\bf  K}}).{{\bf k}})^2}
{|{\bf  P}+\frac12\hbar{\bf  K}-\hbar{\bf k}|^2|{\bf  P}+\frac12\hbar{\bf  K}|^2}
\sin^2\frac{ \theta_{|{\bf  P}+\frac12\hbar{\bf  K}-\hbar{\bf k}|}}2
\sin^2\frac{ \theta_{|{\bf  P}+\frac12\hbar{\bf  K}|}}2
\nonumber\\ &&
+q^2_ec^2
\left(
3\cos^2\frac{ \theta_{|\bf P+\frac12\hbar \bf K|}}2  
\sin^2 \frac{ \theta_{|\bf P+\frac12\hbar \bf K-\hbar{\bf k}|}}2 
+3\cos^2 \frac{ \theta_{|\bf P+\frac12\hbar \bf K-\hbar{\bf k}|}}2 
\sin^2\frac{ \theta_{|\bf P+\frac12\hbar \bf K|}}2  
\right.\nonumber\\&& \left.
-\frac12
\frac {{\bf P}+\frac12\hbar{\bf K}}{|{\bf P}+\frac12\hbar{\bf K}|}.
\frac {{\bf P}+\frac12\hbar{\bf K}-\hbar{\bf k}}
{|{\bf P}+\frac12\hbar{\bf K}-\hbar{\bf k}|}
\sin{ \theta_{|{\bf  P}+\frac12\hbar{\bf  K}-\hbar{\bf k}|}}
\sin{ \theta_{|{\bf  P}+\frac12\hbar{\bf  K}|}} 
\right)
\label{A.a81}
\end{eqnarray}
Let us note that we can write:
\begin{eqnarray}
&&
\hbar^2
\frac{k^2|{\bf  P}+\frac12\hbar{\bf  K}|^2
-|{\bf k}.({\bf  P}+\frac12\hbar{\bf  K})|^2}
{|{\bf  P}+\frac12\hbar{\bf  K}+\hbar{\bf k}|^2|{\bf  P}+\frac12\hbar{\bf  K}|^2}
\nonumber\\ && 
=\frac{|{\bf  P}+\frac12\hbar{\bf  K}+\hbar{\bf k}|^2|{\bf  P}+\frac12\hbar{\bf  K}|^2
-|({\bf  P}+\frac12\hbar{\bf  K}+\hbar{\bf k}).({\bf  P}+\frac12\hbar{\bf  K})|^2}
{|{\bf  P}+\frac12\hbar{\bf  K}+\hbar{\bf k}|^2|{\bf  P}+\frac12\hbar{\bf  K}|^2}
\nonumber\\ &&  =
1-\frac
{|({\bf  P}+\frac12\hbar{\bf  K}+\hbar{\bf k}).({\bf  P}+\frac12\hbar{\bf  K})|^2}
{|{\bf  P}+\frac12\hbar{\bf  K}+\hbar{\bf k}|^2|{\bf  P}+\frac12\hbar{\bf  K}|^2}
\label{A.a82}
\end{eqnarray}
so that we obtain:
\begin{eqnarray}
&&S^s_{\ref{C.2b}/\ref{C.4}}=-q^2_ec^2 \left(
\cos^2\frac{ \theta_{|{\bf  P}+\frac12\hbar{\bf  K}|}}2
\cos ^2\frac{ \theta_{|{\bf  P}+\frac12\hbar{\bf  K}-\hbar{\bf k}|}}2 
\right.\nonumber\\&&\left.
+\sin^2\frac{ \theta_{|{\bf  P}+\frac12\hbar{\bf  K}-\hbar{\bf k}|}}2
\sin^2 \frac{ \theta_{|{\bf  P}+\frac12\hbar{\bf  K}|}}2
\right)
\nonumber\\&&
+q^2_ec^2
\left(
3\cos^2\frac{ \theta_{|\bf P+\frac12\hbar \bf K|}}2  
\sin^2 \frac{ \theta_{|\bf P+\frac12\hbar \bf K-\hbar{\bf k}|}}2 
+3\cos^2 \frac{ \theta_{|\bf P+\frac12\hbar \bf K-\hbar{\bf k}|}}2 
\sin^2\frac{ \theta_{|\bf P+\frac12\hbar \bf K|}}2  
\right.\nonumber\\&& \left.
-
\frac {{\bf P}+\frac12\hbar{\bf K}}{|{\bf P}+\frac12\hbar{\bf K}|}.
\frac {{\bf P}+\frac12\hbar{\bf K}-\hbar{\bf k}}
{|{\bf P}+\frac12\hbar{\bf K}-\hbar{\bf k}|}
\sin{ \theta_{|{\bf  P}+\frac12\hbar{\bf  K}-\hbar{\bf k}|}}
\sin{ \theta_{|{\bf  P}+\frac12\hbar{\bf  K}|}} 
\right)
\label{A.a83}
\end{eqnarray}
We now consider the contribution of the second diagram which can be obtained
by changing the sign of ${\bf  K}$  in the previous expression: 
\begin{eqnarray}
&&S^s_{\ref{C.5b}/\ref{C.1}}=-q^2_ec^2 \left(
\cos^2\frac{ \theta_{|{\bf  P}-\frac12\hbar{\bf  K}|}}2
\cos ^2\frac{ \theta_{|{\bf  P}-\frac12\hbar{\bf  K}-\hbar{\bf k}|}}2 
\right.\nonumber\\&&\left.
+\sin^2\frac{ \theta_{|{\bf  P}-\frac12\hbar{\bf  K}-\hbar{\bf k}|}}2
\sin^2 \frac{ \theta_{|{\bf  P}-\frac12\hbar{\bf  K}|}}2
\right)
\nonumber\\&&
+q^2_ec^2
\left(
3\cos^2\frac{ \theta_{|\bf P-\frac12\hbar \bf K|}}2  
\sin^2 \frac{ \theta_{|\bf P-\frac12\hbar \bf K-\hbar{\bf k}|}}2 
+3\cos^2 \frac{ \theta_{|\bf P-\frac12\hbar \bf K-\hbar{\bf k}|}}2 
\sin^2\frac{ \theta_{|\bf P-\frac12\hbar \bf K|}}2  
\right.\nonumber\\&& \left.
-
\frac {{\bf P}-\frac12\hbar{\bf K}}{|{\bf P}-\frac12\hbar{\bf K}|}.
\frac {{\bf P}-\frac12\hbar{\bf K}-\hbar{\bf k}}
{|{\bf P}-\frac12\hbar{\bf K}-\hbar{\bf k}|}
\sin{ \theta_{|{\bf  P}-\frac12\hbar{\bf  K}-\hbar{\bf k}|}}
\sin{ \theta_{|{\bf  P}-\frac12\hbar{\bf  K}|}} 
\right)
\label{A.a84}
\end{eqnarray}
We now consider the contribution of the third diagram which can be obtained
from (\ref{A.a46}), (\ref{A.a68}) and  (\ref{A.a72}):
\begin{eqnarray}
&&S^s_{\ref{C.20}/\ref{C.28}}=-q_e^2c^2\left( 
\cos^2\frac{ \theta_{|{\bf  P}+\frac12\hbar{\bf  K}|}}2
\sin ^2\frac{ \theta_{|{\bf  P}+\frac12\hbar{\bf  K}+\hbar{\bf k}|}}2 
\right.\nonumber\\&&\left.
+\cos^2\frac{ \theta_{|{\bf  P}+\frac12\hbar{\bf  K}+\hbar{\bf k}|}}2
\sin^2 \frac{ \theta_{|{\bf  P}+\frac12\hbar{\bf  K}|}}2
\right.\nonumber\\&&\left.
-\frac12\frac {{\bf  P}+\frac12\hbar{\bf  K}+\hbar{\bf k}}
{|{\bf  P}+\frac12\hbar{\bf  K}+\hbar{\bf k}|}
.\frac{{\bf  P}+\frac12\hbar{\bf  K}}{|{\bf  P}+\frac12\hbar{\bf  K}|} 
\sin  \theta_{|{\bf  P}+\frac12\hbar{\bf  K}|} 
\sin  \theta_{|{\bf  P}+\frac12\hbar{\bf  K}+\hbar{\bf k}|} 
\right)
\nonumber\\&&
+q_e^2c^2\hbar^2
\frac{k^2|{\bf  P}+\frac12\hbar{\bf  K}|^2-|{\bf k}.({\bf  P}+\frac12\hbar{\bf  K})|^2}
{|{\bf  P}+\frac12\hbar{\bf  K}+\hbar{\bf k}|^2
|{\bf  P}+\frac12\hbar{\bf  K}|^2}
\nonumber\\&&
\times
\sin^2 \frac{ \theta_{|\bf P+\frac12\hbar \bf K|}}2 
\sin^2 \frac{ \theta_{|\bf P+\frac12\hbar \bf K+\hbar{\bf k}|}}2 
\nonumber\\&&
+3q_e^2c^2 \cos^2\varphi_{{\bf P+\frac12\hbar\bf K} , 
{\bf P-\frac12\hbar\bf K+\hbar\bf k}} 
\nonumber\\&& 
-4q_e^2c^2 
\cos\varphi_{{\bf P+\frac12\hbar\bf K} , 
{\bf P+\frac12\hbar\bf K+\hbar\bf k}} 
\left(
\frac {(\bf P+\frac12\hbar\bf K+\hbar\bf k)}{|\bf P+\frac12\hbar\bf K+\hbar\bf k|} .
\frac {\bf P+\frac12\hbar\bf K}{|\bf P+\frac12\hbar\bf K|}
\right)
\nonumber\\&& \times
\sin \frac{ \theta_{|\bf P+\frac12\hbar\bf K|}}2 
\sin \frac{ \theta_{|\bf P+\frac12\hbar\bf K+\hbar\bf k|}}2 
\nonumber\\&& 
+2 q_e^2c^2
 \left\vert
1+
\left(
\frac {\bf P+\frac12\hbar\bf K}{|\bf P+\frac12\hbar\bf K|}.
\frac {\bf P+\frac12\hbar\bf K+\hbar\bf k}
{|\bf P+\frac12\hbar\bf K+\hbar\bf k|}
\right)^2 \right\vert
\nonumber\\&&\times
\sin^2 \frac{ \theta_{|\bf P+\frac12\hbar\bf K|}}2 
\sin^2 \frac{ \theta_{|\bf P+\frac12\hbar\bf K+\hbar\bf k|}}2 
\label{A.a85}
\end{eqnarray}
Using the explicit form of $\cos \varphi$ and the relation (\ref{A.a82}), we obtain
\begin{eqnarray}
&&S^s_{\ref{C.20}/\ref{C.28}}=
-q_e^2c^2
\left( 
\cos^2\frac{ \theta_{|{\bf  P}+\frac12\hbar{\bf  K}|}}2
\sin ^2\frac{ \theta_{|{\bf  P}+\frac12\hbar{\bf  K}+\hbar{\bf k}|}}2 
\right.\nonumber\\&&\left.
+\cos^2\frac{ \theta_{|{\bf  P}+\frac12\hbar{\bf  K}+\hbar{\bf k}|}}2
\sin^2 \frac{ \theta_{|{\bf  P}+\frac12\hbar{\bf  K}|}}2
\right)
\nonumber\\&&
+3q_e^2c^2\left( 
\cos^2\frac{ \theta_{|{\bf  P}+\frac12\hbar{\bf  K}|}}2
\cos ^2\frac{ \theta_{|{\bf  P}+\frac12\hbar{\bf  K}+\hbar{\bf k}|}}2 
\right.\nonumber\\&&\left.
+\sin^2\frac{ \theta_{|{\bf  P}+\frac12\hbar{\bf  K}+\hbar{\bf k}|}}2
\sin^2 \frac{ \theta_{|{\bf  P}+\frac12\hbar{\bf  K}|}}2
\right)
\nonumber\\&&
+q_e^2c^2 
\frac {{\bf  P}+\frac12\hbar{\bf  K}+\hbar{\bf k}}
{|{\bf  P}+\frac12\hbar{\bf  K}+\hbar{\bf k}|}
.\frac{{\bf  P}+\frac12\hbar{\bf  K}}{|{\bf  P}+\frac12\hbar{\bf  K}|} 
\nonumber\\&&\times
\sin  \theta_{|{\bf  P}+\frac12\hbar{\bf  K}|} 
\sin  \theta_{|{\bf  P}+\frac12\hbar{\bf  K}+\hbar{\bf k}|} 
\label{A.a87}
\end{eqnarray}
We now consider the contribution of the fourth diagram which can be obtained
by changing the sign of ${\bf  K}$  in the previous expression:
\begin{eqnarray}
&&S^s_{\ref{C.27}/\ref{C.21}}=
-q_e^2c^2
\left( 
\cos^2\frac{ \theta_{|{\bf  P}-\frac12\hbar{\bf  K}|}}2
\sin ^2\frac{ \theta_{|{\bf  P}-\frac12\hbar{\bf  K}+\hbar{\bf k}|}}2 
\right.\nonumber\\&&\left.
+\cos^2\frac{ \theta_{|{\bf  P}-\frac12\hbar{\bf  K}+\hbar{\bf k}|}}2
\sin^2 \frac{ \theta_{|{\bf  P}-\frac12\hbar{\bf  K}|}}2
\right)
\nonumber\\&&
+3q_e^2c^2\left( 
\cos^2\frac{ \theta_{|{\bf  P}-\frac12\hbar{\bf  K}|}}2
\cos ^2\frac{ \theta_{|{\bf  P}-\frac12\hbar{\bf  K}+\hbar{\bf k}|}}2 
\right.\nonumber\\&&\left.
+\sin^2\frac{ \theta_{|{\bf  P}-\frac12\hbar{\bf  K}+\hbar{\bf k}|}}2
\sin^2 \frac{ \theta_{|{\bf  P}-\frac12\hbar{\bf  K}|}}2
\right)
\nonumber\\&&
+q_e^2c^2 
\frac {{\bf  P}-\frac12\hbar{\bf  K}+\hbar{\bf k}}
{|{\bf  P}-\frac12\hbar{\bf  K}+\hbar{\bf k}|}
.\frac{{\bf  P}-\frac12\hbar{\bf  K}}{|{\bf  P}-\frac12\hbar{\bf  K}|} 
\nonumber\\&&\times
\sin  \theta_{|{\bf  P}-\frac12\hbar{\bf  K}|} 
\sin  \theta_{|{\bf  P}-\frac12\hbar{\bf  K}+\hbar{\bf k}|} 
\label{A.a88}
\end{eqnarray}
We know therefore all the required elements of the operator $\Theta^{(2)}$.
The presence of a saclar part only for describing the spin dependence of the computed elements 
of that operator shows that they are also diagonal in the spin indices, and not only in their variables 
$\bf K$ and $\bf P$.

\section{The logarithmic divergence of the collision operator $\Theta^{(2)}$}
\def\theequation{\thesection.\arabic{equation}}
\setcounter{equation}{0}

The dominant part of the integrand in these expressions are (for the part that does not vanish through
angular integration):
\begin{equation}
\left.I(\omega_k)_{M\ref{C.2b}/\ref{C.4}}\right\vert_{dom} 
=
-i\frac{q^2_ec^2}{2\varepsilon_0(2\pi)^3 \omega_{{\bf k}}} 
\frac1{2\hbar\omega_k} 
\label{5.229}
\end{equation}
\begin{equation}
\left.I(\omega_k)_{M\ref{C.5b}/\ref{C.1}}\right\vert_{dom} 
=
i\frac{q^2_ec^2}{2\varepsilon_0(2\pi)^3 \omega_{{\bf k}}} 
\frac1{2\hbar\omega_k} 
\label{5.230}
\end{equation}
\begin{equation}
\left.I(\omega_k)_{M\ref{C.20}/\ref{C.28}}\right\vert_{dom} 
=
i\frac{q^2_ec^2}{2\varepsilon_0(2\pi)^3 \omega_{{\bf k}}} 
\frac1{2\hbar\omega_k} 
\label{5.231}
\end{equation}
\begin{equation}
\left.I(\omega_k)_{M\ref{C.27}/\ref{C.21}}\right\vert_{dom} 
=
-i\frac{q^2_ec^2}{2\varepsilon_0(2\pi)^3 \omega_{{\bf k}}} 
\frac1{2\hbar\omega_k} 
\label{5.232}
\end{equation}
The sum of these contributions clearly vanishes.
We now consider the correction due to the propagator.
\begin{eqnarray}
&&\left.I(\omega_k)_{M\ref{C.2b}/\ref{C.4}}\right\vert_{Pr} 
=-i\frac{q^2_ec^2}{2\varepsilon_0(2\pi)^3 \omega_{{\bf k}}} 
\frac1{4\hbar^2\omega^2_{k}}
\left(E_{|{\bf  P}+\frac12\hbar {\bf  K}|}
\right.\nonumber\\&&\left.
-\hat{{\bf k}}.\frac{{\bf  P}+\frac12\hbar{\bf  K}}{|{\bf  P}+\frac12\hbar{\bf  K}|} 
(-1)c({\bf  P}+\frac\hbar2{\bf  K}).\hat{{\bf k}}
\sin  \theta_{|{\bf  P}+\frac12\hbar{\bf  K}|} \right) 
\label{5.233}
\end{eqnarray}
\begin{eqnarray}
&&\left.I(\omega_k)_{M\ref{C.5b}/\ref{C.1}}\right\vert_{Pr} 
=i\frac{q^2_ec^2}{2\varepsilon_0(2\pi)^3 \omega_{{\bf k}}} 
\frac1{4\hbar^2\omega^2_{k}}
\left(E_{|{\bf  P}-\frac12\hbar {\bf  K}|}
\right.\nonumber\\&&\left.
+\hat{{\bf k}}.\frac{{\bf  P}-\frac12\hbar{\bf  K}}{|{\bf  P}-\frac12\hbar{\bf  K}|} 
c({\bf  P}-\frac\hbar2{\bf  K}).\hat{{\bf k}}
\sin  \theta_{|{\bf  P}-\frac12\hbar{\bf  K}|} \right) 
\label{5.234}
\end{eqnarray}
\begin{eqnarray}
&&\left.I(\omega_k)_{M\ref{C.20}/\ref{C.28}}\right\vert_{Pr} 
=i\frac{q^2_ec^2}{2\varepsilon_0(2\pi)^3 \omega_{{\bf k}}} 
\frac1{4\hbar^2\omega^2_{k}}
\left(-E_{|{\bf  P}+\frac12\hbar {\bf  K}|}
\right.\nonumber\\&&\left.
+\hat{{\bf k}}.\frac{{\bf  P}+\frac12\hbar{\bf  K}}{|{\bf  P}+\frac12\hbar{\bf  K}|} 
c({\bf  P}+\frac\hbar2{\bf  K}).(-\hat{{\bf k}})
\sin  \theta_{|{\bf  P}+\frac12\hbar{\bf  K}|} \right) 
\label{5.235}
\end{eqnarray}
\begin{eqnarray}
&&\left.I(\omega_k)_{M\ref{C.27}/\ref{C.21}}\right\vert_{Pr} 
=-i\frac{q^2_ec^2}{2\varepsilon_0(2\pi)^3 \omega_{{\bf k}}} 
\frac1{4\hbar^2\omega^2_{k}}
\left(-E_{|{\bf  P}-\frac12\hbar {\bf  K}|}
\right.\nonumber\\&&\left.
+\hat{{\bf k}}.\frac{{\bf  P}-\frac12\hbar{\bf  K}}{|{\bf  P}-\frac12\hbar{\bf  K}|} 
(-1)c({\bf  P}-\frac\hbar2{\bf  K}).(-\hat{{\bf k}})
\sin  \theta_{|{\bf  P}-\frac12\hbar{\bf  K}|} \right) 
\label{5.236}
\end{eqnarray}
Therefore,
\begin{eqnarray}
&&\left.I(\omega_k)_{M}\right\vert_{Pr} 
=-i\frac{q^2_ec^2}{2\varepsilon_0(2\pi)^3 \omega_{{\bf k}}} 
\frac1{2\hbar^2\omega^2_{k}}
\left(E_{|{\bf  P}+\frac12\hbar {\bf  K}|}-E_{|{\bf  P}-\frac12\hbar {\bf  K}|}
\right.\nonumber\\&&\left.
+c\frac{(\hat{{\bf k}}.({\bf  P}+\frac12\hbar{\bf  K}))^2}{|{\bf  P}+\frac12\hbar{\bf  K}|} 
\sin  \theta_{|{\bf  P}+\frac12\hbar{\bf  K}|} 
-c\frac{(\hat{{\bf k}}.({\bf  P}-\frac12\hbar{\bf  K}))^2}{|{\bf  P}-\frac12\hbar{\bf  K}|} 
\sin  \theta_{|{\bf  P}-\frac12\hbar{\bf  K}|} 
\right) 
\nonumber\\&&
\label{5.237}
\end{eqnarray}
We now consider the correction due to the $k$ dependence in the trigonometric
functions.  We use (\ref{A.100}) and (\ref{A.101}).The last terms, 
in $\sin  \theta_{|{\bf  P}\pm\frac12\hbar{\bf  K}|} $do not contribute.
\begin{eqnarray}
&&\left.I(\omega_k)_{M\ref{C.2b}/\ref{C.4}}\right\vert_{Tr} 
=
i\frac{q^2_ec^2}{2\varepsilon_0(2\pi)^3 \omega_{{\bf k}}} 
\frac{mc^2}{2\hbar \omega_{\bf k}}
\nonumber\\&&\times
\left(
4\cos^2 \frac{ \theta_{|{\bf  P}+\frac12\hbar{\bf  K}|}}2
-4\sin^2 \frac{ \theta_{|{\bf  P}+\frac12\hbar{\bf  K}|}}2
\right)
\frac1{-2\hbar\omega_k} 
\label{5.238}
\end{eqnarray}
\begin{eqnarray}
&&\left.I(\omega_k)_{M\ref{C.5b}/\ref{C.1}}\right\vert_{Tr} 
=
i\frac{q^2_ec^2}{2\varepsilon_0(2\pi)^3 \omega_{{\bf k}}} 
\frac{mc^2}{2\hbar \omega_{\bf k}}
\nonumber\\&&\times
\left(
4\cos^2 \frac{ \theta_{|{\bf  P}-\frac12\hbar{\bf  K}|}}2
-4\sin^2 \frac{ \theta_{|{\bf  P}-\frac12\hbar{\bf  K}|}}2
\right)
\frac1{2\hbar\omega_k} 
\label{5.239}
\end{eqnarray}
\begin{eqnarray}
&&\left.I(\omega_k)_{M\ref{C.20}/\ref{C.28}}\right\vert_{Tr} 
=
-i\frac{q^2_ec^2}{2\varepsilon_0(2\pi)^3 \omega_{{\bf k}}} 
\frac{mc^2}{2\hbar \omega_{\bf k}}
\nonumber\\&&\times
\left(
4\cos^2 \frac{ \theta_{|{\bf  P}+\frac12\hbar{\bf  K}|}}2
-4\sin^2 \frac{ \theta_{|{\bf  P}+\frac12\hbar{\bf  K}|}}2
\right)
\frac1{2\hbar\omega_k} 
\label{5.240}
\end{eqnarray}
\begin{eqnarray}
&&\left.I(\omega_k)_{M\ref{C.27}/\ref{C.21}}\right\vert_{Tr} 
=
-i\frac{q^2_ec^2}{2\varepsilon_0(2\pi)^3 \omega_{{\bf k}}} 
\frac{mc^2}{2\hbar \omega_{\bf k}}
\nonumber\\&&\times
\left(
4\cos^2 \frac{ \theta_{|{\bf  P}-\frac12\hbar{\bf  K}|}}2
-4\sin^2 \frac{ \theta_{|{\bf  P}-\frac12\hbar{\bf  K}|}}2
\right)
\frac1{-2\hbar\omega_k} 
\label{5.241}
\end{eqnarray}

We consider the corrections due to the $k$-dependent factors.
We have
\begin{eqnarray}
&&({\bf  P}+\frac12\hbar{\bf  K}-\hbar{\bf  k})^2
=\hbar^2k^2-2({\bf  P}+\frac12\hbar{\bf  K}).\hbar{\bf  k}+\dots
\nonumber\\&&
=\hbar^2k^2\left(1-2\frac1{\hbar k}({\bf  P}+\frac12\hbar{\bf  K}).\hat{\bf  k}\right)+\dots
\label{A.103}
\end{eqnarray}
Therefore,
\begin{eqnarray}
&&|{\bf  P}+\frac12\hbar{\bf  K}-\hbar{\bf  k}|
=\hbar k\left(1-\frac1{\hbar k}({\bf  P}+\frac12\hbar{\bf  K}).\hat{\bf  k}\right)+\dots
\nonumber\\&&
\frac1{|{\bf  P}+\frac12\hbar{\bf  K}-\hbar{\bf  k}|}=\frac1{\hbar k}
\left(1+\frac1{\hbar k}({\bf  P}+\frac12\hbar{\bf  K}).\hat{\bf  k}\right)+\dots
\label{A.104}
\end{eqnarray}
 Using these relations, we have:
\begin{eqnarray}
&&
\left.I(\omega_k)_{M\ref{C.2b}/\ref{C.4}}\right\vert_{F} 
=
i\frac{q^2_ec^2}{2\varepsilon_0(2\pi)^3 \omega_{{\bf k}}} 
\frac{|{\bf  P}+\frac12\hbar{\bf  K}|}
{\hbar k}
\sin  \theta_{|{\bf  P}+\frac12\hbar{\bf  K}|} 
\frac1{-2\hbar\omega_k} 
\nonumber\\&&
i\frac{q^2_ec^2}{2\varepsilon_0(2\pi)^3 \omega_{{\bf k}}} 
\frac1{\hbar k}
\left(\hat{\bf k}.({\bf  P}+\frac12\hbar{\bf  K})\right)
\left((-\hat{\bf k}).
\frac{{\bf  P}+\frac12\hbar{\bf  K}}
{|{\bf  P}+\frac12\hbar{\bf  K}|}\right)
\sin  \theta_{|{\bf  P}+\frac12\hbar{\bf  K}|} 
\frac1{-2\hbar\omega_k} 
\nonumber\\&&
\label{5.242}
\end{eqnarray}
\begin{eqnarray}
&&
\left.I(\omega_k)_{M\ref{C.5b}/\ref{C.1}}\right\vert_{F} 
=
i\frac{q^2_ec^2}{2\varepsilon_0(2\pi)^3 \omega_{{\bf k}}} 
\frac{|{\bf  P}-\frac12\hbar{\bf  K}|}
{\hbar k}
\sin  \theta_{|{\bf  P}-\frac12\hbar{\bf  K}|} 
\frac1{2\hbar\omega_k} 
\nonumber\\&&
i\frac{q^2_ec^2}{2\varepsilon_0(2\pi)^3 \omega_{{\bf k}}} 
\frac1{\hbar k}
\left(\hat{\bf k}.({\bf  P}-\frac12\hbar{\bf  K})\right)
\left((-\hat{\bf k}).
\frac{{\bf  P}-\frac12\hbar{\bf  K}}
{|{\bf  P}-\frac12\hbar{\bf  K}|}\right)
\sin  \theta_{|{\bf  P}-\frac12\hbar{\bf  K}|} 
\frac1{2\hbar\omega_k} 
\nonumber\\&&
\label{5.243}
\end{eqnarray}
\begin{eqnarray}
&&
\left.I(\omega_k)_{M\ref{C.20}/\ref{C.28}}\right\vert_{F} 
=-i\frac{q^2_ec^2}{2\varepsilon_0(2\pi)^3 \omega_{{\bf k}}} 
\frac{|{\bf  P}+\frac12\hbar{\bf  K}|}
{\hbar k}
\sin  \theta_{|{\bf  P}+\frac12\hbar{\bf  K}|} 
\frac1{2\hbar\omega_k} 
\nonumber\\&&
-i\frac{q^2_ec^2}{2\varepsilon_0(2\pi)^3 \omega_{{\bf k}}} 
\frac1{\hbar k}
\left((-\hat{\bf k}).({\bf  P}+\frac12\hbar{\bf  K})\right)
\left((\hat{\bf k}).
\frac{{\bf  P}+\frac12\hbar{\bf  K}}
{|{\bf  P}+\frac12\hbar{\bf  K}|}\right)
\sin  \theta_{|{\bf  P}+\frac12\hbar{\bf  K}|} 
\frac1{2\hbar\omega_k} 
\nonumber\\&&
\label{5.244}
\end{eqnarray}
\begin{eqnarray}
&&
\left.I(\omega_k)_{M\ref{C.27}/\ref{C.21}}\right\vert_{F} 
=-i\frac{q^2_ec^2}{2\varepsilon_0(2\pi)^3 \omega_{{\bf k}}} 
\frac{|{\bf  P}-\frac12\hbar{\bf  K}|}
{\hbar k}
\sin  \theta_{|{\bf  P}-\frac12\hbar{\bf  K}|} 
\frac1{-2\hbar\omega_k} 
\nonumber\\&&
-i\frac{q^2_ec^2}{2\varepsilon_0(2\pi)^3 \omega_{{\bf k}}} 
\frac1{\hbar k}
\left((-\hat{\bf k}).({\bf  P}-\frac12\hbar{\bf  K})\right)
\left((\hat{\bf k}).
\frac{{\bf  P}-\frac12\hbar{\bf  K}}
{|{\bf  P}-\frac12\hbar{\bf  K}|}\right)
\sin  \theta_{|{\bf  P}-\frac12\hbar{\bf  K}|} 
\frac1{-2\hbar\omega_k} 
\nonumber\\&&
\label{5.245}
\end{eqnarray}
Combining these expressions with (\ref{5.237}), we obtain
\begin{eqnarray}
&&\left.I(\omega_k)_{M}\right\vert_{Pr} 
+\left.I(\omega_k)_{M}\right\vert_{F} 
=i\frac{q^2_ec^2}{2\varepsilon_0(2\pi)^3 \omega_{{\bf k}}} 
\frac1{2\hbar^2\omega^2_{k}}
\left(E_{|{\bf  P}+\frac12\hbar {\bf  K}|}-E_{|{\bf  P}-\frac12\hbar {\bf  K}|}
\right.\nonumber\\&&\left.
+3c\frac{(\hat{{\bf k}}.({\bf  P}+\frac12\hbar{\bf  K}))^2}{|{\bf  P}+\frac12\hbar{\bf  K}|} 
\sin  \theta_{|{\bf  P}+\frac12\hbar{\bf  K}|} 
-3c\frac{(\hat{{\bf k}}.({\bf  P}-\frac12\hbar{\bf  K}))^2}{|{\bf  P}-\frac12\hbar{\bf  K}|} 
\sin  \theta_{|{\bf  P}-\frac12\hbar{\bf  K}|} 
\right.\nonumber\\&&\left.
-2|{\bf  P}+\frac12\hbar{\bf  K}|c
\sin  \theta_{|{\bf  P}+\frac12\hbar{\bf  K}|} 
+2|{\bf  P}-\frac12\hbar{\bf  K}|c
\sin  \theta_{|{\bf  P}-\frac12\hbar{\bf  K}|}
\right)  
\label{5.246}
\end{eqnarray}
The angular integration with $(\hat{{\bf k}}.({\bf  P}+\frac12\hbar{\bf  K}))^2$ will provide the $\frac13$
of the angular integration with $\vert{\bf  P}+\frac12\hbar{\bf  K}\vert^2$.
A simplification is thus possible and we have:
\begin{eqnarray}
&&\left.I(\omega_k)_{M}\right\vert_{Pr} 
+\left.I(\omega_k)_{M}\right\vert_{F} 
=i\frac{q^2_ec^2}{2\varepsilon_0(2\pi)^3 \omega_{{\bf k}}} 
\frac1{2\hbar^2\omega^2_{k}}
\left(E_{|{\bf  P}+\frac12\hbar {\bf  K}|}-E_{|{\bf  P}-\frac12\hbar {\bf  K}|}
\right.\nonumber\\&&\left.
-|{\bf  P}+\frac12\hbar{\bf  K}|c
\sin  \theta_{|{\bf  P}+\frac12\hbar{\bf  K}|} 
+|{\bf  P}-\frac12\hbar{\bf  K}|c
\sin  \theta_{|{\bf  P}-\frac12\hbar{\bf  K}|}
\right)  
\nonumber\\&&
=i\frac{q^2_ec^2}{2\varepsilon_0(2\pi)^3 \omega_{{\bf k}}} 
\frac1{2\hbar^2\omega^2_{k}}
\left(E_{|{\bf  P}+\frac12\hbar {\bf  K}|}-E_{|{\bf  P}-\frac12\hbar {\bf  K}|}
\right.\nonumber\\&&\left.
-\frac{|{\bf  P}+\frac12\hbar{\bf  K}|^2c^2}
{E_{|{\bf  P}+\frac12\hbar {\bf  K}|}}
+\frac{|{\bf  P}-\frac12\hbar{\bf  K}|^2c^2}
{E_{|{\bf  P}-\frac12\hbar {\bf  K}|}}
\right)  
\nonumber\\&&
=i\frac{q^2_ec^2}{2\varepsilon_0(2\pi)^3 \omega_{{\bf k}}} 
\frac1{2\hbar^2\omega^2_{k}}
\left(
\frac{m^2c^4}
{E_{|{\bf  P}+\frac12\hbar {\bf  K}|}}
-\frac{m^2c^4}
{E_{|{\bf  P}-\frac12\hbar {\bf  K}|}}
\right)  
\label{5.247}
\end{eqnarray}

\section{Expliciting the second  order contribution $ (\chi^{-1})^{(2)}$ to the dressing operator
$(\chi^{-1})$}

\setcounter{equation}{0}
\def\theequation{\thesection.\arabic{equation}}

In order to evaluate the right hand side of (\ref{4.a10}), we have to compute the $ (\chi^{-1})^{(2)}$
operator. Anew, we do not consider the presence of an incident field and the expression of $
(\chi^{-1})^{(2)}$ can be deduced from (\ref{3.a7}).
We call $f_{\alpha\beta}^{W(2)}(-{\bf K},-{\bf  X})$ the second order contribution. 
We write explicitly only the contribution arising from $\tilde f_{110000}^{\mu\mu'}$.
The creation operator $ C^{(1)}$ requires one matrix element of the coupling between the Dirac and
electromagnetic fields.
We take into account the vanishing of a matrix element when more than two matter indices are
changed and the vanishing of two elements of $\gamma_{\alpha\beta}$ to obtain:
\begin{eqnarray}
&&f_{\alpha\beta}^{W(2)}(-{\bf K},-{\bf  X})
\\&&
=\hbar^3\int d^3p\int d^3p'\sum_{\mu\mu'}\sum_{\mu_1\mu'_1}\sum_{\lambda_1}
\int d^3k_1\,
\gamma_{\alpha\beta}^{1\mu_11 \mu'_110}({\bf p},{\bf p}',k_1) 
\nonumber\\&&\times
<110010\mu_1 \mu'_1\vert C^{(1)}\vert 110000\mu\mu'> 
 \tilde f_{110000}^{\mu\mu'}
\nonumber\\&&
+\hbar^3\int d^3p\int d^3p'\sum_{\mu\mu'}\sum_{\mu_1\mu'_1}\sum_{\lambda'_1}
\int d^3k'_1\,
\gamma_{\alpha\beta}^{1\mu_11\mu'_101}({\bf p},{\bf p}',k'_1) 
\nonumber\\&&\times
<110001\mu_1 \mu'_1\vert C^{(1)}\vert 110000\mu\mu'> 
 \tilde f_{110000}^{\mu\mu'}
\nonumber\\&&
+\hbar^3\int d^3p'\int d^3\tilde p'\sum_{\mu\mu'}\sum_{\mu_1\mu'_1}\sum_{\lambda'_1}
\int d^3k'_1\,
\gamma_{\alpha\beta}^{1\mu_1\bar1 \mu'_101}({\bf p}',\tilde{{\bf p}}',k'_1)
\nonumber\\&&\times 
<010101\mu_1 \mu'_1\vert C^{(1)}\vert 110000\mu\mu'> 
 \tilde f_{110000}^{\mu\mu'}
\nonumber\\&&
+\hbar^3\int d^3p\int d^3\tilde p\sum_{\mu\mu'}\sum_{\mu_1\mu'_1}\sum_{\lambda_1}
\int d^3k_1\,
\gamma_{\alpha\beta}^{\bar1\tilde\mu 1\mu  10}({\bf p},\tilde{{\bf p}},k_1) 
\nonumber\\&&\times
<101010\mu_1 \mu'_1\vert C^{(1)}\vert 110000\mu\mu'> 
 \tilde f_{110000}^{\mu\mu'}
\label{4.a26}
\end{eqnarray}
The values of the first order $\gamma$, determined by the Weyl's rule of correspondence, have been
given previously (\ref{5.30}-\ref{5.36}).

In order to use the form of $\chi^{(0)\xi\mu\xi'\mu'}_{\alpha\beta}$ (\ref{4.a23}), we have to write
(\ref{4.a26}) in the appropriate variables.
We write the  four terms in (\ref{4.a26}) as the sum of four contributions affected by the indices
corresponding to the nature of the Dirac and the e. m. fields.
We replace in each term the $\gamma$ by its expression and we
take the Fourier transform with respect to the variable ${\bf  X}$.
We change ${\bf K}\to -{\bf K}$.
From our definition of Fourier transform, we have to multiply both sides by $\frac1{(2\pi)^3}
e^{\frac{-i{\bf P}.{\bf  X}}{\hbar} }$ and integrate over
${\bf  X}$.
That procedure provides a function of the variable $\frac{{\bf P}}{\hbar}$.
In order to obtain a normalisable function of ${\bf P}$, that function has to be multiplied by the factor
$\frac1{\hbar^3}$ 
\begin{eqnarray}
&&f_{\alpha\beta110010}^{W(2)}({\bf K},{\bf P})
=\frac1{(2\pi\hbar)^3}\frac{-iq_e}\hbar\sqrt{\frac{\hbar}{2\varepsilon_0  (2\pi)^3}}
 \int d^3X\,e^{\frac{-i{\bf P}.{\bf  X}}{\hbar} }
\nonumber\\&&\times
\int d^3p\int d^3p'\sum_{\mu\mu'}\sum_{\mu_1\mu'_1}\sum_{\lambda_1}
\int d^3k_1\,
\nonumber\\&&\times
\delta({\bf p}-{\bf p}'-\hbar{\bf K}+\hbar{\bf k}_1)
e^{-i\frac12{\bf K}.{\bf  X}}
\frac1{\sqrt{\omega_{{\bf k}_1}}}{\bf e}_{{\bf k}_1 \lambda_1}.
{\bf  X} 
e^{ i{\bf k}_1.{\bf  X}}
\nonumber\\&&\times
e^{\frac{i{\bf p}.{\bf  X}}{\hbar} }
\delta_{\alpha\beta}^{1\mu'_11\mu_1}({\bf p}', {\bf p})
<110010\mu_1 \mu'_1\vert C^{(1)}\vert 110000\mu\mu'> 
 \tilde f_{110000}^{\mu\mu'}
\label{4.a36}
\end{eqnarray}
\begin{eqnarray}
&&f_{\alpha\beta110001}^{W(2)}({\bf K},{\bf P})
=
\frac1{(2\pi\hbar)^3}\frac{-iq_e}\hbar\sqrt{\frac{\hbar}{2\varepsilon_0  (2\pi)^3}}
 \int d^3X\,e^{\frac{-i{\bf P}.{\bf  X}}{\hbar} }
\nonumber\\&&\times
\int d^3p\int d^3p'\sum_{\mu\mu'}\sum_{\mu_1\mu'_1}\sum_{\lambda'_1}
\int d^3k'_1\,
\nonumber\\&&\times
\delta({\bf p}-{\bf p}'-\hbar{\bf K}-\hbar{\bf k}'_1)
e^{-i\frac12{\bf K}.{\bf  X}}
\frac1{\sqrt{\omega_{{\bf k}'_1}}}{\bf e}_{{\bf k}'_1 \lambda_1}.
{\bf  X} 
\nonumber\\&&\times
e^{\frac{i{\bf p}.{\bf  X}}{\hbar} }
\delta_{\alpha\beta}^{1\mu'_11\mu_1}({\bf p}', {\bf p})
<110001\mu_1 \mu'_1\vert C^{(1)}\vert 110000\mu\mu'> 
 \tilde f_{110000}^{\mu\mu'}
\label{4.a37}
\end{eqnarray}
\begin{eqnarray}
&&f_{\alpha\beta010101}^{W(2)}({\bf K},{\bf P})
=\frac1{(2\pi\hbar)^3}\frac{-iq_e}\hbar\sqrt{\frac{\hbar}{2\varepsilon_0  (2\pi)^3}}
 \int d^3X\,e^{\frac{-i{\bf P}.{\bf  X}}{\hbar} }
\nonumber\\&&\times
\int d^3p'\int d^3\tilde p'\sum_{\mu\mu'}\sum_{\mu_1\mu'_1}\sum_{\lambda'_1}
\int d^3k'_1\,
\nonumber\\&&\times
\delta(-{\bf p}'-\tilde{{\bf p}}'-\hbar{\bf K}-\hbar{\bf k}'_1)
e^{-i\frac12{\bf K}.{\bf  X}}
\frac1{\sqrt{\omega_{{\bf k}'_1}}}{\bf e}_{{\bf k}'_1 \lambda'_1}.
{\bf  X} 
(1- e^{ -i{\bf k}'_1.{\bf  X}})
\nonumber\\&&\times
e^{-\frac{i\tilde{{\bf p}}'.{\bf  X}}{\hbar} }
\delta_{\alpha\beta}^{1\mu_1\bar1\bar\mu'_1}({\bf p}', -\tilde{{\bf p}}')
<010101\mu_1 \mu'_1\vert C^{(1)}\vert 110000\mu\mu'> 
 \tilde f_{110000}^{\mu\mu'}
\nonumber\\&&
\label{4.a39}
\end{eqnarray}
\begin{eqnarray}
&&f_{\alpha\beta101010}^{W(2)}({\bf K},{\bf P})
=\frac1{(2\pi\hbar)^3}\frac{-iq_e}\hbar\sqrt{\frac{\hbar}{2\varepsilon_0  (2\pi)^3}}
 \int d^3X\,e^{\frac{-i{\bf P}.{\bf  X}}{\hbar} }
\nonumber\\&&\times
\int d^3p\int d^3\tilde p\sum_{\mu\mu'}\sum_{\mu_1\mu'_1}\sum_{\lambda_1}
\int d^3k_1\,
\nonumber\\&&\times
\delta({\bf p}+\tilde{{\bf p}}-\hbar{\bf K}+\hbar{\bf k}_1)
e^{-i\frac12{\bf K}.{\bf  X}}
\frac1{\sqrt{\omega_{{\bf k}_1}}}{\bf e}_{{\bf k}_1 \lambda_1}.
{\bf  X} 
(e^{ i{\bf k}_1.{\bf  X}}-1)
\nonumber\\&&\times
e^{\frac{i{{\bf p}}.{\bf  X}}{\hbar} }
\delta_{\alpha\beta}^{\bar1\bar\mu'_11\mu_1}(-\tilde{{\bf p}},{\bf p} )
<101010\mu_1 \mu'_1\vert C^{(1)}\vert 110000\mu\mu'> 
 \tilde f_{110000}^{\mu\mu'}
\label{4.a40}
\end{eqnarray}

Trough usual manipulations, we have successsively for the action of any $g({\bf p}, {\bf p}')$ function:
\begin{eqnarray}
&&\int d^3p\int d^3p' \int d^3X\,e^{\frac{-i{\bf P}.{\bf  X}}{\hbar} }
\delta({\bf p}-{\bf p}'-\hbar{\bf K}+\hbar{\bf k}_1)
\nonumber\\&&\times
e^{-i\frac12{\bf K}.{\bf  X}}{\bf e}_{{\bf k}_1 \lambda_1}.
{\bf  X} 
e^{ i{\bf k}_1.{\bf  X}}
e^{\frac{i{\bf p}.{\bf  X}}{\hbar} }
g({\bf p}, {\bf p}')
\nonumber\\&&=
(2\pi)^3\hbar^3
\frac\hbar i
\int d^3p\int d^3p'
\delta({\bf p}-{\bf P}-\frac12\hbar{\bf K}+\hbar{\bf k}_1)
\delta({\bf p}'-{\bf P}+\frac12\hbar{\bf K})
\nonumber\\&&\times
{\bf e}_{{\bf k}_1 \lambda_1}.
[-\nabla_{{\bf p}}-\nabla_{{\bf p}'}]
g({\bf p}, {\bf p}')
\label{4.a42}
\end{eqnarray}
\begin{eqnarray}
&&\int d^3p\int d^3p'\int d^3X\,e^{\frac{-i{\bf P}.{\bf  X}}{\hbar} }
\delta({\bf p}-{\bf p}'-\hbar{\bf K}-\hbar{\bf k}'_1)
\nonumber\\&&\times
e^{-i\frac12{\bf K}.{\bf  X}}{\bf e}_{{\bf k}'_1 \lambda_1}.{\bf  X} 
e^{\frac{i{\bf p}.{\bf  X}}{\hbar} }
g({\bf p}, {\bf p}')
\nonumber\\&&=
(2\pi)^3\hbar^3
\frac\hbar i
\int d^3p\int d^3p'
\delta({\bf p}-{\bf P}-\frac12\hbar{\bf K})
\delta({\bf p}'-{\bf P}+\frac12\hbar{\bf K}+\hbar{\bf k}'_1)
\nonumber\\&&\times
{\bf e}_{{\bf k}'_1 \lambda'_1}.
[-\nabla_{{\bf p}}-\nabla_{{\bf p}'}]
g({\bf p}, {\bf p}')
\label{4.a43}
\end{eqnarray}
\begin{eqnarray}
 &&\int d^3p'\int d^3\tilde p'\int d^3X\,e^{\frac{-i{\bf P}.{\bf  X}}{\hbar} }
\delta(-{\bf p}'-\tilde{{\bf p}}'-\hbar{\bf K}-\hbar{\bf k}'_1)
\nonumber\\&&\times
e^{-i\frac12{\bf K}.{\bf  X}}{\bf e}_{{\bf k}'_1 \lambda_1}.{\bf  X} 
(1- e^{ -i{\bf k}'_1.{\bf  X}})
e^{-\frac{i\tilde{{\bf p}}'.{\bf  X}}{\hbar} }
g({\bf p}',\tilde{{\bf p}}')
\nonumber\\&&=
(2\pi)^3\hbar^3
\frac\hbar i
 \int d^3p'\int d^3\tilde p'
\left[\delta(\tilde{{\bf p}}'+{\bf P}+\frac12\hbar{\bf K})
\delta({\bf p}'-{\bf P}+\frac12\hbar{\bf K}+\hbar{\bf k}'_1)
\right.\nonumber\\ &&\left.
-\delta(\tilde{{\bf p}}'+{\bf P}+\frac12\hbar{\bf K}+\hbar{\bf k}'_1)
\delta({\bf p}'-{\bf P}+\frac12\hbar{\bf K})
\right]{\bf e}_{{\bf k}'_1 \lambda'_1}.
[-\nabla_{{\bf p}'}+\nabla_{\tilde{{\bf p}}'}]
g({\bf p}',\tilde{{\bf p}}')
\nonumber\\&&
\label{4.a45}
\end{eqnarray}
\begin{eqnarray}
&&\int d^3p\int d^3\tilde p \int d^3X\,e^{\frac{-i{\bf P}.{\bf  X}}{\hbar} }
\delta({\bf p}+\tilde{{\bf p}}-\hbar{\bf K}+\hbar{\bf k}_1)
\nonumber\\&&\times
e^{-i\frac12{\bf K}.{\bf  X}}{\bf e}_{{\bf k}_1 \lambda_1}.{\bf  X} 
(e^{ i{\bf k}_1.{\bf  X}}-1)
e^{\frac{i{{\bf p}}.{\bf  X}}{\hbar} }
g({\bf p}, \tilde{{\bf p}})
\nonumber\\&&=
(2\pi)^3\hbar^3
\frac\hbar i
 \int d^3p\int d^3\tilde p
\left[\delta({\bf p}-{\bf P}-\frac12\hbar{\bf K}+\hbar{\bf k}_1)
\delta(\tilde{{\bf p}}+{\bf P}-\frac12\hbar{\bf K})
\right.\nonumber\\&&\left.
-\delta({\bf p}-{\bf P}-\frac12\hbar{\bf K})
\delta(\tilde{{\bf p}}+{\bf P}-\frac12\hbar{\bf K}+\hbar{\bf k}_1)
\right]{\bf e}_{{\bf k}_1 \lambda_1}.
[-\nabla_{{\bf p}}+\nabla_{\tilde{{\bf p}}}]
g({\bf p},\tilde{{\bf p}})
\nonumber\\&&
\label{4.a46}
\end{eqnarray}

\section{Computation of $[\chi^{(0)} (\chi^{-1})^{(2)},\Theta^{(0)}]$}

\setcounter{equation}{0}
\def\theequation{\thesection.\arabic{equation}}

In our investigation, we are interested whether that the second term in (\ref{4.a10}) may compensate
the divergent contribution of the first one.
We focus on one kind of contribution, namely the connection between 
$ \tilde f_{110000}^{\mu\mu'}({\bf p},{\bf p}')$ and itself.
From (\ref{4.a21}), this corresponds also to the connection of $\tilde f^{W1\mu1\mu'}({\bf K},{\bf P})$
to itself. 
When the argument is changed by a photon wave number, no divergence occurs due to the
assumed behaviour of $ \tilde f_{110000}^{\mu\mu'}({\bf p},{\bf p}')$ for large values of its arguments.
We have to restrict to cases with the same value for the argument.
The possible existence of such diagonal contributions can be easily realized. 
Indeed, the change of argument
due to the photon in the Dirac delta functions in (\ref{4.a42})-(\ref{4.a46}) can be
compensated by the change of argument arising from the creation operator in
(\ref{4.a36})-(\ref{4.a40}). 
The operator $\Theta^{(0)}$ is purely diagonal.
In order that the diagonal contribution of the commutator provides a non vanishing
result,  it will be necessary that the gradients present in (\ref{4.a42})-(\ref{4.a46}) acts on the
diagonal operator
$\Theta^{(0)}$.

To obtain the matrix elements of $\chi^{(0)}(\chi^{-1})^{(2)}$, we acts with $\chi^{(0)}$ on both
sides of (\ref{4.a36})-(\ref{4.a40}), after use of the expressions (\ref{4.a42})-(\ref{4.a46}).
We introduce lower indices in expressions such as 
$[\chi^{(0)}f_{\alpha\beta}^{W(2)}]^{1\mu1\mu'}_{110010}({\bf K},{\bf P})$ to keep trace of the origin of
each contribution: the indices ${110010}$ are the indices of the intermediate state present in the
creation operator.
The intermediate steps are given only for that contribution since the other ones are computed in a
completely similar way.
\begin{eqnarray}
&&[\chi^{(0)}f_{\alpha\beta}^{W(2)}]^{1\mu1\mu'}_{110010}({\bf K},{\bf P})
=
\frac14\sum_{\alpha\beta}
\delta_{\alpha\beta}^{1\mu1\mu'}({\bf P}+\frac{\hbar}{2}{\bf K}, 
{\bf P}-\frac{\hbar}{2}{\bf K})
\nonumber\\&&\times
\left(\frac{-iq_e}{\hbar}\sqrt{\frac{\hbar}{2\varepsilon_0  (2\pi)^3}}\right)
\sum_{\mu_1\mu'_1}\sum_{\mu_2\mu'_2}\sum_{\lambda_1}
\int d^3k_1
 \int d^3p\int d^3p'
\nonumber\\&&\times
\delta({\bf p}-{\bf P}-\frac\hbar2{\bf K}+\hbar{\bf k}_1)
\delta({\bf p}'-{\bf P}+\frac\hbar2{\bf K}))
\nonumber\\&&\times
\frac1{\sqrt{\omega_{k_1}}}
\frac\hbar i{\bf e}_{{\bf k}_1 \lambda_1}. [-\nabla_{{\bf p}} -\nabla_{{\bf p}'}]
\delta_{\alpha\beta}^{1\mu'_11\mu_1}({\bf p}', {\bf p})
\nonumber\\&&\times
<110010\mu_1 \mu'_1\vert C^{(1)}\vert 110000\mu_2\mu'_2> 
 \tilde f_{110000}^{\mu_2\mu'_2}
\label{4.a48}
\end{eqnarray}

We replace also the function $ \tilde f_{110000}^{\mu\mu'}({\bf p},{\bf p}')$ by its expression in
terms of $\tilde f^{1\mu1\mu'}({\bf K},{\bf P})$ (\ref{4.a20}) in order that both sides of the expressions
are in similar variables.
\begin{eqnarray}
&&[\chi^{(0)}f_{\alpha\beta}^{W(2)}]^{1\mu1\mu'}_{110010}({\bf K},{\bf P})
=
\frac14\frac1{\hbar^3}\sum_{\alpha\beta}
\delta_{\alpha\beta}^{1\mu1\mu'}({\bf P}+\frac{\hbar}{2}{\bf K}, 
{\bf P}-\frac{\hbar}{2}{\bf K})
\nonumber\\&&\times
\left(\frac{-iq_e}{\hbar}\sqrt{\frac{\hbar}{2\varepsilon_0  (2\pi)^3}}\right)
\sum_{\mu_1\mu'_1}\sum_{\mu_2\mu'_2}\sum_{\lambda_1}
\int d^3k_1
 \int d^3p\int d^3p'
\nonumber\\&&\times
\delta({\bf p}-{\bf P}-\frac\hbar2{\bf K}+\hbar{\bf k}_1)
\delta({\bf p}'-{\bf P}+\frac\hbar2{\bf K}))
\nonumber\\&&\times
\frac1{\sqrt{\omega_{k_1}}}
\frac\hbar i{\bf e}_{{\bf k}_1 \lambda_1}. [-\nabla_{{\bf p}} -\nabla_{{\bf p}'}]
\delta_{\alpha\beta}^{1\mu'_11\mu_1}({\bf p}', {\bf p})
\nonumber\\&&\times
<110010\mu_1 \mu'_1\vert C^{(1)}\vert 110000\mu_2\mu'_2> 
\tilde f^{1\mu_21\mu'_2}(\frac1\hbar({\bf p}-{\bf p}'),\frac12({\bf p}+{\bf p}'))
\nonumber\\&&
\label{4.a48a}
\end{eqnarray}

The relevant first order creation operator has been determined in appendix D 
and its expression (\ref{4.a63}) can be reported in previous
formula (\ref{4.a48a}) that becomes:
\begin{eqnarray}
&&[\chi^{(0)}f_{\alpha\beta}^{W(2)}]^{1\mu1\mu'}_{110010}({\bf K},{\bf P})
=
\frac14\frac1{\hbar^3}\sum_{\alpha\beta}
\delta_{\alpha\beta}^{1\mu1\mu'}({\bf P}+\frac{\hbar}{2}{\bf K}, 
{\bf P}-\frac{\hbar}{2}{\bf K})
\nonumber\\&&\times
\left(\frac{-iq_e}{\hbar}\sqrt{\frac{\hbar}{2\varepsilon_0  (2\pi)^3}}\right)
\sum_{\mu_1\mu'_1}\sum_{\mu_2\mu'_2}\sum_{\lambda_1}
\int d^3k_1
 \int d^3p\int d^3p'
\nonumber\\&&\times
\delta({\bf p}-{\bf P}-\frac\hbar2{\bf K}+\hbar{\bf k}_1)
\delta({\bf p}'-{\bf P}+\frac\hbar2{\bf K}))
\nonumber\\&&\times
\frac1{\sqrt{\omega_{k_1}}}
\frac\hbar i{\bf e}_{{\bf k}_1 \lambda_1}. [-\nabla_{{\bf p}} -\nabla_{{\bf p}'}]
\delta_{\alpha\beta}^{1\mu'_11\mu_1}({\bf p}', {\bf p})
\nonumber\\&&\times
(-\hbar^3)
\sqrt{\frac{\hbar}{2\varepsilon_0(2\pi)^3 \omega_{{\bf k}_1} }}
\alpha^{11}_{\lambda_1+ \mu_1 \mu_2}
({\bf p}+\frac12\hbar {\bf k}_1,{\bf k}_1)
\delta^{Kr}_{\mu'_1,\mu'_2}
\nonumber\\&&\times
\frac1{E_{{\bf p}+\hbar {\bf k}_1}-E_{{\bf p}}-\hbar\omega_{k_1}}
S({\bf p}+\hbar {\bf k}_1;{\bf p})
\tilde f^{1\mu_21\mu'_2}(\frac1\hbar({\bf p}-{\bf p}'),\frac12({\bf p}+{\bf p}'))
\nonumber\\&&
\label{4.a67a}
\end{eqnarray}

Using the relations (\ref{3.a7}) between the $\alpha$'s and the $\delta$'s,
we obtain, for all contributions:
\begin{eqnarray}
&&[\chi^{(0)}f_{\alpha\beta}^{W(2)}]^{1\mu1\mu'}_{110010}({\bf K},{\bf P})
=
\frac14\sum_{\alpha\beta}
\delta_{\alpha\beta}^{1\mu1\mu'}({\bf P}+\frac{\hbar}{2}{\bf K}, 
{\bf P}-\frac{\hbar}{2}{\bf K})
\nonumber\\&&\times
\left(\frac{-iq_e}{\hbar}\sqrt{\frac{\hbar}{2\varepsilon_0  (2\pi)^3}}\right)
\sum_{\mu_1\mu'_1}\sum_{\mu_2\mu'_2}\sum_{\lambda_1}
\int d^3k_1
 \int d^3p\int d^3p'
\nonumber\\&&\times
\delta({\bf p}-{\bf P}-\frac\hbar2{\bf K}+\hbar{\bf k}_1)
\delta({\bf p}'-{\bf P}+\frac\hbar2{\bf K}))
\nonumber\\&&\times
\frac1{\sqrt{\omega_{k_1}}}
\frac\hbar i{\bf e}_{{\bf k}_1 \lambda_1}. [-\nabla_{{\bf p}} -\nabla_{{\bf p}'}]
\delta_{\alpha\beta}^{1\mu'_11\mu_1}({\bf p}', {\bf p})
\nonumber\\&&\times
(-\hbar^3)
\sqrt{\frac{\hbar}{2\varepsilon_0(2\pi)^3 \omega_{{\bf k}_1} }}
q_e c\frac1{\hbar^3}\sum_{i=1,3}
({\bf e}_{{\bf k}_1\lambda_1})_i
\delta_{1i}^{1\mu_11\mu_2}({\bf p}, {\bf p}+\hbar {\bf k}_1)
\delta^{Kr}_{\mu'_1,\mu'_2}
\nonumber\\&&\times
\frac1{E_{{\bf p}+\hbar {\bf k}_1}-E_{{\bf p}}-\hbar\omega_{k_1}}
S({\bf p}+\hbar {\bf k}_1;{\bf p})
 \tilde f^{1\mu_21\mu'_2}(\frac1\hbar({\bf p}-{\bf p}'),\frac12({\bf p}+{\bf p}'))
\nonumber\\&&
\label{4.a74}
\end{eqnarray}
\begin{eqnarray}
&&[\chi^{(0)}f_{\alpha\beta}^{W(2)}]^{1\mu1\mu'}_{110001}({\bf K},{\bf P})
=
\frac14\sum_{\alpha\beta}
\delta_{\alpha\beta}^{1\mu1\mu'}({\bf P}+\frac{\hbar}{2}{\bf K}, 
{\bf P}-\frac{\hbar}{2}{\bf K})
\nonumber\\&&\times
\left(\frac{-iq_e}{\hbar}\sqrt{\frac{\hbar}{2\varepsilon_0  (2\pi)^3}}\right)
\sum_{\mu_1\mu'_1}\sum_{\mu_2\mu'_2}\sum_{\lambda'_1}
\int d^3k'_1\,
 \int d^3p\int d^3p'
\nonumber\\&&\times
\delta({\bf p}-{\bf P}-\frac\hbar2{\bf K})
\delta({\bf p}'-{\bf P}+\frac\hbar2{\bf K}+\hbar{\bf k}'_1))
\nonumber\\&&\times
\frac1{\sqrt{\omega_{k'_1}}}
\frac\hbar i{\bf e}_{{\bf k}'_1 \lambda'_1}. [-\nabla_{{\bf p}} -\nabla_{{\bf p}'}]
\delta_{\alpha\beta}^{1\mu'_11\mu_1}({\bf p}', {\bf p})
\nonumber\\&&\times
\hbar^3
\sqrt{\frac{\hbar}{2\varepsilon_0(2\pi)^3 \omega_{{\bf k}_1} }}
q_e c\frac1{\hbar^3}\sum_{i=1,3}
({\bf e}_{-{\bf k}'_1\lambda'_1})_i
 \delta_{1i}^{1\mu'_2 1\mu'_1}({\bf p}'+\hbar{\bf k}'_1,
{\bf p}')
\delta^{Kr}_{\mu_1,\mu_2}
\nonumber\\&&\times
\frac1{-E_{{\bf p}'+\hbar {\bf k}'_1}+E_{{\bf p}'}+\hbar\omega_{k'_1}}
S({\bf p}'+\hbar {\bf k}'_1;{\bf p}')
 \tilde f^{1\mu_21\mu'_2}(\frac1\hbar({\bf p}-{\bf p}'),\frac12({\bf p}+{\bf p}'))
\nonumber\\&&
\label{4.a75}
\end{eqnarray}
\begin{eqnarray}
&&[\chi^{(0)}f_{\alpha\beta}^{W(2)}]^{1\mu1\mu'}_{010101}({\bf K},{\bf P})
=
\frac14\sum_{\alpha\beta}
\delta_{\alpha\beta}^{1\mu1\mu'}({\bf P}+\frac{\hbar}{2}{\bf K}, 
{\bf P}-\frac{\hbar}{2}{\bf K})
\nonumber\\&&\times
\left(\frac{-iq_e}{\hbar}\sqrt{\frac{\hbar}{2\varepsilon_0  (2\pi)^3}}\right)
\sum_{\mu_1\bar\mu'_1}\sum_{\mu_2\mu'_2}\sum_{\lambda'_1}
\int d^3k'_1
\int d^3p'\int d^3\tilde p'
\nonumber\\&&\times
\left[\delta(\tilde{{\bf p}}'+{\bf P}+\frac12\hbar{\bf K})
\delta({\bf p}'-{\bf P}+\frac12\hbar{\bf K}+\hbar{\bf k}'_1)
\right.\nonumber\\ &&\left.
-\delta(\tilde{{\bf p}}'+{\bf P}+\frac12\hbar{\bf K}+\hbar{\bf k}'_1)
\delta({\bf p}'-{\bf P}+\frac12\hbar{\bf K})
\right]
\nonumber\\&&\times
\frac1{\sqrt{\omega_{k'_1}}}
\frac\hbar i{\bf e}_{-{\bf k}'_1 \lambda'_1}. [-\nabla_{{\bf p}'} +\nabla_{\tilde{{\bf p}}'}]
\delta_{\alpha\beta}^{1\mu_1\bar1\bar\mu'_1}({\bf p}', -\tilde{{\bf p}}')
\nonumber\\&&\times
\hbar^3
\sqrt{\frac{\hbar}{2\varepsilon_0(2\pi)^3 \omega_{{\bf k}'_1} }}
q_e c\frac1{\hbar^3}\sum_{i=1,3}
({\bf e}_{{{\bf k}}'_1\lambda_1})_i
\delta_{1i}^{\bar1\bar\mu'_1\,1\mu_2}
(-\tilde{{\bf p}}',- \tilde{{\bf p}}'-\hbar{{\bf k}}'_1) \delta^{Kr}_{\mu_1,\mu'_2}
\nonumber\\&&\times
\frac1{E_{-\tilde{{\bf p}}'-\hbar{{\bf k}}'_1}+ E_{\tilde{{\bf p}}'}+\hbar\omega_{k'_1}}
S(-\tilde{{\bf p}}'-\hbar{{\bf k}}'_1;{{\bf p}})  
 \tilde f^{1\mu_21\mu'_2}(\frac1\hbar({\bf p}-{\bf p}'),\frac12({\bf p}+{\bf p}'))
\nonumber\\&&
\label{4.a76}
\end{eqnarray}
\begin{eqnarray}
&&[\chi^{(0)}f_{\alpha\beta}^{W(2)}]^{1\mu1\mu'}_{101010}({\bf K},{\bf P})
=
\frac14\sum_{\alpha\beta}
\delta_{\alpha\beta}^{1\mu1\mu'}({\bf P}+\frac{\hbar}{2}{\bf K}, 
{\bf P}-\frac{\hbar}{2}{\bf K})
\nonumber\\&&\times
\left(\frac{-iq_e}{\hbar}\sqrt{\frac{\hbar}{2\varepsilon_0  (2\pi)^3}}\right)
\sum_{\mu_1\mu'_1}\sum_{\mu_2\mu'_2}\sum_{\lambda_1}
\int d^3k_1
\int d^3p\int d^3\tilde p
\nonumber\\&&\times
\left[\delta({\bf p}-{\bf P}-\frac12\hbar{\bf K}+\hbar{\bf k}_1)
\delta(\tilde{{\bf p}}+{\bf P}-\frac12\hbar{\bf K})
\right.\nonumber\\&&\left.
-\delta({\bf p}-{\bf P}-\frac12\hbar{\bf K})
\delta(\tilde{{\bf p}}+{\bf P}-\frac12\hbar{\bf K}+\hbar{\bf k}_1)
\right]
\nonumber\\&&\times
\frac1{\sqrt{\omega_{k_1}}}
\frac\hbar i{\bf e}_{{\bf k}_1 \lambda_1}. [-\nabla_{{\bf p}} +\nabla_{\tilde{{\bf p}}}]
\delta_{\alpha\beta}^{\bar1\bar\mu'_11\mu_1}(-\tilde{{\bf p}}, {\bf p})
\nonumber\\&&\times
(-\hbar^3)
\sqrt{\frac{\hbar}{2\varepsilon_0(2\pi)^3 \omega_{{\bf k}_1} }}
q_e c\frac1{\hbar^3}\sum_{i=1,3}
({\bf e}_{{\bf k}_1\lambda_1})_i
\delta_{1i}^{1\mu'_2\,\bar1\bar\mu'_1}
(-\tilde{{\bf p}}-\hbar{\bf k}_1, -\tilde{{\bf p}}) 
\delta^{Kr}_{\mu_1,\mu_2}
\nonumber\\&&\times
\frac1{-E_{-\tilde{{\bf p}}-\hbar{{\bf k}}_1}-E_{\tilde{{\bf p}}}-\hbar\omega_{k_1}}
S(-\tilde{{\bf p}}-\hbar{{\bf k}}_1;{{\bf p}}')  
 \tilde f^{1\mu_21\mu'_2}(\frac1\hbar({\bf p}-{\bf p}'),\frac12({\bf p}+{\bf p}'))
\nonumber\\&&
\label{4.a77}
\end{eqnarray}
These expressions (\ref{4.a74}-\ref{4.a77}) provide the four contributions to the matrix elements 
$<1\mu1\mu'\vert\chi^{(0)}(\chi^{-1})^{(2)}\vert1\mu_21\mu_2'>$.
In particular, we have the link between $[f_{\alpha\beta}^{W(2)}]^{1\mu1\mu'}({\bf K},{\bf P})$
 and its Fourier transform $\tilde f_{110000}^{\mu_2\mu'_2}$.
We can evaluate therefore the  diagonal action of $[\chi^{(0)}
(\chi^{-1})^{(2)},\Theta^{(0)}]$ on 
$\tilde f_{110000}^{\mu\mu'}$.
Indeed, the operator $\Theta^{(0)}$ is nothing but ${\cal L}_D$ that is
diagonal in the basis of states $\vert110000\mu_2\mu'_2>$.
In the evaluation of the commutator $[\chi^{(0)} (\chi^{-1})^{(2)},\Theta^{(0)}]$, at least one of the
operators present in 
$\chi^{(0)} (\chi^{-1})^{(2)}$ have to act on $\Theta^{(0)}$ to provide a non vanishing result.
These operators are the  substitution operators and the
gradients $\nabla_{{\bf p}}$,  $\nabla_{{\bf p}'}$, $\nabla_{\tilde{{\bf p}}}$ and
$\nabla_{\tilde{{\bf p}'}}$.

\subsection{The first contribution  with the intermediate state {110010} }

Let us examine more closely the contribution arising from (\ref{4.a74}) and compute the commutator
with ${\cal L}_D$. We have:
\begin{equation}
<110000\mu_2\mu'_2\vert{\cal L}_D\vert110000\mu_2\mu'_2>=
\frac1{i\hbar}\left(E_{p}-E_{p'}\right)
\label{4.a78}
\end{equation}
\begin{eqnarray}
&&<1\mu1\mu'\vert{\cal L}_D\chi^{(0)} (\chi^{-1})^{(2)}\vert110000>_{110010}=
\frac1{i\hbar}\left(E_{{\bf P}+\frac{\hbar}{2}{\bf K}}-E_{{\bf P}-\frac{\hbar}{2}{\bf K}}\right)
\nonumber\\&&\times
\frac14\sum_{\alpha\beta}
\delta_{\alpha\beta}^{1\mu1\mu'}({\bf P}+\frac{\hbar}{2}{\bf K}, 
{\bf P}-\frac{\hbar}{2}{\bf K})
\left(\frac{-iq_e}{\hbar}\sqrt{\frac{\hbar}{2\varepsilon_0  (2\pi)^3}}\right)
\sum_{\mu_1\mu'_1}\sum_{\mu_2\mu'_2}\sum_{\lambda_1}
\nonumber\\&&\times
\int d^3k_1
 \int d^3p\int d^3p'
\delta({\bf p}-{\bf P}-\frac\hbar2{\bf K}+\hbar{\bf k}_1)
\delta({\bf p}'-{\bf P}+\frac\hbar2{\bf K}))
\nonumber\\&&\times
\frac1{\sqrt{\omega_{k_1}}}
\frac\hbar i{\bf e}_{{\bf k}_1 \lambda_1}. [-\nabla_{{\bf p}} -\nabla_{{\bf p}'}]
\delta_{\alpha\beta}^{1\mu'_11\mu_1}({\bf p}', {\bf p})
\nonumber\\&&\times
(-\hbar^3)
\sqrt{\frac{\hbar}{2\varepsilon_0(2\pi)^3 \omega_{{\bf k}_1} }}
q_e c\frac1{\hbar^3}\sum_{i=1,3}
({\bf e}_{{\bf k}_1\lambda_1})_i
\delta_{1i}^{1\mu_11\mu_2}({\bf p}, {\bf p}+\hbar {\bf k}_1)
\delta^{Kr}_{\mu'_1,\mu'_2}
\nonumber\\&&\times
\frac1{E_{{\bf p}+\hbar {\bf k}_1}-E_{{\bf p}}-\hbar\omega_{k_1}}
S({\bf p}+\hbar {\bf k}_1;{\bf p})
\label{4.a79}
\end{eqnarray}
\begin{eqnarray}
&&<1\mu1\mu'\vert\chi^{(0)} (\chi^{-1})^{(2)}{\cal L}_D\vert110000>_{110010}=
\frac14\sum_{\alpha\beta}
\delta_{\alpha\beta}^{1\mu1\mu'}({\bf P}+\frac{\hbar}{2}{\bf K}, 
{\bf P}-\frac{\hbar}{2}{\bf K})
\nonumber\\&&\times
\left(\frac{-iq_e}{\hbar}\sqrt{\frac{\hbar}{2\varepsilon_0  (2\pi)^3}}\right)
\sum_{\mu_1\mu'_1}\sum_{\mu_2\mu'_2}\sum_{\lambda_1}
\int d^3k_1
 \int d^3p\int d^3p'
\nonumber\\&&\times
\delta({\bf p}-{\bf P}-\frac\hbar2{\bf K}+\hbar{\bf k}_1)
\delta({\bf p}'-{\bf P}+\frac\hbar2{\bf K}))
\nonumber\\&&\times
\frac1{\sqrt{\omega_{k_1}}}
\frac\hbar i{\bf e}_{{\bf k}_1 \lambda_1}. [-\nabla_{{\bf p}} -\nabla_{{\bf p}'}]
\delta_{\alpha\beta}^{1\mu'_11\mu_1}({\bf p}', {\bf p})
\nonumber\\&&\times
(-\hbar^3)
\sqrt{\frac{\hbar}{2\varepsilon_0(2\pi)^3 \omega_{{\bf k}_1} }}
q_e c\frac1{\hbar^3}\sum_{i=1,3}
({\bf e}_{{\bf k}_1\lambda_1})_i
\delta_{1i}^{1\mu_11\mu_2}({\bf p}, {\bf p}+\hbar {\bf k}_1)
\delta^{Kr}_{\mu'_1,\mu'_2}
\nonumber\\&&\times
\frac1{E_{{\bf p}+\hbar {\bf k}_1}-E_{{\bf p}}-\hbar\omega_{k_1}}
S({\bf p}+\hbar {\bf k}_1;{\bf p})
\frac1{i\hbar}\left(E_{p}-E_{p'}\right)
\label{4.a79a}
\end{eqnarray}
\begin{eqnarray}
&&<1\mu1\mu'\vert\chi^{(0)} (\chi^{-1})^{(2)}{\cal L}_D\vert110000>_{110010}=
\frac14\sum_{\alpha\beta}
\delta_{\alpha\beta}^{1\mu1\mu'}({\bf P}+\frac{\hbar}{2}{\bf K}, 
{\bf P}-\frac{\hbar}{2}{\bf K})
\nonumber\\&&\times
\left(\frac{-iq_e}{\hbar}\sqrt{\frac{\hbar}{2\varepsilon_0  (2\pi)^3}}\right)
\sum_{\mu_1\mu'_1}\sum_{\mu_2\mu'_2}\sum_{\lambda_1}
\int d^3k_1
 \int d^3p\int d^3p'
\nonumber\\&&\times
\delta({\bf p}-{\bf P}-\frac\hbar2{\bf K}+\hbar{\bf k}_1)
\delta({\bf p}'-{\bf P}+\frac\hbar2{\bf K}))
\nonumber\\&&\times
\frac1{\sqrt{\omega_{k_1}}}
\frac\hbar i{\bf e}_{{\bf k}_1 \lambda_1}. [-\nabla_{{\bf p}} -\nabla_{{\bf p}'}]
\left(E_{{\bf p}+\hbar {\bf k}_1}-E_{p'}\right)
\delta_{\alpha\beta}^{1\mu'_11\mu_1}({\bf p}', {\bf p})
\nonumber\\&&\times
i\hbar^2
\sqrt{\frac{\hbar}{2\varepsilon_0(2\pi)^3 \omega_{{\bf k}_1} }}
q_e c\frac1{\hbar^3}\sum_{i=1,3}
({\bf e}_{{\bf k}_1\lambda_1})_i
\delta_{1i}^{1\mu_11\mu_2}({\bf p}, {\bf p}+\hbar {\bf k}_1)
\delta^{Kr}_{\mu'_1,\mu'_2}
\nonumber\\&&\times
\frac1{E_{{\bf p}+\hbar {\bf k}_1}-E_{{\bf p}}-\hbar\omega_{k_1}}
S({\bf p}+\hbar {\bf k}_1;{\bf p})
\label{4.a80}
\end{eqnarray}

\subsection{Diagonality of the contribution  for {110010} in the variables ${\bf K}$-${\bf P}$ }

In order to have a possible compensation, the operator $[\chi^{(0)} (\chi^{-1})^{(2)},\Theta^{(0)}]$ should
be diagonal in the variables ${\bf K}$-${\bf P}$, as the diverging contributions of
$\Theta^{(2)}$. 

We consider first the contribution for which no gradient bears on the energy difference.
We have to substract the energy difference (\ref{4.a78}) for the initial and final states involved.
The left state is for values  ${\bf K}$-${\bf P}$ while the right state involves $\left(E_{p}-E_{p'}\right)$.
After use of the substitution operator $S({\bf p}+\hbar {\bf k}_1;{\bf p})$ and the Dirac delta
functions, we note that the energy difference vanishes identically.

For the terms where the gradient bears on the energy difference, we need:
\begin{equation}
\nabla_{{\bf p}}E_{{\bf p}}=\frac{{\bf p} c^2}{E_{{\bf p}}}
\qquad
\nabla_{{\bf p}}E_{{\bf p}+\hbar {\bf k}_1}=
\frac{({\bf p} +\hbar {\bf k}_1)c^2}{E_{{\bf p}+\hbar {\bf k}_1}}
\label{4.a81}
\end{equation}
The commutator involves the computation of
\begin{equation}{\bf e}_{{\bf k}_1 \lambda_1}. [-\nabla_{{\bf p}} -\nabla_{{\bf p}'}]
[E_{{\bf p}+\hbar {\bf k}_1}-E_{{\bf p}'}]
\label{4.a82}
\end{equation}
Combining (\ref{4.a80}) and this last expression, we
have the structure:
\begin{eqnarray}
&&
\int d^3p\int d^3p'
\delta({\bf p}-{\bf P}-\frac\hbar2{\bf K}+\hbar{\bf k}_1)
\delta({\bf p}'-{\bf P}+\frac\hbar2{\bf K}))
\dots
\nonumber\\&&\times
\tilde f_{110000}^{\mu\mu'}({\bf p} +\hbar {\bf k}_1,{\bf p}')
\to
\tilde f_{110000}^{\mu\mu'}({\bf P}+\frac\hbar2{\bf K} ,
{\bf P}-\frac\hbar2{\bf K})
\label{4.a85}
\end{eqnarray}
In the variables ${\bf K}$, ${\bf P}$, the argument of the last function is  indeed ${\bf K} $,
${\bf P}$ and the term could compensate the problems due to $\Theta^{(2)}$ that is
diagonal in both  ${\bf K}$ and ${\bf P}$. \footnote{ A permutation of the position
of $\bar{{\bf A}}_{\bot l} ({\bf r})$ and $\bar{{\bf A}}_{\bot r} ({\bf r})$  in (\ref{4.37a}) would lead to a
vanishing diagonal contribution}

Therefore,  our generalisation of Weyl's form for the quantum observables  provide a possible
compensation mechanism.  We have to look at the other contributions arising from
(\ref{4.a75}-\ref{4.a77}).

\subsection{Diagonality of the contribution  for ${110001}$ in the variables ${\bf K}$-${\bf P}$ }

The contribution with the intermediate state {110001}  provides {\it a priori} a similar contribution,
with the same property regarding its diagonality in ${\bf K}$-${\bf P}$.
Indeed, instead of (\ref{4.a85}), we have:
\begin{eqnarray}
&&
\int d^3p\int d^3p'
\delta({\bf p}-{\bf P}-\frac\hbar2{\bf K})
\delta({\bf p}'-{\bf P}+\frac\hbar2{\bf K}+\hbar{\bf k}'_1))
\dots
\nonumber\\&&\times
\tilde f_{110000}^{\mu\mu'}({\bf p} ,{\bf p}'+\hbar {\bf k}'_1)
\to
\tilde f_{110000}^{\mu\mu'}({\bf P}+\frac\hbar2{\bf K},
{\bf P}-\frac\hbar2{\bf K})
\label{4.a87a}
\end{eqnarray}
that is diagonal, as expected.

\subsection{The contribution  with the intermediate state ${010101}$ }

Instead of (\ref{4.a80}), we have now
\begin{eqnarray}
&&<1\mu1\mu'\vert\chi^{(0)} (\chi^{-1})^{(2)}{\cal L}_D\vert110000>_{010101}=
\frac14\sum_{\alpha\beta}
\delta_{\alpha\beta}^{1\mu1\mu'}({\bf P}+\frac{\hbar}{2}{\bf K}, 
{\bf P}-\frac{\hbar}{2}{\bf K})
\nonumber\\&&\times
\left(\frac{-iq_e}{\hbar}\sqrt{\frac{\hbar}{2\varepsilon_0  (2\pi)^3}}\right)
\sum_{\mu_1\mu'_1}\sum_{\mu_2\mu'_2}\sum_{\lambda_1}
\int d^3k'_1
\int d^3p'\int d^3\tilde p'
\nonumber\\&&\times
\left[\delta(\tilde{{\bf p}}'+{\bf P}+\frac12\hbar{\bf K})
\delta({\bf p}'-{\bf P}+\frac12\hbar{\bf K}+\hbar{\bf k}'_1)
\right.\nonumber\\&&\left.
-\delta(\tilde{{\bf p}}'+{\bf P}+\frac12\hbar{\bf K}+\hbar{\bf k}'_1)
\delta({\bf p}'-{\bf P}+\frac12\hbar{\bf K})
\right]
\nonumber\\&&\times
\frac1{\sqrt{\omega_{k'_1}}}{\bf e}_{-{\bf k}'_1 \lambda'_1}. [-\nabla_{{\bf p}'}
+\nabla_{\tilde{{\bf p}}'}]
\left(E_{-\tilde{{\bf p}}'-\hbar{{\bf k}}'_1}-E_{p'}\right)
\nonumber\\&&\times
\delta_{\alpha\beta}^{1\mu_1\bar\mu'_1}({\bf p}', -\tilde{{\bf p}}')
(-i)\hbar^2
\sqrt{\frac{\hbar}{2\varepsilon_0(2\pi)^3 \omega_{{\bf k}'_1} }}
q_e c\frac1{\hbar^3}
\nonumber\\&&\times
\sum_{i=1,3}
({\bf e}_{{{\bf k}}'_1\lambda_1})_i
\delta_{1i}^{1\mu'_2\,\bar1\bar\mu'_1}
(\tilde{{\bf p}}, \tilde{{\bf p}}+\hbar{{\bf k}}'_1) \delta^{Kr}_{\mu_1,\mu'_2}
\nonumber\\&&\times
\frac1{E_{-\tilde{{\bf p}}'-\hbar{{\bf k}}'_1}+ E_{\tilde{{\bf p}}'}+\hbar\omega_{k'_1}}
S(-\tilde{{\bf p}}'-\hbar{{\bf k}}'_1;{{\bf p}})  
\label{4.a88}
\end{eqnarray}
The commutator involves the computation of
\begin{equation}
{\bf e}_{{\bf k}'_1 \lambda'_1}. [-\nabla_{{\bf p}'} +\nabla_{\tilde{{\bf p}}'}]
\left[E_{-\tilde{{\bf p}}'-\hbar{{\bf k}}'_1}-E_{p'}\right]
\label{4.a89}
\end{equation}
where the gradient bears on the energy difference.

For that term and all the terms where all the gradients act on the energy difference, we have the
structure:
\begin{eqnarray}
&&
\int d^3p'\int d^3\tilde p'\,
\left[\delta(\tilde{{\bf p}}'+{\bf P}+\frac12\hbar{\bf K})
\delta({\bf p}'-{\bf P}+\frac12\hbar{\bf K}+\hbar{\bf k}'_1)
\right.\nonumber\\&&\left.
-\delta(\tilde{{\bf p}}'+{\bf P}+\frac12\hbar{\bf K}+\hbar{\bf k}'_1)
\delta({\bf p}'-{\bf P}+\frac12\hbar{\bf K})
\right]
\dots
\nonumber\\&&\times
\tilde f_{110000}^{\mu\mu'}(-\tilde{{\bf p}}'-\hbar{{\bf k}}'_1,{\bf p}')
\to
\tilde f_{110000}^{\mu\mu'}({\bf P}+\frac\hbar2{\bf K} -\hbar{{\bf k}}'_1,
{\bf P}-\frac\hbar2{\bf K}-\hbar{{\bf k}}'_1)
\nonumber\\&&
-\tilde f_{110000}^{\mu\mu'}({\bf P}+\frac\hbar2{\bf K} ,
{\bf P}-\frac\hbar2{\bf K})
\label{4.a90}
\end{eqnarray}
In the variables ${\bf K}$, ${\bf P}$, the argument of the last function is ${\bf K} $,
${\bf P}$ and the term can compensate the problems due to $\Theta^{(2)}$ that is
diagonal in both  ${\bf K}$ and ${\bf P}$.

\subsection{The contribution  with the intermediate state {101010} }

It is easily realised that we obtain a contribution diagonal in the variables ${\bf K} $,
${\bf P}$ and the contribution when no gradient acts on the energy difference vanishes identically.
Indeed, instead of (\ref{4.a90}), we obtain
\begin{eqnarray}
&&
\int d^3p'\int d^3\tilde p'\,
\left[\delta({\bf p}-{\bf P}-\frac12\hbar{\bf K}+\hbar{\bf k}_1)
\delta(\tilde{{\bf p}}+{\bf P}-\frac12\hbar{\bf K})
\right.\nonumber\\&&\left.
-\delta({\bf p}-{\bf P}-\frac12\hbar{\bf K})
\delta(\tilde{{\bf p}}+{\bf P}-\frac12\hbar{\bf K}+\hbar{\bf k}_1)
\right]
\nonumber\\&&\times
\dots
S(-\tilde{{\bf p}}-\hbar{{\bf k}}_1;{{\bf p}}')  
 \tilde f_{110000}^{\mu_2\mu'_2}({\bf p},{\bf p}')
\nonumber\\&&=
\int d^3p'\int d^3\tilde p'\,
\delta({\bf p}-{\bf P}-\frac12\hbar{\bf K}+\hbar{\bf k}_1)
\delta(\tilde{{\bf p}}+{\bf P}-\frac12\hbar{\bf K})
\dots
\nonumber\\&&\times
\tilde f_{110000}^{\mu_2\mu'_2}({\bf p},-\tilde{{\bf p}}-\hbar{{\bf k}}_1)
\to
\tilde f_{110000}^{\mu\mu'}({\bf P}+\frac\hbar2{\bf K}-\hbar{{\bf k}}_1,
{\bf P}-\frac\hbar2{\bf K}-\hbar{{\bf k}}_1)
\nonumber\\&&
-\tilde f_{110000}^{\mu\mu'}({\bf P}+\frac\hbar2{\bf K},
{\bf P}-\frac\hbar2{\bf K})
\label{4.a91}
\end{eqnarray}

\section{Contribution to the $\phi_M^{(2)}$}
\setcounter{equation}{0}
\def\theequation{\thesection.\arabic{equation}}

We now turn to the explicit evaluation of the diagonal singular part of
$(\chi^{(0)}\Phi^{(2)}(\chi^{-1})^{(0)}$ =$\Theta^{(2)}+[\chi^{(0)} (\chi^{-1})^{(2)},\Theta^{(0)}]$.
The diagonality has to be considered with respect to three criterions.
In the first one, we consider in $\Theta^{(2)}$ and $ (\chi^{-1})^{(2)}$ the diagonal connection between
the state ${110000}$ and  itself.
We expect indeed that eventual divergences in that connection are independent of the other
contributions.
In the second one, we consider the contribution diagonal with respect to the value of the argument of
the functions. 
The third criterion refers to the spin variables.
We will not compute the matrix elements for all spin indices but introduce usual  functions
describing the spin properties of one particle.
It is useful at this stage to introduce the distribution function 
$f({\bf K},{\bf P},t)$, describing the electron independently of its spin orientation:
\begin{equation}
f({\bf K},{\bf P},t) =
\sum_{\sigma} f_{110000}^{\sigma \sigma} ({\bf K}, {\bf P}, t)
\label{5.a1}
\end{equation}
and the  spin orientation distribution function (in short SODF) (more precisely,
it is a vector but we keep the terminology of the litterature \cite{LL82}, \cite{dH90b} )
$\pmb{$ \cal M$}({\bf K},{\bf P},t)$  
defined in terms of the Pauli matrices trough a partial trace over the 
spin variables:
\begin{equation}
\pmb{$ \cal M$}({\bf K},{\bf P},t) =
{\rm tr}_{spins}\quad \pmb{$\sigma$} 
\left(f_{110000}({\bf K},{\bf P}, t) \right)
\label{5.a2}
\end{equation}
where $f_{110000}({\bf K},{\bf P}, t) $ is the one electron 
Wdf considered as a matrix in spin variables.

These relations (\ref{5.a1}-\ref{5.a2}) can be inverted to provide:
\begin{equation}
f_{110000}^{\sigma \sigma'} ({\bf K}, {\bf P}, t) 
=\frac12 [ f({\bf K}, {\bf P}, t) I
+ \pmb{$ \cal M$}({\bf K},{\bf P},t). \pmb{$\sigma$}]_{\sigma \sigma'} 
\label{5.a3}
\end{equation}
The diagonal kinetic operator acting on this new set of functions receives a $M$ index.
The operator $\phi^{(2)}_{M}$ is  the operator acting on those 
functions $f({\bf K}, {\bf P}, t)$ that contributes to their time derivative. 
Vectorial (in spin space) and tensorial operators would  also be required to complete the kinetic
equations, that involve also
 $\pmb{$ \cal M$}({\bf K},{\bf P},t)$. 
Their explicit form  is irrelevant here.
We focus here on the diagonal contributions arising from $[\chi^{(0)} (\chi^{-1})^{(2)},\Theta^{(0)}]$.
We have the elements to compute the diagonal contributions leading from $110000$ to itself.
We place an index $a$ to notify the fact that only that contribution is included and we still note the
intermediary state explicitly.
For the first contribution, the required
expressions in the
${\bf K}$ and ${\bf P}$ variables  can be obtained from 
(\ref{4.a80}), (\ref{4.a79}) and (\ref{4.a74}):
\begin{eqnarray}
&&[\chi^{(0)} (\chi^{-1})^{(2)},\Theta^{(0)}]_{Ma110010}
=\sum_{\mu,\mu_1,\mu_2}\frac12
\frac14\sum_{\alpha\beta}
\delta_{\alpha\beta}^{1\mu1\mu}({\bf P}+\frac{\hbar}{2}{\bf K}, 
{\bf P}-\frac{\hbar}{2}{\bf K})
\nonumber\\&&\times
\left(\frac{-iq_e}{\hbar}\sqrt{\frac{\hbar}{2\varepsilon_0  (2\pi)^3}}\right)
\sum_{\lambda_1}
\int d^3k_1
\frac1{\sqrt{\omega_{k_1}}}
\frac\hbar i{\bf e}_{{\bf k}_1 \lambda_1}. \left[-\nabla_{{\bf P}}
\left(E_{{\bf P}+\frac\hbar2{\bf K}}-E_{{\bf P}-\frac\hbar2{\bf K}}\right)\right]
\nonumber\\&&\times
\delta_{\alpha\beta}^{1\mu_21\mu_1}
({\bf P}-\frac\hbar2{\bf K}, {\bf P}+\frac\hbar2{\bf K}-\hbar{\bf k}_1)
i\hbar^2
\sqrt{\frac{\hbar}{2\varepsilon_0(2\pi)^3 \omega_{{\bf k}_1} }}
\nonumber\\&&\times
q_e c\frac1{\hbar^3}\sum_{i=1,3}
({\bf e}_{{\bf k}_1\lambda_1})_i
\delta_{1i}^{1\mu_11\mu_2}({\bf P}+\frac\hbar2{\bf K}-\hbar{\bf k}_1, {\bf P}+\frac\hbar2{\bf K})
\nonumber\\&&\times
\frac1{E_{{\bf P}+\frac\hbar2{\bf K}}-E_{{\bf P}+\frac\hbar2{\bf K}-\hbar{\bf k}_1}
-\hbar\omega_{k_1}} 
\label{5.a5}
\end{eqnarray}
The other contributions are determied in a similar way. 
From  (\ref{4.a75}), we obtain:
\begin{eqnarray}
&&[\chi^{(0)} (\chi^{-1})^{(2)},\Theta^{(0)}]_{Ma110001}
=\sum_{\mu,\mu'_1,\mu_2}
\frac12\frac14\sum_{\alpha\beta}
\delta_{\alpha\beta}^{1\mu1\mu}({\bf P}+\frac{\hbar}{2}{\bf K}, 
{\bf P}-\frac{\hbar}{2}{\bf K})
\nonumber\\&&\times
\left(\frac{-iq_e}{\hbar}\sqrt{\frac{\hbar}{2\varepsilon_0  (2\pi)^3}}\right)
\sum_{\lambda'_1}
\int d^3k'_1\,
\frac1{\sqrt{\omega_{k'_1}}}
\frac\hbar i{\bf e}_{{\bf k}'_1 \lambda'_1}. \left[-\nabla_{{\bf P}}
\left(E_{{\bf P}+\frac\hbar2{\bf K}}-E_{{\bf P}-\frac\hbar2{\bf K}}\right)\right]
\nonumber\\&&\times
\delta_{\alpha\beta}^{1\mu'_11\mu_2}
({\bf P}-\frac\hbar2{\bf K}-\hbar{\bf k}'_1, {\bf P}+\frac\hbar2{\bf K})
(-i\hbar^2)
\sqrt{\frac{\hbar}{2\varepsilon_0(2\pi)^3 \omega_{{\bf k}_1} }}
q_e c\frac1{\hbar^3}\sum_{i=1,3}
({\bf e}_{-{\bf k}'_1\lambda'_1})_i
\nonumber\\&&\times
 \delta_{1i}^{1\mu_2 1\mu'_1}({\bf P}-\frac\hbar2{\bf K},
{\bf P}-\frac\hbar2{\bf K}-\hbar{\bf k}'_1)
\frac1{-E_{{\bf P}-\frac\hbar2{\bf K}}+E_{{\bf P}-\frac\hbar2{\bf K}-\hbar{\bf k}'_1}
+\hbar\omega_{k'_1}}
\label{5.a7}
\end{eqnarray}
From (\ref{4.a76}) (second contribution with the $\delta\delta$, hence a $(-1)$ factor):
\begin{eqnarray}
&&[\chi^{(0)} (\chi^{-1})^{(2)},\Theta^{(0)}]_{Ma010101}
=\sum_{\mu,\mu'_1,\mu_2} \frac12
\frac14\sum_{\alpha\beta}
\delta_{\alpha\beta}^{1\mu1\mu}({\bf P}+\frac{\hbar}{2}{\bf K}, 
{\bf P}-\frac{\hbar}{2}{\bf K})
\nonumber\\&&\times
\left(\frac{-iq_e}{\hbar}\sqrt{\frac{\hbar}{2\varepsilon_0  (2\pi)^3}}\right)
\sum_{\lambda'_1}
\int d^3k'_1
(-1)
\frac1{\sqrt{\omega_{k'_1}}}
\frac\hbar i{\bf e}_{{\bf k}'_1 \lambda'_1}. 
\left[-\nabla_{{\bf P}}
\left(E_{{\bf P}+\frac\hbar2{\bf K}}-E_{{\bf P}-\frac\hbar2{\bf K}}\right)\right]
\nonumber\\&&\times
\delta_{\alpha\beta}^{1\mu_2\bar1\bar\mu'_1}({\bf P}-\frac12\hbar{\bf K}, {\bf P}+\frac12\hbar{\bf K}+\hbar{\bf k}'_1)
(-i\hbar^2)
\sqrt{\frac{\hbar}{2\varepsilon_0(2\pi)^3 \omega_{{\bf k}'_1} }}
q_e c\frac1{\hbar^3}\sum_{i=1,3}
({\bf e}_{{-{\bf k}}'_1\lambda_1})_i
\nonumber\\&&\times
\delta_{1i}^{\bar1\bar\mu'_1\,1\mu_2}
({\bf P}+\frac12\hbar{\bf K}+\hbar{\bf k}'_1, {\bf P}+\frac12\hbar{\bf K})
\frac1{E_{{\bf P}+\frac12\hbar{\bf K}}+ E_{-{\bf P}-\frac12\hbar{\bf K}-\hbar{\bf k}'_1}
+\hbar\omega_{k'_1}}
\label{5.a9}
\end{eqnarray}
From (\ref{4.a77}), (second contribution with the $\delta\delta$, hence a $(-1)$ factor):
\begin{eqnarray}
&&[\chi^{(0)} (\chi^{-1})^{(2)},\Theta^{(0)}]_{Ma101010}
=\sum_{\mu,\bar\mu'_1,\mu_2}\frac12
\frac14\sum_{\alpha\beta}
\delta_{\alpha\beta}^{1\mu1\mu}({\bf P}+\frac{\hbar}{2}{\bf K}, 
{\bf P}-\frac{\hbar}{2}{\bf K})
\nonumber\\&&\times
\left(\frac{-iq_e}{\hbar}\sqrt{\frac{\hbar}{2\varepsilon_0  (2\pi)^3}}\right)
\sum_{\lambda_1}
\int d^3k_1
(-1)
\frac1{\sqrt{\omega_{k_1}}}
\frac\hbar i{\bf e}_{{\bf k}_1 \lambda_1}.
\left[-\nabla_{{\bf P}}
\left(E_{{\bf P}+\frac\hbar2{\bf K}}-E_{{\bf P}-\frac\hbar2{\bf K}}\right)\right]
\nonumber\\&&\times
\delta_{\alpha\beta}^{\bar1\bar\mu'_11\mu_2}({\bf P}-\frac12\hbar{\bf K}+\hbar{\bf k}_1, {\bf P}+\frac12\hbar{\bf K})
i\hbar^2
\sqrt{\frac{\hbar}{2\varepsilon_0(2\pi)^3 \omega_{{\bf k}_1} }}
q_e c\frac1{\hbar^3}\sum_{i=1,3}
({\bf e}_{{\bf k}_1\lambda_1})_i
\nonumber\\&&\times
\delta_{1i}^{1\mu_2\,\bar1\bar\mu'_1}
({\bf P}-\frac12\hbar{\bf K}, {\bf P}-\frac12\hbar{\bf K}+\hbar{\bf k}_1) 
\frac1{-E_{{\bf P}-\frac12\hbar{\bf K}}-E_{{\bf P}-\frac12\hbar{\bf K}+\hbar{\bf k}_1}
-\hbar\omega_{k_1}}
\label{5.a11}
\end{eqnarray}
Elementary but lengthy algebra provides the relations 
($\hat{\bf v}$ is an arbitrary unit vector, $\hat{\bf v}_s$ its $s$ component)
\begin{eqnarray}
&&\sum_{\alpha,\beta}\sum_{s=1,3}\sum_{\mu_1,\mu_2}\hat{\bf v}_s
\sum_{\mu}\delta_{\alpha\beta}^{1\mu1\mu}({\bf p}, {\bf p}')
\delta_{\alpha\beta}^{1\mu_21\mu_1}({\bf p}', {\bf p}")
\delta_{1s}^{1\mu_11\mu_2}({\bf p}", {\bf p})
\nonumber\\&&=
8
\sin\frac{ \theta_{p}}2\cos\frac{ \theta_{p}}2
(\hat {{\bf p}}.\hat{\bf v})
+8
\sin\frac{ \theta_{p"}}2\cos\frac{ \theta_{p"}}2
( \hat {{\bf p}}".\hat{\bf v})
\label{5.a200}
\end{eqnarray}
\begin{eqnarray}
&&\sum_{\alpha,\beta}\sum_{s=1,3}\sum_{\mu_1,\mu_2}\hat{\bf v}_s
\sum_{\mu}\delta_{\alpha\beta}^{1\mu1\mu}({\bf p}, {\bf p}')
\delta_{\alpha\beta}^{1\mu_2\bar1\mu_1}({\bf p}', {\bf p}")
\delta_{1s}^{\bar1\mu_11\mu_2}({\bf p}", {\bf p})
\nonumber\\&&
=8
\sin\frac{ \theta_{p}}2\cos\frac{ \theta_{p}}2
(\hat {{\bf p}}.\hat{\bf v})
-8
\sin\frac{ \theta_{p"}}2\cos\frac{ \theta_{p"}}2
( \hat {{\bf p}}".\hat{\bf v})
\label{5.a126}
\end{eqnarray}
For the expression (\ref{5.a5}), we have to consider $\hat{\bf v}={\bf e}_{{\bf k}_1 \lambda_1}$.
We have also 
\begin{equation}
\nabla_{{\bf P}}E_{{\bf P}+\frac\hbar2{\bf K}}
=c^2\frac{{\bf P}+\frac\hbar2{\bf K}}{E_{{\bf P}+\frac\hbar2{\bf K}}}
\qquad
\nabla_{{\bf P}}E_{{\bf P}-\frac\hbar2{\bf K}}
=c^2\frac{{\bf P}-\frac\hbar2{\bf K}}{E_{{\bf P}-\frac\hbar2{\bf K}}}
\label{5.a201}
\end{equation}
We deduce for (\ref{5.a5}):
\begin{eqnarray}
&&[\chi^{(0)} (\chi^{-1})^{(2)},\Theta^{(0)}]_{Ma110010}
=
-i\left({\frac{q^2_e c}{2\varepsilon_0  (2\pi)^3}}\right)
\sum_{\lambda_1}
\int d^3k_1
\frac1{\omega_{k_1}}
\nonumber\\&&\times
\left[\sin\frac{ \theta_{\vert{\bf P}+\frac{\hbar}{2}{\bf K} \vert}}2
\cos\frac{ \theta_{\vert{\bf P}+\frac{\hbar}{2}{\bf K} \vert}}2
\left(\frac{{\bf P}+\frac{\hbar}{2}{\bf K} }{\vert{\bf P}+\frac{\hbar}{2}{\bf K} \vert}.{\bf e}_{{\bf k}_1\lambda_1}\right) 
\right.\nonumber\\&&\left.
+\sin\frac{ \theta_{\vert{\bf P}+\frac{\hbar}{2}{\bf K} -\hbar{\bf k}_1\vert}}2
\cos\frac{ \theta_{\vert{\bf P}+\frac{\hbar}{2}{\bf K} -\hbar{\bf k}_1\vert}}2
\left(\frac{{\bf P}+\frac{\hbar}{2}{\bf K}-\hbar{\bf k}_1 }
{\vert{\bf P}+\frac{\hbar}{2}{\bf K} -\hbar{\bf k}_1\vert}.{\bf e}_{{\bf k}_1\lambda_1}\right)\right ]
\nonumber\\&&\times
{\bf e}_{{\bf k}_1\lambda_1}.\left[-c^2\frac{{\bf P}+\frac\hbar2{\bf K}}{E_{{\bf P}+\frac\hbar2{\bf K}}}
+c^2\frac{{\bf P}-\frac\hbar2{\bf K}}{E_{{\bf P}-\frac\hbar2{\bf K}}}
\right]
\frac1{E_{{\bf P}+\frac\hbar2{\bf K}}-E_{{\bf P}+\frac\hbar2{\bf K}-\hbar{\bf k}_1}
-\hbar\omega_{k_1}} 
\nonumber\\&&
\label{5.a202}
\end{eqnarray}
and the summation over the polarisation index $\lambda_1$ can be performed easily, using
(only the transverse components play a role):
\begin{equation}
\sum_{\lambda_1=\epsilon, \epsilon'} (\bf a.{\bf e}_{{\bf k}_1 \lambda_1})(\bf b.{\bf e}_{{\bf k}_1
\lambda_1}) =(\bf a.\bf b)-(\bf a.{\hat{\bf k}}_1)({\hat{\bf k}}_1.\bf b)
\label{5.a203}
\end{equation}
\begin{eqnarray}
&&[\chi^{(0)} (\chi^{-1})^{(2)},\Theta^{(0)}]_{Ma110010}
=
-i\left({\frac{q^2_e c^3}{2\varepsilon_0  (2\pi)^3}}\right)
\int d^3k_1
\frac1{\omega_{k_1}}
\nonumber\\&&\times
\left\{
\sin\frac{ \theta_{\vert{\bf P}+\frac{\hbar}{2}{\bf K} \vert}}2
\cos\frac{ \theta_{\vert{\bf P}+\frac{\hbar}{2}{\bf K} \vert}}2
\right.\nonumber\\&&\left.\times
\left(\frac{{\bf P}+\frac{\hbar}{2}{\bf K} }{\vert{\bf P}+\frac{\hbar}{2}{\bf K} \vert}.
\left[-\frac{{\bf P}+\frac\hbar2{\bf K}}{E_{{\bf P}+\frac\hbar2{\bf K}}} +\frac{{\bf P}-\frac\hbar2{\bf K}}{E_{{\bf P}-\frac\hbar2{\bf K}}}
\right]
\right.\right.\nonumber\\&&\left.\left.
-
\left({\hat{\bf k}}_1.\frac{{\bf P}+\frac{\hbar}{2}{\bf K} }{\vert{\bf P}+\frac{\hbar}{2}{\bf K} \vert}
\right)
\left[-\frac{{\bf P}+\frac\hbar2{\bf K}}{E_{{\bf P}+\frac\hbar2{\bf K}}} 
+\frac{{\bf P}-\frac\hbar2{\bf K}}{E_{{\bf P}-\frac\hbar2{\bf K}}}
\right].{\hat{\bf k}}_1
\right) 
\right.\nonumber\\&&\left.
+\sin\frac{ \theta_{\vert{\bf P}+\frac{\hbar}{2}{\bf K} -\hbar{\bf k}_1\vert}}2
\cos\frac{ \theta_{\vert{\bf P}+\frac{\hbar}{2}{\bf K} -\hbar{\bf k}_1\vert}}2
\right.\nonumber\\&&\left.\times
\left(\frac{{\bf P}+\frac{\hbar}{2}{\bf K}-\hbar{\bf k}_1 }
{\vert{\bf P}+\frac{\hbar}{2}{\bf K} -\hbar{\bf k}_1\vert}.
\left[-\frac{{\bf P}+\frac\hbar2{\bf K}}{E_{{\bf P}+\frac\hbar2{\bf K}}} 
+\frac{{\bf P}-\frac\hbar2{\bf K}}{E_{{\bf P}-\frac\hbar2{\bf K}}}
\right]
\right.\right.\nonumber\\&&\left.\left.
-\left({\hat{\bf k}}_1.
\frac{{\bf P}+\frac{\hbar}{2}{\bf K}-\hbar{\bf k}_1 }
{\vert{\bf P}+\frac{\hbar}{2}{\bf K} -\hbar{\bf k}_1\vert}\right)
\left[-\frac{{\bf P}+\frac\hbar2{\bf K}}{E_{{\bf P}+\frac\hbar2{\bf K}}} 
+\frac{{\bf P}-\frac\hbar2{\bf K}}{E_{{\bf P}-\frac\hbar2{\bf K}}}
\right].{\hat{\bf k}}_1.
\right) 
\right\}
\nonumber\\&&\times
\frac1{E_{{\bf P}+\frac\hbar2{\bf K}}-E_{{\bf P}+\frac\hbar2{\bf K}-\hbar{\bf k}_1}
-\hbar\omega_{k_1}} 
\label{5.a204}
\end{eqnarray}
For large $\omega_{k_1}$, taking into account that angular integrations may vanish, the integrand
behaves as
\begin{eqnarray}
&&\left.I(\omega_{k_1})_{Ma110010}\right\vert_{dom}
=
-i\left({\frac{q^2_e c^3}{2\varepsilon_0  (2\pi)^3}}\right)
\frac1{\omega_{k_1}}\frac1{-2\hbar\omega_{k_1}} \frac23
\nonumber\\&&\times
\sin\frac{ \theta_{\vert{\bf P}+\frac{\hbar}{2}{\bf K} \vert}}2
\cos\frac{ \theta_{\vert{\bf P}+\frac{\hbar}{2}{\bf K} \vert}}2
\left(\frac{{\bf P}+\frac{\hbar}{2}{\bf K} }{\vert{\bf P}+\frac{\hbar}{2}{\bf K} \vert}.
\left[-\frac{{\bf P}+\frac\hbar2{\bf K}}{E_{{\bf P}+\frac\hbar2{\bf K}}} +\frac{{\bf P}-\frac\hbar2{\bf K}}{E_{{\bf P}-\frac\hbar2{\bf K}}}
\right]
\right) 
\nonumber\\&&
\label{5.a205}
\end{eqnarray}
From (\ref{4.a76}), we obtain:
\begin{eqnarray}
&&[\chi^{(0)} (\chi^{-1})^{(2)},\Theta^{(0)}]_{Ma010101}
=-i\left({\frac{q^2_e c^3}{2\varepsilon_0  (2\pi)^3}}\right)
\int d^3k'_1
\frac1{\omega_{k'_1}}
\nonumber\\&&\times
\left\{
\sin\frac{ \theta_{\vert{\bf P}+\frac{\hbar}{2}{\bf K} \vert}}2
\cos\frac{ \theta_{\vert{\bf P}+\frac{\hbar}{2}{\bf K} \vert}}2
\right.\nonumber\\&&\left.\times
\left(\frac{{\bf P}+\frac{\hbar}{2}{\bf K} }{\vert{\bf P}+\frac{\hbar}{2}{\bf K} \vert}.
\left[-\frac{{\bf P}+\frac\hbar2{\bf K}}{E_{{\bf P}+\frac\hbar2{\bf K}}} +\frac{{\bf P}-\frac\hbar2{\bf K}}{E_{{\bf P}-\frac\hbar2{\bf K}}}
\right]
\right.\right.\nonumber\\&&\left.\left.
-
\left({\hat{\bf k}}'_1.\frac{{\bf P}+\frac{\hbar}{2}{\bf K} }{\vert{\bf P}+\frac{\hbar}{2}{\bf K} \vert}
\right)
\left[-\frac{{\bf P}+\frac\hbar2{\bf K}}{E_{{\bf P}+\frac\hbar2{\bf K}}} 
+\frac{{\bf P}-\frac\hbar2{\bf K}}{E_{{\bf P}-\frac\hbar2{\bf K}}}
\right].{\hat{\bf k}}'_1
\right) 
\right.\nonumber\\&&\left.
-\sin\frac{ \theta_{\vert{\bf P}+\frac{\hbar}{2}{\bf K} +\hbar{\bf k}'_1\vert}}2
\cos\frac{ \theta_{\vert{\bf P}+\frac{\hbar}{2}{\bf K} +\hbar{\bf k}'_1\vert}}2
\right.\nonumber\\&&\left.\times
\left(\frac{{\bf P}+\frac{\hbar}{2}{\bf K}-\hbar{\bf k}_1 }
{\vert{\bf P}+\frac{\hbar}{2}{\bf K} -\hbar{\bf k}_1\vert}.
\left[-\frac{{\bf P}+\frac\hbar2{\bf K}}{E_{{\bf P}+\frac\hbar2{\bf K}}} 
+\frac{{\bf P}-\frac\hbar2{\bf K}}{E_{{\bf P}-\frac\hbar2{\bf K}}}
\right]
\right.\right.\nonumber\\&&\left.\left.
-\left({\hat{\bf k}}_1.
\frac{{\bf P}+\frac{\hbar}{2}{\bf K}+\hbar{\bf k}'_1 }
{\vert{\bf P}+\frac{\hbar}{2}{\bf K} +\hbar{\bf k}'_1\vert}\right)
\left[-\frac{{\bf P}+\frac\hbar2{\bf K}}{E_{{\bf P}+\frac\hbar2{\bf K}}} 
+\frac{{\bf P}-\frac\hbar2{\bf K}}{E_{{\bf P}-\frac\hbar2{\bf K}}}
\right].{\hat{\bf k}}'_1
\right) 
\right\}
\nonumber\\&&\times
\frac1{E_{{\bf P}+\frac12\hbar{\bf K}}+ E_{-{\bf P}-\frac12\hbar{\bf K}-\hbar{\bf k}'_1}
+\hbar\omega_{k'_1}}
\label{5.a206}
\end{eqnarray}
and
\begin{eqnarray}
&&\left.I(\omega_{k_1})_{Ma010101}\right\vert_{dom}
=
-i\left({\frac{q^2_e c^3}{2\varepsilon_0  (2\pi)^3}}\right)
\frac1{\omega_{k_1}}\frac1{2\hbar\omega_{k_1}} 
\frac23
\nonumber\\&&\times
\sin\frac{ \theta_{\vert{\bf P}+\frac{\hbar}{2}{\bf K} \vert}}2
\cos\frac{ \theta_{\vert{\bf P}+\frac{\hbar}{2}{\bf K} \vert}}2
\left(\frac{{\bf P}+\frac{\hbar}{2}{\bf K} }{\vert{\bf P}+\frac{\hbar}{2}{\bf K} \vert}.
\left[-\frac{{\bf P}+\frac\hbar2{\bf K}}{E_{{\bf P}+\frac\hbar2{\bf K}}} 
+\frac{{\bf P}-\frac\hbar2{\bf K}}{E_{{\bf P}-\frac\hbar2{\bf K}}}
\right]
\right) 
\nonumber\\&&
\label{5.a207}
\end{eqnarray}
The linear divergence cancels out from the two expressions  (\ref{5.a205}) and (\ref{5.a207}).

From (\ref{5.a200}) and (\ref{5.a126}), we obtain, using the hermiticity of the $\delta$ matrices:
\begin{eqnarray}
&&\sum_{\alpha,\beta}\sum_{s=1,3}\sum_{\mu_1,\mu_2}\hat{\bf v}_s
\sum_{\mu}\delta_{\alpha\beta}^{1\mu1\mu}({\bf p}', {\bf p})
\delta_{\alpha\beta}^{1\mu_11\mu_2}({\bf p}", {\bf p}')
\delta_{1s}^{1\mu_21\mu_1}({\bf p}, {\bf p}")
\nonumber\\&&=
8
\sin\frac{ \theta_{p}}2\cos\frac{ \theta_{p}}2
(\hat {{\bf p}}.\hat{\bf v})
+8
\sin\frac{ \theta_{p"}}2\cos\frac{ \theta_{p"}}2
( \hat {{\bf p}}".\hat{\bf v})
\label{5.a208}
\end{eqnarray}
\begin{eqnarray}
&&\sum_{\alpha,\beta}\sum_{s=1,3}\sum_{\mu_1,\mu_2}\hat{\bf v}_s
\sum_{\mu}\delta_{\alpha\beta}^{1\mu1\mu}({\bf p}', {\bf p})
\delta_{\alpha\beta}^{\bar1\mu_11\mu_2}({\bf p}", {\bf p}')
\delta_{1s}^{1\mu_2\bar1\mu_1}({\bf p}, {\bf p}")
\nonumber\\&&
=8
\sin\frac{ \theta_{p}}2\cos\frac{ \theta_{p}}2
(\hat {{\bf p}}.\hat{\bf v})
-8
\sin\frac{ \theta_{p"}}2\cos\frac{ \theta_{p"}}2
( \hat {{\bf p}}".\hat{\bf v})
\label{5.a209}
\end{eqnarray}
In (\ref{4.a76}), we can use (\ref{5.a208}) 
with the correspondence ${\bf p}'\to{\bf P}+\frac\hbar2{\bf
K}$,
${\bf p}\to{\bf P}-\frac\hbar2{\bf K}$, ${\bf p}"\to{\bf P}-\frac\hbar2{\bf K}-\hbar{\bf k}'_1$.
Therefore,
\begin{eqnarray}
&&[\chi^{(0)} (\chi^{-1})^{(2)},\Theta^{(0)}]_{Ma110001}
=i\left({\frac{q^2_e c^3}{2\varepsilon_0  (2\pi)^3}}\right)
\int d^3k'_1
\frac1{\omega_{k'_1}}
\nonumber\\&&\times
\left\{
\sin\frac{ \theta_{\vert{\bf P}-\frac{\hbar}{2}{\bf K} \vert}}2
\cos\frac{ \theta_{\vert{\bf P}-\frac{\hbar}{2}{\bf K} \vert}}2
\right.\nonumber\\&&\left.\times
\left(\frac{{\bf P}-\frac{\hbar}{2}{\bf K} }{\vert{\bf P}-\frac{\hbar}{2}{\bf K} \vert}.
\left[-\frac{{\bf P}+\frac\hbar2{\bf K}}{E_{{\bf P}+\frac\hbar2{\bf K}}} 
+\frac{{\bf P}-\frac\hbar2{\bf K}}{E_{{\bf P}-\frac\hbar2{\bf K}}}
\right]
\right.\right.\nonumber\\&&\left.\left.
-\left({\hat{\bf k}}'_1.\frac{{\bf P}-\frac{\hbar}{2}{\bf K} }{\vert{\bf P}-\frac{\hbar}{2}{\bf K} \vert}
\right)
\left[-\frac{{\bf P}+\frac\hbar2{\bf K}}{E_{{\bf P}+\frac\hbar2{\bf K}}} 
+\frac{{\bf P}-\frac\hbar2{\bf K}}{E_{{\bf P}-\frac\hbar2{\bf K}}}
\right].{\hat{\bf k}}'_1
\right) 
\right.\nonumber\\&&\left.
+\sin\frac{ \theta_{\vert{\bf P}-\frac{\hbar}{2}{\bf K} +\hbar{\bf k}'_1\vert}}2
\cos\frac{ \theta_{\vert{\bf P}-\frac{\hbar}{2}{\bf K} +\hbar{\bf k}'_1\vert}}2
\right.\nonumber\\&&\left.\times
\left(\frac{{\bf P}-\frac{\hbar}{2}{\bf K}-\hbar{\bf k}'_1 }
{\vert{\bf P}-\frac{\hbar}{2}{\bf K} -\hbar{\bf k}'_1\vert}.
\left[-\frac{{\bf P}+\frac\hbar2{\bf K}}{E_{{\bf P}+\frac\hbar2{\bf K}}} 
+\frac{{\bf P}-\frac\hbar2{\bf K}}{E_{{\bf P}-\frac\hbar2{\bf K}}}
\right]
\right.\right.\nonumber\\&&\left.\left.
-\left({\hat{\bf k}}'_1.
\frac{{\bf P}-\frac{\hbar}{2}{\bf K}-\hbar{\bf k}'_1 }
{\vert{\bf P}-\frac{\hbar}{2}{\bf K} -\hbar{\bf k}'_1\vert}
\right)
\left[-\frac{{\bf P}+\frac\hbar2{\bf K}}{E_{{\bf P}+\frac\hbar2{\bf K}}} 
+\frac{{\bf P}-\frac\hbar2{\bf K}}{E_{{\bf P}-\frac\hbar2{\bf K}}}
\right].{\hat{\bf k}}'_1
\right) 
\right\} 
\nonumber\\&&\times
\frac1{-E_{{\bf P}-\frac\hbar2{\bf K}}+E_{{\bf P}-\frac\hbar2{\bf K}-\hbar{\bf k}'_1}
+\hbar\omega_{k'_1}}
\label{5.a210}
\end{eqnarray}
This expression is obtained from (\ref{5.a204}) by replacing 
${\bf K}\to -{\bf K}$, ${\bf k}_1\to {\bf k}'_1$.
\newline
\noindent
The bracket $\left[-\frac{{\bf P}+\frac\hbar2{\bf K}}{E_{{\bf P}+\frac\hbar2{\bf K}}} 
+\frac{{\bf P}-\frac\hbar2{\bf K}}{E_{{\bf P}-\frac\hbar2{\bf K}}}\right]$ generates a change of sign,
analogous to the change of sign required in the correspondence in the elements of $\Theta^{(2)}$. 

In (\ref{4.a77}), we can use (\ref{5.a209}) with the correspondence ${\bf p}'\to{\bf P}+\frac\hbar2{\bf
K}$,
${\bf p}\to{\bf P}-\frac\hbar2{\bf K}$, ${\bf p}"\to{\bf P}-\frac\hbar2{\bf K}+\hbar{\bf k}_1$.
We then obtain 
\begin{eqnarray}
&&[\chi^{(0)} (\chi^{-1})^{(2)},\Theta^{(0)}]_{Ma101010}
=i\left({\frac{q^2_e c^3}{2\varepsilon_0  (2\pi)^3}}\right)
\int d^3k_1
\frac1{\omega_{k_1}}
\nonumber\\&&\times
\left\{
\sin\frac{ \theta_{\vert{\bf P}-\frac{\hbar}{2}{\bf K} \vert}}2
\cos\frac{ \theta_{\vert{\bf P}-\frac{\hbar}{2}{\bf K} \vert}}2
\right.\nonumber\\&&\left.\times
\left(\frac{{\bf P}-\frac{\hbar}{2}{\bf K} }{\vert{\bf P}-\frac{\hbar}{2}{\bf K} \vert}.
\left[-\frac{{\bf P}+\frac\hbar2{\bf K}}{E_{{\bf P}+\frac\hbar2{\bf K}}}
+\frac{{\bf P}-\frac\hbar2{\bf K}}{E_{{\bf P}-\frac\hbar2{\bf K}}} 
\right]
\right.\right.\nonumber\\&&\left.\left.
-\left(
{\hat{\bf k}}_1.\frac{{\bf P}-\frac{\hbar}{2}{\bf K} }{\vert{\bf P}-\frac{\hbar}{2}{\bf K} \vert}
\right) 
\left[-\frac{{\bf P}+\frac\hbar2{\bf K}}{E_{{\bf P}+\frac\hbar2{\bf K}}}
+\frac{{\bf P}-\frac\hbar2{\bf K}}{E_{{\bf P}-\frac\hbar2{\bf K}}} 
\right].{\hat{\bf k}}_1
\right) 
\right.\nonumber\\&&\left.
-\sin\frac{ \theta_{\vert{\bf P}-\frac{\hbar}{2}{\bf K} +\hbar{\bf k}_1\vert}}2
\cos\frac{ \theta_{\vert{\bf P}-\frac{\hbar}{2}{\bf K} +\hbar{\bf k}_1\vert}}2
\right.\nonumber\\&&\left.\times
\left(\frac{{\bf P}-\frac{\hbar}{2}{\bf K}+\hbar{\bf k}_1 }
{\vert{\bf P}-\frac{\hbar}{2}{\bf K} +\hbar{\bf k}_1\vert}.
\left[-\frac{{\bf P}+\frac\hbar2{\bf K}}{E_{{\bf P}+\frac\hbar2{\bf K}}}
+\frac{{\bf P}-\frac\hbar2{\bf K}}{E_{{\bf P}-\frac\hbar2{\bf K}}} 
\right]
\right.\right.\nonumber\\&&\left.\left.
-\left(
{\hat{\bf k}}_1.
\frac{{\bf P}-\frac{\hbar}{2}{\bf K}+\hbar{\bf k}_1 }
{\vert{\bf P}-\frac{\hbar}{2}{\bf K} +\hbar{\bf k}_1\vert}
\right)
\left[-\frac{{\bf P}+\frac\hbar2{\bf K}}{E_{{\bf P}+\frac\hbar2{\bf K}}}
+\frac{{\bf P}-\frac\hbar2{\bf K}}{E_{{\bf P}-\frac\hbar2{\bf K}}} 
\right].{\hat{\bf k}}_1
\right) 
\right\} 
\nonumber\\&&\times
\frac1{-E_{{\bf P}-\frac12\hbar{\bf K}}- E_{-{\bf P}+\frac12\hbar{\bf K}-\hbar{\bf k}_1}
-\hbar\omega_{k_1}}
\label{5.a211}
\end{eqnarray}
For the dominant part, linearly divergent, we have the same cancellation for the sum of 
(\ref{5.a210}) and
(\ref{5.a211}) as for (\ref{5.a204}) and (\ref{5.a206}).

Therefore, the remaining contribution diverges logarithmically.

\bibliographystyle{unsrt}

\end{document}